\documentclass[a4paper,11pt]{article}
\usepackage[margin=1in,footskip=0.5in]{geometry}
 \usepackage[bottom]{footmisc}%
 \usepackage{setspace}
\usepackage{amsmath}
\usepackage{graphicx}  
\usepackage{extarrows}
\usepackage{float}  
\usepackage{subfigure}  
\usepackage[utf8]{inputenc} % allow utf-8 input
\usepackage[T1]{fontenc}    % use 8-bit T1 fonts
\usepackage{hyperref}       % hyperlinks
\usepackage{url}            % simple URL typesetting
\usepackage{booktabs}       % professional-quality tables
\usepackage{amsfonts}       % blackboard math symbols
\usepackage{nicefrac}       % compact symbols for 1/2, etc.
\usepackage{microtype}      % microtypography
\usepackage{lipsum}		% Can be removed after putting your text content
\usepackage[dvipsnames]{xcolor}
\usepackage{amsthm}
 \usepackage{appendix}
 \usepackage{amssymb}
  \usepackage{natbib}
  \usepackage{mathrsfs}  
   \usepackage{algorithmicx}
 \usepackage{multirow}
  \usepackage{natbib}
  \usepackage[ruled,vlined]{algorithm2e}
  \newtheorem{theorem}{Theorem}

\newtheorem{remark}{Remark}
\newtheorem{corollary}{Corollary}
\newtheorem{lemma}{Lemma}
\newtheorem{proposition}{Proposition}
\newtheorem{definition}{Definition}
\newtheorem*{definition*}{Definition}
 \DeclareMathOperator*{\argmin}{arg\,min}

\usepackage[markup=underlined]{changes}
\definechangesauthor[color=magenta]{ab}
\usepackage{caption}
\def\m{\mathcal}
\def\mb{\mathbb}

\def\ms{\mathscr}
 \def\dd{{\rm d}}

\def\wt{\widetilde}
\def\wh{\widehat}
\def\ov{\overline}

\newcommand{\mnorm}[1]{{\vert\kern-0.25ex\vert\kern-0.25ex\vert #1 
    \vert\kern-0.25ex\vert\kern-0.25ex\vert}}
\newcommand{\bmnorm}[1]{{\big\vert\kern-0.25ex\big\vert\kern-0.25ex\big\vert #1 
    \big\vert\kern-0.25ex\big\vert\kern-0.25ex\big\vert}}
\newcommand{\Bmnorm}[1]{{\Big\vert\kern-0.25ex\Big\vert\kern-0.25ex\Big\vert #1 
    \Big\vert\kern-0.25ex\Big\vert\kern-0.25ex\Big\vert}}
\newcommand{\bbmnorm}[1]{{\bigg\vert\kern-0.25ex\bigg\vert\kern-0.25ex\bigg\vert #1 
    \bigg\vert\kern-0.25ex\bigg\vert\kern-0.25ex\bigg\vert}}
\newcommand{\BBmnorm}[1]{{\Bigg\vert\kern-0.25ex\Bigg\vert\kern-0.25ex\Bigg\vert #1 
    \Bigg\vert\kern-0.25ex\Bigg\vert\kern-0.25ex\Bigg\vert}}

\definecolor{cobalt}{rgb}{0.0, 0.28, 0.67}
\def\ccb{\color{black}}

\title {Robust Bayesian Inference on Riemannian Submanifold}
 
\author{Rong Tang\textsuperscript{$\ast$}, Anirban Bhattacharya\textsuperscript{\S}, Debdeep Pati\textsuperscript{\S}, and Yun Yang\textsuperscript{$\dagger$}}

\date{\textsuperscript{$\ast$}Department of Mathematics, The Hong Kong University of Science and Technology\\
\textsuperscript{\S}Department of Statistics, Texas A\&M University\\
\textsuperscript{$\dagger$}Department of Statistics, University of Illinois Urbana-Champaign}

\begin{document}
\maketitle
\begin{abstract}
Manifold-valued parameters routinely arise in modern statistical applications such as in medical imaging, robotics, and computer vision, to name a few. While traditional Bayesian approaches are applicable to such  settings by considering an ambient Euclidean space as the parameter space, we demonstrate the benefits of integrating manifold structure into the  Bayesian  framework, both theoretically and computationally. Moreover, existing Bayesian approaches which are designed specifically for manifold-valued parameters are primarily model-based, which are typically subject to inaccurate uncertainty quantification under model misspecification. In this article, we propose a robust model-free Bayesian inference for  parameters defined on a Riemannian submanifold, which is shown to provide valid uncertainty quantification from a frequentist perspective. Computationally, we propose a Markov chain Monte Carlo to sample from the posterior on the Riemannian submanifold, where the mixing time, in the large sample regime, is shown to depend only on the intrinsic dimension of the parameter space instead of the potentially much larger ambient dimension. Our numerical results demonstrate the effectiveness of our approach on a variety of problems, such as multiple quantile regression, reduced-rank regression, and Fr\'{e}chet mean estimation.
\end{abstract}

\tableofcontents
 
\section{Introduction}

There has been a growing interest in statistical inference from complex data that are non-Euclidean, such as covariance matrices (diffusion tensor) in diffusion tensor imaging~\citep{LEE2017152,https://doi.org/10.1002/cjs.11601}, column orthogonal matrices (the Stiefel manifold) in orbit data analysis~\citep{10.1214/18-AOS1692,lin2017bayesian} , shape objects in medical vision~\citep{guo2022statistical,kendall1984shape}, and  Grassmannian-supported data in computer vision~\citep{chakraborty2015recursive}.  
Typically, the target parameters of interest are population summary measures such as the Fr\'{e}chet mean~\citep{frechet1948elements} or the extrinsic mean~\citep{bhattacharya2003large}, both of which reside on the same manifold as the data. Non-Euclidean parameter spaces also arise when the target parameters are subject to certain constraints (e.g., low-rank, fixed sum, etc.). In such cases, one can cast the constrained parameter space as a manifold without the constraint. For instance, Grassmannian and Stiefel manifolds are widely explored in various dimensionality
reduction problems~\citep{holbrook2016bayesian,zhang2018grassmannian,pmlr-v9-suzuki10a}. Other constraints may be related to the
rank of the optimal solutions~\citep{markovsky2012low,zhou2015rank}, which arise in various areas such as signal and image processing~\citep{ji2011robust}, computational finance~\citep{zhang2003optimal} and multivariate reduced-rank regression~\citep{izenman1975reduced,reinsel2023multivariate}.

With the rapid growth in data acquisition devices, algorithms for performing statistical inference on non-Euclidean spaces encounter challenges associated with increasing sample size, complexity of the manifold, and the intrinsic dimension of the manifold.  To that end, techniques for optimization on a Riemannian manifold have gained significant attention over the past decade~\citep{hosseini2015matrix,ferreira2019gradient,zhou2021manifold,boumal2022intromanifolds}.  While Euclidean-based optimization typically uses Euclidean (stochastic) gradient descent, the Riemannian gradient, also known as the {\it natural} gradient~\citep{martens2020new}, exploits the underlying data geometry to form a steepest descent direction of the cost function relative to the induced (Riemannian) metric. This leads to stability and substantial speed-up in comparison with the Euclidean stochastic gradient descent.  Despite the rich development in 
 Riemannian optimization, extending the techniques to obtain uncertainty quantification in the form of credible or confidence sets along with point estimates remains a comparatively less explored direction.

In a likelihood-based paradigm, a Bayesian approach offers a natural way to quantify uncertainty through a posterior distribution. Numerous studies have investigated Bayesian inference on specific manifolds using both parametric and more flexible semi-parametric approaches. For instance, ~\cite{https://doi.org/10.48550/arxiv.1710.09443,lin2017bayesian} focus on Bayesian inference on the Stiefel manifold. Meanwhile, ~\cite{holbrook2016bayesian,mao2010supervised,thomas2022learning} utilize Bayesian inference on the Grassmannian manifolds to perform dimensionality reduction.  Moreover, Bayesian approaches have been proven useful in dealing with circular and directional statistics~\citep{BINETTE2022118,mcvinish2008semiparametric,ravindran2011bayesian}. Bayesian techniques have also been effectively utilized in manifold regression~\citep{10.1214/15-AOS1390, 10.1214/18-BA1135} and density estimation on manifold~\citep{bhattacharya2010nonparametric,berenfeld2022estimating}. 
Despite the existing literature, most of the current methods and associated sampling algorithms are tailored to specific manifolds. On the theoretical side, posterior consistency results are available~\citep{castillo2014thomas,bhattacharya2010nonparametric} when the observations sit on a compact manifold. However, the impact of crucial geometric properties, such as the {\it intrinsic dimension} and the {\it curvature}, on the limiting posterior distribution remains largely unexplored. Therefore, despite the significant benefits of Bayesian methods in providing a unified framework for point estimation and uncertainty quantification, the theoretical and computational advantages of incorporating manifold structures into a Bayesian posterior remains unclear.

To address this gap, we first consider a Bayesian approach that incorporates manifold information through the prior distribution, and study the asymptotic properties of the resulting Bayesian posterior supported on a Riemannian submanifold $\m M$ embedded in $\mb R^D$. A central question in Bayesian asymptotics is the limiting shape of a posterior distribution.  In a {\it regular} finite-dimensional model, the classical Bernstein von-Mises (BvM; \cite[Chapter 1]{ghosh2003bayesian}) theorem establishes that the posterior distribution for a $D$-dimensional parameter $\theta$ %is 
approximately {\ccb assumes} a normal {\ccb shape} %distribution 
as %$n$ 
{\ccb the sample size increases} to infinity. However, the same result does not apply when the posterior distribution is defined on a low-dimensional submanifold, which is singular with respect to the Lebesgue measure on $\mb R^D$.
To address the challenges,  we introduce a general theoretical framework analogous to the BvM theorem,  called the {\it manifold Bernstein von-Mises theorem}, to characterize the asymptotic shape of such manifold-supported posteriors.  Rather than working directly with the singular posterior, we analyze the {\it projection} of the posterior onto the tangent space $T_{\theta^\ast}\mathcal{M}$ at the target parameter $\theta^\ast$, in  which  $T_{\theta^*}\m M$ serves as a first-order approximation of $\m M$ locally around $\theta^*$.  {Moreover,  we allow for non-differentiable likelihoods that relax classical  smoothness assumptions, and develop our theory using gradient‑like estimating functions defined on the manifold, thereby substantially broadening the scope of applicability of our theorems.} Our results (c.f. Theorem~\ref{th:classicbvm}) 
%split into 
{\ccb encompass}
two {\ccb important} regimes. In {\it well-specified models}, where the parameter $\theta$ indexes a true likelihood $p(x|\theta^\ast)$ with $\theta^*\in \m M$, the posterior {\ccb arising from} a manifold-supported prior, 
%with a manifold-supported prior, 
after projection onto
$T_{\theta^\ast}\mathcal{M}$, is asymptotically normal {\ccb under mild assumptions}. {\ccb In particular, it possesses} a ``correct'' covariance, in the sense that it matches the limiting frequentist covariance of the posterior's center. Consequently, credible sets derived from the posterior {\ccb asymptotically} attain {\it valid} frequentist coverage, {\ccb akin to the regular Euclidean case}. Moreover, {\ccb we show that} these credible sets can be {\it shorter} than those from the classical ambient-space posterior that ignores the manifold structure (c.f., Corollary~\ref{co:classicbvm_region}), providing quantitative statistical justification for exploiting the manifold constraint. However, in {\it misspecified models},  where either the manifold assumption $\theta^*\in \m M$ is violated or the likelihood function $p(x|\theta^*)$ is incorrect, the posterior generally fails to deliver correct uncertainty quantification, {\ccb once again generalizing similar conclusions from the Euclidean framework} \citep{CHERNOZHUKOV2003293,kleijn2012bernstein}.

 % Despite the success of Bayesian approaches for well-specified models, there is an increasing recognition of their limitations in misspecified settings, particularly concerning the frequentist coverage of credible sets \citep{kleijn2012bernstein}. Unsurprisingly, this continues to be the case for manifold-valued parameters; see Table~\ref{Table:DTI1} for an illustration. 
 
 % The main challenge lies in accurately specifying a distribution family that captures the true data generating distribution $\m P^*$ (or likelihood function).
 
To address the challenge of accurately specifying a distribution family that captures the true data generating distribution,~\cite{lin2020robust} propose the use of the ``median-of-mean'' approach  for robust and scalable inference on manifolds for problems that can be formalized as optimization problems over non-Euclidean
spaces;
\cite{bhattacharya2017omnibus} derive a central limit theorem for the empirical Fr\'{e}chet mean of manifold-valued data,  which serves as the foundation for nonparametric inference based on the Fr\'{e}chet mean.  
  Another possible direction for robust inference on manifolds is to enhance the flexibility of the model by considering  mixtures of parametric models~\citep{bhattacharya2010nonparametric}. However, this approach often leads to over-parameterization and requires the usage of a multitude of nuisance parameters, leading to potential loss of statistical and computational efficiency when the focus is on simple population summary measures. Additionally, when dealing with manifold-valued parameters, Bayesian inference requires specific constructions and prior elicitation tailored to the manifold structure, making it more challenging to develop flexible nonparametric Bayesian models. Furthermore, even with more flexible models, problems associated with misspecification may still persist. 
%   \begin{table}
%     \caption{ \em Coverage probabilities for $95\%$ credible intervals of functions of manifold-valued parameters, obtained from a Wishart-based parametric Bayesian inference in the analysis of a diffusion tensor dataset, are provided. Additional details about the experiment can be found in Section~\ref{Sec:DTI}.
%       }\label{Table:DTI1} 
% \centering
% \begin{tabular}{cccc}
%     \hline
%  &FA&Trace&Maximum eigenvalue\\
%    \hline
% {Extrinsic Mean} &99.4&59.2&81.6\\
%    \hline
%    {Fr\'{e}chet Mean} &>99.9&68.7&95.1\\
%  \hline
%      \end{tabular}
%      \end{table}
One promising approach in the literature involves the use of pseudo-posterior distributions \citep{CHERNOZHUKOV2003293,jiang2008gibbs,chib2018bayesian}, where the parameter is treated as the solution of an {\it estimating equation} and one generates a likelihood targeting that. Specifically, we represent the parameter of interest $\theta(\m P^*)$, viewed as a functional $\theta(\cdot)$ evaluated at $\m P^*$, through a population risk minimizer on a Riemannian submanifold $\m M$:
\begin{equation*}
    \theta(\m P^*)\in \underset{\theta\in \m M}{\arg\min} \,\m R(\theta; \m P^*), \quad\text{with }  \m R(\theta; \m P^*):=\mb{E}_{\m P^*}[\ell (X,\theta)].
\end{equation*}
Here, $\ell :\m X\times \m M\to \mb R$ is a {\it loss function} that evaluates the compatibility of  $\theta$ (the parameter of interest) with
the data point $X$, and $\mb E_{\m P^*}$ represents the expectation under $\m P^*$. Observe that the maximum likelihood estimator (MLE) over a restricted parameter space can also be seen as a risk minimizer on a manifold with the loss function being the negative log-density function.

 Many statistical problems can be cast as risk minimization on a manifold. For example, the Bures-Wasserstein barycenter estimation of a set of multivariate normal distributions $N(u,\Sigma)$ ($\mu\in \mb R^p$ and $\Sigma\in \mb S^p_{+}$\footnote{Here $ \mb S^p_{+}$ denotes the space of $p\times p$ symmetric positive semidefinite matrices.}) can be framed as a risk minimization over $\mb R^p \times \mb S^p_{+}$ using a Wasserstein loss function~\citep{10.1214/20-AAP1618}.  In the context of multiple quantile modeling~\citep{lian2019multiple} where quantile slope coefficients are enforced to share information across quantile levels, the target parameter (matrix with rows corresponding to quantile slope coefficients at various quantile levels) can be seen as the risk minimizer  with respect to a check loss function  on the space of  low-rank matrices. Expressing the parameter as a risk minimizer allows us to partially define a model through the loss function without fully specifying the data-generating distribution. Drawing inspiration from the first-order  optimality condition for identifying a local minimizer on Riemannian submanifold,  we propose a  {\it Riemannian exponentially tilted empirical likelihood} (RETEL) function, which replaces the standard parametric likelihood. Moreover, a penalty term associated with the empirical risk function is added to ensure the concentration of the posterior around the global minimizer. We refer to the resulting posterior as the Bayesian Riemannian penalized exponentially tilted empirical likelihood (RPETEL) posterior, which extends the Euclidean PETEL~\citep{tang2022bayesian}. As we will demonstrate, the Bayesian RPETEL posterior provides three forms of robustness:  (1) It satisfies a  Manifold Bernstein-von Mises theorem (c.f.,  Theorem~\ref{th1}), yielding automatically calibrated uncertainty quantification  without requiring a correctly specified likelihood.
{ Establishing this result necessitates new analytical tools to handle posterior singularity,  to represent the posterior in local coordinates, and to control curvature-induced distortions absent in the Euclidean setting.}
 (2)
Even when the global risk minimizer over the ambient space $\mb R^D$ lies off the manifold or does not exist, the mean of the posterior remains meaningful -- it closely matches the risk minimizer over the manifold,  and the covariance matrix of the posterior matches that of the frequentist sampling distribution of this mean; (3)   In a well-specified setting where both the likelihood  and the manifold assumption are correct, the posterior is asymptotically equivalent to the  standard Bayesian posterior with the same manifold-supported prior. Hence, the proposed Bayesian RPETEL posterior is robust to model misspecification, while incurring no efficiency loss under correct specification.

Beyond the appealing asymptotic properties of the proposed posterior, we also demonstrate computational advantages from the incorporation of low-dimensional manifold structure. Specifically,  we develop a Riemannian Random-walk Metropolis (RRWM) algorithm for sampling from posterior distributions defined on submanifolds, which adapts the classical Random-walk Metropolis (RWM) algorithm to the manifold setting.  We show that, in  the large-sample regime,  the RRWM algorithm can be applied for sampling from the derived posterior with a mixing time that is almost linear in the intrinsic dimension $d$ (c.f.~Corollary~\ref{th:mixingbpetel}). This is different from the conventional RWM algorithm for sampling from the  standard Bayesian posterior in the ambient space $\mb R^D$, where the mixing time is at least linear in the ambient dimension $D$~\citep{10.1214/aoap/1034625254}.

The rest of the paper is organized as follows. Section 1.1 provides a summary of the notation used throughout the paper, while Section 1.2 offers a brief background on Riemannian submanifolds. In Section 2.1, we discuss the non-asymptotic properties of the Bayesian posterior on submanifolds,  highlighting its advantage over the Euclidean counterpart that does not incorporate the manifold structure, as well as its limitation. 
Section 3 presents our proposed Bayesian RPETEL posterior for statistical inference on Riemannian submanifolds without fully specifying the model.
Section 4 presents a Riemannian Random-Walk Metropolis algorithm for posterior sampling and analyzes its computational complexity. Section 5 reports numerical studies, and Section 6 concludes with a discussion. {\ccb Proofs and other technical details are deferred to a Supplementary Material.}

\subsection{Notation}

  We use $0_d$ to denote the $d$-dimensional all zero vector, and may omit the subscript when no ambiguity may arise. We use  $B_r(x)$ to denote the closed ball centered at $x$ with radius $r$ (under the $\ell_2$ distance) in the Euclidean space.
  We use $\|\cdot\|_{p}$  to denote the usual vector $\ell_p$ norm, and reserve $\|\cdot\|$ for the $\ell_2$ norm (that is, suppress the subscript when $p=2$).  For a topological space $\m X$, we use $\m X^n=\{(x_1,x_2,\cdots, x_n)\,:\,\forall i,\, x_i\in \m X\}$ to denote the $n$-fold Cartesian product of $\m X$. We use $\ms \mathbb P(\Omega)$ to denote the set of probability measures on space $\Omega$.  For a measure $\mu\in \ms \mathbb P(\Omega)$ and a map $G: \Omega\to \Omega'$, we use $\nu=G_{\#}\mu\in \ms \mathbb P(\Omega')$ to denote the push-forward measure of $\mu$ by $G$ so that $\nu(\m A)=\mu(G^{-1}(\m A))$
  holds for any measurable set $\m A$ on $\Omega'$.  A $k$-dimensional real random vector $X$ is 
  %called a normal random vector from $\m N(\mu,\Sigma)$ 
  {\ccb said to follow a multivariate normal distribution, denoted $\m N(\mu,\Sigma)$,}
  if there exists {\ccb a full-rank matrix} $A\in \mb R^{k\times l}$ {\ccb (with $l \le k$)} so that $\Sigma=AA^T$ and $X=AZ+\mu$, where $Z$ is an $l$-dimensional standard normal random vector {\ccb -- when $l < k$, the distribution is said to be {\it singular}}.  For a matrix $A\in \mb R^{D\times D}$, we use $A^{\dagger}\in \mb R^{D\times D}$ to denote its pseudo-inverse (Moore-Penrose inverse), and the pseudo-determinant of $A$ is defined as $|A|_{+}=\lim_{\alpha\to 0}\frac{{\rm det}(A+\alpha I_D)}{\alpha^{D-{\rm rank}(A)}}$, where ${\rm det}(A)$ denotes the usual determinant, $I_D$ denotes the $D\times D$ identity matrix, and ${\rm rank}(A)$ denotes the rank of $A$. For a metric space $(\m F,\rho)$, let $\bold{N}(\m F,\rho,\varepsilon)$ denote the $\varepsilon$-covering number of $\m F$ with respect to $\rho$, and define the metric entropy as  $\log {\bold N}(\m F,\rho,\varepsilon)$.
  % For a matrix $B\in \mb R^{p\times k}$, we use ${\rm vec}(B)$ to denote the $pk$-dimensional vector obtained by stacking the columns of $B\in \mb R^{p\times k}$ on top of each other. For a vector $\theta\in \mb R^{pk}$, we use ${\rm mat}_k(\theta)$ to denote the $p\times k$ matrix obtained by unstacking $\theta$ by column.
%   We use symbols $\lesssim$ and $\gtrsim$
% to denote the corresponding inequality up to an $n$-independent constant.  
% Throughout,
% $C, c, C_0, c_0, C_1, c_1,\cdots$ are generically used to denote positive $n$-independent constants whose values might change from one line to another, but are independent from everything else. 

\subsection{Preliminary}\label{sec:optmanifold}

Suppose we observe i.i.d. copies $X^{(n)}=\{X_1,X_2,
\cdots,X_n\}$ of a random variable $X\in \m X$ from an unknown distribution $\m P^*$. {\ccb Henceforth, we use $\mb E$ and $\mb P$ to denote expectation and probability under $\m P^*$}. Our target is to estimate a parameter $\theta^*=\theta(\m P^*)$ associated with the unknown population $\m P^*$ based on the finite observations $X^{(n)}$, and quantify the estimation uncertainty.  In many cases, the functional $\theta(\m P^*)$ can take values in a Riemannian submanifold, such as in Bures-Wasserstein barycenter estimation, Fr\'echet mean estimation, reduced-rank multiple quantile regression, among others. In order to discuss the properties of the parameter space, it is necessary to revisit some {\ccb essential} definitions from manifold theory{\ccb; see \cite{boumal2022intromanifolds} for book-level details}.

 Let $\m M$ be a $d$-dimensional Riemannian submanifold embedded in $\mb R^D$. Intuitively speaking, a manifold is a topological space that locally resembles a Euclidean
space. Formally, the definition is as follows: 
\begin{definition}[Submanifold]
 A subset $\m M$ of $\mathbb{R}^D$ is a $d$-dimensional Riemannian submanifold if for every point $\theta$ in $\m M$, there exists a neighbourhood $U_{\theta}$ of $\theta$ on $\m M$ and an open set $V_{\theta} \subseteq \mathbb{R}^d$, such that there exists a homeomorphism  $\xi_{\theta}$ that maps $V_{\theta}$ to $U_{\theta}$, that is, $\xi_{\theta}: \, V_{\theta}\rightarrow U_{\theta}$ is bijective and both $\xi_{\theta}$ and $\xi^{-1}_{\theta}$ are continuous maps. Moreover,  the  differential $\m D \xi_{\theta}(z)[\cdot]$ of $\xi_{\theta}(\cdot)$ at $z$ exists and is injective for every $z\in V_{\theta}$.
 % \footnote{Here, the differential of $\xi(\cdot)$ at $y$, denoted as $D_y \xi$, is a linear map defined by $D_y \xi [v] = \lim_{t \to 0} \frac{\xi(y + tv) - \xi(y)}{t} = J_{\xi}(y)v$ for $v \in \mb R^d$. The injectiveness of $D_y \xi$ is equivalent to the Jacobian matrix $J_{\xi}(y)$ having full rank.} 
The pair $(U_{\theta},\xi_{\theta}^{-1})$ is called a local coordinate chart near $\theta$,  with $\xi_{\theta}^{-1}$ the coordinate map and $\xi_{\theta}$ a local parameterization. We refer to $D$ as the ambient dimension and $d$ as the intrinsic dimension of $\m M$. 
\end{definition}
The tangent space of $\m M$ at a point $\theta\in \m M$, denoted as $T_{\theta}\m M$, is a {\it linearization} of $\m M$ locally around $\theta$, which contains possible directions (tangent vector) in which one can tangentially pass through $\theta$. This linearization forms a foundation for extending many notions and techniques in Euclidean space to the Riemannian submanifold.  For example, for a smooth function $f$ %action 
{\ccb acting}
on $\m M$,  by considering smooth extension $\ov f$ of $f$ to a neighborhood of $\m M$ in $\mb R^D$,  the so-called {\it Riemannian gradient} ${\rm grad} f(\theta)$ of $f$ at $\theta\in \m M$ can be thought of as the orthogonal projection of the (Euclidean) gradient of $\ov f$ at $\theta$ onto $T_{\theta}\m M$. Therefore, the Riemannian gradient gives the steepest ascent tangent direction for $f$ along the manifold, and we have the so-called \emph{first-order necessary optimality condition} that any local minimizer $\theta$ of a smooth function $f:\m M\to \mb R$ satisfies ${\rm grad}f(\theta)=0$. This first-order necessary optimality condition serves as a main motivation for our developed method in Section~\ref{sec:LFBBI}.

The detailed definitions of all the notions mentioned above are available in Appendix A. Throughout, we assume the manifold is  $C^3$-smooth locally around the risk minimizer $\theta^*$, as specified below. This smoothness guarantees that, near $\theta^*$, the manifold is  well-approximated by its tangent space $T_{\theta^*}\m M$. Such assumptions are standard in manifold-based inference~\cite{10.1214/18-AOS1685,divol2022measure} where $\m M$ is typically assumed to be at least $C^2$-smooth. Our stronger  $C^3$ smoothness assumption enables us to establish the desired BvM result with a root-$n$ convergence rate.

 \begin{definition}[Local $C^3$-Smoothness]
 \label{def:C3smooth}
 Let $\m M$ is a $d$-dimensional submanifold embedded in $\mb R^D$ and let $\theta\in \m M$. We call $\m M$ is  locally $C^3_{r,L}$-smooth at $\theta$ if there exist constants $L>0$  and neighborhoods $U_{\theta}\subset \m M$, $V_{\theta}\subset T_{\theta}\mathcal{M}$ such that 
 \begin{enumerate}
     \item $B_{r}(\theta)\cap \m M\subset U_{\theta}$ and $B_{r}(0_D)\cap T_{\theta}\mathcal{M}\subset V_{\theta}$;
     \item  the map $\psi_{\theta}: U_{\theta}\to V_{\theta}$ defined by $\psi_{\theta}(x)={\rm Proj}_{T_{\theta}\m M}(x-\theta)$ is a bijection onto $V_{\theta}$;
     \item the inverse $\phi_{\theta}: V_{\theta}\to U_{\theta}$ of $\psi_{\theta}$ is thrice Fr\'{e}chet differentiable, and its Fr\'{e}chet derivatives up to order three have operator norms uniformly bounded by $L$.
 \end{enumerate}

%  satisfying $B_{r}(\theta)\cap \m M\subset U\subset \m M$ and $B_{r}(0_D)\cap T_{\theta}\mathcal{M}\subset V\subset T_{\theta}\mathcal{M}$, such that
%  the projection $\psi_{\theta}(\cdot)={\rm Proj}_{T_{\theta^*}\m M}(\cdot-\theta^*)$ maps $U\to V$, and is a bijective.  Its inverse, $\phi_{\theta}$ satisfied that the differentials of $\phi_{\theta}(v)$, up to order three,  have operator norms uniformly bounded by $L$ for $v\in B_r(0_D) \cap T_{\theta}\mathcal{M}$.\footnote{That means, for any $\eta_1,\eta_2,\eta_3\in T_{\theta}\m M$ with unit norms and $v\in B_r(0_D)\cap T_{\theta}\m M$,  
% $$ \big\|\m D_v \phi_{\theta}(v)[\eta_1]\big\|\leq L,\,\big\|\m D_v(\m D_v \phi_{\theta^*}(v)[\eta_1])[\eta_2]\big\|\leq L,\,\big\|\m D_v(\m D_v(\m D_v \phi_{\theta^*}(v)[\eta_1])[\eta_2])[\eta_3]\big\|\leq L,$$
%     where the subscript $v$ means the differential is with respect to $v$. }
\end{definition}

\section{Bayesian Inference with Manifold-supported Priors}\label{sec:LBBI}
In the classical Bayesian framework, we model the unknown data distribution $\m P^*$ by a parametric family $\{p(x|\theta)\,:\,\theta\in \mb R^D\}$. Prior knowledge about $\theta$ is incorporated through a prior distribution, which is updated to a posterior using Bayes' theorem.  {\ccb If the model is correctly specified, i.e., if $\m P^* = p(x | \theta^*)$ for some $\theta^* \in \mb R^D$, then the} celebrated Bernstein-von Mises (BvM) theorem ensures {\ccb under mild regularity conditions} that credible regions derived from the posterior will asymptotically be confidence sets of the same level.  
In many application, however, the parameter of interest is subject to certain constraint -- for example, it may have unit norm, sum to one, or be low-rank.   Such constraints can be represented by viewing the parameter as living on a $d$-dimensional Riemannian submanifold ${\m {M}}$ embedded in the ambient space $\mb R^D$. We can then place a prior directly on ${\m M}$ and perform Bayesian inference.
% % that has a density $\pi_{\m M}(\cdot)$ with respect to the volume measure $\mu_{\m M}$ of $\ov{\m M}$, 
For instance, in many models for multivariate and relational data, the posterior takes the form of a matrix Bingham-von Mises-Fisher distribution under a uniform prior on the Stiefel manifold~\citep{hoff2009simulation}. In Bayesian estimation of a covariance matrix, a natural conjugate prior for the normal sampling model is the inverse Wishart distribution~\citep{alvarez2014bayesian}, which is defined on the space of real-valued positive-definite matrices. The inverse Wishart prior also provides conjugacy for diffusion tensor data sets modeled by the Wishart distribution. A key difference from the unconstrained Euclidean setting is that the posterior, which is supported on a lower-dimensional Riemannian submanifold, is singular with respect to the Lebesgue measure on $\mathbb{R}^D$. As a result, the usual Bernstein-von Mises theorem does not apply in its standard form.  On the frequentist side, several studies have investigated the asymptotic behavior of the empirical risk minimizer over the Riemannian manifold $\mathcal{M}$, under local coordinates in $\mathbb{R}^d$ specified by a chart around the target parameter $\theta^*$. For example,~\cite{10.1214/009053605000000093,https://doi.org/10.48550/arxiv.1306.5806} show that the local coordinate representation of the empirical Fr\'{e}chet mean is asymptotically normal around that of the population Fr\'{e}chet mean. However, only limited general theory has been developed to study the theoretical properties of Bayesian inference in the manifold setting.

To bridge this gap, we analyze the shape of the posterior, denoted by $\Pi(\cdot|X^{(n)})$, in two complementary ways. First, we study $\Pi(\cdot|X^{(n)})$ after projecting it onto the tangent space $T_{\theta^*}\m M$ of $\m M$ at $\theta^*$. The tangent space provides a first-order Euclidean approximation to the manifold $\m M$ near $\theta^*$, which enables the use of conventional analytical tools in Euclidean space while preserving the local geometric structure of the manifold. Moreover, in the vicinity of $\theta^*$, points within $\m M$ have a one-to-one correspondence with tangent vectors via the projection map onto $T_{\theta^*}\m M$. Consequently, this projection can be viewed as a local chart around $\theta^*$ by identifying tangent vectors as $d$-dimensional coordinates using an orthonormal basis of the tangent space. This leads to a clean normal approximation for the projected posterior with an asymptotic covariance matrix that is independent of the choice of local chart. 
Second, we investigate one-dimensional summaries $f(\theta)$ with {\ccb $f: \m M \to \mb R$ smooth, and} $\theta \sim \Pi(\cdot|X^{(n)})$, and study their asymptotic normality. This allows us to directly compare the variance of the push-forward posterior $f_{\#}\Pi(\cdot|X^{(n)})$ with that of the frequentist sampling distribution of its mean, thereby assessing whether credible intervals for $f(\theta)$ remain valid in large samples.

To make our theoretical results broadly applicable, we adopt a general {\it loss-function-based} framework. In the classical Bayesian setting, the loss function $\ell(X,\theta)$ corresponds to the negative log-likelihood, $-\log p(X|\theta)$, and the posterior is likelihood-based. However, for complex data, specifying a flexible and accurate probabilistic model with parameters constrained to a manifold can be challenging. For example, in diffusion tensor imaging, inference on the Fr\'{e}chet or extrinsic mean requires a well-specified distribution on the space of positive semidefinite matrices, which is often difficult to formulate. 
To mitigate strong modeling assumptions, we define the {\it target parameter} as a {\it risk minimizer} -- that is, a minimizer of {\ccb an} expected loss -- and adopt a Gibbs posterior~\citep{bhattacharya2020gibbs,Syring_2018,jiang2008gibbs,CHERNOZHUKOV2003293} constructed from the {\ccb corresponding} empirical loss. This approach preserves the Bayesian updating mechanism while decoupling inference from full likelihood specification, and it naturally accommodates manifold-valued parameters. Our theoretical results are developed within this general risk-minimization and Gibbs posterior framework, with the classical Bayesian case recovered by taking the loss function to be the negative log-likelihood.

% Define $\m M$ to be the support of the prior $\Pi$, and the risk function $\m R:\m M\to \mb R$ with $\m R(\theta)=\mb{E}[\ell(X,\theta)]$,  and 
% \begin{equation*}
%     \theta^*={\arg\min}_{\theta\in \m M} \m R(\theta). 
% \end{equation*}

% \vspace{0.5em}

Let ${\m M}$ be a $d$-dimensional Riemannian submanifold, and let $S_{\Pi}\subseteq {\m M}$ denote the support of the prior $\Pi$. Let $\ell:\m X\times {\m M}\to \mb R$ be the loss function, and define the population risk $\m R(\theta)=\mb{E}_{\m P^*}[\ell(X,\theta)]$. The (Gibbs) posterior is defined as
\begin{equation*}
    \Pi(\dd\theta|X^{(n)})=\frac{\exp(-\sum_{i=1}^n \ell(X_i,\theta))\,\Pi(\dd \theta)}{\int_{\m M} \exp(-\sum_{i=1}^n \ell(X_i,\theta))\,\Pi(\dd \theta)} 
    =\frac{\exp(-\sum_{i=1}^n \ell(X_i,\theta))\,\Pi(\dd \theta)}{\int_{S_{\Pi}} \exp(-\sum_{i=1}^n \ell(X_i,\theta))\,\Pi(\dd \theta)}.
\end{equation*}
Assume the risk minimizer $\theta^*={\arg\min}_{\theta\in \m M}\m R(\theta)$ exists and is unique. Fix positive constants $L, r, \beta_1 \in (0,\infty)$, $\beta_2 \in (0,1]$, and positive integers $d, D$ with $d \leq D$. We impose the following four assumptions. 

\vspace{0.5em}
\noindent\textbf{Assumption 1 (Parameter space regularity).}   
\emph{$\m M$ is a $d$-dimensional Riemannian submanifold of $\mb R^D$ and is locally $C^3_{r,L}$-smooth at $\theta^*$.}

This assumption requires the manifold ${\m M}$ to be locally $C^3$-smooth around $\theta^*$ so that the tangent space $T_{\theta^*}{\m M}$ provides a valid local approximation of ${\m M}$ near $\theta^*$. 

\vspace{0.5em}
\noindent\textbf{Assumption 2 (Prior regularity).}   
 \emph{The support $S_{\Pi}$ of $\Pi$ satisfies that $B_{r}(\theta^*)\cap \m M\subset S_{\Pi} \subset\mb B_{L}(0_D)\cap \m M$.
Moreover, $\Pi$ admits a density  $\pi$ with respect to the volume measure of $\m M$, such that $\pi(\theta^*)>1/L$ and, for all $\theta\in B_{1/L}(\theta^*)\cap S_{\Pi}$, $|\pi(\theta)-\pi(\theta^*)|\leq L\,\|\theta-\theta^*\|$.}

This assumption requires the prior to assign positive mass to a fixed-radius neighborhood of $\theta^*$ on ${\m M}$ and have a Lipschitz continuous density with respect to the volume measure on ${\m M}$ in a neighborhood of $\theta^*$.

\vspace{0.5em}
\noindent\textbf{Assumption 3 (Risk regularity).}
  \emph{For all $\theta\in S_{\Pi}$, $|\mathcal{R}(\theta)-\mathcal{R}(\theta^*)|\geq \frac{1}{L}\|\theta-\theta^*\|^2$. There exists $\ov{\m R}:B_{r}(\theta^*)\to \mb R$ that coincides with $\m R$ on $B_{r}(\theta^*)\cap S_{\Pi}$, where $\ov{\m R}$ has uniformly $L$-bounded partial derivatives up to order three.   }

This assumption requires the risk function $\m R$ to increase locally in a quadratic fashion away from its minimizer $\theta^*$ and admit a smooth extension to a neighborhood of $\theta^*$ in the ambient space.

\vspace{0.5em}
\noindent\textbf{Assumption 4 (Regularity of the loss and gradient-like proxy).} \emph{There exists a function  $b: \m X\to \mb R_{\geq 0}$ with  $\mb{E}[\exp\big((\frac{b(X)}{L})^{\beta_1}\big)]\leq 1$  such that for  all $x\in \m X$ and $\theta,\theta'\in S_{\Pi}$, 
       $|\ell(x,\theta)-\ell(x,\theta')|\leq b(x)\|\theta-\theta'\|$. Moreover, there exists  a map $g: \m X \times   S_{\Pi}\to \mb R^D$  with $g(x,\theta)\in T_{\theta}{\m M}$ and $\|g(x,\theta)\|\leq b(x)$ for all $(x,\theta)\in \m X\times S_{\Pi}$, such that: }
    \begin{enumerate}
   \item (Gradient-like behavior and population-level smoothness).      \emph{For any $\theta\in \mb B_{r}(\theta^*)\cap S_{\Pi}$,  $\mb{E}[  g(X,\theta)]={\rm Proj}_{T_{\theta}{\m M}}(\nabla{\ov {\m R}}(\theta))$, and $\mathbb{E}[\|  g(X,\theta)- g(X,\theta^*)\|^2]\leq L\,\|\theta-\theta^*\|^{2\beta_2}$. Moreover, for any $\theta,\theta'\in \mb B_{r}(\theta^*)\cap S_{\Pi}$, $\mb{E}[(\ell(X,\theta)-\ell(X,\theta')-  g(X,\theta')(\theta-\theta'))^2]\leq L\|\theta-\theta'\|^{2+2\beta_2}.$  For any $\eta\in T_{\theta^*}{{\m M}}$ with unit norm, $\mb{E}[(\eta^T g(X,\theta^*))^2]\geq \frac{1}{L}$. }
  
 \item      \emph{(Metric entropy bounds  for $ {g}(x, \cdot)$). For any $n\in \mb N^{+}$  and sample $\{X_1,X_2,\cdots,X_n\}\in \m X^{n}$, define the  pseudo-metric $d_n^{  g}(\theta,\theta')=\sqrt{\frac{1}{n}\sum_{i=1}^n \|  g(X_i,\theta)-  g(X_i,\theta')\|^2}$. Then for any $ \varepsilon>0$, $\log \bold N(B_{r}(\theta^*)\cap S_{\Pi},d_n^{g},\varepsilon)\leq \max(0,L\log n +L\log (\frac{\sqrt{n^{-1} \sum_{i=1}^n b(X_i)^2}}{\varepsilon}))$.}
 \end{enumerate}

% This assumption requires the loss function $\ell$ to be Lipschitz continuous and admit a gradient-like proxy that behaves smoothly near $\theta^*$ in expectation. Specifically, fix positive constants $L, r, \beta_1 \in (0,\infty)$, $\beta_2 \in (0,1]$, and positive integers $d, D$ with $d \leq D$; we impose the following assumptions on $({\m M}, \ell, \m R, \Pi)$.

Similar assumptions also appear in~\cite{JMLR:v25:23-0783,tang2022bayesian}. Notably, we do not require the loss function $\ell$ to be differentiable. Instead, we introduce a {\it gradient-like proxy} $g(\cdot,\cdot)$ whose expectation $\mb{E}[g(X,\cdot)]$ coincides with the Riemannian gradient of the population risk $\m R(\cdot)$. In particular, for each fixed $X\in \m X$, if $\ell(X, \cdot)$ is differentiable on $\m M$, we may take $g(X,\cdot)$ to be its Riemannian gradient. When $\ell(X, \cdot)$ is not differentiable but admits a Lipschitz extension $\ov\ell(X,\cdot)$ to the ambient space, we define $g(X,\theta)$ as the projection onto $T_{\theta}{\m M}$ of any subgradient of $\theta\mapsto\ov\ell(X,\theta)$. 
The parameter $\beta_2$ characterizes the average smoothness of $g(X, \cdot)$. Specifically, $\mathbb{E}[\| g(X,\theta)-g(X,\theta^*)\|^2]$ controls the Lipschitz behavior of $g(X,\cdot)$, while $\mb{E}[(\ell(X,\theta)-\ell(X,\theta')-g(X,\theta')(\theta-\theta'))^2]$ bounds the mean squared error of the first-order approximation to $\ell(X,\cdot)$ with $g$ in place of the gradient. When $\theta\mapsto\ell(X, \theta)$ is twice differentiable on $\m M$ for every $X$, we have $\beta_2=1$. If $\theta\mapsto\ell(X, \theta)$ is not everywhere differentiable, $\beta_2$ can be smaller than one. For example,~\cite{JMLR:v25:23-0783} shows that in Bayesian quantile regression, where $\ell$ is the non-smooth check loss, the assumptions hold with $\beta_2=\frac{1}{2}$. 
As shown in Theorem~\ref{th:classicbvm}, a larger value of $\beta_2$ implies a faster convergence rate of the posterior $\Pi(\cdot|X^{(n)})$ to its limiting distribution. Finally, the metric-entropy condition ensures uniform control of the random fluctuations of $n^{-1}\sum_{i=1}^n g(X_i,\theta)$ around the Riemannian gradient of the population risk $\m R(\theta)$. 

We now state the main result. Let $P_{\theta^*}\in \mb R^{D\times D}$ be the projection matrix onto $T_{\theta^*}\m M$, and let $\wt{\m H}_{\theta^*}$ denote the Jacobian matrix of the map $\theta \mapsto {\rm Proj}_{T_{\theta}{\m M}}(\nabla\ov{\m R}(\theta))$ evaluated at $\theta=\theta^*$, where $\ov{\m R}(\cdot)$ is the ambient-space extension of $\m R(\cdot)$. Then we define $\m H_{\theta^*}=P_{\theta^*}\wt{\m H}_{\theta^*}P_{\theta^*}$ as the Riemannian Hessian matrix of $\m R(\cdot)$ at $\theta^*$. Also, set $\Delta_{\theta^*}=\mb{E}[g(X,\theta^*)g(X,\theta^*)^T]$. We use $\phi_{\theta^*}\colon V_{\theta^*}\to U_{\theta^*}$ for the local inverse of ${\rm Proj}_{T_{\theta^*}\m M}(\cdot-\theta^*)$, as defined in Definition~\ref{def:C3smooth}.

\begin{theorem}[Manifold BvM Theorem for Gibbs Posterior]\label{th:classicbvm}
Suppose Assumptions 1-4 hold. Let $\wh\theta:\m X^n\to S_{\Pi}$ denote the empirical risk minimizer, that is, $\wh\theta(X^{(n)})\in {\arg\min}_{\theta\in S_{\Pi}}\frac{1}{n}\sum_{i=1}^n\ell(X_i,\theta)$. Then, there exists a set $\m A\subset \m X^{n}$ with $\mathbb P(X^{(n)}\in \m A)\geq 1-n^{-1}$ such that
\begin{enumerate}
\item There exists a constant $C>0$ such that, for every dataset $X^{(n)}\in \m A$,  
 \begin{equation*} 
    \begin{aligned}
    % &\qquad\Pi^{(n)}_{\m M}\Big(\|\theta-\theta^*\|\geq C\sqrt{\frac{\log n}{n}}\,\Big)\leq \frac{1}{n},\text{ and }\\
         &{\rm TV}\Big(\Pi(\cdot|X^{(n)}), \,[\phi_{\theta^*}]_{\#}\big[\m N\big( {\rm Proj}_{T_{\theta^*}\m M}(\wh\theta(X^{(n)})-\theta^*),\, n^{-1}   {\m H}_{\theta^*}^{\dagger} \big)\big|_{V_{\theta^*}}\big]\Big)
 \leq C\, \frac{(\log n)^{\frac{1}{\beta_1}+1}}{n^{\frac{\beta_2}{2}}}.
    \end{aligned}
    \end{equation*}
%  Moreover, $\sqrt{n}\cdot {\rm Proj}_{T_{\theta^*}\m M}(\wh\theta_p^{\m M}-\theta^*)\to \m N(0,  \ov{\m H}_{\theta^*}^{\dagger}\ov\Delta_{\theta^*}\ov{\m H}_{\theta^*}^{\dagger})$ in distribution as $n\to \infty$.   \footnote{We say a distribution sequence $\{\mu_1,\mu_2,\cdots\}$ on  $T_{\theta^*}\m M$ converges in distribution to some $\mu$ on  $T_{\theta^*}\m M$ if
%  \begin{equation*}
%      \lim_{n\to\infty} \mu_n(A)=\mu(A)
%  \end{equation*}
% for every measurable set $A\subset T_{\theta^*}\m M$ with $\mu(\partial A)=0$, where $\partial A$ denotes the topological boundary of $A$.}
where recall that $ {\m H}_{\theta^*}^{\dagger}$ denotes the Moore-Penrose inverse of $ {\m H}_{\theta^*}$.  Moreover, when $n$ is suffiiciently large,  the extrinsic posterior mean $\wh\theta_p(X^{(n)}):\,={\arg \min}_{y\in \mathcal{M}}  \big\|y-\int_{\m M} \theta\,\Pi(\dd\theta|X^{(n)})\big\|^2$  is uniquely defined and  satisfies  $\|\wh\theta_p(X^{(n)})-\wh\theta(X^{(n)})\|\leq C\, \frac{(\log n)^{\frac{1}{\beta_1}+1}}{n^{\frac{\beta_2+1}{2}}}$. 
 
\item  For any fixed positive constants $L_1$ and $r_1$, there exists a constant $C > 0$ such that, for every function 
$f:\mathbb{R}^D \to \mathbb{R}$ satisfying that (1) all partial derivatives of $f$ up to second order are bounded in absolute value by $L_1$ on $\mathbb{B}_{r_1}(\theta^*)$, and (2) $\|{\rm Proj}_{T_{\theta^*}\mathcal{M}}(\nabla f(\theta^*))\| \ge \frac{1}{L_1}$, the following holds:
\begin{enumerate}
    \item for every dataset $X^{(n)}\in \m A$,
$$ {\rm TV}\Big(f_{\#}\Pi(\cdot|X^{(n)}), \,\mathcal{N}\big(f(\wh\theta(X^{(n)})),\, n^{-1} \nabla f(\theta^*)^T  {\m H}_{\theta^*}^{\dagger} \nabla f(\theta^*)\big)\Big) \leq C\, \frac{(\log n)^{\frac{1}{\beta_1}+1}}{n^{\frac{\beta_2}{2}}}.$$
 
    \item  $\sqrt{n}\cdot\big(f\big(\wh\theta(X^{(n)})\big)-f(\theta^*)\big)\to \m N\big(0,   \nabla f(\theta^*)^T  {\m H}_{\theta^*}^{\dagger} \Delta_{\theta^*} {\m H}_{\theta^*}^{\dagger} \nabla f(\theta^*)\big)$ in distribution as $n\to \infty$.
\end{enumerate}

 \end{enumerate}
 
\end{theorem}
The proof of Theorem~\ref{th:classicbvm} is provided in Appendix~\ref{proof:well-specified}. In the proof, we introduce a one-to-one change of variables $y = W_{\theta^*}^T {\rm Proj}_{T_{\theta^*}\mathcal{M}}(\theta - \theta^*) \in \mathbb{R}^d$ locally around $\theta^*$, where $W_{\theta^*}$ is an orthonormal basis of the tangent space $T_{\theta^*}\mathcal{M}$. This transformation allows us to convert the analysis to the Euclidean space $\mathbb{R}^d$. By exploiting the regularity properties of both the manifold and the loss function, we derive a Bernstein--von Mises type result for the posterior distribution after this variable transformation. 

To interpret the first result, note that the tangent space $T_{\theta^*}\mathcal{M}$ locally approximates the manifold $\mathcal{M}$ as a flat space. The map $\theta \mapsto {\rm Proj}_{T_{\theta^*}\mathcal{M}}(\theta - \theta^*)$ projects points from the manifold onto this tangent space, and $V_{\theta^*}$ denotes a neighborhood on the tangent space where this projection is one-to-one, so that each tangent vector in $V_{\theta^*}$ corresponds to a unique point on the manifold. The map $\phi_{\theta^*}$ is the inverse of this projection, mapping points back from the tangent space to the manifold. The first statement in Theorem~\ref{th:classicbvm} then establishes that, after projecting the posterior distribution onto the tangent space, the distribution of the resulting tangent vectors is well-approximated by a normal distribution centered at the projected empirical risk minimizer. Because the projection is invertible on $V_{\theta^*}$, we can also map these normal vectors back to the manifold through $\phi_{\theta^*}$, thereby obtaining a direct approximation of the posterior distribution on $\mathcal{M}$. Furthermore, the posterior mean projected onto $\mathcal{M}$, denoted by $\widehat{\theta}_p(X^{(n)})$, serves as a point estimator that closely aligns with the empirical risk minimizer. 

The second statement in Theorem~\ref{th:classicbvm} concerns one-dimensional functionals of the parameter, given by mappings $f(\theta)$. Specifically, it states that for any smooth function $f$ with bounded derivatives and a nonzero Riemannian gradient at $\theta^*$, the posterior distribution of $f(\theta)$ given data $X^{(n)}$ is well-approximated by a normal distribution centered at $f(\widehat{\theta}(X^{(n)}))$. An important subtlety is that the variance in the posterior normal approximation of $f(\theta)$, given by $n^{-1}\nabla f(\theta^*)^T \mathcal{H}_{\theta^*}^{\dagger} \nabla f(\theta^*)$, generally differs from the asymptotic variance of the sampling distribution of $f(\widehat{\theta}(X^{(n)}))$, which is $n^{-1}\nabla f(\theta^*)^T \mathcal{H}_{\theta^*}^{\dagger} \Delta_{\theta^*} \mathcal{H}_{\theta^*}^{\dagger} \nabla f(\theta^*)$. Consequently, the posterior typically fails to provide correct uncertainty quantification for $f(\theta)$; that is, posterior credible intervals may not achieve valid frequentist coverage. However, when $\mathcal{H}_{\theta^*} = \Delta_{\theta^*}$, the two variances coincide, and the posterior delivers {\it asymptotically correct} uncertainty quantification. This equivalence {\ccb (see \cite{CHERNOZHUKOV2003293,kleijn2012bernstein} for Euclidean analogues)} holds, for instance, under correct model specification---when the loss function corresponds to the negative log-likelihood, $\ell(x,\theta) = -\log p(x \mid \theta)$, and the data are i.i.d. from $p(x \mid \theta^*)$.
% This entails two requirements: (1) the density class $\{p(X|\theta)\,\:\,\theta\in \mb R^D\}$ contains true data-generating distribution;  (2) the true parameter lies on the manifold $\m M$.

For regular models, both $\mathcal{H}_{\theta^*}$ and $\Delta_{\theta^*}$ coincide with the Fisher information matrix $I_{\theta^*}$ projected onto the tangent space $T_{\theta^*}\mathcal{M}$. To formalize this relationship, we introduce the following assumption on correct model specification and model regularity.

\medskip
\noindent\textbf{Assumption 5 (Correct model specification and model regularity).} 
The true data distribution $\mathcal{P}^*$ admits a density $p(x \mid \theta^*)$. 
Moreover, there exists a function $b: \mathcal{X} \to \mathbb{R}_{\ge 0}$ such that $\mb{E}\big[\exp\big(\big(\frac{b(X)}{L}\big)^{\beta_1}\big)\big]\leq 1$, and the density family $\{p(x \mid \theta): \theta \in \mathbb{R}^D\}$ satisfies the following conditions:
(1) for any $x \in \mathcal{X}$, the function $\log p(x \mid \theta)$ and its partial derivatives up to third order are uniformly bounded in absolute value by $b(x)$ for all $\theta \in \mathbb{B}_r(\theta^*) \cap S_{\Pi}$; 
(2) for any $\theta \in S_{\Pi}$, the Kullback--Leibler divergence satisfies ${\rm KL}(p(x \mid \theta^*) \,\|\, p(x \mid \theta)) \ge \frac{1}{L}\|\theta - \theta^*\|^2$; 
(3) the Fisher information matrix $I_{\theta^*} = \mathbb{E}\big[(\nabla_{\theta}\log p(X \mid \theta^*))(\nabla_{\theta}\log p(X \mid \theta^*))^T\big]$ equals the negative Hessian of $\theta \mapsto \mathbb{E}[\log p(X \mid \theta)]$ at $\theta^*$, and satisfies $I_{\theta^*} \succcurlyeq \frac{1}{L} I_D$.

\begin{corollary}[BvM Result under Correct Model Specification]\label{co:classicbvm_region}
Suppose Assumptions 1,2, and 5 hold.  Consider the posterior distribution $\Pi(\dd\theta|X^{(n)})\propto \prod_{i=1}^n p(X_i|\theta)\Pi(\dd\theta)$.
Let $\wh\theta(X^{(n)})=\arg\min_{\theta\in S_{\Pi}}\big\{ n^{-1}\sum_{i=1}^n -\log p(X_i|\theta)\big\}$ denote the MLE over the prior support and $P_{\theta^*}$ be the projection matrix onto $T_{\theta^*}\m M$. Then there exists a set $\m A\subset \m X^{n}$ with $\mathbb P(X^{(n)}\in \m A)\geq 1-n^{-1}$ so that for any function $f:\mb R^D\to f$ satisfies the condition stated in Theorem~\ref{th:classicbvm}, and any $X^{(n)}\in \m A$,
 \begin{equation*} 
    \begin{aligned}
    % &\qquad\Pi^{(n)}_{\m M}\Big(\|\theta-\theta^*\|\geq C\sqrt{\frac{\log n}{n}}\,\Big)\leq \frac{1}{n},\text{ and }\\
         &{\rm TV}\Big(f_{\#}\Pi(\cdot|X^{(n)}), \mathcal{N}\big(f(\wh\theta(X^{(n)})),\, n^{-1} \nabla f(\theta^*)^T (P_{\theta^*}I_{\theta^*}P_{\theta^*})^\dagger \nabla f(\theta^*)\big)\Big) \leq C\, \frac{(\log n)^{\frac{1}{\beta_1}+1}}{\sqrt{n}},
    \end{aligned}
    \end{equation*}
 where the matrix $(P_{\theta^*}I_{\theta^*}P_{\theta^*})^\dagger$ always satisfies $(P_{\theta^*}I_{\theta^*}P_{\theta^*})^\dagger\preccurlyeq I_{\theta^*}^{-1}$. Moreover, for any  $\alpha\in (0,1)$, let $q_{\alpha}^f(X^{(n)})$ denote the $\alpha$-quantile of $f_{\#}\Pi(\cdot|X^{(n)})$. Then 
    \begin{equation*}
        \begin{aligned}
& \Big|\mathbb P\big( q_{\alpha/2}^f(X^{(n)}) \leq\theta^*\leq q_{1-\alpha/2}^f(X^{(n)}) \big)-(1-\alpha)\Big|\leq  \frac{(\log n)^{\frac{1}{\beta_1}+1}}{\sqrt{n}}.
        \end{aligned}
    \end{equation*}
 \end{corollary}

Corollary~\ref{co:classicbvm_region} shows that, under correct model specification, the Bayesian posterior with a manifold-supported prior can provide credible intervals for $f(\theta)$ that are also valid from a frequentist perspective. Another key implication of this result is that using a prior supported on a lower-dimensional submanifold can lead to more efficient inference. Specifically, the asymptotic variance of the posterior for $f(\theta)$, given by $n^{-1} \nabla f(\theta^*)^T (P_{\theta^*} I_{\theta^*} P_{\theta^*})^\dagger \nabla f(\theta^*)$, is always less than or equal to the variance $n^{-1} \nabla f(\theta^*)^T I_{\theta^*}^{-1} \nabla f(\theta^*)$ obtained when using a conventional prior supported on the full Euclidean space. This implies that, in large samples and under correct model specification, credible intervals based on a posterior with a manifold-supported prior are at least as short as, and possibly shorter than, those from a standard Bayesian posterior, while still maintaining correct coverage. 

An interesting special case arises when the Fisher information matrix $I_{\theta^*}$ is the identity matrix. In this setting, the two variances are equal if and only if the gradient $\nabla f(\theta^*)$ lies entirely within the tangent space $T_{\theta^*}\mathcal{M}$. This indicates that, for functions $f$ that are naturally compatible with the manifold structure, the standard Bayesian posterior may achieve the same efficiency as the manifold-aware approach. However, if $\nabla f(\theta^*)$ is not contained in $T_{\theta^*}\mathcal{M}$, then explicitly accounting for the manifold structure can lead to strictly shorter credible intervals for $f(\theta)$. In contrast, ignoring the manifold structure may result in overestimated uncertainty along directions orthogonal to the manifold, which do not correspond to meaningful variations of the parameter under the model.

It is important to note that Corollary~\ref{co:classicbvm_region} relies on two key assumptions: (1) the model family $\{p(X \mid \theta): \theta \in \mathbb{R}^D\}$ contains the true data-generating distribution, and (2) the true parameter $\theta^*$ lies on the manifold $\mathcal{M}$. If either condition is violated, the posterior may yield incorrect uncertainty quantification. In the following, we present specific examples to illustrate these points. Proofs of the statements in these examples are provided in Appendix~\ref{proof:example}.

\begin{figure}[t]
  \centering

  % Left image
  \begin{minipage}{0.49\linewidth}
    \centering
    \vspace{2pt}
    \includegraphics[width=\linewidth]{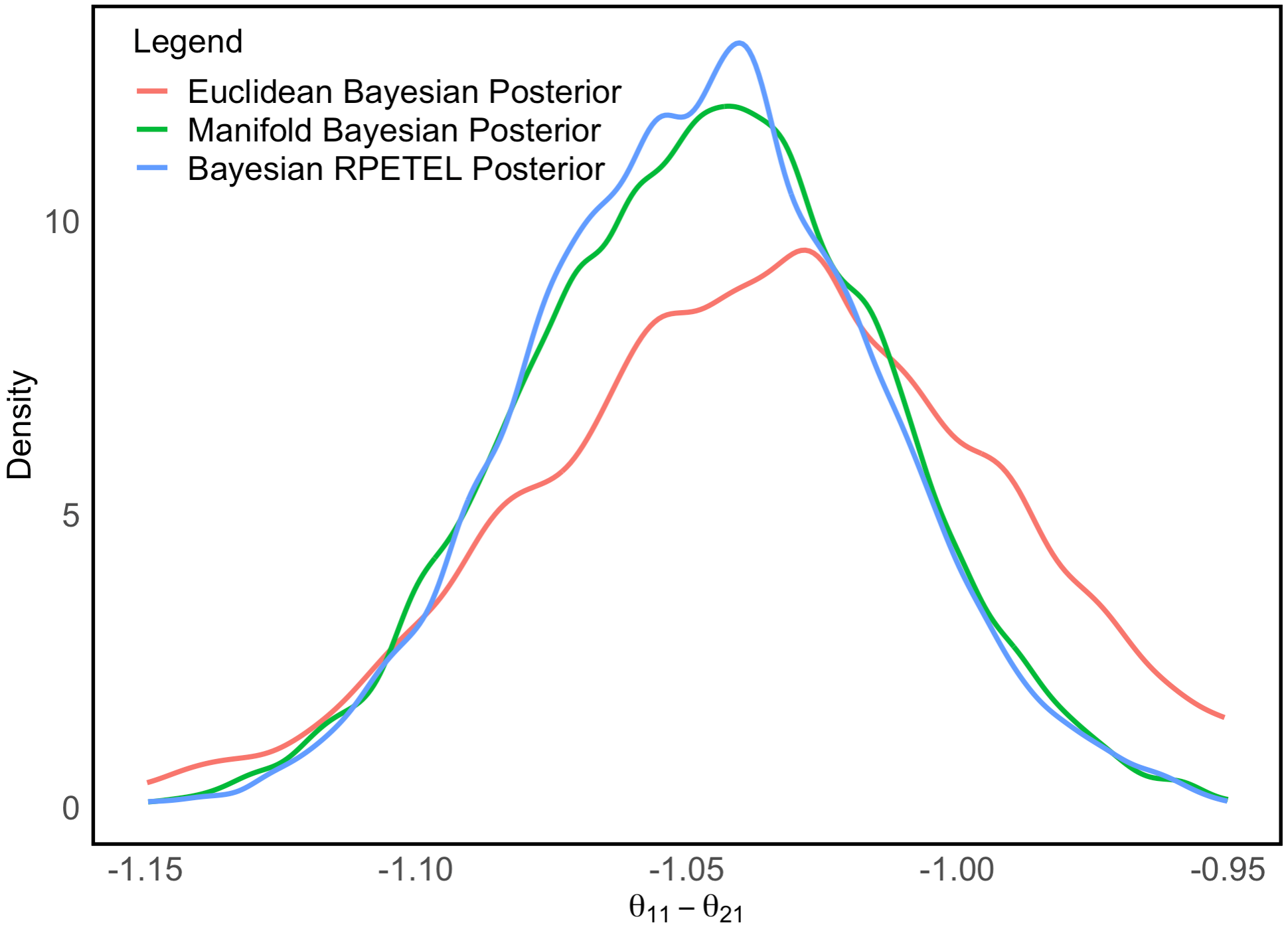}
    \caption*{ {(a) Density plot: Correctly specified likelihood} }
     \end{minipage}
  \hfill
  \begin{minipage}{0.49\linewidth}
    \centering
    \vspace{2pt}
    \includegraphics[width=\linewidth]{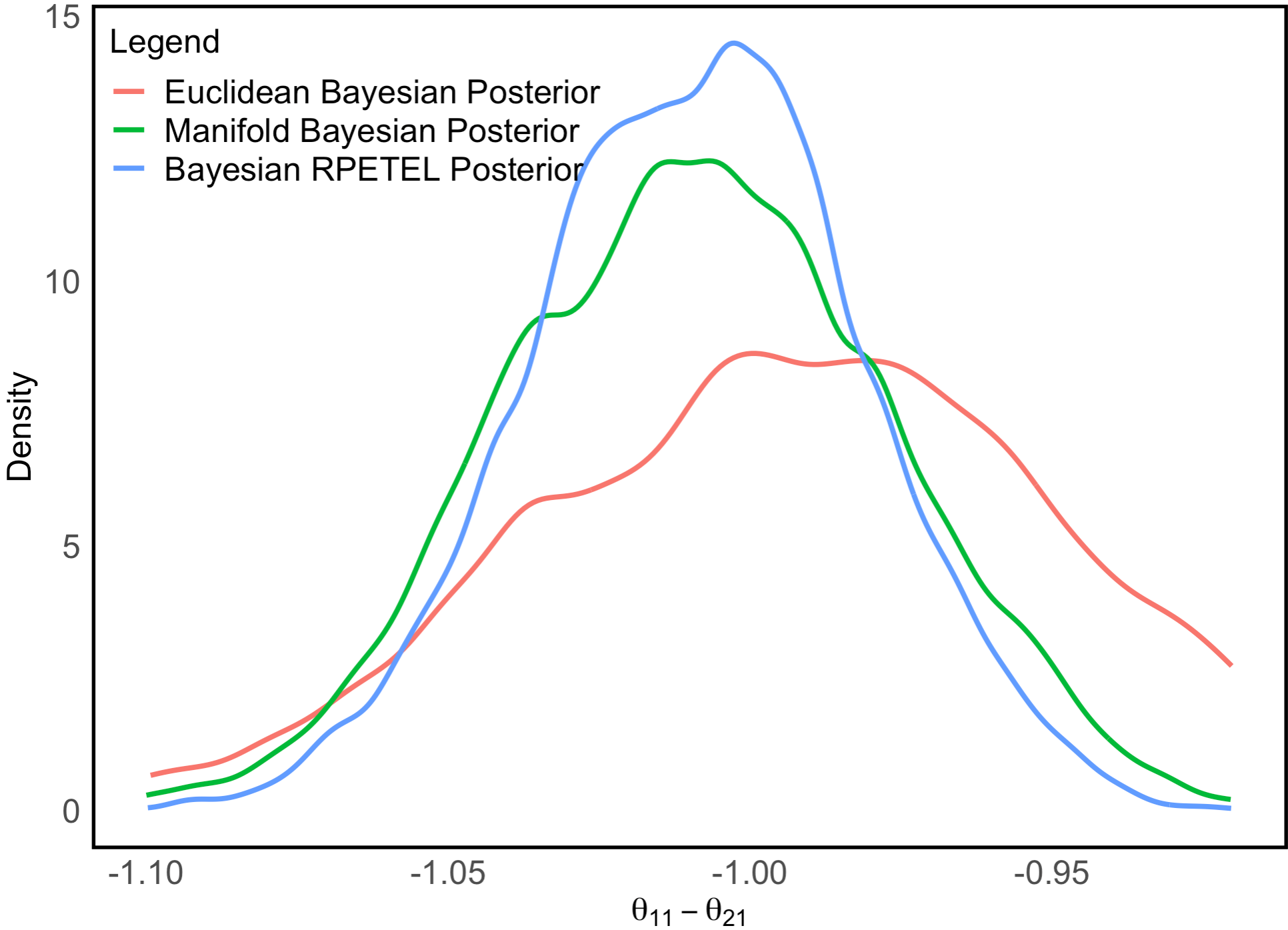}
    \caption*{ {(b) Density plot: Misspecified likelihood}}
 \end{minipage}
  \hfill
  \begin{minipage}{1\linewidth}
    \centering
    \vspace{2pt}
\setlength{\tabcolsep}{6pt}
\renewcommand{\arraystretch}{1.25}
\begin{tabular}{lcccccc}
\hline
& \multicolumn{3}{c}{Correctly specified likelihood} & \multicolumn{3}{c}{Misspecified likelihood} \\
\cline{2-4}\cline{5-7}
& \,\quad$\Pi_{\rm E}$\quad\, & $\Pi_{\rm M}$ & $\Pi_{\rm RP}$& $\,\Pi_{\rm E}\,$ & $\Pi_{\rm M}$ & $\Pi_{\rm RP}$ \\
\hline
Coverage ($\%$, target $90\%$) & 90.3 & 90.1 & 89.4 & 94.3 & 95.6 & 90.6 \\
Interval length ($\times 10^{-1}$)         & 1.47 & 1.15 & 1.14 & 1.47 & 1.15 & 0.98\\
MSE ($\times 10^{-3}$)   & 1.87  & 1.15   & 1.15  & 1.47 & 0.83 & 0.83 \\
Effective sample size & 683  & 944  & 948  & 680  & 942     &   1100   \\
\hline
\end{tabular}
 
\captionof*{table}{{(c) Coverage, interval length, point estimation error and effective sample size (ESS)}}  
  \end{minipage}

  \caption{ {\it Results for Example 1.  Figure (a) and (b) show posterior densities of $\theta_{11}-\theta_{21}$ from a single experiment with $n=1000$, under $\Pi_{\rm E}$ (Bayesian posterior with Euclidean prior: red), $\Pi_{\rm M}$ (Bayesian posterior with manifold-supported prior: green) and $\Pi_{\rm RP}$ (Bayesian RPETEL posterior described in Section~\ref{sec:LFBBI}: blue). Posteriors are estimated from $5000$ samples drawn using the (Riemannian) random-walk Metropolis algorithm  described in Section~\ref{sec:Sampling}. Figure (a) corresponds to the correctly specified  likelihood and  $(b)$ to the misspecified likelihood.   Table (c) reports a quantitative summary based over $1000$ replicated   experiments for inference on $\theta_{11}-\theta_{21}$, including: coverage probability for the nominal $90\%$ credible interval, the average length of that $90\%$ interval, the MSE defined as the average squared Euclidean distance between the posterior mean and the true value, and the effective sample size of the Markov chain for total of $5000$ samples.}
 }
  \label{fig:simutoy1}
\end{figure}

\medskip
\noindent \textbf{Example 1: Reduced-rank multi-response regression.} We consider a linear regression model with multi-dimensional response: $Y = \theta^T X + \varepsilon$, where $Y \in \mathbb{R}^p$ is a $p$-dimensional response vector, $X \in \mathbb{R}^d$ is a covariate, and $\theta = (\beta_1, \beta_2, \cdots, \beta_p) \in \mathbb{R}^{d \times p}$ is the parameter matrix of interest. To share information across different responses, the reduced-rank multi-response regression~\cite{izenman1975reduced} imposes a low-rank constraint ${\rm Rank}(\theta) = r$ with $r < \min(d, p)$. For clarity, we focus on the case $d = p = 2$ and $r = 1$. Suppose the true parameter is given by $\theta^* = (\beta_1^*, \beta_2^*)$ with $\beta_1^* = (1, 1)^T$ and $\beta_2^* = (2, 2)^T$, so indeed $\theta^*$ has rank one. We generate covariates from $\mathcal{N}(0_2, I_2)$ and the noise from $\varepsilon \sim \mathcal{N}(0_2, \Sigma)$, where $\Sigma$ is either the identity or a non-identity full covariance matrix (specified below). Given $n$ i.i.d. samples $\{(X_i, Y_i)\}_{i=1}^n$, we collect the design matrix $\widetilde{X} = (\widetilde{X}_1, \widetilde{X}_2, \cdots, \widetilde{X}_n)^T \in \mathbb{R}^{n \times 2}$ and the response matrix $\widetilde{Y} = (\widetilde{Y}_1, \widetilde{Y}_2, \cdots, \widetilde{Y}_n)^T \in \mathbb{R}^{n \times 2}$. To compare the effect of imposing the low-rank structure, we introduce two priors on the parameter space: $\Pi_{\rm M} = \{\theta \in \mathbb{R}^{2 \times 2}: {\rm Rank}(\theta) = 1, \|\theta\|_F \le 100\}$ and $\Pi_{\rm E} = \{\theta \in \mathbb{R}^{2 \times 2}: \|\theta\|_F \le 100\}$. Specifying the noise distribution as Gaussian with identity covariance (which is misspecified when $\Sigma \neq I_p$), the posteriors take the form $\Pi_{\rm M}(\mathrm{d}\theta \mid (\widetilde{X}, \widetilde{Y})) \propto \Pi_{\rm M}(\mathrm{d}\theta) \exp(-\frac{1}{2} \sum_{i=1}^n \|\widetilde{Y}_i - \theta^T \widetilde{X}_i\|^2)$ and similarly for $\Pi_{\rm E}(\mathrm{d}\theta \mid (\widetilde{X}, \widetilde{Y}))$ with prior $\Pi_{\rm E}$. Let $\widetilde{Y}_{,j}$ ($j=1,2$) denote the $j$-th column of $\widetilde{Y}$ (corresponding to the $j$-th response). Then the ordinary least squares (OLS) estimator is $\widehat{\theta} = (\widehat{\beta}_1, \widehat{\beta}_2)$, where $\widehat{\beta}_j = (\widetilde{X}^T \widetilde{X})^{-1} \widetilde{X}^T \widetilde{Y}_{,j}$ for $j=1,2$. For illustration, we focus inference on $f(\theta) = \theta_{11} - \theta_{21}$, where $\theta_{ij}$ denotes the $(i,j)$ entry of $\theta$. We consider two cases:
\begin{enumerate}
    \item Correctly specified likelihood ($\Sigma=I_2$): the Euclidean posterior  $f_{\#}\Pi_{\rm E}(\cdot|(\wt X,\wt Y))$ approaches $\m N(f(\wh\theta),\frac{2}{n})$,  and $\sqrt{n}(f(\wh\theta)-f(\theta^*))\xrightarrow{d}\m N(0,2)$. On the other hand, the manifold posterior $f_{\#}\Pi_{\rm M}(\cdot|(\wt X,\wt Y))$ approaches $\m N(\wh s(\wh\theta_{11}+\wh\theta_{21}),\frac{1.1}{n})$, where
    % $\wh a=\frac{(\wt Y_{,1}+\wt Y_{,2})^T\wt X\wh\beta_2}{(\wt Y_{,1}+\wt Y_{,2})^T\wt X\wh\beta_1}$
   $\wh s$ is a data-dependent scaling factor, and  $\sqrt{n}(\wh s(\wh\theta_{11}+\wh\theta_{21})-f(\theta^*))\xrightarrow{d}\m N(0,1.1)$. Thus, both posteriors deliver valid uncertainty quantification, with the manifold posterior achieving smaller asymptotic variance.
   
    \item Misspecified likelihood ($\Sigma=\big(\begin{smallmatrix}
  1 & 0.3\\
  0.3 & 1
\end{smallmatrix}\big)$): the Euclidean and manifold posteriors have the same asymptotic Gaussian forms as in the correctly specified case. However, the sampling distributions of the posterior centers differ: $\sqrt{n}(f(\wh\theta)-f(\theta^*))\xrightarrow{d} \m N(0,1.4)$  and $\sqrt{n}(\wh s(\wh\theta_{11}+\wh\theta_{21})-f(\theta^*)) \xrightarrow{d}\m N(0,0.824)$. Hence, although the rank-one manifold assumption is satisfied, the misspecified likelihood prevents both posteriors from providing valid uncertainty quantification.
\end{enumerate}
Figure~\ref{fig:simutoy1} numerically illustrates these findings. In both the correctly specified and misspecified settings, the posterior density of $f(\theta) = \theta_{11} - \theta_{21}$ under the Euclidean Bayesian posterior $\Pi_{\rm E}(\cdot \mid (\widetilde{X}, \widetilde{Y}))$ (red curves) exhibits heavier tails than under the manifold Bayesian posterior $\Pi_{\rm M}(\cdot \mid (\widetilde{X}, \widetilde{Y}))$ (green curves). Moreover, the posterior mean based on $\Pi_{\rm M}$ shows a smaller estimation error compared to that based on $\Pi_{\rm E}$. However, under likelihood misspecification, the empirical coverage of credible intervals for $f(\theta)$ under both $\Pi_{\rm M}$ and $\Pi_{\rm E}$ substantially exceeds the nominal level.

\medskip
\noindent \textbf{Example 2: Mean direction of the von Mises-Fisher distribution.} The von Mises-Fisher distribution models random unit vectors on the $(d-1)$-dimensional unit sphere $\mathbb{S}^{d-1}_1$. Its density takes the form $f(x \mid \theta) = C_d(\|\theta\|)\exp(\theta^T x)$, for $x \in \mathbb{S}^{d-1}_1$ and $\theta \in \mathbb{R}^d$, where the magnitude $\|\theta\|$ controls concentration, the direction of $\theta$ determines the mean direction of the distribution, and $C_d(\|\theta\|)$ is the normalization constant depending only on $\|\theta\|$ and $d$. Consider the case $d = 3$ with true parameter $\theta^* = \kappa^* (1/\sqrt{3}, 1/\sqrt{3}, 1/\sqrt{3})^T$, so that the true mean direction is $\mu^* = (1/\sqrt{3}, 1/\sqrt{3}, 1/\sqrt{3})^T$. Suppose we observe $n$ i.i.d.\ samples $X^{(n)} = \{X_1, X_2, \cdots, X_n\}$ from $f(x \mid \theta^*)$. To infer the mean direction, we take the parameter space $\mathcal{M}$ to be the unit sphere $\mathbb{S}^{d-1}_1$ and place a uniform prior $\Pi$ on $\mathcal{M}$. This yields the posterior distribution $\Pi(\mathrm{d}\theta \mid X^{(n)}) \propto \Pi(\mathrm{d}\theta)\exp\big(\sum_{i=1}^n \theta^T X_i + n\log C_d(\|\theta\|)\big) \propto \Pi(\mathrm{d}\theta)\exp(\sum_{i=1}^n \theta^T X_i)$. The likelihood function is correctly specified, but the manifold assumption $\theta^* \in \mathcal{M}$ holds only if $\kappa^* = 1$. For illustration, consider inference on the first coordinate $f(\theta) = \theta_1$. The posterior distribution $f_{\#}\Pi(\cdot \mid X^{(n)})$ converges to $\mathcal{N}(f(\overline{X} / \|\overline{X}\|), n^{-1}\frac{2}{3A(\kappa^*)})$, where $A(\kappa^*) = \coth(\kappa^*) - 1/\kappa^*$ and $\overline{X} = n^{-1}\sum_{i=1}^n X_i$. On the other hand, the sampling distribution of the posterior center satisfies $\sqrt{n}(f(\overline{X} / \|\overline{X}\|) - f(\mu^*)) \xrightarrow{d} \mathcal{N}(0, \frac{2}{3\kappa^* A(\kappa^*)})$. Thus, unless $\kappa^* = 1$ so that $\theta^*$ lies exactly on the manifold $\mathcal{M}$, the posterior variance fails to match the true sampling variability, and the posterior cannot provide valid uncertainty quantification.

\section{Robust Bayesian Inference on Manifold}\label{sec:LFBBI}
We begin with a likelihood-based formulation in Euclidean space, defining the target parameter as $\theta^*_{\rm E} = {\arg\min}_{\theta \in \mathbb{R}^d} \mathbb{E}[-\log p(X \mid \theta)]$. In practice, however, we often possess additional structural information about the parameter $\theta^*_{\rm E}$—for example, low-rank structure, unit norm, or symmetry—suggesting that $\theta^*$ lies (approximately) on a low-dimensional Riemannian submanifold $\mathcal{M} \subset \mathbb{R}^d$. Such information can be encoded through a prior $\Pi$ supported on $\mathcal{M}$. As discussed in Section~\ref{sec:LBBI}, however, the resulting posterior’s ability to deliver correct uncertainty quantification depends on two strong conditions: the likelihood family must be correctly specified to include the true data-generating distribution, and the Euclidean minimizer $\theta^*_{\rm E}$ must lie exactly on $\mathcal{M}$. Both requirements are often unrealistic. In complex models, specifying a correct likelihood is difficult—for instance, in regression problems, we may primarily care about the regression coefficients, while the noise distribution is difficult to fully characterize and not of primary interest. Moreover, even when $\theta^*_{\rm E}$ does not lie exactly on $\mathcal{M}$, casting the problem as a manifold-constrained optimization can yield substantial computational benefits. In Section~\ref{sec:Sampling}, we show that a manifold-adapted random walk Metropolis algorithm can be used to sample from a manifold-constrained posterior with mixing time nearly linear in the intrinsic dimension $d$ of $\mathcal{M}$, rather than in the ambient dimension $D$. This motivates a deliberate trade-off between small statistical bias and significant computational efficiency gains. For example, in reduced-rank multi-response regression, the true coefficient matrix may not be exactly rank $r$ but can be well approximated by a rank-$r$ matrix, with residual components attributable to weak signals. In such settings, classical Bayesian approaches that rely on fully specified likelihoods and exact manifold adherence may fail to provide valid uncertainty quantification. This motivates the development of robust, manifold-aware pseudo-Bayesian methods that avoid full likelihood specification while retaining valid uncertainty quantification.

 As in Section~\ref{sec:LBBI}, to avoid full likelihood specification, we adopt a loss-function-based framework by introducing a loss function $\ell(X, \theta)$ that quantifies how well a parameter explains the data. Rather than assuming that the Euclidean minimizer $\theta^*_{\rm E}$ lies exactly on $\mathcal{M}$, we define the target parameter directly as the constrained population risk minimizer on $\mathcal{M}$: $$\theta^* = {\arg\min}_{\theta \in \mathcal{M}} \mathcal{R}(\theta), \quad \mathcal{R}(\theta) = \mathbb{E}[\ell(X, \theta)].$$ However, as shown in Section~\ref{sec:LBBI}, treating the scaled empirical risk $\sum_{i=1}^n \ell(X_i, \theta)$ as a surrogate log-likelihood can yield consistent point estimates but generally fails to provide valid uncertainty quantification. To address this limitation, we leverage the first-order optimality condition of $\theta^*$ on $\mathcal{M}$ (see Appendix~\ref{App:manifold} for details): 
\begin{equation}\label{first-order_population}
    {\rm grad}\, \mathcal{R}(\theta^*) = 0_D, \quad \text{where } {\rm grad}\, \mathcal{R}(\theta^*) \text{ denotes the Riemannian gradient of } \mathcal{R}(\cdot) \text{ at } \theta^* \text{ on } \mathcal{M},
\end{equation}
which provides identifying information about $\theta^*$ without requiring a full likelihood specification. In particular, if for any $X$ and $\theta$ the Riemannian gradient ${\rm grad}_{\theta} \, \ell(X, \theta)$ of the loss function $\theta \mapsto \ell(X, \theta)$ exists, then the target parameter can be equivalently characterized by the moment condition $\mathbb{E}[{\rm grad}_{\theta} \, \ell(X, \theta^*)] = {\rm grad} \, \mathcal{R}(\theta^*) = 0_D$. We then incorporate the information from these moment conditions to construct a Bayesian pseudo-posterior.

One choice is to use the exponentially tilted empirical likelihood (ETEL) approach~\citep{schennach2005bayesian,chib2018bayesian,tang2022bayesian}. To do this, we define the Riemannian exponentially tilted empirical likelihood (RETEL) function $L(X^{(n)};\theta)=\prod_{i=1}^n p_i(\theta)$, where $(p_1(\theta),p_2(\theta),\cdots,p_n(\theta))$ is the solution of 
\begin{equation}\label{Eqn:RETEL}
   \begin{aligned}
 &\qquad\quad\max_{(w_1,w_2,\ldots,w_n)} \sum_{i=1}^n \big[-w_i \log (n w_i)\big]\\
\mbox{subject to} \quad & \sum_{i=1}^n w_i=1,\quad \sum_{i=1}^n w_i\,{\rm grad}_{\theta}\,\ell(X_i,\theta)=0_D,\quad w_1,w_2,\ldots,w_n \ge 0.
   \end{aligned}
\end{equation}
We can view $\{p_i(\theta)\}_{i=1}^n$ as the probabilities minimizing the ``backward'' KL divergence between the empirical distribution $(n^{-1},n^{-1},\cdots,n^{-1})$ and the multinomial distribution $(w_1,w_2,\ldots,w_n)$, subject to the constraint that a weighted-sample version of the first-order optimality condition on the Riemannian submanifold is satisfied. It is worth noting that the ETEL function is not the only possible choice for the pseudo-likelihood; for instance, we can consider the Riemannian empirical likelihood (REL) function, which minimizes the ``forward'' KL divergence between the empirical distribution and the multinomial distribution $(w_1,w_2,\ldots,w_n)$ assigned to data points $X^{(n)}$, subject to the same constraint as the RETEL function. Another option is to utilize the generalized method of moments (GMM)~\citep{CHERNOZHUKOV2003293,hall2004generalized,yin2009bayesian}, where the pseudo-likelihood is given by the objective function optimized in GMM using the moment condition~\eqref{first-order_population}, i.e., $-\big(\frac{1}{\sqrt{2n}}\sum_{i=1}^n g(X_i,\theta)\big)^T\big(\frac{1}{n}\sum_{i=1}^n g(X_i,\theta)g(X_i,\theta)^T\big)^{-1}\big(\frac{1}{\sqrt{2n}}\sum_{i=1}^n g(X_i,\theta)\big)$ with $g(X,\theta)={\rm grad}_{\theta}\,\ell(X,\theta)$. However, all these pseudo-likelihoods derived from the moment condition~\eqref{first-order_population} share a common issue: the first-order optimality condition ${\rm grad}\,\m R(\theta)=0_D$ may have multiple solutions, and directly incorporating them as pseudo-likelihoods in a posterior distribution may result in inconsistent estimation.

To address this issue of inconsistent estimation and enforce the concentration of the posterior around the global risk minimizer $\theta^*$, we adopt the approach proposed in~\citep{tang2022bayesian} to exponentially penalize the posterior using the empirical risk function $\m R_n(\theta)=n^{-1}\sum_{i=1}^n \ell(X_i,\theta)$. Given a prior distribution $\Pi$ defined on $\m M$, we define the following Bayesian penalized Riemannian exponentially tilted empirical likelihood (RPETEL) posterior 
\begin{equation}\label{def:RPETEL}
  \Pi_{\rm RP} (\dd\theta|X^{(n)})=\frac{\exp(\sum_{i=1}^n\log p_i(\theta)-\alpha_n \mathcal{R}_n(\theta))\Pi(\dd\theta)}{\int_{\m M}  \exp(\sum_{i=1}^n\log p_i(\theta)-\alpha_n \mathcal{R}_n(\theta)) \Pi(\dd\theta)},
\end{equation}
where $\alpha_n\geq 0$ is an $n$-dependent regularization parameter and $(p_1(\theta),p_2(\theta), \cdots, p_n(\theta))$ is the solution to~\eqref{Eqn:RETEL}. Notice that the RETEL function in~\eqref{def:RPETEL} may be replaced with other moment condition-based pseudo-likelihoods, such as the REL function or the GMM objective discussed earlier.

\begin{figure}[h]

\centering  
\subfigure[Risk function]{
 \includegraphics[ width=0.5\textwidth]{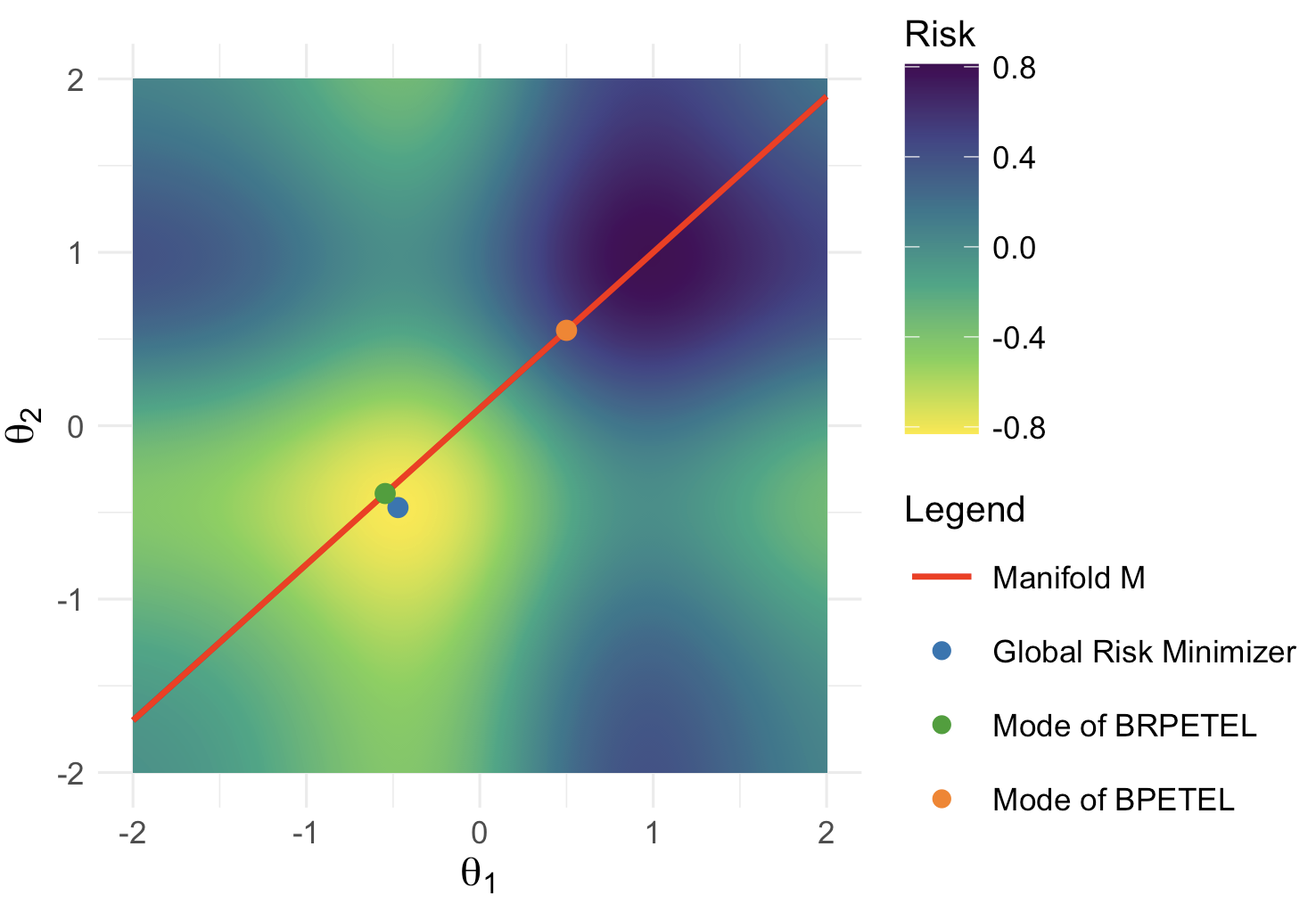}}
\subfigure[Density of posterior distribution]{
 \includegraphics[ width=0.4\textwidth]{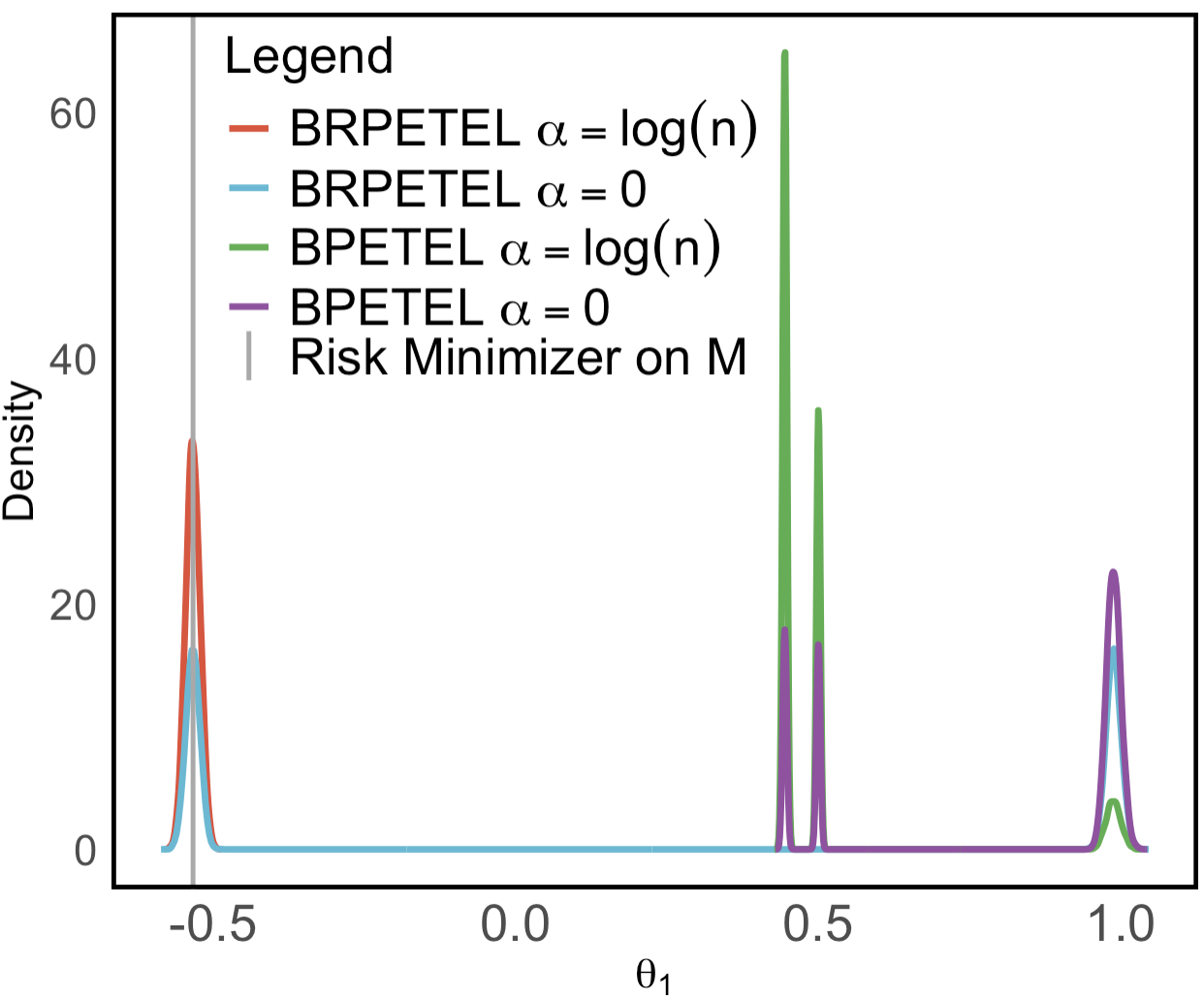}}
\caption{\label{Fig_Density_E_PE} {\em The figures compare Bayesian RPETEL and Bayesian PETEL on a toy example with loss function $\ell(x,\theta)=\sum_{j=1}^2 \exp(-(\theta_j-0.5)^2)\cdot x_j-\exp(-\theta_j^2)$ for $x=(x_1,x_2)^T$ and $\theta=(\theta_1,\theta_2)^T$. The parameter is constrained to the line-shaped manifold $\m M=\{(\theta_1,\theta_2)\in \mb R^2\,:\,\theta_2=0.9\theta_1+0.1\}$. We draw $n=1000$ i.i.d. samples from $\m N(0_2,I_2)$ and use a uniform prior on $\m M\cap[-2,2]^2$. The Bayesian RPETEL posterior (BRPETEL) builds the ETEL function from the Riemannian gradient on $\m M$, while the Bayesian PETEL posterior (BPETEL) uses the Euclidean gradient in $\mb R^2$ to construct the ETEL. Figure~(a) shows the population risk $\m R(\theta)=\mb{E}[\ell(X,\theta)]$ for $\theta\in[-2,2]^2$. The blue dot marks the Euclidean risk minimizer on $[-2,2]^2$, which lies very close to but not exactly on $\m M$. The green and orange dots indicate the posterior modes of BRPETEL and BPETEL, respectively, when $\alpha_n=\log n$; the BRPETEL mode sits close to the true risk minimizer, whereas the BPETEL mode is noticeably farther away. Figure~(b) plots the marginal densities of $\theta_1$ for both posteriors under $\alpha_n=\log n$ and $\alpha_n=0$. The gray line marks $\theta_1^*$, the $\theta_1$-coordinate of the risk minimizer on $\m M$. When $\alpha_n=0$, both posteriors are multimodal because the risk function is nonconvex. In particular, ${\rm grad}_{\theta}\m R(\theta)=0$ has two solutions on $\m M$: one near $(-0.545,0.391)$ corresponding to the risk minimizer on $\m M$, and another near $(0.975,0.977)$ corresponding to a local risk maximizer. With $\alpha_n=0$, the BRPETEL exhibits two comparable modes around these points, while the BPETEL with $\alpha_n=0$ shows a poorly interpretable shape and allocates little mass near $\theta_1^*$. When $\alpha_n=\log n$, the BRPETEL concentrates around $\theta_1^*$, whereas the BPETEL still fails to place meaningful mass there.
  }}
 
\end{figure}

An important design choice in our construction is the use of the Riemannian gradient rather than the full Euclidean gradient. Indeed, if $\theta^*=\theta^*_{\rm E}$, one could instead work with the moment condition $\nabla \m R(\theta)=\mb{E}[\nabla_{\theta}\ell(X,\theta)]=0_D$, in which case our method reduces to the Bayesian PETEL posterior of~\cite{tang2022bayesian} when combined with a manifold-supported prior. However, this approach has several drawbacks. First, it critically relies on the assumption that the global Euclidean minimizer lies exactly on the manifold $\m M$. If this is not the case, the moment condition becomes misspecified. As shown in~\cite{chib2018bayesian}, such misspecification can lead to invalid uncertainty quantification, and the resulting posterior may even fail to concentrate around the risk minimizer $\theta^*$ on the manifold $\m M$, even when the Euclidean minimizer $\theta_{\rm E}^*$ is very close to the manifold (see Figure~\ref{Fig_Density_E_PE} for an illustration). Second, in settings where the parameter is intrinsically defined on a manifold, such as Fr\'{e}chet mean estimation on the sphere, the Euclidean gradient is not a meaningful object, and only the Riemannian gradient correctly encodes the first-order optimality condition.

The Bayesian RPETEL approach also applies when the loss function $\ell(x,\theta)$ is not everywhere differentiable in $\theta$. In such cases, the Riemannian gradient ${\rm grad}_{\theta}\ell(X,\theta)$ can be replaced by any function $g(X,\theta)$ that satisfies Assumption~4. The moment condition underlying $\Pi_{\rm RP}(\cdot|X^{(n)})$ is then specified as $\mb{E}[g(X,\theta)]=0_D$. Under this generalized formulation, we establish a Bernstein–von Mises type theorem for the Bayesian RPETEL posterior. Throughout, we follow the notation of Theorem~\ref{th:classicbvm}, including the Riemannian Hessian matrix $\m H_{\theta^*}$, the Gram matrix $\Delta_{\theta^*}$, the empirical risk minimizer $\wh\theta(X^{(n)})$, and the inverse of the projection map $\phi_{\theta^*}: V_{\theta^*}\to U_{\theta^*}$. We also use $\wh\theta_{p}(X^{(n)})$ to denote the posterior mean of $\Pi_{\rm RP}(\cdot|X^{(n)})$ projected onto $\m M$.

 \begin{theorem}[Manifold BvM theorem for Bayesian RPETEL]\label{th1}
Suppose Assumptions 1-4 hold. Then there exists a set $\m A\subset \m X^{n}$ with $\mathbb P(X^{(n)}\in \m A)\geq 1-n^{-1}$, and positive constants $c_1, c_2$ such that when $c_1\log n\leq \alpha_n\leq c_2 \sqrt{n}$, the followings hold
\begin{enumerate}
\item There exists a constant $C>0$ such that, for every dataset $X^{(n)}\in \m A$, we have
 \begin{equation*} 
    \begin{aligned}
    % &\qquad\Pi^{(n)}_{\m M}\Big(\|\theta-\theta^*\|\geq C\sqrt{\frac{\log n}{n}}\,\Big)\leq \frac{1}{n},\text{ and }\\
         &{\rm TV}\Big(\Pi_{\rm RP}(\cdot|X^{(n)}), \,\phi_{\theta^*}{}_{\#}\big[\m N\big( {\rm Proj}_{T_{\theta^*}\m M}(\wh\theta(X^{(n)})-\theta^*),\, n^{-1} {\m H}_{\theta^*}^{\dagger} \Delta_{\theta^*}{\m H}_{\theta^*}^{\dagger} \big)|_{V_{\theta^*}}\big]\Big)\\
 &\qquad\leq C\, \frac{(\log n)^{\frac{2}{\beta_1}+1}}{n^{\frac{\beta_2}{2}}},
    \end{aligned}
    \end{equation*}
    and when $n$ is sufficiently large,  $\wh\theta_p(X^{(n)})$ is uniquely defined and satisfies $\|\wh\theta_p(X^{(n)})-\wh\theta(X^{(n)})\|\leq C\, \frac{(\log n)^{\frac{1}{\beta_1}+1}}{n^{\frac{\beta_2+1}{2}}}$.
%  Moreover, $\sqrt{n}\cdot {\rm Proj}_{T_{\theta^*}\m M}(\wh\theta_p^{\m M}-\theta^*)\to \m N(0,  \ov{\m H}_{\theta^*}^{\dagger}\ov\Delta_{\theta^*}\ov{\m H}_{\theta^*}^{\dagger})$ in distribution as $n\to \infty$.   \footnote{We say a distribution sequence $\{\mu_1,\mu_2,\cdots\}$ on  $T_{\theta^*}\m M$ converges in distribution to some $\mu$ on  $T_{\theta^*}\m M$ if
%  \begin{equation*}
%      \lim_{n\to\infty} \mu_n(A)=\mu(A)
%  \end{equation*}

\item  There exists a constant $C>0$ such that for any function $f$ satisfies the conditions specified in Theorem~\ref{th:classicbvm} and for every dataset $X^{(n)}\in \m A$, we have
\begin{equation*}
    \begin{aligned}
         {\rm TV}\Big(f_{\#}\Pi_{\rm RP}(\cdot|X^{(n)}), \,\mathcal{N}\big(f(\wh\theta(X^{(n)})),\, n^{-1} \nabla f(\theta^*)^T  {\m H}_{\theta^*}^{\dagger} \Delta_{\theta^*} {\m H}_{\theta^*}^{\dagger} \nabla f(\theta^*)\big)\Big) \leq C\, \frac{(\log n)^{\frac{2}{\beta_1}+1}}{n^{\frac{\beta_2}{2}}}. 
    \end{aligned}
\end{equation*}
\end{enumerate}

\end{theorem}

A key observation is that the variance in the normal approximation of the posterior of $f(\theta)$ coincides with the asymptotic variance of the sampling distribution of $f(\wh\theta(X^{(n)}))$ established in Theorem~\ref{th:classicbvm}. Consequently, credible intervals for $f(\theta)$ obtained from the quantiles of posterior samples from $f_{\#}\Pi_{\rm RP}(\cdot|X^{(n)})$ achieve correct frequentist coverage without requiring the correct model specification assumption (i.e., Assumption~5). Moreover, when we project the posterior onto the tangent space, it is approximately normal with a ``sandwich'' covariance matrix. This makes it natural to construct credible regions for parameters constrained to the manifold: we form a Wald-type region in the tangent space and then map it back to the manifold. Since the target parameter $\theta^*$ and its tangent space $T_{\theta^*}\m M$ are unknown, we instead work around the point estimator $\wh\theta_p=\wh\theta_p(X^{(n)})$. Specifically, define the local projection $\psi_{\wh\theta_p}(\theta)={\rm Proj}_{T_{\wh\theta_p}\m M}(\theta-\wh\theta_p)$. When $\wh\theta_p$ is sufficiently close to $\theta^*$, $\psi_{\wh\theta_p}$ is injective on $\mb B_{r_1}(\wh\theta_p)\cap\m M$ for some small enough radius $r_1>0$ and admits an inverse map $\phi_{\wh\theta_p}$ (c.f., Lemma~\ref{lemmasmootharoundtheta} in Appendix~\ref{proof:main}). Let ${\Sigma}_p={\rm Cov}_{\theta\sim \Pi_{\rm RP}(\cdot|X^{(n)})}\big(\psi_{\wh\theta_p}(\theta)\big)$. For a chosen level of significance $\alpha$, let $q_{\alpha}$ be the $\alpha$-upper quantile of $\psi_{\wh\theta_p}(\theta)^T{\Sigma}_p^\dagger\psi_{\wh\theta_p}(\theta) =(\theta-\wh\theta_p)^T {\Sigma}_p^{\dagger}(\theta-\wh\theta_p)$ under $\theta\sim \Pi_{\rm RP}(\cdot|X^{(n)})$. The Wald-type credible region with level $\alpha$ is then 
\begin{equation}\label{eqn:credible}
\begin{aligned}
        \m E_n&=\{\theta\in \m M\cap B_{r_1}(\wh\theta_p)\,:\, (\theta-\wh\theta_p)^T  \Sigma_p^{\dagger} (\theta-\wh\theta_p)\leq q_{1-\alpha}\}.
\end{aligned}
\end{equation}
The next corollary shows that $\m E_n$ attains valid frequentist coverage.

\begin{corollary}[Validity of Wald-type Credible Region]\label{co1:nonsmooth}
Under the conditions in Theorem~\ref{th1}, there exists a small enough positive constant $r_1$ so that  for any level $\alpha\in (0,1)$, there exists a constant $C$ satisfying
\begin{equation*}
 \big|\mathbb P(\theta^*\in \m E_n)-(1-\alpha)\big|\leq C\,  \frac{(\log n)^{\frac{2}{\beta_1}+1}}{n^{\frac{\beta_2}{2}}}.
\end{equation*}  
\end{corollary}
Next, we return to the likelihood-based setting. A key implication of Theorem~\ref{th1} is that, under correct model specification, the Bayesian RPETEL posterior shares the same asymptotic distribution as the classical Bayesian posterior supported on the manifold. This result is formalized below.
\begin{corollary}[Bayesian RPETEL under Correct Model Specification]\label{co:wellspecified}
 Suppose Assumptions 1,2, and 5 hold, let $\ell(X,\theta)=\log p(X|\theta)$, and take $g(X,\theta)={\rm grad}_{\theta}\ell(X,\theta)$.  Then there exists a set $\m A\subset \m X^{n}$ with $\mathbb P(X^{(n)}\in \m A)\geq 1-n^{-1}$, and positive constants $c_1, c_2$ such that if $c_1\log n\leq \alpha_n\leq c_2 \sqrt{n}$, then for any function $f:\mb R^D\to f$ satisfies the condition stated in Theorem~\ref{th:classicbvm}, and any $X^{(n)}\in \m A$, we have
 \begin{equation*} 
    \begin{aligned}
    % &\qquad\Pi^{(n)}_{\m M}\Big(\|\theta-\theta^*\|\geq C\sqrt{\frac{\log n}{n}}\,\Big)\leq \frac{1}{n},\text{ and }\\
         &{\rm TV}\Big(f_{\#}\Pi_{\rm RP}(\cdot|X^{(n)}), \,\mathcal{N}\big(f(\wh\theta(X^{(n)})),\, n^{-1} \nabla f(\theta^*)^T (P_{\theta^*}I_{\theta^*}P_{\theta^*})^\dagger \nabla f(\theta^*)\big)\Big) \leq C\, \frac{(\log n)^{\frac{2}{\beta_1}+1}}{\sqrt{n}}.
    \end{aligned}
    \end{equation*}  
\end{corollary}
Under the correct model specification in Corollary~\ref{co:wellspecified}, where $\theta^*=\theta^*_{\rm E}$ also satisfies $\mb{E}[\nabla \m R(\theta^*)]=0_D$, the moment condition based on the Riemannian gradient extracts the most ``relevant'' information contained in the full Euclidean gradient. This extraction does not reduce efficiency; rather, it transforms an overidentified problem—where the ambient gradient yields $D$ moment conditions with $D>d$—into an identified one, where the intrinsic dimension $d$ of the manifold matches the effective number of moment conditions. Indeed, with an orthonormal basis $W_{\theta}\in \mb R^{D\times d}$ of the tangent space $T_{\theta}\m M$, the condition $\mb{E}[{\rm grad}_{\theta}\ell(X,\theta)]=0_D$ is equivalent to $\mb{E}[W_{\theta}^T{\rm grad}_{\theta}\ell(X,\theta)]=0_d$. This identifiability provides additional benefits under model misspecification, where $\theta^*_{\rm E}\notin \m M$, preventing spurious overidentification and ensuring that the posterior remains concentrated near $\theta^*$. Overall, these results highlight a threefold robustness of our method: (1) the posterior is automatically calibrated to deliver valid uncertainty quantification without requiring a correctly specified likelihood; (2) even if the unconstrained (Euclidean) risk minimizer lies off the manifold or does not exist, the posterior center remains meaningful—it closely matches the risk minimizer over the manifold, and the posterior covariance aligns with the frequentist sampling covariance of this center; and (3) when both the likelihood model is correctly specified and the manifold assumption holds, the posterior is asymptotically equivalent to the standard Bayesian posterior with the same prior. Hence, there is no loss of efficiency in well-specified models. To conclude this section, we revisit Examples~1 and~2 from Section~\ref{sec:LBBI} to illustrate our method.

\medskip
 \noindent \textbf{Example 1 (revisited): Reduced-rank multi-response regression.}  Take the prior $\Pi=\Pi_{\rm M}$ and recall $f(\theta)=\theta_{11}-\theta_{12}$. When the likelihood is correctly specified  ($\Sigma=I_2$),  $f_{\#}\Pi_{\rm RP}(\cdot|(\wt X,\wt Y))$ approaches to $\m N(\wh s(\wh\theta_{11}+\wh\theta_{21}),\frac{1.1}{n})$, which matches the Gaussian limit of the standard Bayesian posterior $\Pi_{\rm M}(\cdot|X^{(n)})$ with prior $\Pi_{\rm M}$. When the likelihood is misspecified ($\Sigma=\big(\begin{smallmatrix}
  1 & 0.3\\
  0.3 & 1
\end{smallmatrix}\big)$),   $f_{\#}\Pi_{\rm RP}(\cdot|(\wt X,\wt Y))$ approaches to  $\m N(\wh s(\wh\theta_{11}+\wh\theta_{21}),\frac{0.824}{n})$, and recall that $\sqrt{n}(\wh s(\wh\theta_{11}+\wh\theta_{21})-f(\theta^*)) \xrightarrow{d}\m N(0,0.824)$. Thus, even with likelihood misspecification, the Bayesian RPETEL posterior still provides valid uncertainty quantification.

\medskip
\noindent \textbf{Example 2 (revisited): Mean direction of the von Mises-Fisher distribution.} Place a uniform prior $\Pi$ on $\m M=\mb S_1^{2}$ and consider $f(\theta)=\theta_1$. The (pushforward) posterior $f_{\#}\Pi_{\rm RP}(\cdot|X^{(n)})$ converges to $\m N(f({\ov X}/{\|\ov X\|}), n^{-1}\frac{2}{3\kappa^*A(\kappa^*)})$, and recall that $\sqrt{n}(f({\ov X}/{\|\ov X\|})-f(\mu^*))\xrightarrow{d}\m N(0, \frac{2}{3\kappa^*A(\kappa^*)})$. Hence, for any $\kappa^*$, the posterior provides valid uncertainty quantification. Moreover, when $\kappa^*=1$ so that $\theta^*\in \m M$, the Gaussian limits of $f_{\#}\Pi_{\rm RP}(\cdot|X^{(n)})$ and $f_{\#}\Pi(\cdot|X^{(n)})$ coincide, where $\Pi(\cdot|X^{(n)})$ denotes the classical Bayesian posterior with prior $\Pi$.

\section{Posterior Sampling on Riemannian Submanifold}\label{sec:Sampling}
% The posterior $\Pi^{(n)}$ involves a normalization constant expressible as a multi-dimensional integral $\int \pi(\theta) \prod_{i=1}^n p(X_i|\theta)\,d\mu_{\mathcal{M}}(\theta)$, taken with respect to the volume measure of the manifold. Analytically computing this integral is typically infeasible, and its numerical approximation often requires sampling from the volume measure $\mu_{\mathcal{M}}$, a process that can become computationally demanding as the intrinsic or ambient dimension of $\mathcal{M}$ increases. On the the other hand, 
Given that the (pseudo-)posterior has an explicit form up to a normalization constant, we can employ a Markov chain Monte Carlo (MCMC) algorithm for sampling. MCMC transforms the integration problem of computing the normalization constant into a sampling task, thereby circumventing the need to evaluate a high-dimensional integral directly. The main difficulty in implementing MCMC algorithms arises from the fact that the posterior is defined on the manifold $\mathcal{M}$ rather than in a Euclidean space. While several MCMC methods have been developed for sampling from distributions on specific manifolds, such as the sphere~\citep{10.1214/aos/1176342874,doi:10.1080/03610919408813161} and the Stiefel manifold~\citep{jauch2021monte,hoff2009simulation}, these techniques are highly specialized to the particular manifold structures they target. In addition, Hamiltonian Monte Carlo (HMC) algorithms have been adapted for more general manifold sampling. For instance,~\cite{pmlr-v22-brubaker12} introduced a constrained version of HMC for sampling from solution manifolds $\m M_{\bold{q}}=\{\theta\in\mb R^D\,:\,\bold{q}(\theta)=0\}$. However, their method is closely tied to the constraint functions $\bold{q}(\theta)=0$ and does not easily extend beyond such manifolds. Similarly,~\cite{Byrne_2013} explored HMC on general Riemannian submanifolds, but this approach requires the computationally intensive task of evaluating geodesic flows on the submanifold.
~\cite{JMLR:v26:24-0829}  derived error bounds for sampling and estimation using an intrinsically defined Langevin diffusion on a compact Riemannian submanifolds. 
However, both Langevin and HMC-based methods require computing the Riemannian gradient of the log-density, which becomes difficult  when the density is not smooth. Beyond HMC and Langevin methods,~\cite{https://doi.org/10.48550/arxiv.1702.08446} proposed a simpler algorithm for sampling on solution manifolds, where a random vector in the tangent space of the current state is proposed and then projected back onto the manifold using the constraint functions $\bold{q}(\cdot)$ before applying an acceptance–rejection step. Inspired by this idea, we develop a Riemannian Random-Walk Metropolis (RRWM) algorithm that generalizes~\cite{https://doi.org/10.48550/arxiv.1702.08446} for sampling from posteriors supported on general Riemannian submanifolds. In this section, we first introduce the RRWM algorithm and then provide a mixing time analysis for sampling from the Bayesian RPETEL posterior.

  % A special choice of the retraction is the inverse of the projection map ${\rm Proj}_{T_{\theta}\m M}(\cdot-\theta)$ considered in Section~\ref{sec:LBBI} and~\ref{sec:LFBBI}.
  \subsection{Riemannian random-walk Metropolis (RRWM) algorithm}
The key idea of the RRWM algorithm is that, starting from the current state $\theta\in\m M$, we generate a proposal increment $v$ by first drawing an ambient vector from $\m N(0_D,\Sigma)$ and then projecting it onto the tangent space $T_{\theta}\m M$. We then map $v$ back to the manifold using a map $\wt\phi_{\theta}:T_{\theta}\m M\to\m M$, setting the candidate $y=\wt\phi_{\theta}(v)$. The candidate is then accepted or rejected using the standard Metropolis–Hastings acceptance probability. A key design issue is that both the covariance matrix $\Sigma$ and the mapping $\wt\phi_{\theta}$ influence how far the proposal $y$ tends to move from the current state $\theta$. If $\wt\phi_{\theta}$ rescales small tangent steps differently across locations, $\Sigma$ no longer controls the proposal step size consistently. For example, suppose $T_{\theta_1}\m M=T_{\theta_2}\m M$, while $\wt\phi_{\theta_1}(v)\approx v+\theta_1$ and $\wt\phi_{\theta_2}(v)\approx 2v+\theta_2$ for $\|v\|\approx0$. Then, the same tangent increment $v$ produces a move of size $\|v\|$ near $\theta_1$ but $2\|v\|$ near $\theta_2$, leading to location-dependent proposal magnitudes. To eliminate this confounding effect, we require $\wt\phi_{\theta}$ to be a retraction (see Appendix~\ref{App:manifold} for a detailed definition), which guarantees first-order equivalence between $\wt\phi_{\theta}(v)$ and $\theta+v$ when $\|v\|$ is small. This ensures that acceptance-rate tuning via $\Sigma$ is both stable and interpretable. A retraction can be constructed from any local parameterization $\xi_{\theta}(\cdot)$ as 
$$\wt\phi_{\theta}(v)=\xi_{\theta}\Big(\big(W_{\theta}^T J_{\xi_{\theta}}(\xi^{-1}_{\theta}(\theta))\big)^{-1}W_{\theta}^T v+\xi^{-1}_{\theta}(\theta)\Big),$$ 
where $W_{\theta}$ is a $D\times d$ matrix whose columns form an orthonormal basis for $T_{\theta}\m M$, and $\xi^{-1}_{\theta}$ is the inverse of $\xi_{\theta}$.

To describe the RRWM algorithm more precisely, we assume that for any $\theta\in\m M$, there exists a retraction $\wt\phi_{\theta}$ that is injective on an open neighborhood $\wt V_{\theta}$ of $0_D$ in $T_{\theta}\m M$. Let $\wt U_{\theta}=\wt\phi_{\theta}(\wt V_{\theta})$ and define the local inverse $\wt\psi_{\theta}:\wt U_{\theta}\to\wt V_{\theta}$. The set $\wt V_{\theta}$ serves as a ``safety'' region: proposals $y=\wt\phi_{\theta}(v)$ with $v\notin\wt V_{\theta}$ are rejected to prevent the algorithm from stepping outside the domain where $(\wt\psi_{\theta},\wt\phi_{\theta})$ are mutual inverses. The retraction map $\wt\phi_{\theta}$ is used to define proposal states in RRWM, while its inverse $\wt\psi_{\theta}$ plays a crucial role in computing the acceptance probability of those proposals. Given a target density $\mu^*(\theta)$ with respect to the volume measure $\mu_{\m M}$ on the submanifold $\m M$, the RRWM algorithm also involves a step size $\wt h>0$, a symmetric positive definite preconditioning (proposal) covariance matrix $\wt I\in\mb R^{D\times D}$, and an initial distribution $\mu_0$ on $\m M$.
The RRWM algorithm then generates the sequence $\{\theta^k\}_{k\ge 0}$ iteratively as follows: for $k=0,1,2,\ldots$,
 \begin{enumerate}
     \item  \textbf{(Initialization)} If $k=0$, sample $\theta^0$ from initial distribution $\mu_0$;
     \item \textbf{(Proposal)} If $k\geq 1$,
     \begin{enumerate}
         \item {(Generate random vector in tangent space)} sample $\wt v$ from $\mathcal{N}(0,{2\wt h}\wt I)$ and let $v={\rm Proj}_{T_{\theta^{k-1}}\m M}(\wt v)$;
         \item {(Reject proposal if $v$ escape from $\wt V_{\theta^{k-1}}$)} if $ v\notin \wt V_{\theta^{k-1}}$, then set $\theta^{k}= \theta^{k-1}$ and terminate the current iteration;
         \item {(Map back to manifold)} set $y=\wt{\phi}_{\theta^{k-1}}(v)$;
     \end{enumerate}
     \item  \textbf{(Metropolis-Hasting rejection/correction)} 
     \begin{enumerate}
         \item {(Reject proposal if $\theta^{k-1}$ escapes from $\wt U_y$)} if $\theta^{k-1}\notin \wt U_{y}$, then set $\theta^{k}=\theta^{k-1}$ and terminate the current iteration;
         \item (Set acceptance probability) 
let $v'=\wt\psi_y(\theta^{k-1})$, set acceptance probability $A(\theta^{k-1},y)=1\wedge \alpha(\theta^{k-1},y)$ with acceptance ratio statistic:
         \begin{equation*}
            \begin{aligned}
             \alpha(\theta^{k-1},y)&=\frac{\mu^\ast(y)\cdot\exp\big(-v'^T(P_y\wt IP_y)^{\dagger}v'/(4\wt h)\big)}{\mu^\ast(\theta^{k-1})\cdot\exp\big(-v^T(P_{\theta^{k-1}}\wt IP_{\theta^{k-1}})^{\dagger}v/(4\wt h))\big)}\\
    &\qquad\cdot \frac{\big(\big|(\m D{\wt\phi}_y(v')[P_{y}])^T\m D{\wt\phi}_y(v')[P_{y}]\big|_{+}\big)^{-\frac{1}{2}}}{\big(\big|(\m D{\wt{\phi}_{\theta^{k-1}}}(v)[P_{\theta^{k-1}}])^T\m D{\wt{\phi}_{\theta^{k-1}}}(v)[P_{\theta^{k-1}}]\big|_{+}\big)^{-\frac{1}{2}}},
                  \end{aligned}
         \end{equation*}
    where recall that $|\cdot|_{+}$ denotes the pseudo-determinant, $\m D \wt\phi_\theta(v)[\cdot]$ is the differential of $\wt\phi_{\theta}$ at $v$, and we denote $\m D{\wt \phi}_y(v')[ {\m V}]=\big[\m D{\wt \phi}_y(v')[\wt v_1],\m D{\wt \phi}_y(v')[ \wt v_2], \cdots,\m D{\wt\phi}_y(v')[ \wt v_D]\big]$ for $\m V=[\wt v_1,\wt v_2,\cdots,\wt v_D]$.
    \item (Accept/reject the proposal)  flip a coin and accept $y$ with probability $A(\theta^{k-1},y)$ and set $\theta^k = y$; otherwise, set $\theta^k=\theta^{k-1}$.
     \end{enumerate}
 \end{enumerate}

\begin{remark}
The Metropolis-adjusted Langevin algorithm (MALA) is another well-known class of MCMC algorithms that utilizes additional gradient information about the target density to improve its mixing time compared with the Random-Walk Metropolis algorithm~\citep{pmlr-v134-chewi21a}. Let $\mu^*(\theta)\propto\exp(-f(\theta))$ denote the target density with respect to the volume measure $\mu_{\m M}$, where $f(\cdot)$ is the potential function. Similar to the RRWM algorithm, we can develop a Riemannian Metropolis-adjusted Langevin algorithm (RMALA) that leverages Riemannian gradient information for smooth potential functions. In the proposal step of the RMALA algorithm, a random vector $\wt v$ is drawn from $\m N(-\wt h\wt I\cdot{\rm grad}\,f(\theta^{k-1}),2\wt h\wt I)$ and then projected onto the tangent space $T_{\theta^{k-1}}\m M$. The acceptance ratio statistic is given by the following expression:
\begin{equation*}
\begin{aligned}
&\alpha(\theta^{k-1},y)=\frac{\mu^\ast(y)\cdot\big(\big|(\m D{\wt\phi}_y(v')[P_{y}])^T\m D{\wt\phi}_y(v')[P_{y}]\big|_{+}\big)^{-\frac{1}{2}}}{\mu^\ast(\theta^{k-1})\cdot\big(\big|(\m D{{\wt\phi}_{\theta^{k-1}}}(v)[P_{\theta^{k-1}}])^T\m D{{\wt\phi}_{\theta^{k-1}}}(v)[P_{\theta^{k-1}}]\big|_{+}\big)^{-\frac{1}{2}}}\\
&\quad\cdot\frac{\exp\big(-(v'+\wt h\wt I{\rm grad} f(y))^T(P_y\wt I P_y)^{\dagger}(v'+\wt h\wt I{\rm grad} f(y))/(4\wt h)\big)}{\exp\big(-(v+\wt h\wt I{\rm grad}\, f(\theta^{k-1}))^T (P_{\theta^{k-1}}\wt I P_{\theta^{k-1}})^{\dagger}(v+\wt h\wt I{\rm grad}\, f(\theta^{k-1}))/(4\wt h)\big)}.
\end{aligned}
\end{equation*}
Further details about the RMALA algorithm are provided in Appendix~\ref{sec:RMALA}. In particular, RRWM is a zero-th order sampling algorithm (relying only on log-density evaluations), while RMALA is a first-order algorithm that additionally incorporates gradient information of the log-density.
\end{remark}

A crucial component of the RRWM algorithm is the choice of the retraction $\wt\phi_{\theta}(\cdot)$ for the Riemannian submanifold $\m M$. However, in many problems, $\m M$ is not equipped with a ready-made retraction or explicit local parameterization. Instead, it is often defined implicitly through constraints, such as a solution set $\m M_{\bold{q}}=\{\theta\in\mb R^D\,:\,\bold{q}(\theta)=0\}$, or through a structural property, such as the fixed-rank matrix manifold ${\m M}_r=\{B\in\mb R^{p\times k}\,:\,{\rm rank}(B)=r\}$. Therefore, it is necessary to identify a suitable retraction for the specific manifold of interest to ensure that randomly generated samples from the tangent space can be correctly mapped back onto the manifold. One particularly convenient choice is the retraction considered in~\cite{https://doi.org/10.48550/arxiv.1702.08446}, which coincides with the inverse $\phi_{\theta}:V_{\theta}\to U_{\theta}$ of the projection map $\psi_{\theta}:U_{\theta}\to V_{\theta}$, where $\psi_{\theta}(y)={\rm Proj}_{T_{\theta}\m M}(y-\theta)$ as defined in Section~\ref{sec:LBBI}. This retraction $\phi_{\theta}$ has the advantage of eliminating the Jacobian factor in the acceptance ratio:
% \footnote{Conversely, one may start from defining the retraction map as $\wt\phi_\theta(v)={\rm Proj}_{\m M}(\theta+v)$.  However such a retraction may not have the advantage of eliminating the Jacobian factor.}
\begin{equation*}
\big|(\m D{\phi}_y(v')[P_{y}])^T\m D{\phi}_y(v')[P_{y}]\big|_{+}=\big|(\m D{{\phi}_{\theta^{k-1}}}(v)[P_{\theta^{k-1}}])^T\m D{{\phi}_{\theta^{k-1}}}(v)[P_{\theta^{k-1}}]\big|_{+},
\end{equation*}
 where $v'=\psi_y(\theta^{k-1})$, $v=\psi_{\theta^{k-1}}(y)$. As a result,   the acceptance ratio in the RRWM algorithm can be simplified to 
 \begin{equation*}
   \m A(\theta^{k-1},y)=1\wedge  \frac{\mu^\ast(y)\cdot\exp\big(-v'^T(P_y\wt IP_y)^{\dagger}v'/(4\wt h)\big)}{\mu^\ast(\theta^{k-1})\cdot\exp\big(-v^T(P_{\theta^{k-1}}\wt IP_{\theta^{k-1}})^{\dagger}v/(4\wt h))\big)}.
 \end{equation*}

According to~\cite{https://doi.org/10.48550/arxiv.1702.08446}, if the manifold is a solution manifold $\m M_{\bold q}$ for some smooth function $\bold q$, then $y=\phi_{\theta}(v)$ can be obtained by numerically solving the equation
\begin{equation*}
\begin{aligned}
& y=\theta+v+Q_{\theta}^Ta \quad \text{where } Q_{\theta}=\bold{J}_{\bold q}(\theta)\, \text{ and }\, \bold q(\theta+v+Q_{\theta}^Ta)=0_k,
\end{aligned}
\end{equation*}
for example using the Newton–Raphson algorithm, where $\bold{J}_{\bold q}(\theta)$ denotes the $k\times D$ Jacobian matrix of $\bold q$ at $\theta$, i.e., $\big[\bold{J}_{\bold q}(\theta)\big]_{ij}=\frac{\partial q_i(\theta)}{\partial \theta_j}$ for $i\in[k]$ and $j\in[D]$. However, this numerical scheme for computing $\phi_\theta(v)$ does not apply to more general manifolds where the function $\bold q(\cdot)$ does not exist or is difficult to obtain. To address this, we propose a more general numerical scheme for computing $\phi_\theta(v)$ by solving the following optimization problem: given a tangent vector $v\in T_{\theta}\m M$ with sufficiently small norm $\|v\|$, there exists $r>0$ such that $y=\phi_{\theta}(v)$ can be identified as the unique solution to
\begin{equation}\label{eqn:solveproj}
\underset{y\in \m M\cap B_r(\theta)}{\arg\min}\,\|{\rm Proj}_{T_{\theta}\m M}(y-\theta)-v\|^2.
\end{equation}
The optimization problem~\eqref{eqn:solveproj} can be solved using the Riemannian gradient descent method or Riemannian Newton’s method, initialized at $\theta$. Furthermore, the step “reject the proposal if $v\notin V_{\theta^{k-1}}$ or $\theta^{k-1}\notin U_y$” in the RRWM algorithm—used to prevent tangent vectors from escaping the domain of $\phi_{\theta^t}$ or $\phi_y$—can be implemented by rejecting proposals when the algorithms for computing $\phi_{\theta^{k-1}}(v)$ or $\phi_y(v')$ with $v'={\rm Proj}_{T_y\m M}(\theta^{k-1}-y)$ fail to converge. Detailed algorithms are provided in Appendix~\ref{App:detailAlgo}.

 \subsection{Mixing time analysis of RRWM algorithm for Bayesian RPETEL sampling}
It is straightforward to verify that the Markov chain associated with RRWM is time-reversible and has $\mu^*$ as the stationary distribution. Nevertheless, obtaining a quantitative bound on the convergence rate of the algorithm is crucial for guiding its practical design and implementation. Following a common practice in the literature~\citep{JMLR:v21:19-441,3240040402}, we analyze a $\zeta$-lazy version of the RRWM algorithm, where at each iteration a coin is flipped: with probability $1-\zeta$, the algorithm proceeds with the proposal and Metropolis–Hastings acceptance step, and with probability $\zeta$, the chain remains unchanged. We characterize the $\varepsilon$-mixing time in $\chi^2$ divergence for the Markov chain generated by this $\zeta$-lazy version, assuming an $M_0$-warm start $\mu_0$.\footnote{We say $\mu_0\in \ms \mathbb P(\m M)$ is an $M_0$-warm start with respect to the stationary distribution $\Pi_{\rm RP}^{(n)}\in \ms \mathbb P(\m M)$ if $\mu_0(E)\leq M_0\,\Pi_{\rm RP}^{(n)}(E)$ holds for all Borel sets $E\subset \m M$, and we refer to $M_0$ as the warming parameter.} The $\varepsilon$-mixing time is defined as the minimal number of steps required for the chain to reach $\varepsilon^2$-$\chi^2$ divergence from the stationary distribution:
\begin{equation*}
\tau_{\rm mix}(\varepsilon,\mu_0):=\inf\{k\in \mb N\,:\,\sqrt{\chi^2(\mu_k,\mu^*)}\leq \varepsilon\},
\end{equation*}
where $\mu_k$ denotes the probability distribution of the Markov chain after $k$ steps.

 Most existing analyses of MCMC mixing time are conducted in Euclidean settings and rely on restrictive assumptions such as smoothness or strong log-concavity of the target distribution~\citep{JMLR:v21:19-441,https://doi.org/10.48550/arxiv.2109.13055}. However, these assumptions are often violated by Bayesian posterior densities. \cite{10.2307/30243694} shows that, in large-sample regimes where the Bayesian posterior satisfies an asymptotic normality property, the Random-walk Metropolis (RWM) algorithm for sampling from the posterior restricted to a compact subset of $\mb R^D$ has an asymptotic $\varepsilon$-mixing time upper bound of $O(D^2\log(\frac{1}{\varepsilon}))$. This motivates the question of whether the Riemannian Random-walk Metropolis (RRWM) algorithm can achieve faster mixing when the parameter space is a lower-dimensional submanifold.

In the next theorem, we show that for the Bayesian RPETEL posterior $\Pi_{\rm RP}(\cdot|X^{(n)})$, in the large-sample regime, the $\varepsilon$-mixing time of the RRWM algorithm scales as $O\big((d+\log (\frac{1}{\varepsilon}))\cdot\log \frac{1}{\varepsilon}\big)$, given a warm start, an appropriate covariance matrix $\wt I$, and a step size $\wt h\asymp n^{-1}(d+\log (\frac{1}{\varepsilon}))^{-1}$. This behavior differs from that of the conventional RWM algorithm in the ambient space $\mb R^D$, where the mixing time grows at least linearly with the ambient dimension $D$~\citep{10.1214/aoap/1034625254}. The mixing time for sampling from $\Pi_{\rm RP}(\cdot|X^{(n)})$ using RRWM depends only on the intrinsic dimension $d$, demonstrating that incorporating manifold structure into Bayesian inference yields computational advantages in addition to statistical ones.

The main challenge in bounding the mixing time of the RRWM algorithm arises from the irregularity of the density of $\Pi_{\rm RP}(\cdot|X^{(n)})$ with respect to the volume measure on $\m M$: the density may be non-log-concave or even discontinuous. Nevertheless, as shown in Theorem~\ref{th1}, the Bayesian RPETEL posterior $\Pi_{\rm RP}(\cdot|X^{(n)})$ converges to a multivariate normal distribution after projection onto the tangent space $T_{\theta^*}\m M$. Leveraging this result, we obtain the following corollary, which bounds the mixing time of the RRWM algorithm by showing that its convergence behavior is comparable to that of sampling from the corresponding multivariate normal limit. For simplicity, we take the retraction $\wt\phi_{\theta}$ to be $\phi_{\theta}$, the local inverse of the projection map $\psi_{\theta}(y)={\rm Proj}_{T_{\theta}\m M}(y-\theta)$ defined in Section~\ref{sec:LBBI}.

%  \begin{theorem}\label{th:mixingclassic} 
%  Under Assumptions in Theorem~\ref{th:classicbvm}, suppose  there exist $n$-independent constants $\rho_2\geq \rho_1>0$ so that  for any unit vector $\eta\in T_{\theta^*}\m M$, $\rho_1\leq \eta^T\wt I^{\frac{1}{2}}(P_{\theta^*}I_{\theta^*}P_{\theta^*})\wt I^{\frac{1}{2}}\eta\leq \rho_2$. Then we have the following claim: for any positive constant $c$ and ${\varepsilon}\geq \frac{M_0}{n^{c}}$, there exist
%   $(n,d,D)$-independent constants $(c_0,C_1)$  so that for large enough $n$,  the $\zeta$-lazy version ($\zeta\in (0,\frac{1}{2}]$) of the RRWM algorithm for sampling from $\Pi_{\m M}^{(n)}$ with an $M_0$-warm start, $\psi_{\theta}(\cdot)={\rm Proj}_{T_{\theta}\m M}(\cdot-\theta)$, and step size parameter $\wt h=h/n$  where $h=c_0\,\rho_2^{-1}
%  \big(d+\log (\frac{M_0d\rho_2}{\varepsilon\rho_1} )\big)^{-1}$,  with probability at least $1-n^{-1}$,  has  an $\varepsilon$-mixing time in $\chi^2$ divergence bounded as
%  \begin{equation*}
%      \tau_{\rm mix}(\varepsilon,\mu_0)\leq \frac{C_1}{\zeta} \,\bigg\{\bigg[\kappa\cdot \Big(d+\log \big(\frac{M_0d\kappa}{\varepsilon} \big)\Big)\cdot \log \Big(\frac{\log M_0}{\varepsilon}\Big)\bigg]\vee\log \,(M_0)\bigg\}, \quad \text{where}\quad \kappa=\frac{\rho_2}{\rho_1}.
%  \end{equation*}
% \end{theorem}

 \begin{corollary}\label{th:mixingbpetel}%(Mixing time for sampling from Bayesian RPETEL posterior)
 %Consider a tolerance $\varepsilon\in (0,1)$, a covariance matrix $\wt I\in \mb R^{D\times D}$, a lazy-parameter $\zeta\in (0,\frac{1}{2}]$ and a warming parameter $M_0$. 
Assume the conditions of Theorem~\ref{th1} hold, and suppose there exist $n$-independent constants $\rho_2\geq \rho_1>0$ such that $\rho_1\leq \eta^T\wt I^{\frac{1}{2}}\m H_{\theta^*}\Delta_{\theta^*}^{\dagger}\m H_{\theta^*}\wt I^{\frac{1}{2}}\eta\leq \rho_2$ for every unit vector $\eta\in T_{\theta^*}\m M$. Then, for sufficiently large $n$, there exists a set $\m A\subset \m X^{n}$ satisfying $\mathbb P(X^{(n)}\in \m A)\geq 1-n^{-1}$ such that the following holds for every dataset $X^{(n)}\in \m A$:  
let $\mu_0$ be an $M_0$-warm start with respect to $\Pi_{\rm RP}(\cdot|X^{(n)})$. For any positive constant $c$ and accuracy level ${\varepsilon}\geq \frac{M_0}{n^{c}}$, there exist $(n,d,D)$-independent absolute constants $(c_0,C_1)$ such that the $\zeta$-lazy version ($\zeta\in(0,\frac{1}{2}]$) of the RRWM algorithm targeting $\Pi_{\rm RP}(\cdot|X^{(n)})$, using the retraction $\phi_{\theta}(\cdot)$ and step-size parameter $\wt h=h/n$ with $h=c_0\rho_2^{-1}\big(d+\log(\frac{M_0d\rho_2}{\varepsilon\rho_1})\big)^{-1}$, has an $\varepsilon$-mixing time in $\chi^2$ divergence bounded by
\begin{equation*}
\tau_{\rm mix}(\varepsilon,\mu_0)\leq \frac{C_1}{\zeta}\bigg\{\bigg[\kappa\cdot\Big(d+\log\Big(\frac{M_0d\kappa}{\varepsilon}\Big)\Big)\cdot\log\Big(\frac{\log M_0}{\varepsilon}\Big)\bigg]\vee\log(M_0)\bigg\}, 
\quad \text{where}\quad \kappa=\frac{\rho_2}{\rho_1}.
\end{equation*}

\end{corollary}

%  Therefore, the convergence of the Markov chain for sampling from Bayesian RPETEL posterior can be much faster  than from the Bayesian PETEL posterior defined in the ambient space $\mb R^D$.
 \begin{remark}\label{remark:I}
Corollary~\ref{th:mixingbpetel} shows that the mixing time of the RRWM algorithm for sampling from the Bayesian RPETEL posterior is linear in the intrinsic dimension $d$ and independent of the ambient dimension $D$, given a suitable pre-conditioning matrix $\wt I$ and initial distribution. This result also implies that RRWM achieves faster convergence when $\kappa\approx 1$, that is, when the pre-conditioning covariance matrix $\wt I\in\mb R^{D\times D}$ for the proposal state is chosen such that $P_{\theta^*}\wt I P_{\theta^*}\approx \m H_{\theta^*}^{\dagger}\Delta_{\theta^*}\m H_{\theta^*}^{\dagger}$, where $P_{\theta^*}$ denotes the projection matrix onto $T_{\theta^*}\m M$. We suggest two approaches for selecting an appropriate $\wt I$:  
\begin{enumerate}
\item Direct estimation of $\m H_{\theta^*}$ and $\Delta_{\theta^*}$:  
obtain a consistent estimator $\wh\theta$ of $\theta^*$ (for example, via empirical risk minimization), and estimate $\m H_{\theta^*}$ and $\Delta_{\theta^*}$ by replacing $\theta^*$ with $\wh\theta$ and using empirical averages in place of population expectations based on $X^{(n)}$. Then choose $\wt I$ as any symmetric positive definite $D\times D$ matrix satisfying $P_{\wh\theta}\wt I P_{\wh\theta}=\wh{\m H}^{\dagger}\wh\Delta\wh{\m H}^{\dagger}$, where $\wh{\m H}$ and $\wh\Delta$ denote the corresponding estimators of $\m H_{\theta^*}$ and $\Delta_{\theta^*}$.  

\item Estimation of $\m H_{\theta^*}^{\dagger}\Delta_{\theta^*}\m H_{\theta^*}^{\dagger}$ from posterior samples:  
when direct evaluation of the Riemannian Hessian of $\m R(\cdot)$ is infeasible, we can estimate $\m H_{\theta^*}^{\dagger}\Delta_{\theta^*}\m H_{\theta^*}^{\dagger}$ using the covariance of $\psi_{\wh\theta_p}{}_{\#}\Pi_{\rm RP}^{(n)}$. Initialize $\wt I=I_D$ and run the RRWM algorithm for a moderate number of iterations $K$. Let $\wh\theta_p$ be the mean of the resulting samples $\{\theta^k\}_{k=1}^K$, and denote the covariance matrix of $\{\psi_{\wh\theta_p}(\theta^k)\}_{k=1}^K$ by $\wh\Sigma_p$. Then set $\wt I$ such that $P_{\wh\theta_p}^T\wt I P_{\wh\theta_p}=n\cdot \wh\Sigma_p$.  
\end{enumerate}
 \end{remark}
 
More generally, a similar mixing time bound can be established for any target posterior $\mu^*$ that satisfies a manifold Bernstein–von Mises theorem, as stated in Theorem~\ref{th:mixing} of Appendix~\ref{sec:mixgeneral}, with the proof given in Appendix~\ref{Proofmix}. Our analysis builds upon and extends the mathematical techniques, particularly the conductance profile method, developed in~\cite{JMLR:v21:19-441,JMLR:v25:23-0783}. We generalize the mixing time analysis from the Euclidean to the non-Euclidean setting by showing that, under suitable regularity conditions, the conductance profile associated with $\mu^*$ can be lower bounded by that of $\mu^*$ restricted to a high-probability region $K_{\theta}$ around $\theta^*$, denoted $\mu^*|_{K_{\theta}}$, up to higher-order terms. The analysis then reduces to the Euclidean space $\mb R^d$ by establishing a one-to-one correspondence between local coordinates in $\mb R^d$ and points on $K_{\theta}$. Using the manifold Bernstein–von Mises theorem, we further perform a perturbation analysis that connects the Markov transition kernel of the local-coordinate representation of $\theta\sim\mu^*|_{K_{\theta}}$ with that of its Gaussian approximation. Figure~\ref{fig:simutoy1} presents the effective sample size (ESS) for a fixed number of iterations obtained by the RRWM algorithm when sampling from the manifold-supported Bayesian posterior $\Pi_{\rm M}$ and the Bayesian RPETEL posterior $\Pi_{\rm RP}$ in Example 1 (with $d=3<D=4$), as well as by the conventional RWM algorithm when sampling from the Euclidean posterior $\Pi_{\rm E}$. The RRWM ESS values range approximately from $940$ to $1100$, whereas the RWM ESS is about $680$, yielding a ratio close to $D/d$ and demonstrating the computational advantage of the manifold-supported posterior.

\section{Numerical Illustration}\label{sec:simulation}
  
In this section, we evaluate the frequentist operating characteristics of the Bayesian RPETEL method in three problems: (1) multiple-quantile modeling with common slopes; (2) spectral projector estimation; and (3) mean parameter inference for diffusion tensors.
 % We were unable to locate a general purpose competitor that provided uncertainty quantification across all these examples. We thus adapted the calibrated Gibbs (CG) posterior~\citep{bhattacharya2020gibbs,Syring_2018,jiang2008gibbs,CHERNOZHUKOV2003293} from the Euclidean to the manifold setting, and compared against it in all the examples. In particular, we consider the Gibbs posterior defined in Section~\ref{sec:LBBI} plus a tuning learning rate $\beta$
% to get $\Pi_G(\dd\theta|X^{(n)})=\frac{ \exp(-\beta \sum_{i=1}^n \ell(X_i,\theta))}{\int\exp(-\beta \sum_{i=1}^n \ell(X_i,\theta) \pi(\dd\theta)}$. The learning rate is calibrated in a data-driven way using the stochastic approximation strategy proposed in~\citep{Syring_2018} --- $\beta$ is iteratively adjusted until the estimated coverage probability via bootstrap is close to the
% nominal level. Specifically, the coverage probability is estimated by the proportion of times the $(1-\alpha)$ posterior credible region $\{\theta\in\m M\,: \,\pi_{G}(\theta\,|\, X^{(n),b})\geq c_{\alpha})\}$ based on bootstrapping data $X^{(n),b}$, $b\in[B]$ covers the empirical risk minimizer of $\mathcal R_n$ over the manifold $\m M$. 
Unless otherwise specified, for all scenarios considered below, we generate posterior samples $\{\theta^k\}_{k=1}^{K}$ from the Bayesian RPETEL posterior and competitor posteriors using the RRWM algorithm (see Algorithm~\ref{algorithmcom} in Appendix~\ref{App:detailAlgo}), with $K$ specified in each example. The regularization parameter $\alpha_n$ is set to $2\log n$ throughout. For a function $f:\m M\to\mb R$, we construct a $(1-\alpha)$ Bayesian credible interval for $f(\theta)$ using the $\frac{\alpha}{2}$ and $1-\frac{\alpha}{2}$ quantiles of the posterior samples $\{f(\theta^k)\}_{k=1}^{K}$. We also construct a $(1-\alpha)$ credible region for $\theta$ using equation~\eqref{eqn:credible}, where the quantile $q_{\alpha}$ is approximated from the corresponding $\alpha$-quantile of the posterior samples. Coverage probabilities for the resulting credible intervals or credible regions are estimated based on $1000$ simulation replicates. Since closed-form expressions for the population-level global risk minimizers are not available, we approximate them using empirical risk minimizers computed from a large sample of size $5\times10^5$ in all experiments.
% \textcolor{magenta}{(Rong: later on you say 100 replicates in a table caption. pls double check. Also, good to add the default value of $\alpha_n$ for RPETEL here used across all simulation setttings. EDIT: it seems you are using $\log n$ for some sims and $2 
% \log n$ for some others. Can we use a common value across all the sims? Seems like you can easily use $2 \log n$ throughout as you already use it for everything other than 4.1, and the sensitivity analysis suggests that $2 \log n$ works equally well for extrinsic/Frechet.)}
We consider example-specific competitor methods, with details provided in the corresponding sections. These competitors include Gibbs posterior, parametric Bayesian, and Bayesian ETEL approaches, thereby offering a comprehensive set of benchmarks for comparison.

 \subsection{Multiple quantile modeling with common slopes}\label{sec:multiplequantile}

The multiple linear quantile regression problem aims to estimate several quantile regression coefficients simultaneously. At the population level, it solves
\begin{equation}\label{quantile_nonconstraint}
\underset{u,\beta}{\arg\min}\; \sum_{j=1}^{J} \mathbb{E}\big[\rho_{\tau_j}(Y - u_j - \wt X^{T}\beta_j)\big],
\end{equation}
where $u=(u_1,\ldots,u_J)$ are intercepts, $\beta=(\beta_1,\ldots,\beta_J)$ are slope vectors, $(\tau_1,\ldots,\tau_J)$ are the quantile levels, and $\rho_{\tau}(t)=t\{\tau - \mathbf{1}(t\le 0)\}$ is the check loss. Estimating each quantile coefficient separately can be inefficient, so to borrow information across quantile levels, one may impose a common-slope constraint while allowing intercepts to vary~\cite{yang2012bayesian}, restricting $(u,\beta)$ to the hyperplane $\mathcal{M}=\{(u,\beta):\beta_1=\cdots=\beta_J\}$.

In this experiment, we consider quantile levels $(\tau_1,\tau_2,\tau_3)=(0.2,0.4,0.7)$ and simulate data as follows. We generate $n$ i.i.d. samples $\wt X_i\sim \mathcal{N}(0,1)$ and errors $\varepsilon_i\sim \mathrm{Laplace}(0,1)$ independently, and set $Y_i=\wt X_i+\varepsilon_i(1+\epsilon \wt X_i)$ for $i=1,\ldots,n$. Let $q_{\tau}$ denote the $\tau$-th quantile of $\mathrm{Laplace}(0,1)$. Then the population solution to \eqref{quantile_nonconstraint} is $u^{\ast}=(q_{0.2},q_{0.4},q_{0.7})$ and $\beta^{\ast}=1+\epsilon u^{\ast}$. We consider two cases: $\epsilon=0$, where the common-slope model is correctly specified (so $(u^{\ast},\beta^{\ast})\in\mathcal{M}$), and $\epsilon=0.1$, where the model is misspecified. We compare three posteriors:  
(1) BRPETEL: the Bayesian RPETEL posterior with prior $\Pi_{\mathrm{M}}$ uniform on $\{(u,\beta)\in\mathcal{M}:\; |u_j|<100,\; |\beta_j|<100\}$;  
(2) BPETEL$_{\mathrm{M}}$: the Bayesian PETEL posterior (full-gradient ETEL) with the same manifold prior $\Pi_{\mathrm{M}}$;  
(3) BPETEL$_{\mathrm{E}}$: the Bayesian PETEL posterior (full-gradient ETEL) with Euclidean prior $\Pi_{\mathrm{E}}$ uniform on $\{(u,\beta)\in\mathbb{R}^{6}:\; |u_j|<100,\; |\beta_j|<100\}$. Posterior summaries are computed using $K=15{,}000$ Metropolis–Hastings draws, with the posterior mean $(\hat u_p,\hat\beta_p)$ used as the point estimator. For posteriors with prior $\Pi_{\mathrm{M}}$, we use the RRWM algorithm with step size $\wt h=1/n$ and $\wt I=I_6$. For the Euclidean prior $\Pi_{\mathrm{E}}$, we use the RWM algorithm with proposal covariance $(1/n)I_6$.

We report the mean squared error $\mathrm{MSE}=\|(\hat u_p,\hat\beta_p)-(u^{\ast},\beta^{\ast})\|^2$ and sampling efficiency via the effective sample size (ESS), averaged across dimensions. We consider $n=500$ and $n=1000$, with each configuration replicated $1000$ times. Results are summarized in Table~\ref{Table:exp1}. Under correct specification ($\epsilon=0$), the manifold-supported posteriors (BRPETEL, BPETEL$_{\mathrm{M}}$) achieve lower MSE than the Euclidean BPETEL$_{\mathrm{E}}$, with BRPETEL and BPETEL$_{\mathrm{M}}$ having nearly identical MSE. Under misspecification ($\epsilon=0.1$), BRPETEL and BPETEL$_{\mathrm{M}}$ have larger MSE than BPETEL$_{\mathrm{E}}$ due to the bias induced by the common-slope restriction; however, BRPETEL outperforms BPETEL$_{\mathrm{M}}$, showing improved robustness. In terms of computation, the manifold-supported posteriors yield substantially higher ESS than BPETEL$_{\mathrm{E}}$, and BRPETEL additionally shows a slight ESS advantage over BPETEL$_{\mathrm{M}}$.

Table~\ref{Table:exp1.1} reports the sampling coverages of the $95\%$ credible intervals and their average lengths for $u_1$, $u_2$, $u_3$, and $\beta_1$ under the two manifold-supported posteriors (BRPETEL and BPETEL$_\mathrm{M}$), targeting the components of the risk minimizer on $\mathcal{M}$ (not $(u^{\ast},\beta^{\ast})$), based on $1000$ replications. Under correct specification ($\epsilon=0$), both methods achieve coverage close to $95\%$ with nearly identical interval lengths, with BRPETEL’s coverages slightly closer to the nominal level. Increasing $n$ improves coverage for both methods. Under misspecification ($\epsilon=0.1$), BRPETEL maintains coverages near $93$–$95\%$ for all coordinates, whereas BPETEL$_\mathrm{M}$ shows notable undercoverage for $\beta_1$ (e.g., $87.7\%$ when $n=1000$). Increasing $n$ improves the coverage of BRPETEL but has little effect on BPETEL$_\mathrm{M}$. Interval lengths are nearly identical across methods (especially for $n=1000$), indicating that BRPETEL’s superior coverage is due to better calibration rather than wider intervals.

  \begin{table}[]
      \centering
        \caption{MSE (reported as $\times 10^{-2}$) and effective sample size (ESS) for multiple quantile modeling with common slopes. Here MSE is defined as the average of  $\|(\wh u_p, \wh \beta_p)-(u^*,\beta^*)\|^2$ over $1000$ replications, and ESS is computed per parameter dimension of $\theta=(u,\beta)$, averaged across dimensions and then across the $1000$ replications. Results are shown for correctly specified ($\varepsilon=0$) and misspecified ($\varepsilon=0.1$) settings, and for sample sizes $n\in\{500,1000\}$.
   }
       \begin{tabular}{lc|ccc|ccc}
\hline
&& \multicolumn{3}{c|}{Correctly specified ($\varepsilon=0$)} & \multicolumn{3}{c}{Misspecified ($\varepsilon=0.1$)} \\
\cline{3-5}\cline{6-8}
&& BRPETEL &BPETEL$_{\rm M}$ &BPETEL$_{\rm E}$& BRPETEL &BPETEL$_{\rm M}$ &BPETEL$_{\rm E}$ \\
\hline
\multirow{2}{*}{$n=500$}&MSE & $1.144$& $1.140$ &$ 1.418$ & $1.657$ & $1.780$ & $1.437$ \\
&ESS&$695$& $673$ &$ 294$ & $671$ & $629$ & $282$\\
\multirow{2}{*}{$n=1000$}&MSE  & $0.581 $& $0.576$ & $0.724$ & $1.084$ & $1.191$ &$0.730$\\
&ESS&$722$& $709$ &$ 310$ & $702$ & $645$ & $302$\\
\hline
\end{tabular}

      \label{Table:exp1}
  \end{table}

  \begin{table}[]
         \caption{Coverage probabilities and average lengths of $95\%$ credible intervals for multiple-quantile modeling with common slopes, computed over $1000$ replications. Columns labeled $u_i$ and $\beta_1$ report, respectively, the sampling coverage (in $\%$) and average interval length for the $i$th component of $u$ and for $\beta_1$. Coverage targets the corresponding components of the risk minimizer on the manifold $\mathcal M$ (not $(u^\ast,\beta^\ast)$).}
      \centering
     {%
        \begin{tabular}{cc|cccc|cccc}
\hline
&& \multicolumn{4}{c|}{Correctly specified ($\varepsilon=0$)} & \multicolumn{4}{c}{Misspecified ($\varepsilon=0.1$)} \\
\cline{3-6}\cline{7-10}
&& $u_1$ &$u_2$  &$u_3$& $\beta_1$ & $u_1$ &$u_2$ &$u_3$& $\beta_1$ \\
\hline
\multirow{4}{*}{BRPETEL}&{Coverage ($n=500$)}& $94.5$& $94.1$ &$93.8$ & $93.0$ & $93.1$ & $94.0$&$ 94.2$ &$ 92.9$  \\
&Length ($n=500$)&$0.255$& $0.155$ &$0.193$ & $0.153$ & $0.254$ & $0.157$&$ 0.194$ &$ 0.153$ \\
&Coverage ($n=1000$) &$94.6$& $94.4$ &$ 94.8$ & $93.6$ & $94.2$ & $94.2$&$ 94.3$ &$ 93.9$ \\
&Length ($n=1000$) &$0.177$& $0.109$ &$ 0.136$ & $0.107$ & $0.177$ & $0.109$&$ 0.136$ &$ 0.107$ \\
\hline
\multirow{4}{*}{BPETEL$_{\rm M}$}&{Coverage ($n=500$)}& $94.0$& $93.6$ &$ 93.2$ & $92.3$ & $91.2$ & $92.8$&$ 93.5$ &$ 88.0$  \\
&Length ($n=500$)&$0.253$& $0.154$ &$ 0.192$ & $0.146$ & $0.252$ & $0.154$&$ 0.190$ &$ 0.144$ \\
&Coverage ($n=1000$) &$94.1$& $93.9$ &$ 94.1$ & $93.1$ & $91.9$ & $92.5$&$ 93.3$ &$ 87.7$ \\
&Length ($n=1000$) &$0.177$& $0.109$ &$ 0.136$ & $0.103$ & $0.176$ & $0.109$&$ 0.135$ &$ 0.102$ \\
\hline

\end{tabular}
}

\label{Table:exp1.1}
  \end{table}

\subsection{Spectral projectors of covariance matrices}
Let $X\in \mathbb{R}^p$ be a zero-mean random vector with covariance matrix $\Sigma^*$. Consider a direct sum of the eigenspaces associated with the first $k$ largest eigenvalues of $\Sigma^*$ and the corresponding projectors $\mathcal{P}_k(\Sigma^*)$, defined as $\mathcal{P}_k(\Sigma^*)=\sum_{i=1}^k u^*_i{u^*_i}^T$, where $(u^*_1,u^*_2,\cdots,u^*_k)$ are the eigenvectors associated with the first $k$ largest eigenvalues of $\Sigma^*$. Our goal is to quantify the uncertainty in recovering the true spectral projector $P^*_k=\mathcal{P}_k(\Sigma^*)$~\cite{10.1214/18-EJS1451}. We view $P^*_k$ as the minimizer of the risk function $\mathbb{E}[-{\rm tr}(P XX^T)]$  on the Grassmannian manifold $\mathcal{M}=\{P\in \mathbb{R}^{p\times p}\,|\,P^2=P,\, P^T=P,\, {\rm tr}(P)=k\}$. It is noteworthy that $\mathcal{M}$ can be regarded as a solution manifold in $\mathbb{R}^{p^2}$ by flattening matrices in the Grassmannian manifold into $p^2$-dimensional vectors.

We generate $n$ i.i.d.\ samples $X_i\sim \frac{1}{2}\mathcal{N}(0,\Sigma_0)+\frac{1}{2}{\rm Uniform}([-1,1]^3)$ where 
\[
\Sigma_0=\left[\begin{array}{ccc}
1 & 0.1 & 0.1 \\
0.1 & 1.2 & 0.1\\
0.1 & 0.1 & 0.3
\end{array}\right].
\]
Letting $\Sigma^*$ denote the population covariance matrix for the generated data—computed analytically under this model—we focus on inferring the projector $\mathcal{P}_2(\Sigma^*)$. The prior $\Pi$ has density proportional to $\mathbf{1}(\|\!P\!\|_{\mathrm F}^2<100)$ with respect to the volume measure on $\mathcal{M}$, where $\|\cdot\|_{\mathrm F}$ denotes the Frobenius norm.
In addition to our Bayesian RPETEL posterior, we consider two competitors.  
(i) A Gibbs posterior with learning rate $\beta$: 
\[
\Pi_G(\mathrm{d}P\mid X^{(n)}) \propto \exp\!\Big(-\beta\sum_{i=1}^n \ell(X_i,P)\Big)\,\Pi(\mathrm{d}P),\qquad 
\ell(X,P)=\operatorname{tr}(P XX^\top),
\]
where $\beta$ is calibrated via the stochastic approximation method of~\citep{Syring_2018} to match nominal coverage based on bootstrap estimation.\footnote{Coverage is estimated as the proportion of times the $(1-\alpha)$ credible region $\{P\in\mathcal{M}:\,\pi_G(P\mid X^{(n),b})\ge c_\alpha\}$ computed on bootstrap samples $X^{(n),b}$, $b\in[B]$, where $\pi_G(P\mid X^{(n),b})$ is the density function of $\Pi_G(\mathrm{d}P\mid X^{(n),b})$, contains the empirical risk minimizer on $\mathcal{M}$.} The selected value is approximately $0.95$, and we fix $\beta=0.95$ across replications.
(ii) A misspecified parametric Bayesian method assuming the data follow $\mathcal{N}(0,\Sigma)$, with an inverse-Wishart prior $IW(p+1,I_p)$ on $\Sigma$. Under this model, the posterior is $IW(n+d+1,I+\sum_{i=1}^n X_iX_i^T)$. We construct credible intervals by taking spectral projectors of posterior draws of $\Sigma$; we refer to this method as MVN-IW.

The coverage probabilities of credible intervals for the diagonal elements of $\mathcal{P}_2(\Sigma^*)$ and coverage of credible regions for different sample sizes are presented in Table~\ref{Table:PCA}. When $n=500$, Bayesian RPETEL exhibits slightly better performance than both the Gibbs posterior and MVN-IW. For example, for the target $95\%$ coverage, the Gibbs posterior deviates by $2.7\%$ for $\theta_5$, MVN-IW by $2.0\%$, whereas Bayesian RPETEL deviates by at most $1.8\%$. However, all three methods tend to underestimate the precision of $\theta_5$ when $n=500$. With $n=1000$, both Bayesian RPETEL and MVN-IW show substantial improvement, while the Gibbs posterior does not improve and continues to underestimate the precision of $\theta_5$. Overall, Bayesian RPETEL outperforms MVN-IW even at larger sample sizes.

\begin{table}
          \caption{ \label{Table:PCA} \em Coverage probabilities for inference on the spectral projectors of the covariance matrix: ``$\theta_k$'' represents the credible interval  for the $k$th dimension of the parameter $\theta$ obtained by flattening the matrix $P\in \mb R^{d\times d}$ into a vector (by column), and ``Credible region'' refers to the credible region given in~\eqref{eqn:credible}.}    
\centering
\begin{tabular}{cccccccc}
 \hline
&  &Target coverage ($\%$)&$\theta_1$&$\theta_5$&$\theta_9$&Credible region \\
   \hline
  \multirow{4}{*}{$n=500$}&  \multirow{2}{*}{Bayesian RPETEL}&95&94.8&96.8&94.0&94.1\\
 &&90&89.2&91.5&88.4&88.8\\
    
  & \multirow{2}{*}{Gibbs posterior}& 95&95.0&97.7&95.4&94.1\\
 &&90&88.7&93.9&90.9&89.1\\

  & \multirow{2}{*}{MVN-IW}& 95&97.7&97.0&96.1&96.1\\
 &&90&92.4&93.7&91.9&90.9\\
   \hline
   \hline
   \multirow{4}{*}{$n=1000$}&  \multirow{2}{*}{Bayesian RPETEL}&95&95.0&95.9&95.0&94.1\\
 &&90&90.1&90.9&89.1&88.9\\
 
  & \multirow{2}{*}{Gibbs posterior}& 95&96.5&97.9&95.6&95.7\\
 &&90&90.4&94.3&91.0&90.6\\
 
  & \multirow{2}{*}{MVN-IW}& 95&95.9&95.6&96.3&96.4\\
 &&90&91.8&90.7&91.1&91.7\\
   \hline
     \end{tabular}

\end{table}

\subsection{Mean parameter inference for diffusion tensors}\label{Sec:DTI}
The superiority of the Bayesian RPETEL posterior over the competitors in the simulation examples supports our theoretical conclusion that the Bayesian RPETEL posterior can provide valid uncertainty quantification under certain regularity conditions. To examine how well these conditions hold in practice, we analyze an areal diffusion tensor imaging (DTI) dataset and compare Bayesian RPETEL with a parametric competitor. We employ a ``subsampling’’ strategy to empirically assess inferential validity: we repeatedly sample $n=    1000$ observations with replacement from the original dataset, perform this procedure $1000$ times, and compute the coverage probabilities of the $95\%$ Bayesian credible intervals produced by the Bayesian RPETEL posterior and the competitor posterior for covering the empirical risk minimizer $\hat\theta$ computed from the full dataset.

Diffusion tensor data~\citep{LEE2017152,https://doi.org/10.1002/cjs.11601} consist of symmetric, positive definite $3\times 3$ covariance matrices at each voxel, representing the anisotropic diffusion of water molecules. Such data arise not only in diffusion imaging but also in other contexts, including brain connectivity~\citep{ingalhalikar2014sex} and portfolio covariance estimation~\citep{fan2016overview}. Our goal is to infer the mean parameter $\theta$ of a random covariance matrix $X\sim \mu_\theta$ supported on $\mathbb{S}_+^3=\{\Sigma\in\mathbb{R}^{3\times3}:\,\Sigma \text{ is positive semidefinite}\}$. Two common notions of the ``mean’’ of a distribution $\mu$ on $\mathbb{S}_+^3$ are: (1) the extrinsic mean, defined as the minimizer over $\theta\in\mathbb{S}_+^3$ of $\int\|\theta-X\|_2^2\,\mu(\dd X)$, and (2) the Bures–Wasserstein (BW) barycenter~\citep{10.1214/20-AAP1618}, which is also the Fr\'echet mean:
\begin{equation}\label{eqn:bary}
\theta^\ast \in \arg\min_{\theta\in\mathbb{S}_+^3} \int_{\mathbb{S}_+^3} d_{\mathrm{BW}}^2(\theta,X)\,\mu(\dd X),
\end{equation}
where for $Q,S\in\mathbb{S}_+^3$ the BW distance~\citep{BHATIA2019165} is
\[
d_{\mathrm{BW}}^2(Q,S)=\operatorname{tr}(Q)+\operatorname{tr}(S)-2\,\operatorname{tr}\big(Q^{1/2}SQ^{1/2}\big)^{1/2},
\]
which coincides with the $2$-Wasserstein distance between $\mathcal{N}(0,Q)$ and $\mathcal{N}(0,S)$.

We conduct a real data analysis using the diffusion tensor dataset from a $148\times 190\times 160$ scan of Gordon Kindlmann’s brain.\footnote{Brain dataset courtesy of Gordon Kindlmann at the Scientific Computing and Imaging Institute, University of Utah, and Andrew Alexander, W. M. Keck Laboratory for Functional Brain Imaging and Behavior, University of Wisconsin–Madison.} Our focus is a spatial patch $\mathcal{D}$ of neighboring voxels around voxel $(75,95,80)$ with neighborhood size $20$: specifically, we extract the voxels $\{(x,y,z):\, x\in[55,95],\, y\in[75,115],\, z\in[60,100]\}$ and study the extrinsic/Fr\'echet mean of the diffusion tensors within this region. This large neighborhood provides a sufficiently large effective population to assess inferential validity under subsampling. We investigate three scientifically meaningful functionals of the mean diffusion tensor: fractional anisotropy (FA),\footnote{For eigenvalues $(\lambda_1,\lambda_2,\lambda_3)$, the FA is ${\rm FA}= \sqrt{\frac{3}{2}} \frac{\sqrt{\left(\lambda_1-\hat{\lambda}\right)^2+\left(\lambda_2-\hat{\lambda}\right)^2+\left(\lambda_3-\hat{\lambda}\right)^2}}{\sqrt{\lambda_1^2+\lambda_2^2+\lambda_3^2}}$, where $\hat{\lambda}$ is their arithmetic mean.} trace, and maximum eigenvalue. Alongside Bayesian RPETEL, we consider a parametric Bayesian approach that assumes each $3\times 3$ diffusion tensor follows a Wishart distribution ${\rm WI}_3(\theta,v)$ with (extrinsic) mean $\theta\in\mathbb{S}_+^3$ and degrees of freedom $v>4$. The extrinsic mean $\theta$ is given an inverse-Wishart prior with mean $I_3$ and degrees of freedom $5$, and $v$ is assigned a uniform prior over $[5,50]$. Posterior samples of the BW barycenter $\theta'$ are obtained as a functional of posterior draws of $(\theta,v)$, where $\theta'$ solves
\[
\theta'=\arg\min_{\theta'\in\mathbb{S}_+^3}\mathbb{E}_{{\rm WI}_3(\theta,v)}\!\left[d_{\rm BW}^2(\theta',X)\right],
\]
assuming uniqueness. We refer to this approach as \textit{Wishart Modeling}.

% %\footnote{In Wishart Modeling, the posterior samples for the Wasserstein barycenter $\theta'$ are obtained as a function of the  posterior samples for the (extrinsic) mean $\theta$ and the degree of freedom $v$, where $\theta'={\arg\min}_{\theta'\in \mb S_{+}^p} \mb{E}_{ {\rm WI}_p(\theta,v)} \big[d^2_{\rm BW}(\theta', X)\big]$ assuming a unique minimizer exists.} 

\begin{figure}
     \centering
     \includegraphics[width=0.6\textwidth]{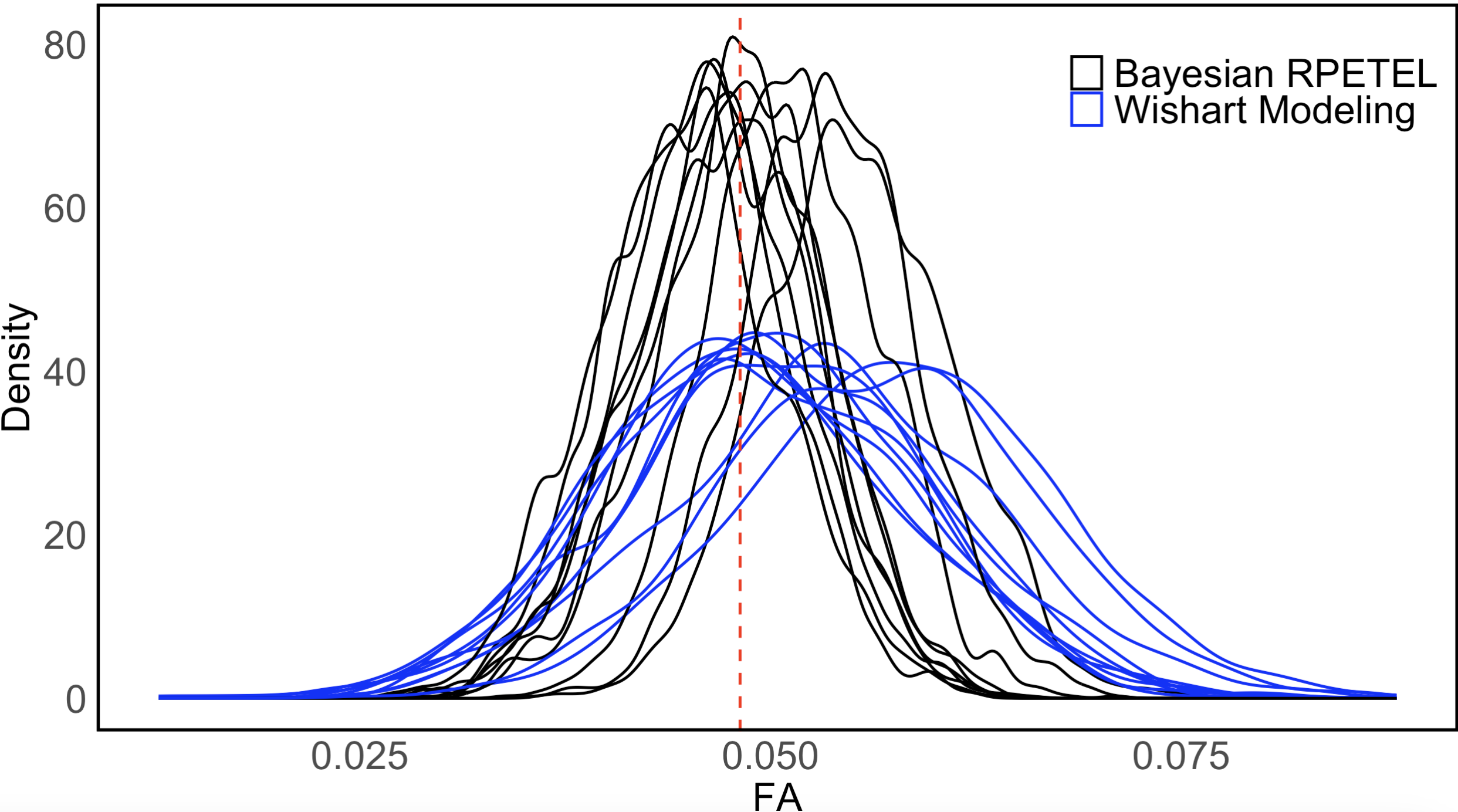}
     \caption{\em Density plots of the fractional anisotropy (FA) computed from the posterior samples obtained using Bayesian RPETEL and Wishart Modeling in the extrinsic mean example.
      The plot includes overlays from ten runs of experiments. The red vertical line indicates the FA of the empirical risk minimizer $\wh \theta$ computed from the original dataset. }
     \label{fig:FA}
 \end{figure}

The coverage probabilities computed by subsampling are given in Table~\ref{Table:DTI}.  We can observe that the Bayesian RPETEL posterior performs significantly better than the Wishart Modeling, especially for the inference of FA and Trace. For instance, Figure~\ref{fig:FA} displays the density plot of the FA for the posterior samples obtained from Bayesian RPETEL and Wishart Modeling in the extrinsic mean example. Notably, the posterior distribution from Wishart Modeling exhibits a much heavier tail compared to Bayesian RPETEL. As a result, the Wishart Modeling approach noticeably underestimates the precision of the inference for FA.

\begin{table}
    \caption{ \label{Table:DTI}\em Coverage probabilities ($\%$) for inference of the fractional anisotropy (FA), trace and maximum eigenvalues of the extrinsic mean and Fr\'{e}chet mean of the diffusion tensors (the target coverage is given by $95\%$).}
\centering
\begin{tabular}{ccccc}
    \hline
% &&\multicolumn{4}{c}{Extrinsic Mean}&\multicolumn{4}{c}{Fr\'{e}chet Mean}\\
%     \hline
 &&FA&Trace&Maximum eigenvalue\\
   \hline
 \multirow{2}{*}{Extrinsic Mean} &Bayesian RPETEL&93.7&94.9&94.5\\
 &Wishart Modeling&99.4&59.2&81.6\\
   \hline
   \hline
 \multirow{2}{*}{Fr\'{e}chet Mean} &Bayesian RPETEL&94.1&93.8&93.6\\
 &Wishart Modeling&>99.9&73.1&87.2\\
 \hline
     \end{tabular}

\end{table}

\section{Conclusion and Discussion}
In this paper, we have developed a general Bayesian inference framework for parameters constrained to a Riemannian submanifold, highlighting the benefits of incorporating manifold structure into Bayesian analysis. To further address potential model misspecification, we proposed a novel Bayesian method—Bayesian Riemannian Penalized Exponentially Tilted Empirical Likelihood (RPETEL)—for robust statistical inference with manifold-valued parameters. This approach avoids the need for a correctly specified likelihood and provides accurate uncertainty quantification with provable asymptotic guarantees. Numerical studies demonstrate that our method outperforms state-of-the-art alternatives in terms of inferential precision.

Our theoretical analysis in this work focuses on the mixing time of the Riemannian random-walk Metropolis algorithm. Future work may extend this analysis to the Riemannian Metropolis-adjusted Langevin and Hamiltonian Monte Carlo algorithms, which leverage higher-order information from the likelihood or loss function. Another promising direction is to extend the current methodology to incorporate additional inequality constraints on the parameters. Finally, although our development centers on parameter spaces with trivial or explicit embeddings in an ambient Euclidean space, it would be of substantial interest to explore more general Riemannian manifolds without explicit Euclidean embeddings, as well as settings where the underlying manifold structure is itself unknown, such as in text, network, and graph data.
\bibliography{ref}

@book{ghosh2003bayesian,
  title={Bayesian nonparametrics},
  author={Ghosh, Jayanta K and Ramamoorthi, RV},
  year={2003},
  publisher={Springer}
}

@article{bhattacharya2003large,
  title={Large sample theory of intrinsic and extrinsic sample means on manifolds},
  author={Bhattacharya, Rabi and Patrangenaru, Vic},
  journal={The Annals of Statistics},
  volume={31},
  number={1},
  pages={1--29},
  year={2003}
}

@article{kleijn2012bernstein,
  title={The Bernstein-Von-Mises theorem under misspecification},
  author={Kleijn, BJK and van der Vaart, AW},
  journal={Electronic Journal of Statistics},
  volume={6},
  pages={354--381},
  year={2012}
}

@article{izenman1975reduced,
  title={Reduced-rank regression for the multivariate linear model},
  author={Izenman, Alan Julian},
  journal={Journal of multivariate analysis},
  volume={5},
  number={2},
  pages={248--264},
  year={1975},
  publisher={Elsevier}
}

@article{tang2022bayesian,
  title={Bayesian inference for risk minimization via exponentially tilted empirical likelihood},
  author={Tang, Rong and Yang, Yun},
  journal={Journal of the Royal Statistical Society. Series B: Statistical Methodology},
  year={2022},
  publisher={Wiley-Blackwell}
}

@book{vershynin_2018, place={Cambridge}, series={Cambridge Series in Statistical and Probabilistic Mathematics}, title={High-Dimensional Probability: An Introduction with Applications in Data Science}, DOI={10.1017/9781108231596}, publisher={Cambridge University Press}, author={Vershynin, Roman}, year={2018}, collection={Cambridge Series in Statistical and Probabilistic Mathematics}}

@book{wainwright_2019, place={Cambridge}, series={Cambridge Series in Statistical and Probabilistic Mathematics}, title={High-Dimensional Statistics: A Non-Asymptotic Viewpoint}, DOI={10.1017/9781108627771}, publisher={Cambridge University Press}, author={Wainwright, Martin J.}, year={2019}, collection={Cambridge Series in Statistical and Probabilistic Mathematics}}

@article{JMLR:v21:19-441,
  author  = {Yuansi Chen and Raaz Dwivedi and Martin J. Wainwright and Bin Yu},
  title   = {Fast mixing of Metropolized Hamiltonian Monte Carlo: Benefits of multi-step gradients},
  journal = {Journal of Machine Learning Research},
  year    = {2020},
  volume  = {21},
  number  = {92},
  pages   = {1-72},
  url     = {http://jmlr.org/papers/v21/19-441.html}
}

@book{Eldering2013,
author="Eldering, Jaap",
title="Normally Hyperbolic Invariant Manifolds: The Noncompact Case",
year="2013",
publisher="Atlantis Press",
address="Paris",
}

@article{Rai__2019,
   title="{A multivariate Berry–Esseen theorem with explicit constants}",
   volume={25},
   ISSN={1350-7265},
   url={http://dx.doi.org/10.3150/18-BEJ1072},
   DOI={10.3150/18-bej1072},
   number={4A},
   journal={Bernoulli},
   publisher={Bernoulli Society for Mathematical Statistics and Probability},
   author={Raič, Martin},
   year={2019},
   month={Nov},
   pages={2824–2853}
}

@Book{boumal2022intromanifolds,
  title     = {An introduction to optimization on smooth manifolds},
  author    = {Boumal, Nicolas},
  publisher = {Cambridge University Press},
  year      = {2023},
  url       = {https://www.nicolasboumal.net/book},
  doi       = {10.1017/9781009166164}
}

@article{JMLR:v26:24-0829,
  author  = {Karthik Bharath and Alexander Lewis and Akash Sharma and Michael V. Tretyakov},
  title   = {Sampling and Estimation on Manifolds using the Langevin Diffusion},
  journal = {Journal of Machine Learning Research},
  year    = {2025},
  volume  = {26},
  number  = {71},
  pages   = {1--50},
  url     = {http://jmlr.org/papers/v26/24-0829.html}
}

@article{10.1214/18-EJS1451,
author = {Igor Silin and Vladimir Spokoiny},
title = {{Bayesian inference for spectral projectors of the covariance matrix}},
volume = {12},
journal = {Electronic Journal of Statistics},
number = {1},
publisher = {Institute of Mathematical Statistics and Bernoulli Society},
pages = {1948 -- 1987},
year = {2018},
doi = {10.1214/18-EJS1451},
URL = {https://doi.org/10.1214/18-EJS1451}
}

@article{https://doi.org/10.48550/arxiv.1702.08446,
author = {Zappa, Emilio and Holmes-Cerfon, Miranda and Goodman, Jonathan},
title = {Monte Carlo on Manifolds: Sampling Densities and Integrating Functions},
journal = {Communications on Pure and Applied Mathematics},
volume = {71},
number = {12},
pages = {2609-2647},
doi = {https://doi.org/10.1002/cpa.21783},
url = {https://onlinelibrary.wiley.com/doi/abs/10.1002/cpa.21783},
eprint = {https://onlinelibrary.wiley.com/doi/pdf/10.1002/cpa.21783},
year = {2018}
}

@article{3240040402,
author = {Lovász, L. and Simonovits, M.},
title = {Random walks in a convex body and an improved volume algorithm},
journal = {Random Structures \& Algorithms},
volume = {4},
number = {4},
pages = {359-412},
doi = {https://doi.org/10.1002/rsa.3240040402},
url = {https://onlinelibrary.wiley.com/doi/abs/10.1002/rsa.3240040402},
eprint = {https://onlinelibrary.wiley.com/doi/pdf/10.1002/rsa.3240040402},
year = {1993}
}

@article{10.1214/aos/1176342874,
author = {Christopher Bingham},
title = {{An antipodally symmetric distribution on the sphere}},
volume = {2},
journal = {The Annals of Statistics},
number = {6},
publisher = {Institute of Mathematical Statistics},
pages = {1201 -- 1225},
year = {1974},
doi = {10.1214/aos/1176342874},
URL = {https://doi.org/10.1214/aos/1176342874}
}

@article{Supplement2291,
author = {Rong Tang and Yun Yang},
title = {Supplement to ``Minimax Rate of Distribution Estimation on Unknown Submanifolds under Adversarial Losses''},
year={2023},
url={https://doi.org/10.1214/23-AOS2291SUPP}
}

@article{doi:10.1080/03610919408813161,
author = { Andrew T.A   Wood },
title = {Simulation of the von mises fisher distribution},
journal = {Communications in Statistics - Simulation and Computation},
volume = {23},
number = {1},
pages = {157-164},
year  = {1994},
publisher = {Taylor & Francis},
doi = {10.1080/03610919408813161},

URL = { 
        https://doi.org/10.1080/03610919408813161
    
},
eprint = { 
        https://doi.org/10.1080/03610919408813161
    
}

}

@InProceedings{pmlr-v22-brubaker12,
  title = 	 {A Family of MCMC Methods on Implicitly Defined Manifolds},
  author = 	 {Brubaker, Marcus and Salzmann, Mathieu and Urtasun, Raquel},
  booktitle = 	 {Proceedings of the Fifteenth International Conference on Artificial Intelligence and Statistics},
  pages = 	 {161--172},
  year = 	 {2012},
  volume = 	 {22},
  month = 	 {21--23 Apr},
  publisher =    {PMLR},
  url = 	 {https://proceedings.mlr.press/v22/brubaker12.html}
}

@article{Byrne_2013,
	doi = {10.1111/sjos.12036},
  
	url = {https://doi.org/10.1111%2Fsjos.12036},
  
	year = 2013,
	month = {sep},
  
	publisher = {Wiley},
  
	volume = {40},
  
	number = {4},
  
	pages = {825--845},
  
	author = {Simon Byrne and Mark Girolami},
  
	title = {Geodesic Monte Carlo on Embedded Manifolds},
  
	journal = {Scandinavian Journal of Statistics}
}

@article{10.1214/18-AOS1685,
author = {Eddie Aamari and Clément Levrard},
title = {{Nonasymptotic rates for manifold, tangent space and curvature estimation}},
volume = {47},
journal = {The Annals of Statistics},
number = {1},
publisher = {Institute of Mathematical Statistics},
pages = {177 -- 204},
year = {2019},
doi = {10.1214/18-AOS1685},
URL = {https://doi.org/10.1214/18-AOS1685}
}

@article{CHERNOZHUKOV2003293,
title = {{An MCMC approach to classical estimation}},
journal = "Journal of Econometrics",
volume = "115",
number = "2",
pages = "293 - 346",
year = "2003",
issn = "0304-4076",
doi = "https://doi.org/10.1016/S0304-4076(03)00100-3",
url = "http://www.sciencedirect.com/science/article/pii/S0304407603001003",
author = "Victor Chernozhukov and Han Hong",
}

@misc{https://doi.org/10.48550/arxiv.1710.09443,
  doi = {10.48550/ARXIV.1710.09443},
  
  url = {https://arxiv.org/abs/1710.09443},
  
  author = {Pourzanjani, Arya A and Jiang, Richard M and Mitchell, Brian and Atzberger, Paul J and Petzold, Linda R},
  
  title = {Bayesian Inference over the Stiefel Manifold via the Givens Representation},
  
  publisher = {arXiv},
  
  year = {2017},
  
  copyright = {arXiv.org perpetual, non-exclusive license}
}

@article{bhattacharya2020gibbs,
  title={Gibbs posterior inference on multivariate quantiles},
  author={Bhattacharya, Indrabati and Martin, Ryan},
  journal={arXiv preprint arXiv:2002.01052},
  year={2020}
}

@article{Syring_2018,
   title={Calibrating general posterior credible regions},
   volume={106},
   ISSN={1464-3510},
   url={http://dx.doi.org/10.1093/biomet/asy054},
   DOI={10.1093/biomet/asy054},
   number={2},
   journal={Biometrika},
   publisher={Oxford University Press (OUP)},
   author={Syring, Nicholas and Martin, Ryan},
   year={2018},
   month={Dec},
   pages={479–486}
}

@misc{https://doi.org/10.48550/arxiv.2109.13055,
  doi = {10.48550/ARXIV.2109.13055},
  
  url = {https://arxiv.org/abs/2109.13055},
  
  author = {Wu, Keru and Schmidler, Scott and Chen, Yuansi},
  
  title = {Minimax Mixing Time of the Metropolis-Adjusted Langevin Algorithm for Log-Concave Sampling},
  
  publisher = {arXiv},
  
  year = {2021},
  
  copyright = {Creative Commons Attribution 4.0 International}
}

@article{10.2307/30243694,
 ISSN = {00905364, 21688966},
 URL = {http://www.jstor.org/stable/30243694},
 author = {Alexandre Belloni and Victor Chernozhukov},
 journal = {The Annals of Statistics},
 number = {4},
 pages = {2011--2055},
 publisher = {Institute of Mathematical Statistics},
 title = {On the Computational Complexity of MCMC-Based Estimators in Large Samples},
 urldate = {2022-08-07},
 volume = {37},
 year = {2009}
}

@misc{https://doi.org/10.48550/arxiv.1306.5806,
  doi = {10.48550/ARXIV.1306.5806},
  
  url = {https://arxiv.org/abs/1306.5806},
  
  author = {Bhattacharya, Rabi and Lin, Lizhen},
  
  title = {Omnibus CLTs for Fréchet means and nonparametric inference on non-Euclidean spaces},
  
  publisher = {arXiv},
  
  year = {2013},
  
  copyright = {arXiv.org perpetual, non-exclusive license}
}

@article{10.1214/009053605000000093,
author = {Rabi Bhattacharya and Vic Patrangenaru},
title = {{Large sample theory of intrinsic and extrinsic sample means on manifolds—II}},
volume = {33},
journal = {The Annals of Statistics},
number = {3},
publisher = {Institute of Mathematical Statistics},
pages = {1225 -- 1259},
year = {2005},
doi = {10.1214/009053605000000093},
URL = {https://doi.org/10.1214/009053605000000093}
}

@article{10.1214/20-AAP1618,
author = {Alexey Kroshnin and Vladimir Spokoiny and Alexandra Suvorikova},
title = {{Statistical inference for Bures–Wasserstein barycenters}},
volume = {31},
journal = {The Annals of Applied Probability},
number = {3},
publisher = {Institute of Mathematical Statistics},
pages = {1264 -- 1298},
year = {2021},
doi = {10.1214/20-AAP1618},
URL = {https://doi.org/10.1214/20-AAP1618}
}

@article{BHATIA2019165,
title = {On the Bures–Wasserstein distance between positive definite matrices},
journal = {Expositiones Mathematicae},
volume = {37},
number = {2},
pages = {165-191},
year = {2019},
issn = {0723-0869},
doi = {https://doi.org/10.1016/j.exmath.2018.01.002},
url = {https://www.sciencedirect.com/science/article/pii/S0723086918300021},
author = {Rajendra Bhatia and Tanvi Jain and Yongdo Lim}
}

@article{hillen2017moments,
  title={Moments of von Mises and Fisher distributions and applications},
  author={Hillen, Thomas and Painter, Kevin J and Swan, Amanda C and Murtha, Albert D},
  journal={Mathematical biosciences and engineering},
  volume={14},
  number={3},
  pages={673--694},
  year={2017},
  publisher={Arizona State University}
}

@article{JMLR:v25:23-0783,
  author  = {Rong Tang and Yun Yang},
  title   = {On the Computational Complexity of Metropolis-Adjusted Langevin Algorithms for Bayesian Posterior Sampling},
  journal = {Journal of Machine Learning Research},
  year    = {2024},
  volume  = {25},
  number  = {157},
  pages   = {1--79},
  url     = {http://jmlr.org/papers/v25/23-0783.html}
}

@InProceedings{pmlr-v134-chewi21a,
  title = 	 {Optimal dimension dependence of the Metropolis-Adjusted Langevin Algorithm},
  author =       {Chewi, Sinho and Lu, Chen and Ahn, Kwangjun and Cheng, Xiang and Gouic, Thibaut Le and Rigollet, Philippe},
  booktitle = 	 {Proceedings of Thirty Fourth Conference on Learning Theory},
  pages = 	 {1260--1300},
  year = 	 {2021},
  volume = 	 {134},
  month = 	 {15--19 Aug},
  publisher =    {PMLR},
  url = 	 {https://proceedings.mlr.press/v134/chewi21a.html}
}

@article{LEE2017152,
title = {Inference for eigenvalues and eigenvectors in exponential families of random symmetric matrices},
journal = {Journal of Multivariate Analysis},
volume = {162},
pages = {152-171},
year = {2017},
issn = {0047-259X},
doi = {https://doi.org/10.1016/j.jmva.2017.08.006},
url = {https://www.sciencedirect.com/science/article/pii/S0047259X17305444},
author = {Han Na Lee and Armin Schwartzman},
}

@article{https://doi.org/10.1002/cjs.11601,
author = {Lan, Zhou and Reich, Brian J. and Bandyopadhyay, Dipankar},
title = {A spatial Bayesian semiparametric mixture model for positive definite matrices with applications in diffusion tensor imaging},
journal = {Canadian Journal of Statistics},
volume = {49},
number = {1},
pages = {129-149},
doi = {https://doi.org/10.1002/cjs.11601},
url = {https://onlinelibrary.wiley.com/doi/abs/10.1002/cjs.11601},
eprint = {https://onlinelibrary.wiley.com/doi/pdf/10.1002/cjs.11601},
year = {2021}
}

@article{ingalhalikar2014sex,
  title={Sex differences in the structural connectome of the human brain},
  author={Ingalhalikar, Madhura and Smith, Alex and Parker, Drew and Satterthwaite, Theodore D and Elliott, Mark A and Ruparel, Kosha and Hakonarson, Hakon and Gur, Raquel E and Gur, Ruben C and Verma, Ragini},
  journal={Proceedings of the National Academy of Sciences},
  volume={111},
  number={2},
  pages={823--828},
  year={2014},
  publisher={National Acad Sciences}
}

@article{fan2016overview,
  title={An overview of the estimation of large covariance and precision matrices},
  author={Fan, Jianqing and Liao, Yuan and Liu, Han},
  journal={The Econometrics Journal},
  volume={19},
  number={1},
  pages={C1--C32},
  year={2016},
  publisher={Oxford University Press Oxford, UK}
}

@article{10.1214/18-AOS1692,
author = {Rudrasis Chakraborty and Baba C. Vemuri},
title = {{Statistics on the Stiefel manifold: Theory and applications}},
volume = {47},
journal = {The Annals of Statistics},
number = {1},
publisher = {Institute of Mathematical Statistics},
pages = {415 -- 438},
year = {2019},
doi = {10.1214/18-AOS1692},
URL = {https://doi.org/10.1214/18-AOS1692}
}

@article{guo2022statistical,
  title={Statistical shape analysis of brain arterial networks (ban)},
  author={Guo, Xiaoyang and Bal, Aditi Basu and Needham, Tom and Srivastava, Anuj},
  journal={The Annals of Applied Statistics},
  volume={16},
  number={2},
  pages={1130--1150},
  year={2022},
  publisher={Institute of Mathematical Statistics}
}

@article{kendall1984shape,
  title={Shape manifolds, procrustean metrics, and complex projective spaces},
  author={Kendall, David G},
  journal={Bulletin of the London mathematical society},
  volume={16},
  number={2},
  pages={81--121},
  year={1984},
  publisher={Wiley Online Library}
}

@inproceedings{chakraborty2015recursive,
  title={Recursive frechet mean computation on the grassmannian and its applications to computer vision},
  author={Chakraborty, Rudrasis and Vemuri, Baba C},
  booktitle={Proceedings of the IEEE International Conference on Computer Vision},
  pages={4229--4237},
  year={2015}
}

@inproceedings{frechet1948elements,
  title={Les {\'e}l{\'e}ments al{\'e}atoires de nature quelconque dans un espace distanci{\'e}},
  author={Fr{\'e}chet, Maurice},
  booktitle={Annales de l'institut Henri Poincar{\'e}},
  volume={10},
  number={4},
  pages={215--310},
  year={1948}
}

@article{lian2019multiple,
  title={Multiple quantile modeling via reduced-rank regression},
  author={Lian, Heng and Zhao, Weihua and Ma, Yanyuan},
  journal={Statistica Sinica},
  volume={29},
  number={3},
  pages={1439--1464},
  year={2019},
  publisher={JSTOR}
}

@article{jiang2008gibbs,
  title={Gibbs posterior for variable selection in high-dimensional classification and data mining},
  author={Jiang, Wenxin and Tanner, Martin A},
  journal={The Annals of Statistics},
  volume={36},
  number={5},
  pages={2207--2231},
  year={2008},
  publisher={Institute of Mathematical Statistics}
}

@article{lin2020robust,
  title={Robust optimization and inference on manifolds},
  author={Lin, Lizhen and Lazar, Drew and Sarpabayeva, Bayan and Dunson, David B},
  journal={arXiv preprint arXiv:2006.06843},
  year={2020}
}

@article{bhattacharya2017omnibus,
  title={Omnibus CLTs for Fr{\'e}chet means and nonparametric inference on non-Euclidean spaces},
  author={Bhattacharya, Rabi and Lin, Lizhen},
  journal={Proceedings of the American Mathematical Society},
  volume={145},
  number={1},
  pages={413--428},
  year={2017}
}

@Book{boumal2020introduction,
  title     = {An introduction to optimization on smooth manifolds},
  author    = {Boumal, Nicolas},
  publisher = {Cambridge University Press},
  year      = {2023},
  url       = {https://www.nicolasboumal.net/book},
  doi       = {10.1017/9781009166164}
}

@book{lee2013smooth,
  author={Lee, John M},
  title={Introduction to smooth manifolds},
  year={2013},
  publisher={Springer}
}

@article{chib2018bayesian,
  title={Bayesian estimation and comparison of moment condition models},
  author={Chib, Siddhartha and Shin, Minchul and Simoni, Anna},
  journal={Journal of the American Statistical Association},
  volume={113},
  number={524},
  pages={1656--1668},
  year={2018},
  publisher={Taylor \& Francis}
}

@article{yang2012bayesian,
  title={Bayesian empirical likelihood for quantile regression},
  author={Yang, Yunwen and He, Xuming},
journal={The Annals of Statistics},
  year={2012}
}

@article{holbrook2016bayesian,
  title={Bayesian inference on matrix manifolds for linear dimensionality reduction},
  author={Holbrook, Andrew and Vandenberg-Rodes, Alexander and Shahbaba, Babak},
  journal={arXiv preprint arXiv:1606.04478},
  year={2016}
}

@article{zhang2018grassmannian,
  title={Grassmannian learning: Embedding geometry awareness in shallow and deep learning},
  author={Zhang, Jiayao and Zhu, Guangxu and Heath Jr, Robert W and Huang, Kaibin},
  journal={arXiv preprint arXiv:1808.02229},
  year={2018}
}

@InProceedings{pmlr-v9-suzuki10a,
  title = 	 {Sufficient Dimension Reduction via Squared-loss Mutual Information Estimation},
  author = 	 {Suzuki, Taiji and Sugiyama, Masashi},
  booktitle = 	 {Proceedings of the Thirteenth International Conference on Artificial Intelligence and Statistics},
  pages = 	 {804--811},
  year = 	 {2010},
  volume = 	 {9},
  address = 	 {Chia Laguna Resort, Sardinia, Italy},
  month = 	 {13--15 May},
  publisher =    {PMLR},
  url = 	 {https://proceedings.mlr.press/v9/suzuki10a.html}
}

@book{markovsky2012low,
  title={Low rank approximation: algorithms, implementation, applications},
  author={Markovsky, Ivan},
  volume={906},
  year={2012},
  publisher={Springer}
}

@article{ji2011robust,
  title={Robust video restoration by joint sparse and low rank matrix approximation},
  author={Ji, Hui and Huang, Sibin and Shen, Zuowei and Xu, Yuhong},
  journal={SIAM Journal on Imaging Sciences},
  volume={4},
  number={4},
  pages={1122--1142},
  year={2011},
  publisher={SIAM}
}

@phdthesis{zhou2015rank,
  title={Rank-constrained optimization: A Riemannian manifold approach},
  author={Zhou, Guifang},
  year={2015},
  school={The Florida State University}
}

@book{reinsel2023multivariate,
  title={Multivariate Reduced-Rank Regression: Theory, Methods and Applications},
  author={Reinsel, Gregory C and Velu, Raja P and Chen, Kun},
  volume={225},
  year={2023},
  publisher={Springer Nature}
}

@article{zhang2003optimal,
  title={Optimal low-rank approximation to a correlation matrix},
  author={Zhang, Zhenyue and Wu, Lixin},
  journal={Linear algebra and its applications},
  volume={364},
  pages={161--187},
  year={2003},
  publisher={Elsevier}
}

@article{hosseini2015matrix,
  title={Matrix manifold optimization for Gaussian mixtures},
  author={Hosseini, Reshad and Sra, Suvrit},
  journal={Advances in Neural Information Processing Systems},
  volume={28},
  year={2015}
}

@article{ferreira2019gradient,
  title={Gradient method for optimization on Riemannian manifolds with lower bounded curvature},
  author={O. P. Ferreira and M. S. Louzeiro and L. F. Prudente},
  journal={SIAM Journal on Optimization},
  volume={29},
  number={4},
  pages={2517--2541},
  year={2019},
  publisher={SIAM}
}

@article{zhou2021manifold,
  title={Manifold optimization-assisted gaussian variational approximation},
  author={Zhou, Bingxin and Gao, Junbin and Tran, Minh-Ngoc and Gerlach, Richard},
  journal={Journal of Computational and Graphical Statistics},
  volume={30},
  number={4},
  pages={946--957},
  year={2021},
  publisher={Taylor \& Francis}
}

@article{martens2020new,
  title={New insights and perspectives on the natural gradient method},
  author={Martens, James},
  journal={The Journal of Machine Learning Research},
  volume={21},
  number={1},
  pages={5776--5851},
  year={2020},
  publisher={JMLRORG}
}

@article{schennach2005bayesian,
  title={Bayesian exponentially tilted empirical likelihood},
  author={Schennach, Susanne M},
  journal={Biometrika},
  volume={92},
  number={1},
  pages={31--46},
  year={2005},
  publisher={Oxford University Press}
}

@book{hall2004generalized,
  title={Generalized method of moments},
  author={Hall, Alastair R},
  year={2004},
  publisher={OUP Oxford}
}

@article{yin2009bayesian,
  title={Bayesian generalized method of moments},
  author={Yin, Guosheng},
journal={Bayesian Analysis},
  year={2009}
}

@article{divol2022measure,
  title={Measure estimation on manifolds: an optimal transport approach},
  author={Divol, Vincent},
  journal={Probability Theory and Related Fields},
  volume={183},
  number={1-2},
  pages={581--647},
  year={2022},
  publisher={Springer}
}

@article{10.1214/aoap/1034625254,
author = {A. Gelman and W. R. Gilks and G. O. Roberts},
title = {{Weak convergence and optimal scaling of random walk Metropolis algorithms}},
volume = {7},
journal = {The Annals of Applied Probability},
number = {1},
publisher = {Institute of Mathematical Statistics},
pages = {110 -- 120},
year = {1997},
doi = {10.1214/aoap/1034625254},
URL = {https://doi.org/10.1214/aoap/1034625254}
}

@article{hoff2009simulation,
  title={Simulation of the matrix Bingham--von Mises--Fisher distribution, with applications to multivariate and relational data},
  author={Hoff, Peter D},
  journal={Journal of Computational and Graphical Statistics},
  volume={18},
  number={2},
  pages={438--456},
  year={2009},
  publisher={Taylor \& Francis}
}

@article{jauch2021monte,
  title={Monte Carlo simulation on the Stiefel manifold via polar expansion},
  author={Jauch, Michael and Hoff, Peter D and Dunson, David B},
  journal={Journal of Computational and Graphical Statistics},
  volume={30},
  number={3},
  pages={622--631},
  year={2021},
  publisher={Taylor \& Francis}
}

@article{lin2017bayesian,
  title={Bayesian nonparametric inference on the Stiefel manifold},
  author={Lin, Lizhen and Rao, Vinayak and Dunson, David},
  journal={Statistica Sinica},
  pages={535--553},
  year={2017},
  publisher={JSTOR}
}

@inproceedings{mao2010supervised,
  title={Supervised dimension reduction using Bayesian mixture modeling},
  author={Mao, Kai and Liang, Feng and Mukherjee, Sayan},
  booktitle={Proceedings of the thirteenth international conference on artificial intelligence and statistics},
  pages={501--508},
  year={2010},
  organization={JMLR Workshop and Conference Proceedings}
}

@article{thomas2022learning,
  title={Learning Subspaces of Different Dimensions},
  author={Thomas, Brian St and You, Kisung and Lin, Lizhen and Lim, Lek-Heng and Mukherjee, Sayan},
  journal={Journal of Computational and Graphical Statistics},
  volume={31},
  number={2},
  pages={337--350},
  year={2022},
  publisher={Taylor \& Francis}
}

@article{castillo2014thomas,
  title={Thomas Bayes’ walk on manifolds},
  author={Castillo, Isma{\"e}l and Kerkyacharian, G{\'e}rard and Picard, Dominique},
  journal={Probability Theory and Related Fields},
  volume={158},
  number={3-4},
  pages={665--710},
  year={2014},
  publisher={Springer}
}

@article{bhattacharya2010nonparametric,
  title={Nonparametric Bayesian density estimation on manifolds with applications to planar shapes},
  author={Bhattacharya, Abhishek and Dunson, David B},
  journal={Biometrika},
  volume={97},
  number={4},
  pages={851--865},
  year={2010},
  publisher={Oxford University Press}
}

@article{10.1214/15-AOS1390,
author = {Yun Yang and David B. Dunson},
title = {{Bayesian manifold regression}},
volume = {44},
journal = {The Annals of Statistics},
number = {2},
publisher = {Institute of Mathematical Statistics},
pages = {876 -- 905},
year = {2016},
doi = {10.1214/15-AOS1390},
URL = {https://doi.org/10.1214/15-AOS1390}
}

@article{10.1214/18-BA1135,
author = {Lizhen Lin and Niu Mu and Pokman Cheung and David Dunson},
title = {{Extrinsic Gaussian Processes for Regression and Classification on Manifolds}},
volume = {14},
journal = {Bayesian Analysis},
number = {3},
publisher = {International Society for Bayesian Analysis},
pages = {887 -- 906},
year = {2019},
doi = {10.1214/18-BA1135},
URL = {https://doi.org/10.1214/18-BA1135}
}

@article{berenfeld2022estimating,
  title={Estimating a density near an unknown manifold: a Bayesian nonparametric approach},
  author={Berenfeld, Cl{\'e}ment and Rosa, Paul and Rousseau, Judith},
  journal={arXiv preprint arXiv:2205.15717},
  year={2022}
}

@article{BINETTE2022118,
title = {Bayesian nonparametrics for directional statistics},
journal = {Journal of Statistical Planning and Inference},
volume = {216},
pages = {118-134},
year = {2022},
issn = {0378-3758},
doi = {https://doi.org/10.1016/j.jspi.2021.05.007},
url = {https://www.sciencedirect.com/science/article/pii/S0378375821000586},
author = {Olivier Binette and Simon Guillotte}
}

@article{mcvinish2008semiparametric,
  title={Semiparametric Bayesian circular statistics},
  author={McVinish, Ross and Mengersen, Kerrie},
  journal={Computational statistics \& data analysis},
  volume={52},
  number={10},
  pages={4722--4730},
  year={2008},
  publisher={Elsevier}
}

@article{ravindran2011bayesian,
  title={Bayesian analysis of circular data using wrapped distributions},
  author={Ravindran, Palanikumar and Ghosh, Sujit K},
  journal={Journal of Statistical Theory and Practice},
  volume={5},
  number={4},
  pages={547--561},
  year={2011},
  publisher={Taylor \& Francis}
}

@article{alvarez2014bayesian,
  title={Bayesian inference for a covariance matrix},
  author={Alvarez, Ignacio and Niemi, Jarad and Simpson, Matt},
  journal={arXiv preprint arXiv:1408.4050},
  year={2014}
}
\bibliographystyle{plain} 

\newpage
\appendix
  \begin{center}
{\bf\Large Appendix}
\end{center}

\textbf{Notations}:  We adopt the notations in the manuscript, and further introduce the following additional notations for technical proofs.  For any function $f:\,\m X\times \mb R^d\to \mb R$, we use $ \nabla_{y}f(x,y)$ to denote the gradient vector of $f(x,y)$ with respect to $y$ for $x\in\m X$ and $y\in \mb R^d$. For any function $f:\mb R^d\to \mb R^k$, we use $\bold{J}_f(y)$ to denote the $k\times d$ Jacobian matrix of $f$ at $y$, i.e., $\bold{J}_f(y)_{ij}=\frac{\partial f_i(y)}{\partial y_j}$ with $f(\cdot)=(f_1(\cdot),f_2(\cdot),\cdots, f_k(\cdot))$. 
  We use $\bold{1}_A(x)$ to denote the indicator function of a set $A$ so that $\bold{1}_A(x)=1$ if $x\in A$ and zero otherwise. We use $\chi^2_{\alpha}(d)$ to denote the $\alpha$-th quantile of $\chi^2$ distribution with $d$ degrees of freedom. For a probability measure $\mu$ and measurable set $A$, the (unnormalized) restriction of $\mu$ to $A$ is defined by $\mu|_A(B)=\mu(B\cap A)$ for  any measurable set $B$, and the normalized restriction of $\mu$ to $A$ is defined by $\mu|_A(B)=\frac{\mu(B\cap A)}{\mu(A)}$. When no ambiguity may arise, we will use the same symbol $\mu^*(\cdot)$ to denote the density function of the probability measure $\mu^*$. If $X$ is an random variable with law $\mu$,  we write $P(X\in A)=\mu(A)$ to denote the probability of $A$ under $\mu$. For any two positive integers $D,d$ with $D\geq d$, we use $\mb O(D,d)$ to denote the set of all orthogonal $d$-frames in $\mb R^D$, i.e., $\mb O(D,d)=\{U\in \mb R^{D\times d}\,:\, U^TU=I_d\}$, and when $D=d$, we write $\mb O(d)=\mb O(d,d)$. We use the notation $\mnorm{\cdot}_{\rm F}$ and $\mnorm{\cdot}_{\rm op}$ to denote the matrix Frobenius norm and operator norm respectively. We use $\mb S_1^d$ to denote the $d$-dimensional $1-$sphere, i.e., $\mb S^d_1=\{\lambda\in \mb R^{d+1}\,:\,\|\lambda\|=1\}$. For two numbers $a,b$, we denote $a\vee b=\max(a,b)$.
The symbols $\lesssim$ and $\gtrsim$
mean the corresponding inequality up to an $n$-independent constant.  We write $A=B+O(a_n)$ if $|A-B|\lesssim a_n$. Throughout, $C$, $c$, $C_0$, $c_0$, $C_1$, $c_1$, $C_2$, $c_2$,\ldots are generically used to denote positive constants whose values might change from one line to another, but are independent from everything else.

 \section{Notions in Riemannian Submanifold}\label{App:manifold}
This subsection provides an introduction to Riemannian submanifolds and their relevance in optimization problems for determining the parameter of interest $\theta^*$. We draw upon the works of~\cite{boumal2020introduction,lee2013smooth} for this purpose. Understanding optimization tools for Riemannian submanifolds is crucial for constructing appropriate posterior distributions for Bayesian inference and for efficient sampling. We begin with a review of first-order embedded geometry and present a popular first-order optimization algorithm for Riemannian submanifolds. Subsequently, we delve into second-order embedded geometry and discuss second-order optimization algorithms.

 \medskip
\noindent\textbf{1. \underline{Definition of manifold.}}  We first recall the definition of manifold.
 \begin{definition}[Submanifold]
 A subset $\m M$ of $\mathbb{R}^D$ is a $d$-dimensional Riemannian submanifold if for every point $\theta$ in $\m M$, there exists a neighbourhood $U_{\theta}$ of $\theta$ on $\m M$ and an open set $V_{\theta} \subseteq \mathbb{R}^d$, such that that there exists a homeomorphism  $\xi_{\theta}$ that maps $V_{\theta}$ to $U_{\theta}$, that is, $\xi_{\theta}: \, V_{\theta}\rightarrow U_{\theta}$ is bijective and both $\xi_{\theta}$ and $\xi^{-1}_{\theta}$ are continuous maps. Moreover,  the  differential $\m D \xi_{\theta}(z)[\cdot]$ of $\xi_{\theta}(\cdot)$ at $z$ exists and is injective for every $z\in V_{\theta}$.
 % \footnote{Here, the differential of $\xi(\cdot)$ at $y$, denoted as $D_y \xi$, is a linear map defined by $D_y \xi [v] = \lim_{t \to 0} \frac{\xi(y + tv) - \xi(y)}{t} = J_{\xi}(y)v$ for $v \in \mb R^d$. The injectiveness of $D_y \xi$ is equivalent to the Jacobian matrix $J_{\xi}(y)$ having full rank.} 
The pair $(U_{\theta},\xi_{\theta}^{-1})$ is called a local coordinate chart near $\theta$,  with $\xi_{\theta}^{-1}$ the coordinate map and $\xi_{\theta}$ a local parameterization. We refer to $D$ as the ambient dimension and $d$ as the intrinsic dimension of $\m M$. 
\end{definition}
Next, we define the atlas.
 \begin{definition}[Atlas]
     A collection of $d$-dimensional charts $\ms A = \{(U_\lambda, \varphi_\lambda)\}_{\lambda\in \Lambda}$ is called an atlas on $\m M$ if 1. $\m M = \bigcup_{\lambda\in\Lambda} U_\lambda$. 2. Each chart $(U_{\lambda}, \varphi_{\lambda})$ in atlas $\ms A$ consists of a homeomorphism $\varphi_{\lambda}:\, U_{\lambda} \to \widetilde U_{\lambda}$,  from an open set $U_{\lambda} \subset \m M$ to an open set $\widetilde U_{\lambda} \subset \mb R^d$. 3. Any two charts $(U, \varphi)$ and $(V, \psi)$ in atlas $\mathscr{A}$ are compatible, meaning that the transition map $\varphi \circ \psi^{-1}: \psi(U \cap V) \rightarrow \varphi(U \cap V)$ is a diffeomorphism.
\end{definition}

%  satisfying $B_{r}(\theta)\cap \m M\subset U\subset \m M$ and $B_{r}(0_D)\cap T_{\theta}\mathcal{M}\subset V\subset T_{\theta}\mathcal{M}$, such that
%  the projection $\psi_{\theta}(\cdot)={\rm Proj}_{T_{\theta^*}\m M}(\cdot-\theta^*)$ maps $U\to V$, and is a bijective.  Its inverse, $\phi_{\theta}$ satisfied that the differentials of $\phi_{\theta}(v)$, up to order three,  have operator norms uniformly bounded by $L$ for $v\in B_r(0_D) \cap T_{\theta}\mathcal{M}$.\footnote{That means, for any $\eta_1,\eta_2,\eta_3\in T_{\theta}\m M$ with unit norms and $v\in B_r(0_D)\cap T_{\theta}\m M$,  
% $$ \big\|\m D_v \phi_{\theta}(v)[\eta_1]\big\|\leq L,\,\big\|\m D_v(\m D_v \phi_{\theta^*}(v)[\eta_1])[\eta_2]\big\|\leq L,\,\big\|\m D_v(\m D_v(\m D_v \phi_{\theta^*}(v)[\eta_1])[\eta_2])[\eta_3]\big\|\leq L,$$
%     where the subscript $v$ means the differential is with respect to $v$. }
% \end{definition}

\noindent  A function $f:M\to\mathbb{R}$ is called $C^k$-smooth if there exists an open neighborhood $U\subset\mathbb{R}^D$ of $M$ and a function $\ov f:U\to\mathbb{R}$ such that $f=\ov f|_M$ and $\ov f$ has continuous partial derivatives up to order $k$ on $U$.

\medskip
\noindent\textbf{2. \underline{First-order embedded geometry.}}
We start by introducing the following concepts of the tangent bundle and the differential of a smooth map, which enable us to define gradients on submanifolds.
\begin{definition}
(Tangent bundle)  For a submanifold $\m M$ embedded in $\mb R^D$, we denote the tangent space of $\m M$ at $\theta$ as $T_{\theta}\m M=\{c'(0)\,|\, c: I\to \m M \text{ is differentiable} \text{ and } c(0)=\theta\}$, where $I$ is any open interval containing $t=0$; we use $T \m M=\{(\theta,v)\,:\, \theta\in \m M, v\in T_{\theta} \m M\}$  to denote the tangent bundle of $\m M$.
\end{definition}
\begin{definition}
(Differential of a smooth map) The differential $f:\m M\to \m M'$ at $\theta$ is a linear operator $\m D f(\theta):T_{\theta}\m M\to T_{f(\theta)}\m M'$ defined by 
\begin{equation*}
    \m D f(\theta)[v]=\frac{\,\dd}{\,\dd t}f(c(t))\big|_{t=0},
\end{equation*}
 where $c$ is a differentiable curve on $\m M$ passing through $\theta$ at $t = 0$ with velocity $c'(0)=v$. 
\end{definition}

\noindent Now, we are prepared to define the Riemannian gradient of a smooth function on a submanifold. 

% \textcolor{magenta}{(should we explicitly say below what a vector field is in the present context, i.e., a map from $\m M \to T\m M$?)}
\begin{definition}
(Riemannian gradient/natural gradient) Let $f: \m M \to \mb R$ be a $C^1$-smooth function on a Riemannian submanifold $\m M$ embedded in $\mb R^D$. The Riemannian gradient ${\rm grad} f$ of $f$ is the vector field  on $\m M$ (i.e., an assignment of a tangent vector to each point in $\m M$)  uniquely defined by the following identities:
  \begin{equation*}
  \begin{aligned}
       & \forall\, (\theta,v)\in T\m M: \quad \m D f(\theta)[v]=\langle v, {\rm grad}f(\theta)\rangle_{\theta},
    \end{aligned}
  \end{equation*}
     where $\langle\cdot, \cdot\rangle_{\theta}: T_{\theta}\m M\times T_{\theta}\m M\to \mb R$ is the Riemannian metric defined by the restriction of the $D$-dimensional Euclidean inner product on $T_{\theta}\m M\times T_{\theta}\m M$.
\end{definition}
  If we consider any $C^1$-smooth extension $\ov f$ of $f$ to a neighborhood of $\m M$ in $\mb R^D$, the Riemannian gradient of $f$ is given by 
 \begin{equation*}
     {\rm grad} f(\theta)={\rm Proj}_{T_{\theta}\m M}(\nabla \ov f(\theta)),
 \end{equation*}
 where ${\rm Proj}_{T_{\theta}\m M}(v)=\underset{v'\in T_{\theta}\m M}{\argmin} \|v'-v\|$ denote the projection of $v$ onto $T_{\theta}\m M$.

 \medskip

\noindent\textbf{3. \underline{First-order algorithm for optimization on a Riemannian submanifold.}} Consider a generic optimization problem
\begin{equation}\label{prob:optimization}
     \underset{\theta\in \m M}{\min} \,f(\theta),
 \end{equation}
  where $\m M$ is a Riemannian submanifold embedded in $\mb R^D$, and $f$ is a smooth function called the objective function. Optimizaing $f$ over $\mathcal{M}$ can be viewed as a ``constrained'' optimization problem, where $\theta$ is not allowed to move freely in the ambient space $\mathbb{R}^D$, but instead must remain on $\mathcal{M}$.    To generalize unconstrained optimization algorithms such as gradient descent, one can utilize the  following \emph{first-order necessary optimality condition} for identifying a local minimizer of $f$ on $\m M$:
 \begin{proposition} (Propositions 4.5 and 4.6 of~\cite{boumal2022intromanifolds} )\label{prop1}
 Any local minimizer $\theta$ of a smooth function $f:\m M\to \mb R$ satisfies ${\rm grad}f(\theta)=0$.
 \end{proposition}
The first-order optimality condition outlined in Proposition~\ref{prop1} enables us to convert the problem of identifying a (local) optimizer for $f$ into the task of solving a system of (first-order) equations. Therefore, one can apply the Riemannian gradient descent method (RGD) to solve problem~\eqref{prob:optimization}. The  Riemannian gradient determines the direction of movement in each iteration, and a retraction  is used to  ensure that the point stays on the manifold. The retraction allows us to  move
away from a point $\theta\in \m M$ along the direction $v\in T_{\theta}\m M$ while remaining on the manifold, which is the
basic operation of  essentially all optimization
algorithms on manifolds.  The retraction is formally defined as follows.
 \begin{definition} (Retraction)
A retraction on a manifold $\m M$ is a map:
\begin{equation*}
    {\rm R}: T\m M\to \m M: (\theta,v)\longmapsto {\rm R}_{\theta}(v)
\end{equation*}
such that each curve $c(t)={\rm R}_{\theta}(tv)$ satisfies $c(0)=\theta$ and $c'(0)=v$;  and if in addition $c''(0)=0$ for each curve $c$, then ${\rm R}$ is called a second-order retraction on $\m M$.
\end{definition}
 Given a retraction ${\rm R}_{\theta}(v)$, the RGD algorithm is as follows: select an initial point $\theta_0\in \m M$, choose step sizes $\alpha_k>0$, and iterate 
\begin{equation*}
    \theta_{k+1}={\rm R}_{\theta_k}(-\alpha_k{\rm grad}f(\theta_k)) \quad k=0,1,2,\cdots
\end{equation*}
% A second-order optimization algorithm  can also be considered using second-order embedded geometry; further details are available in Appendix~\ref{App:manifold}. 

\medskip

\noindent\textbf{4. \underline{Second-order embedded geometry.}} Based on the notion of Riemannian gradient for smooth functions on submanifolds, we can consider the following concept of Riemannian Hessian.
\begin{definition}
Let $\m M$ be a Riemannian submanifold embedded in $\mb R^D$. The Riemannian Hessian of a function $f:\m M\to \mb R$ at point $\theta\in \m M$ is the linear map ${\rm Hess} f(\theta): T_{\theta} \m M\to T_{\theta} \m M$ defined as follows~\citep{boumal2022intromanifolds}:
\begin{equation*}
     {\rm Hess} f(\theta)[u]={\rm Proj}_{T_{\theta} \m M} \Big(\lim_{t \to 0}\frac{{\rm grad} f(c(t))-{\rm grad} f(c(0))}{t}\Big),
\end{equation*}
where $c$ is a differentiable curve on $\m M$ so that $c(0)=\theta$ and $c'(0)=u$.
\end{definition}

Let  $\ov{G}$ be a $C^1$-smooth extension of ${\rm grad}f$ defined on a neighborhood of $\m M$ in $\mb R^D$, then we have
\begin{equation*}
       {\rm Hess} f(\theta)[u]={\rm Proj}_{T_{\theta} \m M}(\m D\ov G(\theta)[u]).
\end{equation*}

The following proposition states the \emph{second-order necessary optimality condition} for identifying local minimizer of $f$ on $\m M$.
 \begin{proposition} (Propositions 6.2 and 6.3 of~\cite{boumal2022intromanifolds} )\label{prop2}
 Any local minimizer $\theta$ of a smooth function $f:\m M\to \mb R$ satisfies ${\rm grad}f(\theta)=0$ and ${\rm Hess} f(\theta)\succcurlyeq 0$.\footnote{We call ${\rm Hess} f(\theta)\succcurlyeq 0$ if $u^T{\rm Hess} f(\theta)[u]\geq 0 $ for all $u \in T_{\theta}\m M$. }
 \end{proposition}

We can consider second-order optimization algorithms based on the second-order necessary optimality condition described in Proposition~\ref{prop2}. The most popular one is the Riemannian Newton's method where the Riemannian Hessian and Riemannian gradient are used to decide the moving direction.  Given a second-order retraction ${\rm R}$ on manifold, the Riemannian Newton's algorithm is given by: select initial point $\theta_0\in \m M$, for $k=0,1,2,\cdots$, iterate
 \begin{equation*}
 \begin{aligned}
       &\text{solve}\quad {\rm Hess} f(\theta_k)[s_k]=-{\rm grad}f(\theta_k) \quad\text{for} \quad s_k\in T_{\theta_k}\m M;\\
       & \qquad\quad\theta_{k+1}={\rm R}_{\theta_k}(s_k).
 \end{aligned}
 \end{equation*}

\noindent\textbf{5. \underline{Riemannian volume measure.}} The Riemannian volume measure is a commonly-used base measure for defining density function on a Riemannian submanifold. Consider a $d$-dimensional Riemannian submanifold $\m M$ embedded in $\mb R^D$, and with atlas  $\ms A=\{(U_{\lambda},\varphi_{\lambda})\}_{\lambda \in \Lambda}$. We first consider the following mathematical technique of \emph{partition of unity}, so that one can glue constructions in the local charts to form a global construction on the manifold. 
\begin{definition}(partition of unity)
 A partition of unity subordinate to altas $\ms A=\{(U_{\lambda},\varphi_{\lambda})\}_{\lambda \in \Lambda}$ is a collection of  functions $\{\rho_\lambda\}_{\lambda\in \Lambda}$ on $\mathcal M$ so that \begin{enumerate} 
     \item $0\leq \rho_\lambda\leq 1$ for all $\lambda\in \Lambda$, and $\sum_{\lambda\in \Lambda}\rho_\lambda(x)=1$ for all $x\in \mathcal{M}$.
     \item ${\rm supp}(\rho_\lambda)\subset U_{\lambda}$ for any $\lambda\in\Lambda$.
     \item Each point $x\in \mathcal{M}$ has a neighborhood which intersects ${\rm supp}(\rho_\lambda)$ for only finitely many $\lambda\in \Lambda$.
 \end{enumerate}
 \end{definition}
 
Then given a partition of unity $\{\rho_\lambda\}_{\lambda\in \Lambda}$  subordinate to altas $\ms A$,  the Riemannian volume measure $\mu_{\m M}$ is defined as
 \begin{equation*}
     \,\dd \mu_{\m M}=\sum_{\lambda\in \Lambda} \rho_{\lambda}(\varphi_\lambda^{-1}(z))\sqrt{{\rm det}(J_{\varphi_\lambda^{-1}}(z)^TJ_{\varphi_\lambda^{-1}}(z))}\,\dd z.
 \end{equation*}
 
 A measure $\mu$  on $\m M$ is said to have a density $f$ (with respect to the volume measure $\mu_{\m M}$) if for any measurable subset $A \subset \m M$, 
 \begin{equation*}
     \mu(A)=\int_{A} f \,\dd  \mu_{\m M}=\sum_{\lambda\in \Lambda}\int_{\varphi_\lambda(U_{\lambda}\cap A)} \rho_{\lambda}(\varphi_\lambda^{-1}(z))\cdot f(\varphi_\lambda^{-1}(z))\sqrt{{\rm det}(J_{\varphi_\lambda^{-1}}(z)^TJ_{\varphi_\lambda^{-1}}(z))}\,\dd z.
 \end{equation*}
 
\section{Mixing time analysis}\label{sec:mixgeneral}

The goal of this section is to characterize the computational complexity of the RRWM algorithm for sampling from a Bayesian posterior $\mu^*$ defined on a submanifold $\m M$.  We impose a condition on the target distribution $\mu^*$ that requires the existence of a  point $\wt \theta\in \m M$ so that ${\psi_{\wt \theta}}_{\#}\mu^*$ is uniformly close to a zero-centered multivariate normal distribution in a high probability set of $\mu^*$. This is formalized as the following Assumptions.

\vspace{1em}
\noindent{\bf Assumption B.1 (Conditions for the retraction at $\theta$): }
 There exist positive constants $r$, $L$ so that  the retraction $\wt \phi_{\theta}:\wt V_{\theta}\to \wt U_{\theta}$, and its inverse $\wt \psi_{\theta}:\wt U_{\theta}\to \wt V_{\theta}$ satisfies that:
 \begin{enumerate}
\item  $ B_{r}(\theta)\cap\mathcal{M}\subset \wt U_{\theta}$ and $B_r(0_D)\cap T_{\theta}\mathcal{M}\subset \wt V_{\theta}$;
\item $\underset{\theta'\in B_{r}(\theta)\cap\mathcal{M}}{\sup}\frac{\|\wt\psi_{\theta}(\theta')-({\theta}'-\theta)\|_2}{\|\theta'-{\theta}\|_2^2}\leq L$ and $\underset{v\in B_{r}(0_D)\cap T_{\theta}\mathcal{M}}{\sup}\frac{\|\wt\phi_{\theta}(v)-(v+\theta)\|_2}{\|v\|_2^2}\leq L$;
    \item for any $v,v'\in  B_r(\bold{0}_D)\cap T_{\theta}\mathcal{M}$ and any unit vector $\eta\in T_{\theta}\m M$,   $\|\m D \wt\phi_{\theta}(v)[\eta]-\m D \wt\phi_{\theta}(v')[\eta]\|_2\leq L\|v-v'\|_2$.
\end{enumerate}

Assumption B.1 requires that the retraction employed in RRWM algorithm is at least $C^2$-smooth. For a submanifold that is locally $C^3$-smooth around $\theta$, there are multiple viable choices for the retraction that meets Assumption B.1. For instance, a special choice of $\wt\psi_{\theta}(\cdot)$ is the projection map $\psi_{\theta}(\cdot) = {\rm Proj}_{T_{\theta}\m M}(\cdot-\theta)$, which has been applied in~\cite{https://doi.org/10.48550/arxiv.1702.08446} for  sampling from a solution manifold. Additionally, the exponential map and logarithmic map pair serves as another example, which is utilized  in~\cite{Byrne_2013} to develop Hamiltonian Monte Carlo algorithm  on a general Riemannian submanifold.

\vspace{0.5em}
\noindent{\bf Assumption B.2 (Conditions for the target distribution $\mu^*$ and hyperparameters in the RRWM algorithm):} There exist a reference point $\wt\theta\in \m M$,  a matrix ${W}_{\wt\theta}\in R^{D\times d}$ whose columns form an orthonormal basis for $T_{\wt\theta}\m M$,  positive numbers $\varepsilon, \varepsilon_1, M_0,n,\rho_1,\rho_2,R,h$ and covariance matrices $\wt I\in \mb R^{D\times D}$ and $J\in \mb R^{d\times d}$, so that
\begin{enumerate}

\item $K_{\theta}=\{x=\wt\phi_{\wt \theta}\big({W}_{\wt\theta}\frac{z}{\sqrt{n}}\Big)\,:\,z\in K\}$ with $K=\{z\in \mb R^d\,:\,\|( W_{\wt \theta}^T \wt I  W_{\wt \theta})^{-\frac{1}{2}}z\|\leq R\}$ is well-defined, that is, $\big\{v= W_{\wt \theta}\frac{z}{\sqrt{n}}\,:\, z\in K\big\}\subset \wt V_{\wt\theta}$;

 \item Let $\mu^*|_{K_{\theta}}$ be the normalized restriction of $\mu^*$ to $K_{\theta}$, and define the push-forward measure  $\mu_{\rm loc}^*=\big[\sqrt{n}\cdot {W}_{\wt \theta}^T\wt\psi_{\wt\theta}(\cdot)\big]_{\#}(\mu^*|_{K_{\theta}})$,  then $\mu_{\rm loc}^*$ is absolute continuous with respect to the Lebesgue measure on $\mb R^d$. Denote its density by the same symbol $\mu^*_{\rm loc}(\cdot)$, then for any $\xi\in K$,  
\begin{equation}\label{eqn:assumptionC}
    \Big|\log(\mu^*_{\rm loc}(\xi))-\log\Big((2\pi{\rm det}(J^{-1}))^{-\frac{d}{2}}\exp(-\frac{1}{2}\xi^TJ\xi)\Big)\Big|\leq \varepsilon_1;
\end{equation}

\item  $J^{\Delta}=(W_{\wt\theta}^T \wt I  W_{\wt\theta})^{\frac{1}{2}}J (W_{\wt\theta}^T \wt I  W_{\wt\theta})^{\frac{1}{2}}$ satisfies $\rho_1I_d\preccurlyeq J^{\Delta}\preccurlyeq \rho_2 I_d$.
\item $\mu^\ast\big(\{x=\wt\phi_{\wt\theta}\big(W_{\wt\theta}\frac{z}{\sqrt{n}}\Big)\,:\,\|(W_{\wt \theta}^T \wt I W_{\wt \theta})^{-\frac{1}{2}}z\|\leq R/2\}\big)\geq 1-\exp(-5\varepsilon_1)\frac{\varepsilon^2{h\rho_1}}{M_0^2}$.
\end{enumerate}

Assumption B.2 is motivated by the manifold BvM theorem.
The first two conditions in Assumption B.2 requires that the target distribution $\mu^*$, when constrained on a neighborhood $K_{\theta}$ around $\wt\theta$, can be approximated by a Gaussian distribution in the local coordinate system characterized by $\wt\psi_{\wt\theta}$.  It is noteworthy that we do not impose any smoothness or convexity constraints on the density of $\mu^*$ with respect to the volume measure of $\m M$, and the deviation characteristic $\varepsilon_1$ can take any value.  The third condition 
 requires the asymptotic covariance matrix $J$, after rescaling by the preconditioning matrix $\wt I$, to have its maximum eigenvalue upper-bounded by $\rho_2$ and its minimum eigenvalue lower-bounded by $\rho_1$. The condition number $\kappa=\frac{\rho_2}{\rho_1}$ serves as an indicator of how well the preconditioning matrix $\wt I$ is chosen to alleviate issues arising from the anisotropy of the target distribution. 
The last condition requires that $\mu^*$ should be concentrated around $\wt\theta$, which ensures that the Markov chain has a low probability of getting stuck in regions far away from $\wt\theta$ and shape constraints on $\mu^*|_{K_{\theta}}$ are sufficient to guarantee fast mixing of the chain.

  \begin{theorem}\label{th:mixing}(Mixing time for sampling from Bayesian posteriors satisfying  Bernstein-von Mises results in local coordinates) Let  $\mu^\ast$ be the target distribution on a submanifold $\m M$. Suppose there exists a reference point $\wt\theta\in \m M$ and a positive radius $r$, so that Assumption B.1 holds uniformly for $\theta\in  B_{r}(\wt \theta)\cap \m M$. Then there exist a small enough absolute $(n,d,D)$-independent constant $c_0$ and an $n$-independent constant $C_1$ so that if Assumption B.2 holds for reference point $\wt\theta$, a tolerance $\varepsilon\in (0,1)$, warming parameter $M_0$, sample size $n$, preconditioning matrix $\wt I$, covariance matrix $J$, eigenvalue constraints $\rho_2\geq \rho_1>0$, a rescaled step size $h$ that can be expressed as 
  \begin{equation*}
    h=c_0\,\rho_2^{-1}
 \Big(d+\log \big(\frac{M_0d\kappa}{\varepsilon} \big)+\varepsilon_1\Big)^{-1}, \quad\text{where}\quad \kappa=\frac{\rho_2}{\rho_1},
\end{equation*}
and radius $R$ satisfying $6\sqrt{d/\rho_1}\leq R\leq C_1 (h\sqrt{n})^{\frac{1}{3}}$. Then the $\zeta$-lazy version of the RRWM Algorithm with an $M_0$-warm start $\mu_0$ and step size $\wt h=\frac{h}{n}$ has  an $\varepsilon$-mixing time in $\chi^2$ divergence bounded as
 \begin{equation*}
 \begin{aligned}
          &\tau_{\rm mix}(\varepsilon,\mu_0)\\
          &\leq \frac{C_2\exp(-2\varepsilon_1)}{\zeta} \,\bigg\{\bigg[\exp(-3\varepsilon_1)\cdot\kappa\cdot \Big(d+\log \big(\frac{M_0d\kappa}{\varepsilon} \big)+\varepsilon_1\Big)\cdot \log \Big(\frac{\log M_0}{\varepsilon}\Big)\bigg]\vee\log \,(M_0)\bigg\},
 \end{aligned}
 \end{equation*}
 where $C_2$ is an $(n,d,D)$-independent constant.

 \end{theorem}

\section{Detailed Algorithms}\label{App:detailAlgo}

\subsection{Algorithm to compute the inverse of the projection map}

In this subsection, we introduce algorithms to compute  the local inverse $\phi_{\theta}(v)$ of the projection map $\phi_{\theta}(\theta')={\rm Proj}_{T_{\theta}\m M}(\theta'-\theta)$.  Notice that this specific choice of the retraction has the advantage of eliminating the Jacobian factor in the acceptance ratio. As a result,   the acceptance ratio in the RRWM algorithm can be simplified to 
 \begin{equation*}
   \m A(\theta^{k-1},y)=1\wedge  \frac{\mu^\ast(y)\cdot\exp\big(-v'^T(P_y\wt IP_y)^{\dagger}v'/(4\wt h)\big)}{\mu^\ast(\theta^{k-1})\cdot\exp\big(-v^T(P_{\theta^{k-1}}\wt IP_{\theta^{k-1}})^{\dagger}v/(4\wt h))\big)}.
 \end{equation*}
  According to~\cite{https://doi.org/10.48550/arxiv.1702.08446},  if the manifold is a solution manifold $\m M_{\bold q}=\{\theta\in \mb R^D\,:\, \bold{q}(\theta)=0\}$ for some smooth function $\bold q$,  then $y=\phi_{\theta}(v)$ can be found by numerically solving the equation
 \begin{equation*}
 \begin{aligned}
   & y=\theta+v+Q_{\theta}^Ta \quad \text{where}\, Q_{\theta}=\bold{J}_{\bold q}(\theta)\,\text{and}\, \bold q(\theta+ v+Q_{\theta}^Ta)=0_k,
 \end{aligned}
 \end{equation*}
 by using the Newton-Raphson algorithm,  where $\bold{J}_{\bold q}(\theta)$ to denote the $k\times D$ Jacobian matrix of $f$ at $\theta$, i.e., $\big[\bold{J}_{\bold q}(\theta)\big]_{ij}=\frac{\partial q_i(\theta)}{\partial \theta_j}$ for $i\in[k]$ and $j\in[D]$.  The detailed algorithm is given as follows.
 
 \begin{algorithm}[H]\label{algorithm.2.1}
\caption{Finding $\phi_{\theta}(v)$: Solution manifold}
\SetAlgoLined
\SetKwRepeat{Do}{do}{while}%
\SetKwFunction{FMain}{\rm project}
 \SetKwProg{Fn}{Function}{:}{}
  \Fn{\FMain{$\mathcal{M}$, $\theta$, $v$}}{
        Initialization: $a=0$, $i=0$, ${\rm flag}=1$, a function $\bold q=(q_1,q_2,\cdots,q_{D-d})$ such that $\mathcal{M}\cap B_r(\theta)=\{x\in B_r(\theta)\,|\,q(x)=0\}$\;
      $Q_{\theta}\leftarrow\bold{J}_{\bold q}(\theta)$\;
      \Repeat{$\|\bold q(\theta+v+Q_{\theta}a)\|_2\leq \varepsilon_0$ \text{ or } $i>{\rm{nmax}}$}{
      Solve $(\bold{J}_{\bold q} (\theta+v+Q_{\theta}a)^T V^{\perp})\Delta a=-\bold q(\theta+v+Q_{\theta}a)$ for  $\Delta a$\;
        $a\leftarrow a+\Delta a$\;
        $i\leftarrow i+1$\;
      }
      \If{$i>{\rm{nmax}}$ \text{ or } $\theta+v+Q_{\theta}a\notin \m M$}
      {${\rm flag}\leftarrow 0$\;}
        \KwRet $[\theta+v+Q_{\theta}a,{\rm flag}]$\;
  }
 \end{algorithm}
  Here ${\rm flag}$ indicates whether  the computation of $\phi_{\theta}(v)$ succeeds (${\rm flag}=1$) or fails $({\rm flag}=0)$.
However, this numerical scheme of computing $\phi_\theta(v)$ does not apply to a more general manifold where the function $\bold q(\cdot)$
 does not exist or is difficult to obtain. Instead, we propose a {more general} numerical scheme of computing $\phi_\theta(v)$ by solving the following optimization problem: given a tangent vector $v\in T_{\theta}\m M$ with a small enough  norm $\|v\|$, there exists $r>0$ so that $y=\phi_{\theta}(v)$ can be identified as the unique solution of 
 \begin{equation}\label{eqn:solveproj}
  \underset{y\in \m M\cap B_r(\theta)}{\arg\min}\,\|{\rm Proj}_{T_{\theta}\m M}(y-\theta)-v\|^2.
  \end{equation}
 The above optimization problem can be solved using the Riemannian gradient descent method or  Riemannian Newton’s method, with the initial point set to $\theta$.  Implementing these methods requires retraction, which can also be used in the RRWM algorithm. However, unlike an arbitrary retraction,  the inverse projection map  eliminates Jacobian terms in the acceptance ratio, which is advantageous when second-order information for the retraction is hard to compute.  The algorithms are detailed  below.

\begin{algorithm}[H]\label{algorithm.2.0}
\caption{Finding $\phi_{\theta}(v)$: Given a retraction (first-order algorithm)}
\SetAlgoLined
\SetKwRepeat{Do}{do}{while}%
\SetKwFunction{FMain}{\rm project}
 \SetKwProg{Fn}{Function}{:}{}
  \Fn{\FMain{$\mathcal{M}$, $\theta$, $v$}}{
        Initialization: $y=\theta$, $k=0$, ${\rm flag}=1$, step size $\{\alpha_k\}_{k\in \mb N}$, a retraction ${\rm R}$ on $\mathcal{M}$\;
        \Repeat{$\|G\|_2\leq \varepsilon_0$ \text{ or } $k>{\rm{nmax}}$ }{
       $G\leftarrow{\rm Proj}_{T_{y}\mathcal{M}}({\rm Proj}_{T_{\theta}\m M}(y-\theta)-v)$\;
         $y\leftarrow {\rm R}_y(-\alpha_k\cdot G)$\;
        $k\leftarrow k+1$\;
      }
      \If{$k>{\rm{nmax}}$ \text{ or } $y\notin  \m M$}
      {${\rm flag}\leftarrow 0$\;}
        \KwRet $[y,{\rm flag}]$\;
  }
 \end{algorithm}

\begin{algorithm}[H]\label{algorithm.2}
\caption{Finding $\phi_{\theta}(v)$: Given a second-order retraction (second-order algorithm)}
\SetAlgoLined
\SetKwRepeat{Do}{do}{while}%
\SetKwFunction{FMain}{\rm project}
 \SetKwProg{Fn}{Function}{:}{}
  \Fn{\FMain{$\mathcal{M}$, $\theta$, $v$}}{
        Initialization: $y=\theta$, $k=0$, ${\rm flag}=1$, a second order retraction ${\rm R}$ on $\mathcal{M}$\;
        \Repeat{$\| \Delta y\|_2\leq \varepsilon_0$ \text{ or } $k>{\rm{nmax}}$ }{
       $G\leftarrow{\rm Proj}_{T_{y}\mathcal{M}}({\rm Proj}_{T_{\theta}\m M}(y-\theta)-v)$\;
     
      Solve ${\rm Hess} f(y) [\Delta y]=-G$ for  $\Delta y\in T_y \mathcal{M}$, where $f: \m M\to \mb R$ is defined as $f(y)=\frac{1}{2}\|{\rm Proj}_{T_{\theta}\m M}(y-\theta)-v\|^2$\;
        $y\leftarrow {\rm R}_y({\Delta y})$\;
        $k\leftarrow k+1$\;
      }
      \If{$k>{\rm{nmax}}$ \text{ or } $y\notin  \m M$}
      {${\rm flag}\leftarrow 0$\;}
        \KwRet $[y,{\rm flag}]$\;
  }
 \end{algorithm}

\subsection{Algorithm for sampling from Bayesian RPETEL}
We now detail the RRWM algorithm for sampling from the Bayesian RPETEL posterior $\Pi_{\rm RP}(\dd\theta|X^{(n)})$, defined as 
 \begin{equation*} 
  \pi_{\rm RP} (\dd\theta|X^{(n)})=\frac{\exp(-\alpha_n \mathcal{R}_n(\theta))\prod_{i=1}^n p_i(\theta)\Pi(\dd\theta)}{\int_{\m M}  \exp(-\alpha_n \mathcal{R}_n(\theta))\prod_{i=1}^n p_i(\theta) \Pi(\dd\theta)},
  \end{equation*}
  where $(p_1(\theta),p_2(\theta), \cdots, p_n(\theta))$ is the solution of 
 \begin{equation*} 
   \begin{aligned}
 \max_{(w_1,w_2,\ldots,w_n)} & \sum_{i=1}^n \big[-w_i \log (n w_i)\big]\\
\mbox{subject to} \quad & \sum_{i=1}^n w_i=1, \sum_{i=1}^nw_i  {\rm grad}_{\theta},\,\ell(X_i,\theta)= 0_D,\quad w_1,w_2,\ldots,w_n \geq 0.
 \end{aligned}
\end{equation*}
Since $ {\rm grad}_{\theta}\, \ell(X,\theta)\in T_{\theta}\m M$, we can reformulate the constraints  $\sum_{i=1}^n w_i\cdot   {\rm grad}_{\theta}\, \ell(X_i,\theta)= 0_D$  by using a 
 matrix $W_{\theta}\in \mb R^{D\times d}$,  whose columns form a basis for $T_{\theta}\mathcal{M}$. This results in  $\sum_{i=1}^n w_i \cdot W_{\theta}^T  {\rm grad}_{\theta}\, \ell(X_i,\theta)=0_d$. 
By introducing Lagrange multipliers, we can rewrite the RETEL function $\prod_{i=1}^n p_i(\theta)$ as
  \begin{equation}\label{Eqn:ETEL2}
L(X^{(n)};\theta)=\prod_{i=1}^n\frac{\exp\left(\lambda(\theta)^T   {W}_{\theta}^T {\rm grad}_{\theta}\, \ell(X_i,\theta)\right)}{\sum_{i=1}^n\exp\left(\lambda(\theta)^T    {W}_{\theta}^T {\rm grad}_{\theta}\, \ell(X_i,\theta)\right)},
 \end{equation}
  where $\lambda(\theta)=\underset{\xi\in \mathbb{R}^d}{\arg\min} \sum_{i=1}^n \exp\big(\xi^T    {W}_{\theta}^T {\rm grad}_{\theta}\, \ell(X_i,\theta)\big)$. Note that the expression~\eqref{Eqn:ETEL2} is invariant to the choice of the basis $W_{\theta}$ and is equivalent to 
  \begin{equation*}
L(X^{(n)};\theta)=\prod_{i=1}^n\frac{\exp\left(\ov\lambda(\theta)^T   {\rm grad}_{\theta}\, \ell(X_i,\theta)\right)}{\sum_{i=1}^n\exp\left(\ov\lambda(\theta)^T  {\rm grad}_{\theta}\, \ell(X_i,\theta)\right)},
  \end{equation*}
  with $\ov\lambda(\theta)=\underset{\xi\in T_{\theta}\m M}{\arg\min}\sum_{i=1}^n \exp(\xi^T {\rm grad}_{\theta}\, \ell(X_i,\theta))$. Therefore, the computation of ETEL function can be simplified by solving an optimization problem on the manifold $T_{\theta}\m M$, which can be performed using  an adjusted Riemannian Newton’s algorithm  outlined  below.
  
\begin{algorithm}\label{algorithm.1}
\caption{Approximate solution to $\ov \lambda(\theta)$}
\SetAlgoLined
\SetKwRepeat{Do}{do}{while}%
     \SetKwFunction{FMain}{$\wh{\lambda}$}
 \SetKwProg{Fn}{Function}{:}{}
  \Fn{\FMain{$\theta$, $\mathcal{M}$, $\ell$}}{
     Define $f(\lambda)\leftarrow \frac{1}{n} \sum_{i=1}^{n} \exp \big(\lambda^{T} {\rm grad}_{\theta}\, \ell(X_i,\theta))\big)$\;
            $k\leftarrow0$\;
            $\ov{\lambda}^0=0$\;
        \Repeat{$\|\Delta \theta\|_2\leq \varepsilon$ }{
      $k\leftarrow k+1$\;
      $\gamma=1$\;
      $H\leftarrow\frac{1}{n}\sum_{i=1}^n\exp\left( {\rm grad}_{\theta}\, \ell(X_i,\theta)^T \ov \lambda^{k-1}\right) {\rm grad}_{\theta}\, \ell(X_i,\theta) {\rm grad}_{\theta}\, \ell(X_i,\theta)^T$\;
 $G\leftarrow\frac{1}{n}\sum_{i=1}^n\exp\left( {\rm grad}_{\theta}\, \ell(X_i,\theta)^T\ov{\lambda}^{k-1}\right) {\rm grad}_{\theta}\, \ell(X_i,\theta)$\;
  Solve $H\Delta\theta=-G$ for $\Delta \theta \in T_{\theta}\mathcal{M}$\;
 \Repeat{$f(\ov{\lambda}^k)\leq f(\ov{\lambda}^{k-1})$}
 { $\ov{\lambda}^k\leftarrow\ov{\lambda}^{k-1}+ \gamma\Delta\theta$\;
$ \gamma=\frac{1}{2}\gamma$;}
      }
   
    \KwRet$ \ov{\lambda}^k$\;
          }
 \end{algorithm}
 Now we can state the RRWM algorithm for sampling from $\Pi_{\rm RP}(\dd\theta|X^{(n)})$.

  \begin{algorithm}[H]\label{algorithmcom}
\caption{RRWM algorithm for sampling from Bayesian RPETEL posterior $\Pi_{\rm RP}(\dd\theta|X^{(n)})$}
\SetAlgoLined
\SetKwRepeat{Do}{do}{while}%
 \textbf{Input}: Number of iteration $L$, step size parameter $\wt h$, covariance matrix $\wt I$, initial distribution $\mu^0$\;
 \textbf{Data}: $X_1,X_2\cdots,X_n$\;
 Sampling $\theta^0$ from $\mu_0$\;
  $\lambda^0=\wh\lambda(\theta^0, \mathcal{M},\ell)$\;
 \For{$t \leftarrow 0 \,\,to\,\, L-1$}{
  Sample $\wt v$ from $\mathcal{N}(0,2\wt h\wt I)$ and let $v= {\rm Proj}_{T_{\theta^t}\mathcal{M}}(\wt v)$\;
  $[y,{\rm flag}]\leftarrow{\rm project}(\mathcal{M},\theta^t,v)$\;
    \eIf{${\rm flag}==0$}
  {$\theta^{t+1}\leftarrow \theta^t$\;}
  { % \eIf{$h_i(y)<0$ for some $i\in [l]$}
%      {$\theta^{t+1}\leftarrow\theta^t$\;}
     { 
     ${v}'\leftarrow {\rm Proj}_{T_{y}\mathcal{M}}(\theta^t-y)$\;
  Generate a uniform random  number $u\in(0,1)$\;
  
  $\lambda=\wh{\lambda}(y,\mathcal{M}, \ell)$\;
  \eIf{
  {$u> \frac{\pi(y)\exp\Big(\sum_{i=1}^n \log \frac{\exp( \lambda^T{\rm grad}_{\theta}\, \ell(X_i,y))}{\sum_{i=1}^n \exp(\lambda^T{\rm grad}_{\theta}\, \ell(X_i,y))}-\alpha_n \mathcal{R}_n(y)\Big)}{\pi(\theta^t)\exp\Big(\sum_{i=1}^n \log \frac{\exp((\lambda^0)^T{\rm grad}_{\theta}\, \ell(X_i,\theta^t))}{\sum_{i=1}^n \exp((\lambda^0)^T{\rm grad}_{\theta}\, \ell(X_i,\theta^t))}-\alpha_n \mathcal{R}_n(\theta^t)\Big)}  \frac{\exp\big(-v'^T(P_y^T\wt IP_y)^{\dagger} v'/(4\wt h)\big)}{\exp\big(-v^T (P_{\theta^{t}}^T\wt IP_{\theta^{t}})^{\dagger} v/(4\wt h))\big)}$}}
  { $\theta^{t+1}\leftarrow \theta^t$\;
   
  }{$[a,{\rm flag}]\leftarrow{\rm project}(\m M, y,{v}')$\;
      \eIf{${\rm flag}==0$}
        {$\theta^{t+1}\leftarrow \theta^t$\;}
  {$\theta^{t+1}\leftarrow y$\;
    $\lambda^{0}\leftarrow \lambda$\;}
  }}}
 }
 \end{algorithm}
In the context of the Bayesian RPETEL posterior with a non-smooth loss function, the term ${\rm grad_{\theta}}\ell(X,\theta)$ can be replaced with $g(X,\theta)\in T_{\theta}\m M$ that satisfies  Assumption 4.

  \subsection{Riemannian MALA algorithm}\label{sec:RMALA}
Let $\mu_{\m M}$ denote the volume measure of $\m M$. Consider a target distribution  that has a density $\mu^*(\theta)=\frac{\exp(-f(\theta))}{\int_{\m M}\exp(-f(\theta))\mu_{\m M}(\dd\theta)}$ with respect to $\mu_{\m M}$, where $f(\theta)$ is the potential of $\mu^*$. The Riemannian Metropolis-adjusted Langevin algorithm (MALA) produces $\{\theta^k\}_{k\geq 0}$ sequentially as follows: for $k=0,1,2,\cdots,$
 \begin{enumerate}
     \item  \textbf{(Initialization)} If $k=0$, sample $\theta^0$ from $\mu_0$;
     \item \textbf{(Proposal)} If $k\geq 1$,
     \begin{enumerate}
         \item {(generate random vector in tangent space)} sample $\wt v$ from $\mathcal{N}(-\wt h\wt I\cdot{\rm grad} f(\theta^{k-1}),{2\wt h}\wt I)$ and let $v={\rm Proj}_{T_{\theta^{k-1}}\m M}(\wt v)$;
         \item {(reject proposal if $v$ escape from $\wt V_{\theta^{k-1}}$)} if $ v\notin \wt V_{\theta^{k-1}}$, then $\theta^{k}= \theta^{k-1}$;
         \item {(``project'' back to manifold)} set $y={\phi}_{\theta^{k-1}}(v)$;
     \end{enumerate}
     \item  \textbf{(Metropolis-Hasting rejection/correction)} 
     \begin{enumerate}
         \item {(reject proposal if $\theta^{k-1}$ escape from $\wt U_y$)} if $\theta^{k-1}\notin \wt U_{y}$, then $\theta^{k}=\theta^{k-1}$;
         \item (set acceptance probability) let $v'=\psi_y(\theta^{k-1})$, set acceptance probability $A(\theta^{k-1},y)=1\wedge \alpha(\theta^{k-1},y)$ with acceptance ratio statistic 
         \begin{equation*}
            \begin{aligned}
          & \alpha(\theta^{k-1},y)=\frac{\mu^\ast(y)\cdot\big(\big|(\m D{\phi}_y(v')[P_{y}])^T\m D{\phi}_y(v')[P_{y}]\big|_{+}\big)^{-\frac{1}{2}}}{\mu^\ast(\theta^{k-1})\cdot\big(\big|(\m D{{\phi}_{\theta^{k-1}}}(v)[P_{\theta^{k-1}}])^T\m D{{\phi}_{\theta^{k-1}}}(v)[P_{\theta^{k-1}}]\big|_{+}\big)^{-\frac{1}{2}}}\\
    &\quad\cdot\frac{\exp\big(-(v'+\wt h\wt I{\rm grad} f(y))^T(P_y\wt I P_y)^{\dagger}(v'+\wt h\wt I{\rm grad} f(y))/(4\wt h)\big)}{\exp\big(-(v+\wt h\wt I{\rm grad}\, f(\theta^{k-1}))^T (P_{\theta^{k-1}}\wt I\wt P_{\theta^{k-1}})^{\dagger}(v+\wt h\wt I{\rm grad}\, f(\theta^{k-1}))/(4\wt h))\big)}.
                  \end{aligned}
         \end{equation*}

    \item (accept/reject the proposal)  flip a coin and accept $y$ with probability $A(\theta^{k-1},y)$ and set $\theta^k = y$; otherwise, set $\theta^k=\theta^{k-1}$.
     \end{enumerate}
 \end{enumerate}
 
\vspace{0.5em}
% \textbf{\noindent Formula for computing the Riemannian gradient of the potential of Bayesian RPETEL}: 
\noindent Then consider the Bayesian RPETEL posterior $  \Pi_{\rm RP} (\dd\theta|X^{(n)})$ whose density with respect to the volume measure $\mu_{\m M}$ is given by 
 \begin{equation*} 
  \pi_{\rm RP} (\theta|X^{(n)})=\frac{\pi(\theta)\exp(-\alpha_n \mathcal{R}_n(\theta))\prod_{i=1}^n p_i(\theta)}{\int \pi(\theta)\exp(-\alpha_n \mathcal{R}_n(\theta))\prod_{i=1}^n p_i(\theta) \,\dd{\mu_{\m M}}}.
  \end{equation*}
The potential $f$ of  $\Pi_{\rm RP}^{(n)}$ is then given by \begin{equation}\label{potential}
      f(\theta)=\alpha_n\m R_n(\theta)-\sum_{i=1}^n \log p_i(\theta)-\log \pi(\theta).
  \end{equation}
 We then have the following lemma about the Riemannian gradient of $f$.
 \begin{lemma}\label{RGradientpotential}
Denote $g(X,\theta)=(g_1(X,\theta),g_2(X,\theta),\cdots,g_D(X,\theta))={\rm grad}_{\theta}\, \ell(X,\theta)$ and  ${\rm grad}_{\theta} \,g(X,\theta)=[{\rm grad}_{\theta}\,g_1(X,\theta),{\rm grad}_{\theta}\,g_2(X,\theta),\cdots,{\rm grad}_{\theta}\, g_D(X,\theta)]$ for $\theta\in \m M$. The Riemannian gardient of $f$ given in~\eqref{potential} on $\m M$ is 
 \begin{equation*}
 \begin{aligned}
   {\rm grad} \,f(\theta)&=\frac{\alpha_n}{n}\sum_{i=1}^n g(X_i,\theta) -\sum_{i=1}^n g_{\ov\lambda}(\theta)^T g(X_i,\theta)\\
   &\,-\sum_{i=1}^n [{\rm grad}_{\theta} \,g(X_i,\theta)]\cdot\ov\lambda(\theta)+n\cdot\frac{\sum_{i=1}^n \exp(\ov\lambda(\theta)^T g(X_i,\theta))[{\rm grad}_{\theta} \,g(X_i,\theta)]\cdot\ov\lambda(\theta)}{\sum_{i=1}^n \exp(\ov\lambda(\theta)^T g(X_i,\theta))},
      \end{aligned}
 \end{equation*}
 where  $\ov\lambda(\theta)={\arg\min}_{\xi\in T_{\theta}\m M}\sum_{i=1}^n \exp(\xi^Tg(X_i,\theta))$ and 
 \begin{equation*}
     \begin{aligned}
       g_{\ov\lambda}(\theta)&=-\left(\sum_{i=1}^n \exp(\ov\lambda(\theta)^T g(X_i,\theta))g(X_i,\theta)g(X_i,\theta)^T\right)^{+}\\
       &\cdot\left(\sum_{i=1}^n \exp(\ov\lambda(\theta)^T g(X_i,\theta))\left(I_D+g(X_i,\theta)\ov\lambda(\theta)^T\right)[{\rm grad}_{\theta} \,g(X,\theta)]^T\right),
     \end{aligned}
 \end{equation*}
 with $A^{+}$ being  the Moore–Penrose inverse of $A$.
 \end{lemma}

\section{Auxiliary and Supporting Lemmas}\label{proof:main}
This appendix contains supporting lemmas  used in the main proofs. We first show that local smoothness at $\theta^*$ implies local smoothness at any $\theta_0$ sufficiently close to $\theta^*$.

\begin{lemma}\label{lemmasmootharoundtheta}
Let $\m M$ be a $d$-dimensional Riemannian submanifold embedded in $\mb R^D$. If $\m M$ is locally $C^3_{r,L}$-smooth at $\theta^*\in \m M$, then  there exist positive constants $r_1,L_1$ such that, for any $\theta_0\in B_{r/2}(\theta^*)\cap \m M$, the manifold $\m M$ is locally $C^3_{r_1,L_1}$-smooth at $\theta_0$.
\end{lemma}

Next, we state a lemma used to prove the manifold Bernstein-von Mises results for all posteriors considered in the main text.
Consider a prior measure $\Pi_{\m M}$ defined on $\m M$ and a point $\theta^*\in \m M$ so that the manifold $\m M$ and the prior $\Pi_{\m M}$ satisfy Assumption 1 and 2 in the main text.  Then since $\m M$ is $C^3_{r,L}$-smooth at $\theta^*$,  there exists $U_{\theta^*},V_{\theta^*}$ with $B_{r}(\theta^*)\cap \m M\subseteq U_{\theta^*}\subseteq \m M$ and $B_{r}(0_D)\cap T_{\theta^*}\m M\subseteq V_{\theta^*}\subset T_{\theta^*}\m M$, so that $\psi_{\theta^*}: U_{\theta^*}\to V_{\theta^*}$ defined by $\psi_{\theta^*}(x)={\rm Proj}_{T_{\theta^*}\m M}(x-\theta^*)$ is a bijective, where the inverse, denoted by $\phi_{\theta^*}$ is thrice Fr\'{e}chet differentiable, and its Fr\'{e}chet derivatives up to order three have operator norms uniformly bounded by $L$.
% Then using Lemma~\ref{lemmasmootharoundtheta}, for any $\theta\in B_{r/2}(\theta^*)$,  there exists $U_{\theta},V_{\theta}$ with $B_{r_1}(\theta)\cap \m M\subseteq U_{\theta}\subseteq \m M$ and $B_{r_1}(0_D)\cap T_{\theta^*}\m M\subseteq V_{\theta}\subset T_{\theta^*}\m M$, so that $\psi_{\theta}: Ui_{\theta}\to V_{\theta}$ defined by $\psi_{\theta}(x)={\rm Proj}_{T_{\theta}\m M}(x-\theta)$ is a bijective, where the inverse, denoted by $\phi_{\theta}$ is thrice Fr\'{e}chet differentiable, and its Fr\'{e}chet derivatives up to order three have operator norms uniformly bounded by $L_1$.
With this notation in place, we can derive the following lemma. We first state the required assumption on low-dimensional summaries of the parameter.

 \medskip
\noindent \textbf{Assumption D (Conditions for function $f$):} There exist some positive constants $(r,C,L,a,b)$ so that the function $f=(f_1,f_2\cdots,f_p): \m M\to \mb R^p$ satisfies that (1) $1\leq p\leq d$; (2) ${\sup}_{\theta\in \m M}\|f(\theta)\|\leq C$; (3) for any $j\in[p]$, $f_j(\cdot)$ has $L$-Lipschitz continuous Riemannian gradient on $B_r(\theta^*)\cap \m M$; (4) denote $J_f=[{\rm grad} f_1(\theta^*),\cdots,{\rm grad} f_p(\theta^*)]^T$,    it holds that $a^2I_p\preccurlyeq J_fJ_f^T \preccurlyeq b^2 I_p$.

\begin{lemma}\label{lemma:generalresult}
Let $\m M$ be a submanifold satisfies Assumption 1. Consider a prior $\Pi=\Pi_{\m M}$ satisfies Assumption 2, and a map $\ms L: \m X^{n}\times \m M\to \mb R$. For any $X^{(n)}=(X_1,X_2,\cdots,X_n)\in \m X^{n}$, define the measure 
    \begin{equation*}
        \Pi^{(n)}(\dd \theta)=\frac{\exp(\ms L(X^{(n)},\theta)) \Pi_{\m M}(\dd \theta)}{\int_{\m M}\exp(\ms L(X^{(n)},\theta)) \Pi_{\m M}(\dd \theta)}.
    \end{equation*}
    Suppose there exist a set $\m A\in \m X^{(n)}$, a $D\times d$ matrix $W_{\theta^*}$ whose columns form an orthonormalbasis of $T_{\theta^*}\m M$, a map $\wh\theta^\diamond: \m A\to \m M$, a positive definite $d\times d$ matrix $\Sigma$, and absolute constants $\delta,\delta_1,c,\gamma_2>0$, $\gamma_1\in (0,\frac{1}{2}]$, $\gamma_0\geq 1$ and $c_1\geq 2$, such that, for any $X^{(n)}\in \m A$, writing $\wh\theta^\diamond=\wh\theta^\diamond(X^{(n)})$, the following hold:
    \begin{enumerate}
        \item $\|\wh\theta^\diamond-\theta^*\|\leq c\sqrt{\frac{\log n}{n}}$, $  \Pi^{(n)}(\|\theta-\wh\theta^\diamond\|\geq \delta)\leq n^{-c_1}$ and  $B_{\delta}(\wh\theta^\diamond)\cap \m M\subseteq U_{\theta^*}$.
        \item For any  $h\in B_{\delta_1(\log n)^{3/2}}(0_d)$, 
        \begin{equation*}
            \begin{aligned}
          & \bigg| \log\ms L\Big(X^{(n)},\phi_{\theta^*}\Big(\frac{W_{\theta^*}h}{\sqrt{n}}+\psi_{\theta^*}(\wh\theta^\diamond)\Big)\Big)+ \frac{1}{2} h^T \Sigma^{-1} h\bigg|\\
          &\qquad\leq C\frac{(\log n)^{\gamma_0}}{n^{\gamma_1}}(\|h\|^{\gamma_2}+1).
            \end{aligned}
        \end{equation*}
        
        \item For any $h\in\mb R^d$ satisfying  $\|h\|\geq \delta_1(\log n)^{3/2}$ and $\frac{W_{\theta^*}h}{\sqrt{n}}+\psi_{\theta^*}(\wh\theta^\diamond)\in V_{\theta^*}$,  it holds that  $    \log\ms L\big(X^{(n)},\phi_{\theta^*}\big(\frac{W_{\theta^*}h}{\sqrt{n}}+\psi_{\theta^*}(\wh\theta^\diamond)\big)\big)\leq -2c_1\,d\, \log n.$
 
    \end{enumerate}
    Then there exist positive constants $C_0,C_1,C_2$ so that for any $X^{(n)}\in \m A$, 
    \begin{enumerate}
        \item $\Pi^{(n)}(\|\theta-\wh\theta^\diamond\|\geq C_0\sqrt{\frac{\log n}{n}})\leq C_1 n^{-c_1}$,
        \item  Consider the projection of the posterior mean to the manifold: $\wh\theta_p={\arg \min}_{y\in \mathcal{M}}  \|y-\mathbb{E}_{\Pi^{(n)}}[\theta]\|^2$, when $n$ is sufficiently large, $\wh\theta_p$ is uniquely defined and it holds that  $\|\wh\theta_p-\wh\theta^\diamond\|\leq C_1\,\frac{(\log n)^{\gamma_0}}{n^{\gamma_1+\frac{1}{2}}}$.
        \item    for any $f: \m M\to \mb R^p$  satisfying Assumption D, it holds that
 $${\rm TV}\Big(f_{\#}\Pi^{(n)},\m N\big(f(\wh\theta^\diamond),n^{-1}J_fW_{\theta^*}\Sigma W_{\theta^*}^TJ_f^T\big)\Big)\leq C_2 \frac{(\log n)^{\gamma_0}}{n^{\gamma_1}}.$$
 \end{enumerate}
    \end{lemma}

\subsection{Proof of Lemma~\ref{lemmasmootharoundtheta}}

We will use the following lemma to show the desired result; its proof   follows directly from the proof of the first statement of Lemma F.4 in~\cite{Supplement2291}.
\begin{lemma}\label{lemmasmootharoundtheta1}
 Let $\m M$ be a $d$-dimensional Riemannian submanifold embedded in $\mb R^D$ and let $\theta_0\in \m M$. Suppose there exist positive constants $(r,L)$,  sets $U\subseteq \m M$ and $\Upsilon\subseteq \mb R^d$,  and a map $\xi: U\to \Upsilon$ such that (1) $\m M\cap B_{r}(\theta_0)\subseteq U$ and $B_{r}(\xi(\theta_0))\subset \Upsilon$; (2) $\xi$ is a bijective, and its inverse $\xi^{-1}$ is three times differentiable, with all mixed partial derivatives up to order three are bounded in absolute values by $L$; (3) $\inf_{z\in \Upsilon} \lambda_{\min}(J_{\xi_{-1}}(z)^TJ_{\xi_{-1}}(z))\geq 1/L$, where $ \lambda_{\min}(\cdot)$ denotes the minimal eigenvalue and $J_{\xi^{-1}}$ is the Jacobian matrix of $\xi^{-1}$. Then there exist positive constants $(r_1,L_1)$ such that $\m M$ is $C^3_{r_1,L_1}$-smooth around $\theta_0$.
\end{lemma}
So we only need to verify the three conditions in Lemma~\ref{lemmasmootharoundtheta1}. Notice that since $\m M$ is locally $C^3_{\tau,r,L}$-smooth around $\theta^*\in \m M$, there exists $U,V$ with $B_{r}(\theta^*)\cap \m M\subseteq U\subseteq \m M$ and $B_{r}(0_D)\cap T_{\theta^*}\m M\subseteq V\subseteq T_{\theta^*}\m M$, so that $\psi_{\theta^*}: U\to V$ defined by $\psi_{\theta^*}(x)={\rm Proj}_{T_{\theta^*}\m M}(x-\theta)$ is a bijective, where the inverse, denoted by $\phi_{\theta^*}$ is thrice Fr\'{e}chet differentiable, and its Fr\'{e}chet derivatives up to order three have operator norms uniformly bounded by $L$.

Then let $W_{\theta^*}\in \mb R^{D\times d}$ be an arbitrary orthonormal basis of $T_{\theta^*}\m M$. Define $\Upsilon=B_{r}(0_d)$ and $U_1=\{x=\phi_{\theta^*}(W_{\theta^*}z)\,:\, z\in \Upsilon\}$. Then notice that for any $x\in B_r(\theta^*)\cap \m M$, it holds that $\|W_{\theta^*}^T\psi_{\theta^*}(x)\|< r$. We have $B_{r}(\theta^*)\cap \m M\subseteq U_1$. Define
\begin{equation*}
    \begin{aligned}
        \xi: U_1\to \Upsilon,\, \xi(\theta)=W_{\theta^*}^T(\theta-\theta^*).
    \end{aligned}
\end{equation*}
Then $\xi$ is a bijective with inverse given by $\xi^{-1}(z)=\phi_{\theta^*}(W_{\theta^*}z)$. Moreover, $\xi^{-1}$ is three-times differentiable, and there exists a constant $L_1$ so that all mixed partial derivatives up to order three are bounded in absolute values by $L_1$, this verifies the second condition  in Lemma~\ref{lemmasmootharoundtheta1}. Now using the fact that, for any $z\in \Upsilon=B_{r}(0_d)$, it holds that
\begin{equation*}
    J_{\xi(\xi^{-1}(\cdot))}(z)=I_d \,\Rightarrow\,  W_{\theta^*}^T J_{\xi^{-1}}(z)=I_d.
\end{equation*}
Therefore,  
\begin{equation*}
    \underset{z\in \Upsilon}{\inf} \lambda_{\min}(J_{\xi^{-1}}(z)^TJ_{\xi^{-1}}(z))\geq 1,
\end{equation*}
this verifies the third condition  in Lemma~\ref{lemmasmootharoundtheta1}.
 Moreover, for any $\theta_0\in B_{r/2}(\theta^*)\cap \m M$,  it holds that $\|\xi(\theta_0)\|< r/2$ and $B_{r/2}(\xi(\theta_0))\subseteq B_{r}(0_d)=\Upsilon$. Moreover, $B_{r/2}(\theta_0)\cap \m M\subseteq  B_{r}(\theta^*)\cap \m M \subseteq U_1$. So the first condition  in Lemma~\ref{lemmasmootharoundtheta1} is also verified, and this completes the proof.

\subsection{Proof of Lemma~\ref{lemma:generalresult}}
By the thrice differentiablity of $\phi_{\theta^*}$, there exists a constant $C$ so that 
 
\begin{equation*}
    \underset{v\in  V_{\theta^*} }{\sup} \frac{\|\phi_{\theta^*}(v)-(v+\theta^*)\|}{\|v\|^2}\leq C,
\end{equation*}
and
\begin{equation*}
\begin{aligned}
      \underset{\theta'\in U_{\theta^*}}{\sup}\frac{\|\psi_{\theta^*}(\theta')-(\theta'-\theta^*)\|}{\|\theta^*-\theta'\|^2}&= \underset{\theta'\in U_{\theta^*}}{\sup}\frac{\|\phi_{\theta^*}(\psi_{\theta^*}(\theta'))-(\theta^*+\psi_{\theta^*}(\theta'))\|}{\|\theta^*-\theta'\|}\\
      &\leq  \underset{v\in  V_{\theta^*} }{\sup} \frac{\|\phi_{\theta^*}(v)-(v+\theta^*)\|}{\|v\|^2}\leq C.
\end{aligned}
 \end{equation*}
Then let's fix an $X^{(n)}\in \m A$ and write $\wh\theta^\diamond=\wh\theta^\diamond(X^{(n)})$. Define $V^\diamond_{\theta^*}=\{W_{\theta^*}^T(y-\psi_{\theta^*}(\wh\theta^\diamond))\,:\, y\in V_{\theta^*}\}$, and the 
function $  \psi^\diamond: \mb R^D\to \mb R^d$ by $\psi^\diamond(\theta)=W_{\theta^*}^T(\theta-\wh\theta^\diamond)$. Its restriction $\psi^\diamond|_{U_{\theta^*}}: U_{\theta^*}\to  V^\diamond_{\theta^*}$ satisfies, for any $\theta\in U_{\theta^*}$, $\psi^\diamond|_{U_{\theta^*}}(\theta)=\psi^\diamond(\theta)=W_{\theta^*}^T(\psi_{\theta^*}(\theta)-\psi_{\theta^*}(\wh\theta^\diamond))$. Then we define the function $\phi^\diamond: V^\diamond_{\theta^*}\to U_{\theta^*}$ by $\phi^\diamond(z)=\phi_{\theta^*}\big(W_{\theta^*}z+\psi_{\theta^*}(\wh\theta^\diamond)\big)$. It is straightforward to verify that $\phi^\diamond$ is the inverse of  $\psi^\diamond|_{U_{\theta^*}}$, and $\phi^\diamond$ is thrice differentiable with $\bold{J}_{\phi^{\diamond}}(-W_{\theta^*}^T\psi_{\theta^*}(\wh\theta^\diamond))=W_{\theta^*}$. 

\medskip
\noindent\textbf{Step 1. We will first show that}
\begin{equation}\label{statement:TVpsi}
    {\rm TV}(\psi^\diamond_{\#}\Pi^{(n)}, \m N(0, n^{-1}\Sigma))\lesssim \frac{(\log n)^{\gamma_0}}{n^{\gamma_1}}.
\end{equation}
Using the assumption that $\|\wh\theta^\diamond-\theta^*\|\leq c\sqrt{\frac{\log n}{n}}$ and $B_{\delta}(\wh\theta^\diamond)\cap \m M\subseteq U_{\theta^*}$, there exist positive constants $C_1,C_2$ so that for any $\theta\in B_{\delta}(\wh\theta^\diamond)\cap \m M$ and $y=\psi^\diamond(\theta)$,
 \begin{equation}\label{eqn:diffmatrix}
 \begin{aligned}
&\Bmnorm{\mathbf{J}_{\phi^{\diamond}}(y)- W_{\theta^*}}_{\rm op}=\Bmnorm{\mathbf{J}_{\phi^{\diamond}}(y)- \bold{J}_{\phi^{\diamond}}(-W_{\theta^*}^T\psi_{\theta^*}(\wh\theta^\diamond))}_{\rm op}\leq C_1\|y+W_{\theta^*}^T\psi_{\theta^*}(\wh\theta^\diamond)\|,\\
 &\mathcal{J}^{\diamond}(y)=\sqrt{\Big|\mathbf{J}_{\phi^{\diamond}}(y)^{T} \mathbf{J}_{\phi^{\diamond}}(y)\Big|}\leq C_2.
      \end{aligned}
 \end{equation}
Moreover, when $n$ is large enough,  there exist positive constants $r_1$ so that for any $\theta\in B_{r_1}(\theta^*)\cap \mathcal{M}$ and $y=\psi^\diamond(\theta)$,
 \begin{equation*} 
  \mathcal{J}^{\diamond}(y)\geq 1/2.
 \end{equation*}
Let $h=\sqrt{n}\cdot\psi^\diamond(\theta)$ be the local coordinate, and define    $\m B_\delta=\{y= \psi^{\diamond}(\theta)\,:\, \theta\in \mathcal{M},\, \|\theta-\wh{\theta}^{\diamond}\|\leq \delta\}$. Let $\pi(\cdot)$ be the density function of $\Pi_{\m M}$ with respect to the volume measure of $\m M$. We will start by showing the following result.

 \medskip
 \noindent\textbf{Step 1.1. Show}
 \begin{equation}\label{eqnstatementTV}
 \begin{aligned}
& \int \bigg|\pi\Big(\phi^{\diamond}\big(\frac{h}{\sqrt{n}}\big)\Big)\textbf{1}\Big(\frac{h}{\sqrt{n}}\in \m B_{\delta}\Big)\mathcal{J}^\diamond\big( \frac{h}{\sqrt{n}}\big)\cdot\exp\Big(\log \ms L\big(X^{(n)},\phi^{\diamond}\big(\frac{h}{\sqrt{n}}\big)\big)\Big)\\&\qquad -\pi(\theta^*)\exp\big(-\frac{1}{2} h^T \Sigma^{-1}h\big)\bigg|\,\dd h\leq C\, \frac{(\log n)^{\gamma_0}}{n^{\gamma_1}}.
 \end{aligned}
 \end{equation}
 To show the desired result, we define the following set of $h\in \mathbb{R}^d$,
\begin{equation}\label{hset}
 \begin{aligned}
 &A_1=\left\{\|h\|_2\leq \delta_2\sqrt{\log n}\right\},\\
 &A_2=\left\{\delta_2\sqrt{\log n}\leq \|h\|_2\leq \delta_1(\log n)^{1.5}\right\},\\
 &A_3=\left\{\|h\|_2\geq \delta_1 (\log n)^{1.5}\right\},\\
   \end{aligned}
 \end{equation}
where $\delta_2$ is a large enough constant that will be chosen later, $\delta_1$ is the one specified in the assumption. Then notice that when $\frac{h}{\sqrt{n}}=\psi^\diamond(\theta)\in \m B_{\delta}$, it holds that $\frac{W_{\theta^*}h}{\sqrt{n}}+\psi_{\theta^*}(\wh\theta^\diamond)=W_{\theta^*}W_{\theta^*}{}^T(\theta-\theta^*)=\psi_{\theta^*}(\theta)$, where $\theta\in B_{\delta}(\wh\theta^\diamond)\cap \m M\subseteq U_{\theta^*}$. Therefore, by the third condition in the assumptions, it holds for any $h\in A_3$ with $\frac{h}{\sqrt{n}}\in \m B_{\delta}$, that  $$\log\ms L\big(X^{(n)},\phi^\diamond\big(\frac{h}{\sqrt{n}}\big)\big)\leq -2c_1\,d\, \log n.$$ We can derive that, when $n$ is large enough,
\begin{equation}\label{eqnA3}
 \begin{aligned}
&  \int_{A_3} \bigg|\pi\Big(\phi^{\diamond}\big(\frac{h}{\sqrt{n}}\big)\Big)\textbf{1}\Big(\frac{h}{\sqrt{n}}\in \m B_{\delta}\Big)\mathcal{J}^\diamond\big( \frac{h}{\sqrt{n}}\big)\cdot\exp\Big(\log \ms L\big(X^{(n)},\phi^{\diamond}\big(\frac{h}{\sqrt{n}}\big)\big)\Big)\\
&\qquad\qquad-  \pi(\theta^*)\exp\big(-\frac{1}{2} h^T  \Sigma^{-1} h\big)\bigg|\,\dd h\\
 &  \leq \int_{A_3} \pi\Big(\phi^{\diamond}\big(\frac{h}{\sqrt{n}}\big)\Big)\textbf{1}\Big(\frac{h}{\sqrt{n}}\in \m B_{\delta}\Big)\mathcal{J}^\diamond\big( \frac{h}{\sqrt{n}}\big)\cdot\exp\Big(\log \ms L\big(X^{(n)},\phi^{\diamond}\big(\frac{h}{\sqrt{n}}\big)\big)\Big) \,\dd h\\
 &\qquad+\int_{A_3} \pi(\theta^*)\exp\big(-\frac{1}{2} h^T  \Sigma^{-1}h\big) \,\dd h\\
&\leq \exp(-2c_1 d\log n) \cdot n^{\frac{d}{2}} \int_{\m B_{\delta}}\pi(\phi^{\diamond}(y)) \m J^{\diamond}(y)\,\dd y+\int_{A_3} \pi(\theta^*)\exp\big(-\frac{1}{2} h^T  \Sigma^{-1}h\big) \,\dd h\\
&\leq n^{-c_1},
 \end{aligned}
 \end{equation}
 where the last inequality  follows from the positive definiteness of $\Sigma$ and the sub-exponential tail bound (see for example,~\cite{wainwright_2019})  to control $\int_{A_3} \pi(\theta^*)\exp\big(-\frac{1}{2} h^T  \Sigma^{-1}h\big) \,\dd h$.
Next we consider $h\in A_2$, by the second condition in the assumptions, there exist  positive constants $C_3,C_4,C_5$ so that when $\delta_2\geq C_3$,
\begin{equation}\label{eqnA2}
 \begin{aligned}
  &\int_{A_2} \pi\Big(\phi^{\diamond}\big(\frac{h}{\sqrt{n}}\big)\Big)\textbf{1}\Big(\frac{h}{\sqrt{n}}\in \m B_{\delta}\Big)\mathcal{J}^\diamond\big( \frac{h}{\sqrt{n}}\big)\cdot\exp\Big(\log \ms L\big(X^{(n)},\phi^{\diamond}\big(\frac{h}{\sqrt{n}}\big)\big)\Big) \,\dd h\\
  &\leq  \int_{A_2} \pi\Big(\phi^{\diamond}\big(\frac{h}{\sqrt{n}}\big)\Big)\textbf{1}\Big(\frac{h}{\sqrt{n}}\in \m B_{\delta}\Big)\mathcal{J}^\diamond\big( \frac{h}{\sqrt{n}}\big)\cdot\exp\Big(-\frac{1}{2} h^T \Sigma^{-1} h+C_4\frac{(\log n)^{\gamma_0+1.5\gamma_2}}{n^{\gamma_1}}\Big) \,\dd h\\
  &\leq C_5 \int_{A_2} \exp(-\frac{1}{2} h^T \Sigma^{-1}h)\,\dd h  \leq n^{-c_1},
  \end{aligned}
 \end{equation}
 and 
 \begin{equation*}
 \begin{aligned}
  \int_{A_2} \pi(\theta^*)\exp(-\frac{1}{2} h^T \Sigma^{-1}h)\,\dd h \leq n^{-c_1}.
  \end{aligned}
 \end{equation*}
Then we consider $h\in A_1$,
\begin{equation*}
 \begin{aligned}
& \int_{A_1} \bigg|\pi\Big(\phi^{\diamond}\big(\frac{h}{\sqrt{n}}\big)\Big)\textbf{1}\Big(\frac{h}{\sqrt{n}}\in \m B_{\delta}\Big)\mathcal{J}^\diamond\big( \frac{h}{\sqrt{n}}\big)\cdot\exp\Big(\log \ms L\big(X^{(n)},\phi^{\diamond}\big(\frac{h}{\sqrt{n}}\big)\big)\Big)-  \pi(\theta^*)\exp\big(-\frac{1}{2} h^T  \Sigma^{-1} h\big)\bigg|\,\dd h\\
&\leq \underbrace{\int_{A_1}\pi\Big(\phi^{\diamond}\big(\frac{h}{\sqrt{n}}\big)\Big)\textbf{1}\Big(\frac{h}{\sqrt{n}}\in \m B_{\delta}\Big)\mathcal{J}^\diamond\big( \frac{h}{\sqrt{n}}\big)\cdot\bigg|\exp\Big(\log \ms L\big(X^{(n)},\phi^{\diamond}\big(\frac{h}{\sqrt{n}}\big)\big)\Big) -\exp(-\frac{1}{2} h^T \Sigma^{-1}h)\bigg|\,\dd h}_{(I_A)}\\
&+\underbrace{\int_{A_1}\exp(-\frac{1}{2} h^T \Sigma^{-1}h)\Big|\pi\Big(\phi^{\diamond}\big(\frac{h}{\sqrt{n}}\big)\Big)\textbf{1}\Big(\frac{h}{\sqrt{n}}\in \m B_{\delta}\Big)\mathcal{J}^\diamond\big( \frac{h}{\sqrt{n}}\big)-\pi(\theta^*)\Big|\,\dd h}_{(I_B)}.
 \end{aligned}
 \end{equation*}
 For the term $(I_A)$, by the second condition in the assumption, we have, for some positive constants $C,C_1$ that,
 \begin{equation*}
 \begin{aligned}
 &\int_{A_1}\pi\Big(\phi^{\diamond}\big(\frac{h}{\sqrt{n}}\big)\Big)\textbf{1}\Big(\frac{h}{\sqrt{n}}\in \m B_{\delta}\Big)\mathcal{J}^\diamond\big( \frac{h}{\sqrt{n}}\big)\cdot\bigg|\exp\Big(\log \ms L\big(X^{(n)},\phi^{\diamond}\big(\frac{h}{\sqrt{n}}\big)\big)\Big) -\exp(-\frac{1}{2} h^T \Sigma^{-1}h)\bigg|\,\dd h\\
&\leq C\,\int_{A_1} \exp(-\frac{1}{2} h^T \Sigma^{-1}h)  \frac{(\log n)^{\gamma_0}}{n^{\gamma_1}}(\|h\|^{\gamma_2}+1) \,\dd h\\
&\leq C_1\,\frac{ (\log n)^{\gamma_0}}{n^{\gamma_1}}.
  \end{aligned}
 \end{equation*}
 For the term $(I_B)$, by using the Lipschitz continuity of the prior and the local $C^3$ smoothness of the manifold around $\theta^*$ (Assumption 1) , it holds for any $h\in A_1$ that,
 \begin{equation*}
 \begin{aligned}
&\Big| \pi\big(\phi^{\diamond}\big(\frac{h}{\sqrt{n}}\big)\big)-\pi(\theta^*)\Big|\leq L\,\big \|\phi^{\diamond}\big(\frac{h}{\sqrt{n}}\big)-\theta^*\|=L\, \big\|\phi_{\theta^*}\big(W_{\theta^*}\frac{h}{\sqrt{n}}+\psi_{\theta^*}(\wh\theta^\diamond)\big)-\phi_{\theta^*}(0)\big\|\\
&\leq L\, \big\|W_{\theta^*} \frac{h}{\sqrt{n}}+\psi_{\theta^*}(\wh\theta^\diamond)\big\|+C\,\big\|W_{\theta^*} \frac{h}{\sqrt{n}}+\psi_{\theta^*}(\wh\theta^\diamond)\big\|^2\leq C_1\,\frac{\|h\|_2+\sqrt{\log n}}{\sqrt{n}}.
\end{aligned}
 \end{equation*}
  \begin{equation*}
 \textbf{1}\Big(\frac{h}{\sqrt{n}}\in \m B_{\delta}\Big)=1,
  \end{equation*}
 and 
 \begin{equation*}
  \begin{aligned}
&\big|\mathcal{J}^\diamond\big( \frac{h}{\sqrt{n}}\big)-1\big|=\bigg|\sqrt{\Big|\mathbf{J}_{\phi^{\diamond}}(\frac{h}{\sqrt{n}})^{T} \mathbf{J}_{\phi^{\diamond}}(\frac{h}{\sqrt{n}})\Big|}-\sqrt{\Big|W_{\theta^*}^TW_{\theta^*}\Big|}\bigg|\leq C_1\,\frac{\|h\|_2+\sqrt{\log n}}{\sqrt{n}}.
  \end{aligned}
  \end{equation*}
  where the last inequality uses equation~\eqref{eqn:diffmatrix}. Therefore, we can get $(I_B)\lesssim \sqrt{\frac{\log n}{n}}$, and for some positive constant $C$,
\begin{equation*}
\begin{aligned}
  &\int_{A_1} \bigg|\pi\Big(\phi^{\diamond}\big(\frac{h}{\sqrt{n}}\big)\Big)\textbf{1}\Big(\frac{h}{\sqrt{n}}\in \m B_{\delta}\Big)\mathcal{J}^\diamond\big( \frac{h}{\sqrt{n}}\big)\cdot\exp\Big(\log \ms L\big(X^{(n)},\phi^{\diamond}\big(\frac{h}{\sqrt{n}}\big)\big)\Big)\\
  &\qquad\qquad-  \pi(\theta^*)\exp\big(-\frac{1}{2} h^T  \Sigma^{-1} h\big)\bigg|\,\dd h \leq C\, \frac{ (\log n)^{\gamma_0}}{n^{\gamma_1}}.  
\end{aligned}
\end{equation*}
Putting together the bounds over $A_1$, $A_2$, and $A_3$ gives the desired result in \eqref{eqnstatementTV}.

 \medskip
 \noindent\textbf{Step 1.2. Show that there exist positive constants $C_0,C_1$ such that}
\begin{equation*}
         \Pi^{(n)}\Big(\|\theta-\wh\theta^\diamond\|\geq C_0\sqrt{\frac{\log n}{n}}\Big)\leq C_1n^{-c_1}.
\end{equation*}
This is precisely the first claim in Lemma~\ref{lemma:generalresult}. 

\medskip

 \noindent Note that since $\Pi^{(n)}(\|\theta-\wh\theta^\diamond\|\geq \delta)\leq n^{-c_1}$, we have
\begin{equation*}
   \begin{aligned}
          \Pi^{(n)}\Big(\|\theta-\wh\theta^\diamond\|\geq C_0\sqrt{\frac{\log n}{n}}\Big)&\leq \Pi^{(n)}\Big(C_0\sqrt{\frac{\log n}{n}}\leq \|\theta-\wh\theta^\diamond\|\leq \delta \Big)+n^{-c_1}\\
          &\leq \Pi^{(n)}\Big(|\|\theta-\wh\theta^\diamond\|\geq C_0\sqrt{\frac{\log n}{n}}\, \Big|\,\|\theta-\wh\theta^\diamond\|\leq \delta \Big)+n^{-c_1}.
             \end{aligned}
   \end{equation*}
Moreover, recall that for any $v\in V_{\theta^*}$, $\|\phi_{\theta^*}(v)-v-\theta^*\|\leq C\, \|v\|^2$ and for any $\theta\in U_{\theta^*}$, $\|\psi_{\theta^*}(\theta)-(\theta-\theta^*)\|\leq C\|\theta-\theta^*\|^2$.  Consider any  $z\in \mb R^d$ with $\|z\|\leq \frac{C_0}{2}\sqrt{\frac{\log n}{n}}$, then for $\theta=\phi^\diamond(z)=\phi_{\theta^*}(W_{\theta^*}z+\psi_{\theta^*}(\wh\theta^\diamond))$, we have for sufficiently large $n$,
\begin{equation*}
\begin{aligned}
        \|\theta-\wh\theta^\diamond\|&=\|\theta-(\theta^*+W_{\theta^*}z+\psi_{\theta^*}(\wh\theta^\diamond))\|+\|\theta^*+\psi_{\theta^*}(\wh\theta^\diamond)-\wh\theta^\diamond\|+\|W_{\theta^*}z\|\\
        &\leq C\|W_{\theta^*}z+\psi_{\theta^*}(\wh\theta^\diamond)\|^2+C\, \|\wh\theta^\diamond-\theta^*\|^2+\frac{C_0}{2}\sqrt{\frac{\log n}{n}}\leq C_0\sqrt{\frac{\log n}{n}}.
\end{aligned}
\end{equation*}
 Therefore, we can derive
\begin{equation*}
   \begin{aligned}
          &\Pi^{(n)}\Big(\|\theta-\wh\theta^\diamond\|\geq C_0\sqrt{\frac{\log n}{n}}\Big)\\
          &\leq \Pi^{(n)}\Big(|\|\theta-\wh\theta^\diamond\|\geq C_0\sqrt{\frac{\log n}{n}}\, \Big|\,\|\theta-\wh\theta^\diamond\|\leq \delta \Big)+n^{-c_1}\\
          &\leq  \Pi^{(n)}\Big( \|\psi^\diamond(\theta)\|\geq C_0\sqrt{\frac{\log n}{n}}\,\Big|\,\|\theta-\wh\theta^\diamond\|\leq \delta  \Big)+ n^{-c_1}.
             \end{aligned}
   \end{equation*}
Moreover, using~\eqref{eqnA3} and~\eqref{eqnA2}, when $C_0\geq \delta_2$,
    \begin{equation*}
 \begin{aligned}
 & \Pi^{(n)}\Big( \|\psi^\diamond(\theta)\|\geq C_0\sqrt{\frac{\log n}{n}}\,\Big|\,\|\theta-\wh\theta^\diamond\|\leq \delta  \Big)\\
&=\frac{ \int_{\frac{h}{\sqrt{n}}\in \m B_{\delta}, \|h\|\geq C_0\sqrt{\frac{\log n}{n}}} \pi\Big(\phi^{\diamond}\big(\frac{h}{\sqrt{n}}\big)\Big)\mathcal{J}^\diamond\big( \frac{h}{\sqrt{n}}\big)\cdot\exp\Big(\log \ms L\big(X^{(n)},\phi^{\diamond}\big(\frac{h}{\sqrt{n}}\big)\big)\Big) \,\dd h}{\int_{\frac{h}{\sqrt{n}}\in \m B_{\delta}} \pi\Big(\phi^{\diamond}\big(\frac{h}{\sqrt{n}}\big)\Big)\mathcal{J}^\diamond\big( \frac{h}{\sqrt{n}}\big)\cdot\exp\Big(\log \ms L\big(X^{(n)},\phi^{\diamond}\big(\frac{h}{\sqrt{n}}\big)\big)\Big) \,\dd h}\\
&\lesssim \bigg(\int_{\frac{h}{\sqrt{n}}\in \m B_{\delta}} \pi\Big(\phi^{\diamond}\big(\frac{h}{\sqrt{n}}\big)\Big)\mathcal{J}^\diamond\big( \frac{h}{\sqrt{n}}\big)\cdot\exp\Big(\log \ms L\big(X^{(n)},\phi^{\diamond}\big(\frac{h}{\sqrt{n}}\big)\big)\Big) \,\dd h\bigg)^{-1}\cdot n^{-c_1}.
     \end{aligned}
 \end{equation*}
Moreover, using~\eqref{eqnstatementTV}, we have 
\begin{equation}\label{eqnstatementTV1}
    \begin{aligned}
  &\bigg|\int_{\frac{h}{\sqrt{n}}\in \m B_{\delta}} \pi\Big(\phi^{\diamond}\big(\frac{h}{\sqrt{n}}\big)\Big)\mathcal{J}^\diamond\big( \frac{h}{\sqrt{n}}\big)\cdot\exp\Big(\log \ms L\big(X^{(n)},\phi^{\diamond}\big(\frac{h}{\sqrt{n}}\big)\big)\Big) \,\dd h-\pi(\theta^*)(2\pi)^{\frac{d}{2}}|\Sigma|^{\frac{1}{2}}\bigg|\\
  &\leq  \int \bigg|\pi\Big(\phi^{\diamond}\big(\frac{h}{\sqrt{n}}\big)\Big)\textbf{1}\Big(\frac{h}{\sqrt{n}}\in \m B_{\delta}\Big)\mathcal{J}^\diamond\big( \frac{h}{\sqrt{n}}\big)\cdot\exp\Big(\log \ms L\big(X^{(n)},\phi^{\diamond}\big(\frac{h}{\sqrt{n}}\big)\big)\Big)\\&\qquad -\pi(\theta^*)\exp\big(-\frac{1}{2} h^T \Sigma^{-1}h\big)\bigg|\,\dd h\leq C\, \frac{(\log n)^{\gamma_0}}{n^{\gamma_1}}.
    \end{aligned}
\end{equation}
Moreover, using the fact that $\pi(\theta^*)$ is lower bounded away from zero and $\Sigma$ is positive definite, we have, for sufficiently large $n$,
    \begin{equation*}
 \begin{aligned}
 & \Pi^{(n)}\Big( \|\psi^\diamond(\theta)\|\geq C_0\sqrt{\frac{\log n}{n}}\Big)\leq \Pi^{(n)}\Big( \|\psi^\diamond(\theta)\|\geq C_0\sqrt{\frac{\log n}{n}}\,\Big|\,\|\theta-\wh\theta^\diamond\|\leq \delta  \Big)+n^{-c_1}\leq C_1n^{-c_1}.
     \end{aligned}
 \end{equation*}

\medskip

\noindent{\textbf{Step 1.3. Show the desired result of~\eqref{statement:TVpsi}.}}

\medskip
 \noindent Given that $\Pi^{(n)}(\|\theta-\wh\theta^\diamond\|\geq \delta)\leq n^{-c_1}$, it holds for any measurable set $ B\subseteq \mathbb{R}^d$ that,
\begin{equation*}
\left|\Pi^{(n)}\big(\psi^{\diamond}(\theta)
\in B\big)-\Pi^{(n)}\big(\psi^{\diamond}(\theta)
\in B\,\big|\,\|\theta-\wh{\theta}^{\diamond}\|_2\leq \delta\big)\right|\leq 2\,n^{-c_1}\leq 2 \,n^{-2}.
 \end{equation*}
Then we have 
   \begin{equation*}
 \begin{aligned}
 & \Pi^{(n)}\Big( \psi^\diamond(\theta)\in B\,\Big|\,\|\theta-\wh\theta^\diamond\|\leq \delta  \Big)\\
&=\frac{ \int_{\frac{h}{\sqrt{n}}\in \m B_{\delta}\cap B} \pi\Big(\phi^{\diamond}\big(\frac{h}{\sqrt{n}}\big)\Big)\mathcal{J}^\diamond\big( \frac{h}{\sqrt{n}}\big)\cdot\exp\Big(\log \ms L\big(X^{(n)},\phi^{\diamond}\big(\frac{h}{\sqrt{n}}\big)\big)\Big) \,\dd h}{\int_{\frac{h}{\sqrt{n}}\in \m B_{\delta}} \pi\Big(\phi^{\diamond}\big(\frac{h}{\sqrt{n}}\big)\Big)\mathcal{J}^\diamond\big( \frac{h}{\sqrt{n}}\big)\cdot\exp\Big(\log \ms L\big(X^{(n)},\phi^{\diamond}\big(\frac{h}{\sqrt{n}}\big)\big)\Big) \,\dd h}.
     \end{aligned}
 \end{equation*}
Let $Z$ be a random variable with distribution $\m N(0,n^{-1}\Sigma)$, then 
 \begin{equation}\label{eqn:BPETELfinal}
 \begin{aligned}
    &\Big| \Pi^{(n)}(\psi^{\diamond}(\theta)\in B)-P(Z\in B)\Big|\\
    &\leq \left|\Pi^{(n)}\big(\psi^{\diamond}(\theta)
\in B\big)-\Pi^{(n)}\big(\psi^{\diamond}(\theta)
\in B\,\big|\,\|\theta-\wh{\theta}^{\diamond}\|_2\leq \delta\big)\right|\\
&\quad+\Big|\Pi^{(n)}\big(\psi^{\diamond}(\theta)
\in B\,\big|\,\|\theta-\wh{\theta}^{\diamond}\|_2\leq \delta\big)-P(Z\in B)\Big|\\
    &\leq 2 \, n^{-2}+\bigg|\Big[\int_{\frac{h}{\sqrt{n}}\in \m B_{\delta}} \pi\Big(\phi^{\diamond}\big(\frac{h}{\sqrt{n}}\big)\Big)\mathcal{J}^\diamond\big( \frac{h}{\sqrt{n}}\big)\cdot\exp\Big(\log \ms L\big(X^{(n)},\phi^{\diamond}\big(\frac{h}{\sqrt{n}}\big)\big)\Big) \,\dd h\Big]^{-1}\\
    &\qquad\qquad\cdot\Big[\int_{\frac{h}{\sqrt{n}}\in \m B_{\delta}\cap B} \pi\Big(\phi^{\diamond}\big(\frac{h}{\sqrt{n}}\big)\Big)\mathcal{J}^\diamond\big( \frac{h}{\sqrt{n}}\big)\cdot\exp\Big(\log \ms L\big(X^{(n)},\phi^{\diamond}\big(\frac{h}{\sqrt{n}}\big)\big)\Big) \,\dd h\Big]\\
&\qquad\quad-\pi(\theta^*)^{-1}\cdot(2\pi)^{-\frac{d}{2}}|\Sigma|^{-\frac{1}{2}}\cdot \int_{\frac{h}{\sqrt{n}}\in B} \pi(\theta^*)\cdot\exp(-\frac{1}{2} h^T \Sigma h)\,\dd h\bigg|\\
&\leq 2 \, n^{-2}+\Big[\int_{\frac{h}{\sqrt{n}}\in \m B_{\delta}} \pi\Big(\phi^{\diamond}\big(\frac{h}{\sqrt{n}}\big)\Big)\mathcal{J}^\diamond\big( \frac{h}{\sqrt{n}}\big)\cdot\exp\Big(\log \ms L\big(X^{(n)},\phi^{\diamond}\big(\frac{h}{\sqrt{n}}\big)\big)\Big) \,\dd h\Big]^{-1}\\
&\qquad \cdot \int \bigg|\pi\Big(\phi^{\diamond}\big(\frac{h}{\sqrt{n}}\big)\Big)\textbf{1}\Big(\frac{h}{\sqrt{n}}\in \m B_{\delta}\Big)\mathcal{J}^\diamond\big( \frac{h}{\sqrt{n}}\big)\cdot\exp\Big(\log \ms L\big(X^{(n)},\phi^{\diamond}\big(\frac{h}{\sqrt{n}}\big)\big)\Big) -\pi(\theta^*)\exp\big(-\frac{1}{2} h^T \Sigma^{-1}h\big)\bigg|\,\dd h\\
&+\int_{\frac{h}{\sqrt{n}}\in B} \pi(\theta^*)\cdot\exp(-\frac{1}{2} h^T \Sigma h)\,\dd h\\
&\cdot\bigg|\Big[\int_{\frac{h}{\sqrt{n}}\in \m B_{\delta}} \pi\Big(\phi^{\diamond}\big(\frac{h}{\sqrt{n}}\big)\Big)\mathcal{J}^\diamond\big( \frac{h}{\sqrt{n}}\big)\cdot\exp\Big(\log \ms L\big(X^{(n)},\phi^{\diamond}\big(\frac{h}{\sqrt{n}}\big)\big)\Big) \,\dd h\Big]^{-1}
-\pi(\theta^*)^{-1}\cdot(2\pi)^{-\frac{d}{2}}|\Sigma|^{-\frac{1}{2}}\bigg|.
     \end{aligned}
 \end{equation}
Now using~\eqref{eqnstatementTV} and~\eqref{eqnstatementTV1}, we have, for any measurable set $B\subseteq \mb R^d$, 
 \begin{equation*} 
 \begin{aligned}
    &\Big| \Pi^{(n)}(\psi^{\diamond}(\theta)\in B)-P(z\in B)\Big|\lesssim \frac{ (\log n)^{\gamma_0}}{n^{\gamma_1}},
     \end{aligned}
 \end{equation*}
which completes the proof of Statement~\eqref{statement:TVpsi}.

\medskip
\noindent\textbf{Step 2. Then we will show  $\|\wh\theta_p-\wh\theta^\diamond\|\leq C_1\,\frac{ (\log n)^{\gamma_0}}{n^{\gamma_1+\frac{1}{2}}}$.} 

\medskip
\noindent This is precisely the second claim in Lemma~\ref{lemma:generalresult}.  Note that 
\begin{equation*} 
 \begin{aligned}
 &\Big\|\mathbb{E}_{\Pi^{(n)}}\big[\psi^\diamond(\theta)\textbf{1}(\|\theta-\wh\theta^\diamond\|\leq \delta)\big]\Big\|\\
 &=\frac{\Pi^{(n)}(\|\theta-\wh\theta^\diamond\|\leq \delta)}{\int_{\frac{h}{\sqrt{n}}\in \m B_{\delta}} \pi\Big(\phi^{\diamond}\big(\frac{h}{\sqrt{n}}\big)\Big)\mathcal{J}^\diamond\big( \frac{h}{\sqrt{n}}\big)\cdot\exp\Big(\log \ms L\big(X^{(n)},\phi^{\diamond}\big(\frac{h}{\sqrt{n}}\big)\big)\Big) \,\dd h}\\
    &\qquad\cdot \Big\|\int_{\frac{h}{\sqrt{n}}\in \m B_{\delta}} \frac{h}{\sqrt{n}} \pi\Big(\phi^{\diamond}\big(\frac{h}{\sqrt{n}}\big)\Big)\mathcal{J}^\diamond\big( \frac{h}{\sqrt{n}}\big)\cdot\exp\Big(\log \ms L\big(X^{(n)},\phi^{\diamond}\big(\frac{h}{\sqrt{n}}\big)\big)\Big) \,\dd h\Big\| \\
    &\lesssim \frac{1}{\sqrt{n}}\cdot\Big\|\int_{\frac{h}{\sqrt{n}}\in \m B_{\delta}} h\cdot \pi\Big(\phi^{\diamond}\big(\frac{h}{\sqrt{n}}\big)\Big)\mathcal{J}^\diamond\big( \frac{h}{\sqrt{n}}\big)\cdot\exp\Big(\log \ms L\big(X^{(n)},\phi^{\diamond}\big(\frac{h}{\sqrt{n}}\big)\big)\Big) \,\dd h\\
    &\qquad\qquad-\int h\cdot \pi(\theta^*) \cdot\exp(-\frac{1}{2}h^T\Sigma^{-1}h)\,\dd h\Big\|.
  \end{aligned}
 \end{equation*}
Then, by an analysis analogous to Step 1.1 (used to prove \eqref{eqnstatementTV}), we obtain
 \begin{equation}\label{eqn:mean}
 \begin{aligned}
 \Big\|\mathbb{E}_{\Pi^{(n)}}\big[\psi^\diamond(\theta)\textbf{1}(\|\theta-\wh\theta^\diamond\|\leq \delta)\big]\Big\|\leq C\,\frac{ (\log n)^{\gamma_0}}{n^{\gamma_1+\frac{1}{2}}}.
  \end{aligned}
 \end{equation}
Similarly,
 \begin{equation*} 
     \begin{aligned}
       &\Bmnorm{\mathbb{E}_{\Pi^{(n)}}\big[\psi^\diamond(\theta)\psi^\diamond(\theta)^T \textbf{1}(\|\theta-\wh\theta^\diamond\|\leq \delta)\big]-n^{-1}\Sigma}_{\rm F} \\
        &=\Big\vert\kern-0.25ex\Big\vert\kern-0.25ex\Big\vert\frac{\Pi^{(n)}(\|\theta-\wh\theta^\diamond\|\leq \delta)}{\int_{\frac{h}{\sqrt{n}}\in \m B_{\delta}} \pi\Big(\phi^{\diamond}\big(\frac{h}{\sqrt{n}}\big)\Big)\mathcal{J}^\diamond\big( \frac{h}{\sqrt{n}}\big)\cdot\exp\Big(\log \ms L\big(X^{(n)},\phi^{\diamond}\big(\frac{h}{\sqrt{n}}\big)\big)\Big) \,\dd h}\\
    &\qquad\cdot  \int_{\frac{h}{\sqrt{n}}\in \m B_{\delta}} \frac{hh^T}{n} \pi\Big(\phi^{\diamond}\big(\frac{h}{\sqrt{n}}\big)\Big)\mathcal{J}^\diamond\big( \frac{h}{\sqrt{n}}\big)\cdot\exp\Big(\log \ms L\big(X^{(n)},\phi^{\diamond}\big(\frac{h}{\sqrt{n}}\big)\big)\Big) \,\dd h-\frac{\Sigma}{n}\Big\vert\kern-0.25ex\Big\vert\kern-0.25ex\Big\vert_{\rm F} \\
    &\lesssim \frac{1}{n}\cdot \Big|\frac{\Pi^{(n)}(\|\theta-\wh\theta^\diamond\|\leq \delta)}{\int_{\frac{h}{\sqrt{n}}\in \m B_{\delta}} \pi\Big(\phi^{\diamond}\big(\frac{h}{\sqrt{n}}\big)\Big)\mathcal{J}^\diamond\big( \frac{h}{\sqrt{n}}\big)\cdot\exp\Big(\log \ms L\big(X^{(n)},\phi^{\diamond}\big(\frac{h}{\sqrt{n}}\big)\big)\Big) \,\dd h}-(2\pi)^{\frac{d}{2}}\pi(\theta^*) |\Sigma|^{\frac{1}{2}}\Big|\\
    &\qquad+\frac{1}{n}
  \cdot\Big\vert\kern-0.25ex\Big\vert\kern-0.25ex\Big\vert\int_{\frac{h}{\sqrt{n}}\in \m B_{\delta}} hh^T\cdot \pi\Big(\phi^{\diamond}\big(\frac{h}{\sqrt{n}}\big)\Big)\mathcal{J}^\diamond\big( \frac{h}{\sqrt{n}}\big)\cdot\exp\Big(\log \ms L\big(X^{(n)},\phi^{\diamond}\big(\frac{h}{\sqrt{n}}\big)\big)\Big) \,\dd h\\
    &\qquad\qquad-\int hh^T\cdot \pi(\theta^*) \cdot\exp(-\frac{1}{2}h^T\Sigma^{-1}h)\,\dd h\Big\vert\kern-0.25ex\Big\vert\kern-0.25ex\Big\vert_{\rm F}.
     \end{aligned}
 \end{equation*}
Then, by an analysis analogous to Step 1.1, we can obtain
 \begin{equation}\label{eqn:cov}
     \begin{aligned}
       &\Bmnorm{n\cdot\mathbb{E}_{\Pi^{(n)}}\big[\psi^\diamond(\theta)\psi^\diamond(\theta)^T \textbf{1}(\|\theta-\wh\theta^\diamond\|\leq \delta)\big]-\Sigma}_{\rm F}\leq C\, \frac{ (\log n)^{\gamma_0}}{n^{\gamma_1}}.
     \end{aligned}
 \end{equation}
Notice that $W_{\theta^*}\psi^\diamond(\theta)=W_{\theta^*}W_{\theta^*}^T(\psi_{\theta^*}(\theta)-\psi_{\theta^*}(\wh\theta^\diamond))= \psi_{\theta^*}(\theta)-\psi_{\theta^*}(\wh\theta^\diamond)$ and 
\begin{equation*}
    \begin{aligned}
     &  \| \psi_{\theta^*}(\theta)-\psi_{\theta^*}(\wh\theta^\diamond)-(\theta-\wh\theta^\diamond)\|\\
     &\leq \| \psi_{\theta^*}(\theta)-(\theta-\theta^*)\|+\|\psi_{\theta^*}(\wh\theta^\diamond)-(\wh\theta^\diamond-\theta^*)\|\\
     &\leq C\, \|\theta-\theta^*\|^2+C \|\wh\theta^\diamond-\theta^*\|^2\\
     &\leq C\, \|\theta-\wh\theta^\diamond\|^2+C_1\frac{\log n}{n}.
    \end{aligned}
\end{equation*}
Then let $r_1=\min(\delta,\frac{1}{2C})$,  for any $\theta\in \m M\cap B_{r_1}(\wh\theta^\diamond)$, we have
\begin{equation*}
    \begin{aligned}
 \frac{1}{2}\|\theta-\wh\theta^\diamond\|\leq \|\theta-\wh\theta^\diamond\|- C\, \|\theta-\wh\theta^\diamond\|^2&\leq       \|\psi_{\theta^*}(\theta)-\psi_{\theta^*}(\wh\theta^\diamond)\|+C_1\frac{\log n}{n}.
    \end{aligned}
\end{equation*}
Therefore,
\begin{equation*}
 \begin{aligned}
& \Big\|\mathbb{E}_{\Pi^{(n)}}\big[W_{\theta^*}\psi^\diamond(\theta)\textbf{1}(\|\theta-\wh\theta^\diamond\|\leq \delta)\big]-\mathbb{E}_{\Pi^{(n)}}\big[(\theta-\wh{\theta}^{\diamond})\textbf{1}(\|\theta-\wh\theta^\diamond\|\leq \delta)\big]\Big\| \\
&\leq C\,\mb{E}_{\Pi^{(n)}}\big[\|\theta-\wh\theta^\diamond\|^2\textbf{1}(\|\theta-\wh\theta^\diamond\|\leq \delta)\big]+C_1\frac{\log n}{n}\\
&\leq  C\,\mb{E}_{\Pi^{(n)}}\big[\|\theta-\wh\theta^\diamond\|^2\textbf{1}(\|\theta-\wh\theta^\diamond\|\leq r_1)\big]+ C\,\mb{E}_{\Pi^{(n)}}\big[\|\theta-\wh\theta^\diamond\|^2\textbf{1}(r_1\leq \|\theta-\wh\theta^\diamond\|\leq \delta)\big]+C_1\frac{\log n}{n}\\
&\leq  4C\,\mb{E}_{\Pi^{(n)}}\bigg[\Big( \|\psi_{\theta^*}(\theta)-\psi_{\theta^*}(\wh\theta^\diamond)\|+C_1\frac{\log n}{n}\Big)^2\textbf{1}(\|\theta-\wh\theta^\diamond\|\leq r_1)\bigg]+ C\delta^2\,\Pi^{(n)}(\|\theta-\wh\theta^\diamond\|\geq r_1)+C_1\frac{\log n}{n}\\
&\leq 4C\,\mathbb{E}_{\Pi^{(n)}}\big[\psi^\diamond(\theta)^T\psi^\diamond(\theta)\textbf{1}(\|\theta-\wh\theta^\diamond\|\leq \delta)\big] +C_2 \frac{\log n}{n}\\
 &\leq C_3\, \frac{\log n}{n},
   \end{aligned}
 \end{equation*}
 and 
 \begin{equation*}
 \begin{aligned}
& \Big\|\mathbb{E}_{\Pi^{(n)}}\big[(\theta-\wh{\theta}^{\diamond})\textbf{1}(\|\theta-\wh\theta^\diamond\|\leq \delta)\big]-\big(\mathbb{E}_{\Pi^{(n)}}[\theta]-\wh{\theta}^{\diamond}\big)\Big\|\\
&\leq\Big \|\mathbb{E}_{\Pi^{(n)}}\big[\theta\cdot\textbf{1}(\|\theta-\wh\theta^\diamond\|\leq \delta)\big]-\mathbb{E}_{\Pi^{(n)}}[\theta]\Big\|+ \Big\|\mathbb{E}_{\Pi^{(n)}}\big[\wh{\theta}^{\diamond}\cdot\textbf{1}(\|\theta-\wh\theta^\diamond\|\leq \delta)\big]-\wh{\theta}^{\diamond}\Big\|\\
&\leq C_4\, \frac{1}{n^2}.
   \end{aligned}
 \end{equation*}
Combined with statements~\eqref{eqn:mean}, we can get
\begin{equation}\label{eqnthetap}
 \begin{aligned}
& \|\mathbb{E}_{\Pi^{(n)}}[\theta]-\wh{\theta}^{\diamond}\| \\
&\leq \Big\| \mathbb{E}_{\Pi^{(n)}}\big[(\theta-\wh{\theta}^{\diamond})\textbf{1}(\|\theta-\wh\theta^\diamond\|\leq \delta)\big]\Big\|+\frac{C_4}{n^2}\\
&\leq \Big\|\mathbb{E}_{\Pi^{(n)}}\big[W_{\theta^*}\psi^\diamond(\theta)\textbf{1}(\|\theta-\wh\theta^\diamond\|\leq \delta)\big]\Big\| +C_3\, \frac{\log n}{n}+\frac{C_4}{n^2}\\
&= \Big\|\mathbb{E}_{\Pi^{(n)}}\big[\psi^\diamond(\theta)\textbf{1}(\|\theta-\wh\theta^\diamond\|\leq \delta)\big]\Big\|+C_3\, \frac{\log n}{n}+\frac{C_4}{n^2}\\
&\leq C_5\,\frac{ (\log n)^{\gamma_0}}{n^{\gamma_1+\frac{1}{2}}}.
\end{aligned}
 \end{equation}
Based on the fact, we can now show the existence of  ${\arg \min}_{y\in \mathcal{M}}  \|y-\mathbb{E}_{\Pi^{(n)}}[\theta]\|^2$.  
Using $\|\wh\theta^\diamond-\theta^*\|\leq c\sqrt{\frac{\log n}{n}}$,  it holds that $\|\mathbb{E}_{\Pi^{(n)}}[\theta]-\theta^*\|\leq c\sqrt{\frac{\log n}{n}}+C_5\,\frac{ (\log n)^{\gamma_0}}{n^{\gamma_1+\frac{1}{2}}}\leq 2c\sqrt{\frac{\log n}{n}}$. So
 for any $y\in \m M$ with  $\|y-\mathbb{E}_{\Pi^{(n)}}[\theta]\|\leq \|\theta^*-\mathbb{E}_{\Pi^{(n)}}[\theta]\|$, it holds that 
\begin{equation*}
    \|y-\theta^*\|\leq \|y-\mathbb{E}_{\Pi^{(n)}}[\theta]\|+\|\mathbb{E}_{\Pi^{(n)}}[\theta]-\theta^*\|\leq  4c\sqrt{\frac{\log n}{n}}.
\end{equation*}
Moreover, for any $y\in  B(\theta^*,4c\sqrt{\frac{\log n}{n}})\cap \m M\subset U_{\theta^*}$, it holds that $\|\psi^\diamond(y)\|=\|W_{\theta^*}^T(y-\wh\theta^\diamond)\|\leq \|y-\wh\theta^\diamond\|\leq 5c\sqrt{\frac{\log n}{n}}.$ So we have $y\in \{\phi^{\diamond}(z)\, :\, \|z\|\leq 5c\sqrt{\frac{\log n}{n}} \}$, and 
\begin{equation*}
    \begin{aligned}
        {\arg \min}_{y\in \mathcal{M}}  \|y-\mathbb{E}_{\Pi^{(n)}}[\theta]\|^2= {\arg \min}_{z\in \mb R^d, \|z\|\leq 5c\sqrt{\frac{\log n}{n}}}  \|\phi^\diamond(z)-\mathbb{E}_{\Pi^{(n)}}[\theta]\|^2.
    \end{aligned}
\end{equation*}
Using the compactness of $\{z\in \mb R^d\,:\,  \|z\|\leq 5c\sqrt{\frac{\log n}{n}}\}$, the  set of ${\arg \min}_{z\in \mb R^d, \|z\|\leq 5c\sqrt{\frac{\log n}{n}}}  \|\phi^\diamond(z)-\mathbb{E}_{\Pi^{(n)}}[\theta]\|^2$ is well defined and non-empty. Then we will show that the set only contains a single point . Suppose there exist $z_1, z_2\in \{z\in \mb R^d\,:\,  \|z\|\leq 5c\sqrt{\frac{\log n}{n}}\}$ with $z_1\neq z_2$ so that  $ \|\phi^\diamond(z_1)-\mathbb{E}_{\Pi^{(n)}}[\theta]\|=\|\phi^\diamond(z_2)-\mathbb{E}_{\Pi^{(n)}}[\theta]\|=\min_{y\in \m M} \|y-\mathbb{E}_{\Pi^{(n)}}[\theta]\|$. Then note that $\phi^{\diamond}(0_d)=\wh\theta^\diamond$, it holds that
\begin{equation*}
    \begin{aligned}
&\|\phi^\diamond(z_1)-\phi^{\diamond}(0_d)\|\leq  \|\phi^\diamond(z_1)-\mathbb{E}_{\Pi^{(n)}}[\theta]\|+\|\wh\theta^\diamond-\mathbb{E}_{\Pi^{(n)}}[\theta]\|\\
&\leq 2\|\wh\theta^\diamond-\mathbb{E}_{\Pi^{(n)}}[\theta]\|\leq 2C_5\,\frac{ (\log n)^{\gamma_0}}{n^{\gamma_1+\frac{1}{2}}}.
    \end{aligned}
\end{equation*}
So $\|z_1\|\leq \|\phi^\diamond(z_1)-\phi^{\diamond}(0_d)\|\leq 2C_5\,\frac{ (\log n)^{\gamma_0}}{n^{\gamma_1+\frac{1}{2}}}$. Similarly, we have $\|z_2\|\leq  2C_5\,\frac{ (\log n)^{\gamma_0}}{n^{\gamma_1+\frac{1}{2}}}$. So both $z_1$ and $z_2$ are interior points of $\{z\in \mb R^d\,:\,  \|z\|\leq 5c\sqrt{\frac{\log n}{n}}\}$, and we have
\begin{equation*}
    \begin{aligned}
       \bold{J}_{\phi^\diamond}(z_1)^T (\mathbb{E}_{\Pi^{(n)}}[\theta]-\phi^\diamond(z_1)) =   \bold{J}_{\phi^\diamond}(z_2)^T (\mathbb{E}_{\Pi^{(n)}}[\theta]-\phi^\diamond(z_2))=0_d.
    \end{aligned}
\end{equation*}
Then we can further write
 \begin{equation*}
     \bold{J}_{\phi^\diamond}(z_2)^T(\phi^\diamond(z_1)-\phi^\diamond(z_2))=      \bold{J}_{\phi^\diamond}(z_2)^T(\phi^\diamond(z_1)-\mathbb{E}_{\Pi^{(n)}}[\theta]).
 \end{equation*}
 For the left hand side, when $n$ is sufficiently large, there exist some constants $C_1,C_2$ so that
\begin{equation*}
    \begin{aligned}
         &\big\|\bold{J}_{\phi^\diamond}(z_2)^T(\phi^\diamond(z_1)-\phi^\diamond(z_2))\big\|\\
         &\geq \|\bold{J}_{\phi^\diamond}(z_2)^T\bold{J}_{\phi^\diamond}(z_2)(z_1-z_2)\big\|-C_1 \|z_1-z_2\|^2\\
         &\geq C_2 \|z_1-z_2\|-C_1 \|z_1-z_2\|^2\\
         &\geq \frac{C_2}{2} \|z_1-z_2\|.
    \end{aligned}
\end{equation*}
For the right hand side, there exists  a constant $C_3$ so that 
\begin{equation*}
    \begin{aligned}
          &\big\|\bold{J}_{\phi^\diamond}(z_2)^T(\phi^\diamond(z_1)-\mathbb{E}_{\Pi^{(n)}}[\theta])\big\|\\
             &\leq \Big\|\big(\bold{J}_{\phi^\diamond}(z_2)^T-\bold{J}_{\phi^\diamond}(z_1)^T\big)(\phi^\diamond(z_1)-\mathbb{E}_{\Pi^{(n)}}[\theta])\Big\|\\
    &\leq \bmnorm{\bold{J}_{\phi^\diamond}(z_2)^T-\bold{J}_{\phi^\diamond}(z_1)^T}_{\rm F}\|\phi^\diamond(z_1)-\mathbb{E}_{\Pi^{(n)}}[\theta]\|\\
    &\leq C_3 \frac{ (\log n)^{\gamma_0}}{n^{\gamma_1+\frac{1}{2}}}\|z_1-z_2\|.
 \end{aligned}
\end{equation*}
So combined with these inequalities, we have 
\begin{equation*}
    C_3 \frac{ (\log n)^{\gamma_0}}{n^{\gamma_1+\frac{1}{2}}}\|z_1-z_2\|\geq  \frac{C_2}{2} \|z_1-z_2\|.
\end{equation*}
So when $n$ is large enough so that  $  C_3 \frac{ (\log n)^{\gamma_0}}{n^{\gamma_1+\frac{1}{2}}}<\frac{C_2}{2}$,  we must have $z_1=z_2$. This shows that  $\wh\theta_p={\arg \min}_{y\in \mathcal{M}}  \|y-\mathbb{E}_{\Pi^{(n)}}[\theta]\|^2$ is well  and uniquely defined, and 
\begin{equation*}
    \|\wh\theta_p-\wh\theta^\diamond\|=\|\phi^\diamond(z_1)-\phi^{\diamond}(0_d)\|\leq 2C_5\,\frac{ (\log n)^{\gamma_0}}{n^{\gamma_1+\frac{1}{2}}}.
\end{equation*}

 \medskip
 \noindent\textbf{Step 3. Show that for any $f: \m M\to \mb R^p$  satisfying Assumption D, it holds that
 ${\rm TV}\Big(f_{\#}\Pi^{(n)},\m N\big(f(\wh\theta^\diamond),n^{-1}J_f\Sigma J_f^T\big)\Big)\leq C_2 \frac{(\log n)^{ \frac{1+\beta}{2}}}{n^{\frac{\beta}{2}}}.$}

 \medskip
This is precisely the third claim in Lemma~\ref{lemma:generalresult}.   Note that it suffices to consider the case where $p=d$. Indeed, when $p<d$,  uses  that each column of $J_f^T$ lies in $T_{\theta^*}\m M$ and that $J_f$ is full rank. There exists a matrix $J^{\perp}=[v_1,v_2,\cdots,v_{d-p}]\in \mb R^{D\times (d-p)}$ whose columns lie in $T_{\theta^*}\m M$, satisfy $(J^{\perp})^TJ^{\perp}=I_{d-p}$, and orthogonal to the columns of $J_f^T$. Define $f_{p+j}(\theta)=v_j^T\theta$ for $j=\{1,2,\cdots,d-p\}$, and set $\ov{f}=({f}_1,{f}_2,\cdots,{f}_d)$. Then Assumption B holds for $\ov f$ with $p=d$. Consequently, the desired conclusion for $f_{\#}\Pi^{(n)}$ follows from the result for $\ov f_{\#}\Pi^{(n)}$ by the data processing inequality. Now let's suppose $d=p$.
% Recall $\wh{\theta}^{\diamond}={\phi}_{\theta^*}(-V_{\theta^*}\m H_0^{-1}\frac{1}{n}\sum_{i=1}^n g(X_i,{0}_d))$, and there exists a constant $C$ so that it holds with  probability at  least $1-\frac{1}{n^2}$ that $\|\wh\theta^\diamond-\theta^\ast\|\leq C\,\sqrt{\frac{\log n}{n}}$ and 
% \begin{equation*} 
%  \begin{aligned}
% {\rm TV}\Big((V_{\wh{\theta}^{\diamond}}^T \psi_{\wh{\theta}^{\diamond}})_{\#}\Pi_{\rm RP}^{(n)}, \mathcal{N}\big(0,\frac{1}{n}\m H_0^{-1}\Delta_0\m H_0^{-1}\big)\Big) \leq C\, \frac{ (\log n)^{1+\frac{\beta}{2}}}{n^{\frac{\beta}{2}}}.
% \end{aligned}
%  \end{equation*}
Since $\bold{J}_{\phi^{\diamond}}(-W_{\theta^*}^T\psi_{\theta^*}(\wh\theta^\diamond))=W_{\theta^*}$, we have 
\begin{equation*}
    \bold{J}_{f\circ \phi^\diamond}(-W_{\theta^*}^T\psi_{\theta^*}(\wh\theta^\diamond))=J_f W_{\theta^*}, 
\end{equation*}
 and 
 \begin{equation*}
     a^2 I_p \preccurlyeq \bold{J}_{f\circ \phi^\diamond}(-W_{\theta^*}^T\psi_{\theta^*}(\wh\theta^\diamond)) \bold{J}_{f\circ \phi^\diamond}(-W_{\theta^*}^T\psi_{\theta^*}(\wh\theta^\diamond))^T=J_fJ_f^T\preccurlyeq b^2 I_p.
 \end{equation*}
 Moreover, it holds for any measurable set $A\subset \mb R^d$ that,
\begin{equation*}
    \begin{aligned}
        &\Big|\Pi^{(n)}(f(\theta)\in A)-\Pi^{(n)}\Big(\psi^{\diamond}(\theta)\in B_{C_0\sqrt{\frac{\log n}{n}}}({0}_d), f\circ \phi^{\diamond}(\psi^{\diamond}(\theta))\in A  \Big)\Big|\\
        &\leq  \Big|\Pi^{(n)}\big( \theta\in B_{C_0\sqrt{\frac{\log n}{n}}}(\wh\theta^\diamond), f(\theta)\in A\big)-\Pi^{(n)}\Big(\psi^{\diamond}(\theta)\in B_{C_0\sqrt{\frac{\log n}{n}}}({0}_d), f\circ \phi^{\diamond}(\psi^{\diamond}(\theta))\in A  \Big)\Big|+\frac{1}{n^2}\\
        &=\Big|\Pi^{(n)}\big(\theta\in B_{C_0\sqrt{\frac{\log n}{n}}}(\wh\theta^\diamond),f\circ{\phi}^\diamond({\psi}^\diamond(\theta))\in A\big)\\
        &\qquad-\Pi^{(n)}\Big(\psi^{\diamond}(\theta)\in B_{C_0\sqrt{\frac{\log n}{n}}}({0}_d), f\circ \phi^{\diamond}(\psi^{\diamond}(\theta))\in A  \Big)\Big|+\frac{1}{n^2}\\
        &\leq \Pi^{(n)}\big(\theta\notin B_{C_0\sqrt{\frac{\log n}{n}}}(\wh\theta^\diamond)\big)+\frac{1}{n^2}\leq \frac{2}{n^2}.
    \end{aligned}
\end{equation*}
Let $Z$ be a random vector with distribution $\m N(0,n^{-1}\Sigma_0)$.
Then by statement~\eqref{statement:TVpsi}, we have 
\begin{equation*}
    \begin{aligned}
        &\Big|\Pi^{(n)}\Big(\psi^{\diamond}(\theta)\in B_{C_0\sqrt{\frac{\log n}{n}}}({0}_d), f\circ \phi^{\diamond}(\psi^{\diamond}(\theta))\in A  \Big)-P\Big(Z\in B_{C_0\sqrt{\frac{\log n}{n}}}({0}_d), f\circ \phi^{\diamond}(Z)\in A  \Big)\Big|\\
        &\leq C\, \frac{ (\log n)^{\gamma_0}}{n^{\gamma_1}},
    \end{aligned}
\end{equation*}
which leads to
 \begin{equation}\label{eqnPIPEftheta}
    \begin{aligned}
        &\Big|\Pi^{(n)}(f(\theta)\in A)-P\Big(Z\in B_{C_0\sqrt{\frac{\log n}{n}}}({0}_d), f\circ \phi^{\diamond}(Z)\in A  \Big)\Big|\leq C_1 \frac{ (\log n)^{\gamma_0}}{n^{\gamma_1}}.
      \end{aligned}
\end{equation}
Furthermore, using the Lipschitz continuous  of the Riemannian gradient of $f$, and the thrice differentiability of $\phi^\diamond$, there exists a positive constant $r$ so that  for any $z\in B_r(0_d)$,
 \begin{equation}\label{JZdiff}
    \bmnorm{{\bold J}_{f\circ {\phi}^{\diamond}}(z)- \bold{J}_{f\circ  \phi^\diamond}(-W_{\theta^*}^T\psi_{\theta^*}(\wh\theta^\diamond))}_{\rm F}\leq C\, \|z+W_{\theta^*}^T\psi_{\theta^*}(\wh\theta^\diamond)\|\leq C\,\|z\|+C\, \|\theta^\diamond-\theta^*\|.
 \end{equation}
Therefore,  using $a^2 I_d \preccurlyeq \bold{J}_{f\circ \phi^\diamond}(-W_{\theta^*}^T\psi_{\theta^*}(\wh\theta^\diamond)) \bold{J}_{f\circ \phi^\diamond}(-W_{\theta^*}^T\psi_{\theta^*}(\wh\theta^\diamond))^T\preccurlyeq b^2 I_d$, when $r$ is small enough, we have  for any $z\in B_{r}({0}_d)$, 
 $\frac{a^2}{2}I_d\preccurlyeq {\bold J}_{f\circ {\phi}^{\diamond}}(z){\bold J}_{f\circ {\phi}^{\diamond}}(z)^T\preccurlyeq \frac{b^2}{2}I_d$. Moreover,  by inverse function theorem for H\"{o}lder space (see for example, Appendix A of ~\citep{Eldering2013}), there exist positive constants $r_1,r_2$ and open sets $\m U\supset B_{r_1}(0_d)$ and $\m V\supset B_{r_2}(f(\wh\theta^\diamond))$, so that $f\circ {\phi}^{\diamond}|_{\m U}: \m U\to \m V$ is a diffeomorphism. Moreover, there exist positive constants $C,C_1$ so that for any $y,y'\in  B_{r_2}(f(\wh\theta^\diamond))$, 
 \begin{equation*}
     |(f\circ {\phi}^{\diamond}|_{\m U})^{-1}(y)- (f\circ {\phi}^{\diamond}|_{\m U})^{-1}(y')|\leq C \|y-y'\|.
 \end{equation*}
and
 \begin{equation}\label{eqnfinverse}
 \begin{aligned}
    &\| (f\circ  {\phi}^{\diamond}|_{\m U})^{-1}(y)- (J_f W_{\theta^*})^{-1}(y-f(\wh\theta^\diamond))\|\\
   &= \| (f\circ  {\phi}^{\diamond}|_{\m U})^{-1}(y)- (f\circ  {\phi}^{\diamond}|_{\m U})^{-1}(f(\wh\theta^\diamond))- \bold{J}_{(f\circ  {\phi}^{\diamond}|_{\m U})^{-1}}(f(\wh\theta^\diamond))(y-f(\wh\theta^\diamond))\|\\
   &\quad+\|\bold{J}_{(f\circ  {\phi}^{\diamond}|_{\m U})^{-1}}(f(\wh\theta^\diamond))(y-f(\wh\theta^\diamond))- ({\bold J}_f W_{\theta^*})^{-1}(y-f(\wh\theta^\diamond))\|\\
   &\leq C\,\|y-f(\wh\theta^\diamond)\|^2+C\|y-f(\wh\theta^\diamond)\|\cdot   \bmnorm{J_{f\circ {\phi}^{\diamond}}(0_d)- \bold{J}_{f\circ \phi^\diamond}(-W_{\theta^*}^T\psi_{\theta^*}(\wh\theta^\diamond))}_{\rm F}\\
   &\leq C_1 \|y-f(\wh\theta^\diamond)\|^2+C_1\frac{\log n}{n}.
     \end{aligned}
 \end{equation}
Therefore, let $Y$ be a random vector with distribution $\m N(f(\wh\theta^\diamond),n^{-1} {J}_f W_{\theta^*}\Sigma W_{\theta^*}^T{J}_f^T)$,  when $C_0$ is large enough, we have 
 \begin{equation*}
 \begin{aligned}
     &0\leq P(Y\in A)-P\big(Y\in A\cap \{y=f\circ {\phi}^\diamond(z)\,:\, z\in B_{C_0\sqrt{\frac{\log n}{n}}}({0}_d)\} \big)\\
     &\leq 1- P\Big(Y\in B_{\frac{C_0}{C}\sqrt{\frac{\log n}{n}}}(f(\wh\theta^\diamond)) \Big)\leq \frac{1}{n}.
    \end{aligned}  
 \end{equation*}
Consider 
 \begin{equation*}
 \begin{aligned}
       &P\Big( Z\in B_{C_0\sqrt{\frac{\log n}{n}}}({0}_d),f\circ {\phi}^{\diamond}(Z)\in A\Big)\\
       &=\int_{y\in A \cap \{y=f\circ {\phi}^{\diamond}(z)\,:\, z\in B_{C_0\sqrt{\frac{\log n}{n}}}({0}_d)\}}\frac{{\rm det}\big(J_{(f\circ {\phi}^{\diamond}|_{\m U})^{-1}}(y)\big)}{(2\pi)^{\frac{d}{2}}{\rm det}(n^{-1}\Sigma)^{\frac{1}{2}}}\exp\Big(-\frac{n}{2}(f\circ {\phi}^{\diamond}|_{\m U})^{-1}(y)^T\Sigma^{-1}(f\circ {\phi}^{\diamond}|_{\m U})^{-1}(y)\Big)\,\dd y.\\
       \end{aligned}
 \end{equation*}
We have
 \begin{equation*}
 \begin{aligned}
       &\Big|P\Big(Z\in B_{C_0\sqrt{\frac{\log n}{n}}}({0}_d),f\circ {\phi}^{\diamond}(Z)\in A \Big)-P(Y\in A)\Big|\\
       &\leq \int_{y\in A \cap \{y=f\circ {\phi}^{\diamond}(z)\,:\, z\in B_{C_0\sqrt{\frac{\log n}{n}}}({0}_d)\}}\bigg|\frac{{\rm det}\big({\bold J}_{(f\circ {\phi}^{\diamond}|_{\m U})^{-1}}(y)\big)}{(2\pi)^{\frac{d}{2}}{\rm det}(n^{-1}\Sigma)^{\frac{1}{2}}}\exp\Big(-\frac{n}{2}(f\circ {\phi}^{\diamond}|_{\m U})^{-1}(y)^T\Sigma^{-1}(f\circ {\phi}^{\diamond}|_{\m U})^{-1}(y)\Big)\\
       &\qquad-\frac{{\rm det}(({J}_f W_{\theta^*})^{-1})}{(2\pi)^{\frac{d}{2}}{\rm det}(n^{-1}\Sigma)^{\frac{1}{2}}}\exp\Big(-\frac{n}{2}(y-f(\wh\theta^\diamond))^T((J_f W_{\theta^*})^{-1})^T\Sigma^{-1}(J_f W_{\theta^*})^{-1}(y-f(\wh\theta^\diamond))\Big)\bigg|\,\dd y+\frac{1}{n}\\
       &\leq \int_{y\in A \cap \{y=f\circ {\phi}^{\diamond}(z)\,:\, z\in B_{C_0\sqrt{\frac{\log n}{n}}}({0}_d)\}}\frac{{\rm det}\big({\bold J}_{(f\circ {\phi}^{\diamond}|_{\m U})^{-1}}(y)\big)}{(2\pi)^{\frac{d}{2}}{\rm det}(n^{-1}\Sigma)^{\frac{1}{2}}}\cdot\bigg|\exp\Big(-\frac{n}{2}(f\circ {\phi}^{\diamond}|_{\m U})^{-1}(y)^T\Sigma^{-1}(f\circ {\phi}^{\diamond}|_{\m U})^{-1}(y)\Big)\\
       &\underbrace{\qquad\qquad\qquad-\exp\Big(-\frac{n}{2}(y-f(\wh\theta^\diamond))^T(( {J}_f W_{\theta^*})^{-1})^T\Sigma^{-1}( {J}_f W_{\theta^*})^{-1}(y-f(\wh\theta^\diamond))\Big)\bigg|\,\dd y}_{(I_C)}\\
       &+ \int_{y\in A \cap \{y=f\circ {\phi}^{\diamond}(z)\,:\, z\in B_{C_0\sqrt{\frac{\log n}{n}}}({0}_d)\}}\frac{\big|{\rm det}\big({\bold J}_{(f\circ {\phi}^{\diamond}|_{\m U})^{-1}}(y)\big)-{\rm det}((\bold{J}_f W_{\theta^*})^{-1})\big|}{(2\pi)^{\frac{d}{2}}{\rm det}(n^{-1}\Sigma_0)^{\frac{1}{2}}}\\
       &\underbrace{\qquad\qquad\cdot\exp\Big(-\frac{n}{2}(y-f(\wh\theta^\diamond))^T(( {J}_f W_{\theta^*})^{-1})^T\Sigma^{-1}( {J}_f W_{\theta^*})^{-1}(y-f(\wh\theta^\diamond))\Big)\,\dd y}_{(I_D)}+\frac{1}{n}.
       \end{aligned}
 \end{equation*}
Furthermore, for term $(I_C)$, using~\eqref{eqnfinverse}, we have for some constants $C,C_1$,
\begin{equation*}
    \begin{aligned}
        (I_C)&\leq C\,\int_{y\in A \cap \{y=f\circ {\phi}^{\diamond}(z)\,:\, z\in B_{C_0\sqrt{\frac{\log n}{n}}}({0}_d)\}}\frac{{\rm det}\big({\bold J}_{(f\circ {\phi}^{\diamond}|_{\m U})^{-1}}(y)\big)}{(2\pi)^{\frac{d}{2}}{\rm det}(n^{-1}\Sigma)^{\frac{1}{2}}}\\
        &\qquad\cdot \exp\Big(-\frac{n}{2}(y-f(\wh\theta^\diamond))^T(( {J}_f W_{\theta^*})^{-1})^T\Sigma^{-1}( {J}_f W_{\theta^*})^{-1}(y-f(\wh\theta^\diamond))\Big) \\
        &\qquad\qquad \cdot n\cdot (\|y-f(\wh\theta^\diamond)\|^2+\frac{\log n}{n})\cdot   \|y-f(\wh\theta^\diamond)\|\,\dd y\\
        &\leq C_1 \frac{\log n}{\sqrt{n}}.
 \end{aligned}
\end{equation*}
For term $(I_D)$, using ${\bold J}_{(f\circ {\phi}^{\diamond}|_{\m U})^{-1}}(y)=({\bold J}_{f\circ {\phi}^\diamond}(z))^{-1}$ with $z=(f\circ {\phi}^{\diamond}|_{\m U})^{-1}(y)$, and 
\begin{equation*}
    \begin{aligned}
        &\|(f\circ {\phi}^{\diamond}|_{\m U})^{-1}(y)\|=  &\|(f\circ {\phi}^{\diamond}|_{\m U})^{-1}(y)- (f\circ {\phi}^{\diamond}|_{\m U})^{-1}(f(\wh\theta^\diamond))\|\leq C\, \|y-f(\wh\theta^\diamond)\|,
    \end{aligned}
\end{equation*}
as well as inequality~\eqref{JZdiff}, we can get for some constants $C,C_1$,
 \begin{equation*}
     \begin{aligned}
         (I_D)&\leq C\, n^{\frac{d}{2}} \int_{y\in A \cap \{y=f\circ {\phi}^{\diamond}(z)\,:\, z\in B_{C_0\sqrt{\frac{\log n}{n}}}({0}_d)\}}   \Big( \|y-f(\wh\theta^\diamond)\|+\sqrt{\frac{\log n}{n}}\Big)\\
         &\qquad\qquad\cdot \exp\Big(-\frac{n}{2}(y-f(\wh\theta^\diamond))^T((\bold{J}_f W_{\theta^*})^{-1})^T\Sigma^{-1}(\bold{J}_f W_{\theta^*})^{-1}(y-f(\wh\theta^\diamond))\Big)\,\dd y\\
         &\leq C_1 \sqrt{\frac{\log n}{n}}.
     \end{aligned}
 \end{equation*}
  Then combined with~\eqref{eqnPIPEftheta}, it holds for $Y\sim\m N(f(\wh\theta^\diamond),n^{-1} {J}_f W_{\theta^*}\Sigma W_{\theta^*}^T{J}_f^T)$ and any measurable set $A\subset\mb R^d$ that, 
 \begin{equation*}
    \Big|P(Y\in A)-\Pi^{(n)}(f(\theta)\in A)\Big|\leq C\,\frac{ (\log n)^{\gamma_0}}{n^{\gamma_1}},
 \end{equation*}
 which leads to the desired result.

\section{Proof for Bayesian RPETEL Posterior}\label{proofRP}
 We will prove Theorem~\ref{th1}, Corollary~\ref{co1:nonsmooth} and Corollary~\ref{co:wellspecified} in this section.

\subsection{Proof of Theorem~\ref{th1}}
We will use Lemma~\ref{lemma:generalresult} to show the desired result. We will start with verifying the conditions of  Lemma~\ref{lemma:generalresult}.

\noindent\textbf{Step 1: Define the set $\m A$.}
Consider   sufficiently large constants $C$ and $C_1$, and define
\begin{equation*} 
    \m A_1=\big\{X^{(n)}\in \m X^{n}\,:\,\forall i\in [n],\,  b(X_i)\leq C(\log n)^{\frac{1}{\beta_1}}\Big\};
\end{equation*}
 \begin{equation*} 
    \m A_2=\Big\{X^{(n)}\in \m X^{n}\,:\,     \Big\|\frac{1}{n}\sum_{i=1}^n  g(X_i,\theta^*)\cdot\bold{1}\big(b(X_i)\leq  C(\log n)^{\frac{1}{\beta_1}}\big)\Big\|\leq C_1\sqrt{\frac{\log n}{n}}
\Big\};
\end{equation*}
\begin{equation*}
   \m A_3=\Big\{X^{(n)}\in \m X^n\,:\, n^{-1}\sum_{i=1}^n \| g(X_i,\theta^*)\|^3\cdot \bold{1}\big(b(X_i)\leq C(\log n)^{\frac{1}{\beta_1}}\big)\leq C_1\big\};
\end{equation*}
\begin{equation*}
   \begin{aligned}
       \m A_4&=\bigg\{X^{(n)}\in \m X^n\,:\, \forall\, \theta\in B_r(\theta^*)\cap \m M, \, \Big\vert\kern-0.25ex\Big\vert\kern-0.25ex\Big\vert n^{-1}\sum_{i=1}^n g(X_i,\theta) g(X_i,\theta)^T\\
      &\qquad\qquad-\mathbb{E} \big[  g(X,\theta)  g(X,\theta)^T\big]\Big\vert\kern-0.25ex\Big\vert\kern-0.25ex\Big\vert_{\rm F} \leq C_1\, \frac{(\log n)^{\frac{2}{\beta_1}+\frac{1}{2}}}{\sqrt{n}}\bigg\};
   \end{aligned}
\end{equation*}
\begin{equation*}
   \begin{aligned}
       \m A_5&=\bigg\{X^{(n)}\in \m X^n\,:\,  \forall\,\theta\in B_r(\theta^*)\cap \m M,  \,\Big\| n^{-1}\sum_{i=1}^n  g(X_i,\theta)- n^{-1}\sum_{i=1}^n    g(X_i,\theta^*)\\
       &\qquad-\mathbb{E} [g(X,\theta)]+\mathbb{E} [ g(X,\theta^*)]\Big\|_2 \leq C_1\,(\log n)^{\frac{1}{\beta_1}}\Big(\sqrt{\frac{\log n}{n}} \, \|\theta-\theta^*\|^{\beta_2}+\frac{\log n}{n}\Big)\bigg\};
 \end{aligned}
\end{equation*}
\begin{equation*}
   \begin{aligned}
       \m A_6&=\bigg\{X^{(n)}\in \m X^n\,:\,\forall\, \theta,\theta'\in S_{\Pi},\,  \Big|n^{-1}\sum_{i=1}^n \ell(X_i,\theta)-n^{-1}\sum_{i=1}^n \ell(X_i,\theta')\\
       &\qquad-\mathbb{E} [\ell(X,\theta)]+\mathbb{E} [\ell(X,\theta')]\Big| \leq C_1\,(\log n)^{\frac{1}{\beta_1}}\Big(\sqrt{\frac{\log n}{n}} \, \|\theta-\theta^*\|_2+\frac{\log n}{n}\Big)\bigg\};
 \end{aligned}
\end{equation*}
\begin{equation*}
   \begin{aligned}
       \m A_7&=\bigg\{X^{(n)}\in \m X^n\,:\,\forall\, \theta,\theta'\in B_r(\theta^*)\cap \m M, \,\Big|n^{-1}\sum_{i=1}^n \ell(X_i,\theta)-n^{-1}\sum_{i=1}^n \ell(X_i,\theta') \\ &\qquad- \frac{1}{n}\sum_{i=1}^n   g(X_i,\theta')^T(\theta-\theta')
     -\mathbb{E} [\ell(X,\theta)]+\mathbb{E} [\ell(X,\theta')]+\mb{E}\big[  g(X,\theta')^T(\theta-\theta')\big]\Big| \\
       &\qquad\qquad\leq C_1\,(\log n)^{\frac{1}{\beta_1}}\Big(\sqrt{\frac{\log n}{n}} \, \|\theta-\theta'\|^{\beta_2+1}+\frac{\log n}{n}\|\theta-\theta'\|+(\frac{\log n}{n})^2\Big)\bigg\}.
\end{aligned}
\end{equation*}
Let $\m A=\m A_1\cap \m A_2\cap\cdots \cap \m A_7$. Then we have the following lemma concerning the probability of $\m A$.
\begin{lemma}\label{lemma1} 
Under Assumptions 1-4,  there exist constants $C$ and $C_1$ such that $P(X^{(n)}\in \m A)\geq 1-n^{-2}$.
\end{lemma}
\noindent Now let's fix an $X^{(n)}\in \m A$ and let $W_{\theta^*}\in \mb R^{D\times d}$ be an arbitrary matrix whose columns form an orthonormal basis of $T_{\theta^*}\m M$.  

\medskip
\noindent\textbf{Step 2: Show $ \|\theta^*-\wh\theta^{\diamond}(X^{(n)})\|\lesssim \sqrt{\frac{\log n}{n}}$.}

Since $\m M$ is locally $C^3_{r,L}$ around $\theta^*$,  there exists $U_{\theta^*},V_{\theta*}$ with $B_{r}(\theta)\cap \m M\subseteq U_{\theta^*}\subseteq \m M$ and $B_{r}(0_D)\cap T_{\theta^*}\m M\subseteq V_{\theta^*}\subset T_{\theta^*}\m M$, so that $\psi_{\theta^*}: U_{\theta^*}\to V_{\theta^*}$ defined by $\psi_{\theta^*}(x)={\rm Proj}_{T_{\theta^*}\m M}(x-\theta^*)$ is a bijective, where the inverse, denoted by $\phi_{\theta^*}$ is thrice Fr\'{e}chet differentiable, and its Fr\'{e}chet derivatives up to order three have operator norms uniformly bounded by $L$.  By the thrice differentiablity of $\phi_{\theta^*}$, there exists a constant $C_2$ so that 
 \begin{equation*}
    \underset{v\in  V_{\theta^*} }{\sup} \frac{\|\phi_{\theta^*}(v)-(v+\theta^*)\|}{\|v\|^2}\leq C_2,
\end{equation*}
and
\begin{equation*}
\begin{aligned}
      \underset{\theta'\in U_{\theta^*}}{\sup}\frac{\|\psi_{\theta^*}(\theta')-(\theta'-\theta^*)\|}{\|\theta^*-\theta'\|^2}&= \underset{\theta'\in U_{\theta^*}}{\sup}\frac{\|\phi_{\theta^*}(\psi_{\theta^*}(\theta'))-(\theta^*+\psi_{\theta^*}(\theta'))\|}{\|\theta^*-\theta'\|}\\
      &\leq  \underset{v\in  V_{\theta^*} }{\sup} \frac{\|\phi_{\theta^*}(v)-(v+\theta^*)\|}{\|v\|^2}\leq C_2.
\end{aligned}
 \end{equation*}
 Note that $\m R(\theta)-\m R(\theta^*)\geq \frac{1}{L}\|\theta-\theta^*\|^2$ holds for any $\theta\in B_{r}(\theta^*)\cap \m M$. Moreover, using second-order Taylor expansion on curves of manifold (see for example (5.26) of~\cite{boumal2020introduction}), it holds that  for any $v\in \mb R^d$ with $\|v\|=1$, and $0<t\leq r$, 
\begin{equation*}
    \begin{aligned}
        \m R(\phi_{\theta^*}(tW_{\theta^*}v))-\m R(\theta^*)\leq \frac{t^2}{2} v^TW_{\theta^*}^T{\m H}_{\theta^*}W_{\theta^*}v +C\, t^3.
    \end{aligned}
\end{equation*}
Moreover when $t$ is small enough so that $\phi_{\theta^*}(tW_{\theta^*}v)\in S_{\Pi}$, it holds that 
\begin{equation*}
          \m R(\phi_{\theta^*}(tW_{\theta^*}v))-\m R(\theta^*)\geq \frac{1}{L}\,\|\phi_{\theta^*}(tW_{\theta^*}v)-\theta^*\|^2\geq \frac{t^2}{L}\|v\|^2.
\end{equation*}
So let $t\to 0$, we have $ v^TW_{\theta^*}^T{\m H}_{\theta^*}W_{\theta^*}v\geq \frac{2}{L}\|v\|^2$ holds for any $v\in \mb R^d$ with $\|v\|=1$, and therefore, $W_{\theta^*}^T{\m H}_{\theta^*}W_{\theta^*}\succcurlyeq \frac{2}{L}I_d$.
Then since  $X^{(n)}\in \m A$, we have
 \begin{equation*}
 \begin{aligned}
& \Big\|W_{\theta^*}(W_{\theta^*}^T{\m H}_{\theta^*}W_{\theta^*})^{-1}\frac{1}{n}\sum_{i=1}^n W_{\theta^*}^T g(X_i,\theta^*)\Big\|\\
&= \Big\|W_{\theta^*}(W_{\theta^*}^T{\m H}_{\theta^*}W_{\theta^*})^{-1}\frac{1}{n}\sum_{i=1}^n W_{\theta^*}^T g(X_i,\theta^*)\cdot\bold{1}\big(b(X_i)\leq  C(\log n)^{\frac{1}{\beta_1}}\big)\Big\|\\
&\lesssim \sqrt{\frac{\log n}{n}}.
 \end{aligned}
 \end{equation*}
When $n$ is sufficiently large, we have $-W_{\theta^*}(W_{\theta^*}^T{\m H}_{\theta^*}W_{\theta^*})^{-1}\frac{1}{n}\sum_{i=1}^n W_{\theta^*}^T g(X_i,\theta^*)\in V_{\theta^*}$. 
Define $$\wh\theta^{\diamond}(X^{(n)})=\phi_{\theta^*}
\big(-W_{\theta^*}(W_{\theta^*}^T{\m H}_{\theta^*}W_{\theta^*})^{-1}\frac{1}{n}\sum_{i=1}^n W_{\theta^*}^T g(X_i,\theta^*)\big).$$
For brevity, write $\wh\theta^{\diamond}=\wh\theta^{\diamond}(X^{(n)})$. Then there exists  a constant $C_3$ so that
\begin{equation}\label{boundthetadia}
\begin{aligned}
  & \|\theta^*-\wh\theta^{\diamond}\|=\Big\|\theta^*-\phi_{\theta^*}
\big(-W_{\theta^*}(W_{\theta^*}^T{\m H}_{\theta^*}W_{\theta^*})^{-1}\frac{1}{n}\sum_{i=1}^n W_{\theta^*}^Tg(X_i,\theta^*)\big)\Big\|\\
   &\leq \Big\|W_{\theta^*}(W_{\theta^*}^T{\m H}_{\theta^*}W_{\theta^*})^{-1}\frac{1}{n}\sum_{i=1}^n W_{\theta^*}^Tg(X_i,\theta^*)\Big\|+ C_2\Big\|W_{\theta^*}(W_{\theta^*}^T{\m H}_{\theta^*}W_{\theta^*})^{-1}\frac{1}{n}\sum_{i=1}^n W_{\theta^*}^Tg(X_i,\theta^*)\Big\|^2 \\
   &\leq C_3\,\sqrt{\frac{\log n}{n}}.
   \end{aligned}
\end{equation}

 \medskip
 \noindent\textbf{Step 3: Let  $\ms L(X^{(n)},\theta)= \frac{L(X^{(n)},\theta)}{(\frac{1}{n})^n}\exp\big(-\alpha_n \cdot(\m R_n(\theta)-\m R_n(\theta^*))\big)$. Show  that for any  $h\in B_{\delta_1(\log n)^{3/2}}(0_d)$, 
        \begin{equation*}
            \begin{aligned}
          & \bigg| \log \ms L\Big(X^{(n)},\phi_{\theta^*}\Big(\frac{W_{\theta^*}h}{\sqrt{n}}+\psi_{\theta^*}(\wh\theta^\diamond)\Big)\Big)+ \frac{1}{2} h^T  \m H_0\Delta_0^{-1} \m H_0 h\bigg| \leq C\,\frac{(\log n)^{\frac{2}{\beta_1}+1}}{n^{\frac{\beta_2}{2}}}\cdot(\|h\|^3+1).
            \end{aligned}
        \end{equation*}}

\medskip
\noindent For any $\theta\in U_{\theta^*}$, set
$W_{{\theta}}= \bold{J}_{{\phi}_{\theta^*}(W_{\theta^*}y)}(y=W_{\theta^*}^T {\psi}_{\theta^*}({\theta}))$. Since $ g(X_i,\theta)\in T_{\theta}\m M$,  we can rewrite the constraint $\sum_{i=1}^n w_i \cdot g(X_i,\theta)=0_D$ as
\begin{equation*}
\sum_{i=1}^n w_i \cdot W_{\theta}^T g(X_i,\theta)=0_d
\end{equation*}
and introducing Lagrange multipliers,  the RETEL function can be rewritten as 
 \begin{equation*}
L(X^{(n)};\theta)=\prod_{i=1}^n\frac{\exp\left(\lambda(\theta)^T W_{\theta}^T g(X_i,\theta) \right)}{\sum_{i=1}^n\exp\left(\lambda(\theta)^T W_{\theta}^T g(X_i,\theta)\right)},
  \end{equation*}
  with $\lambda(\theta)=\underset{\xi\in \mb R^d}{\arg\min}\sum_{i=1}^n \exp(\xi^T W_{\theta}^T g(X_i,\theta))$.  Denote  $\m H_0=W_{\theta^*}^T {\m H}_{\theta^*} W_{\theta^*}$ and  $\Delta_0=W_{\theta^*}^T {\Delta}_{\theta^*} W_{\theta^*}$. We present the following lemma that provides an approximate form to $\lambda(\theta)$.

\begin{lemma}\label{lemma2}
For any $X^{(n)}\in \m A$, define 
\begin{equation*}
    \begin{aligned}
        &\lambda(\theta)=\underset{\xi \in \mathbb{R}^d}{\arg\min}\frac{1}{n}\sum_{i=1}^n \exp(\xi^T  W_{\theta}^T{g}(X_i,\theta));\\
        &\wt{\lambda}(\theta)=-\Delta_{0}^{-1} \m H_{0} W_{\theta^*}^{T} \big(\psi_{\theta^*}(\theta)-\psi_{\theta^*}(\wh\theta^\diamond(X^{(n)}))\big).
    \end{aligned}
\end{equation*}
Then for any positive constant $\delta$, there exists a constant $C$ so that for any $\theta\in \m M$ with $\|\theta-\theta^*\|\leq \delta\frac{(\log n)^{3/2}}{\sqrt{n}}$,
\begin{equation*}
 \|\lambda(\theta)-\wt{\lambda}(\theta)\|_2\leq C\Big( (\log n)^{\frac{2}{\beta_1}} \Big(\frac{\log n}{n}+\sqrt{\frac{\log n}{n}}\|\theta-\theta^*\|_2^{\beta_2}\Big) +\|\theta-\theta^*\|^2\Big).
\end{equation*}
\end{lemma}
\noindent Fix an  arbitrary $h\in B_{\delta_1(\log n)^{3/2}}(0_d)$ and let $\theta=\phi_{\theta^*}\big(\frac{W_{\theta^*}h}{\sqrt{n}}+\psi_{\theta^*}(\wh\theta^\diamond)\big)$. It holds that 
\begin{equation*}
    \begin{aligned}
        \|\theta-\theta^*\|\leq \|\frac{W_{\theta^*}h}{\sqrt{n}}+\psi_{\theta^*}(\wh\theta^\diamond)\|+C \|\frac{W_{\theta^*}h}{\sqrt{n}}+\psi_{\theta^*}(\wh\theta^\diamond)\|^2\lesssim \frac{\|h\|}{\sqrt{n}}+ \sqrt{\frac{\log n}{n}}\lesssim\frac{(\log n)^{\frac{3}{2}}}{\sqrt{n}}.
    \end{aligned}
\end{equation*}
 Then we can write 
 \begin{equation*}
 \log \frac{L(X^{(n)};\theta)}{(\frac{1}{n})^n}=\sum_{i=1}^n \lambda(\theta)^T W_{\theta}^T g(X_i,\theta)-n\cdot\log \bigg(\frac{1}{n}\sum_{i=1}^n\exp\Big(\lambda(\theta)^T W_{\theta}^T g(X_i,\theta)\Big)\bigg).
 \end{equation*}
By Lemma~\ref{lemma2},  we have
\begin{equation}\label{difflambdatheta}
\begin{aligned}
\|\lambda(\theta)+\Delta_{0}^{-1} \m H_{0}\frac{h}{\sqrt{n}}\|&= \|\lambda(\theta)+\Delta_{0}^{-1} \m H_{0} W_{\theta^*}^{T} (\psi_{\theta^*}(\theta)-\psi_{\theta^*}(\wh\theta^\diamond))\|\\
&\lesssim (\log n)^{\frac{2}{\beta_1}} \Big(\frac{\log n}{n}+\sqrt{\frac{\log n}{n}}\|\theta-\theta^*\|_2^{\beta_2}\Big) +\|\theta-\theta^*\|^2\\
&\lesssim (\log n)^{\frac{2}{\beta_1}} \sqrt{\frac{\log n}{n}} \Big(\frac{\|h\|+\sqrt{\log n}}{\sqrt{n}}\Big)^{\beta_2}.
    \end{aligned}
\end{equation}
Therefore,  using $ \|\theta^*-\wh\theta^{\diamond}\|\lesssim \sqrt{\frac{\log n}{n}}$, we can get
\begin{equation}\label{boundlambda}
\|\lambda(\theta)\|\leq \|\lambda(\theta)+\Delta_{0}^{-1} \m H_{0}\frac{h}{\sqrt{n}}\|+\frac{1}{\sqrt n}\|\Delta_0^{-1}\m H_0h\|\lesssim \Big(\frac{(\log n)^{\frac{1}{2}+\frac{2}{\beta_1}}}{\sqrt{n}}\Big)^{1+\beta_2}+\frac{\|h\|_2}{\sqrt{n}}.
\end{equation}
Then we can obtain
\begin{equation*}
\begin{aligned}
 &\Big|\frac{1}{n}\sum_{i=1}^n\exp\big(\lambda(\theta)^T W_{\theta}^T g(X_i,\theta)\big)-1-\frac{1}{n}\sum_{i=1}^n\lambda(\theta)^T W_{\theta}^T g(X_i,\theta)- \frac{1}{2n}\sum_{i=1}^n \big(\lambda(\theta)^T W_{\theta}^T g(X_i,\theta)\big)^2\Big|\\
&\leq \frac{1}{n}\sum_{i=1}^n\big(\lambda(\theta)^T W_{\theta}^T g(X_i,\theta)\big)^3\lesssim \|\lambda(\theta)\|^3 \frac{1}{n}\sum_{i=1}^n  \|g(X_i,\theta)\|^3\lesssim \|h\|^2n^{-\frac{3}{2}}+ \Big(\frac{(\log n)^{\frac{1}{2}+\frac{2}{\beta_1}}}{\sqrt{n}}\Big)^{3+3\beta_2},
 \end{aligned}
 \end{equation*}
and 
\begin{equation*}
\begin{aligned}
 &\bigg|\log \bigg(\frac{1}{n}\sum_{i=1}^n\exp\Big(\lambda(\theta)^T W_{\theta}^T g(X_i,\theta)\Big)\bigg)-\frac{1}{n}\sum_{i=1}^n\lambda(\theta)^T W_{\theta}^T g(X_i,\theta)\\
 &\quad- \frac{1}{2n}\sum_{i=1}^n \big(\lambda(\theta)^T W_{\theta}^T g(X_i,\theta)\big)^2+\frac{1}{2}\Big(\frac{1}{n}\sum_{i=1}^n\lambda(\theta)^T W_{\theta}^T g(X_i,\theta)\Big)^2\bigg|\lesssim  \|h\|^2n^{-\frac{3}{2}}+ \Big(\frac{(\log n)^{\frac{1}{2}+\frac{2}{\beta_1}}}{\sqrt{n}}\Big)^{3+3\beta_2}.
 \end{aligned}
 \end{equation*}
Therefore, it holds for a constant $C$ that,
 \begin{equation*}
 \begin{aligned}
 &\left|\log \frac{L(X^{(n)};\theta)}{(\frac{1}{n})^n}+\frac{1}{2}\sum_{i=1}^n \big(\lambda(\theta)^T W_{\theta}^T g(X_i,\theta)\big)^2-\frac{n}{2}\Big(\frac{1}{n}\sum_{i=1}^n\lambda(\theta)^T W_{\theta}^T g(X_i,\theta)\Big)^2\right|\\
 &\leq C\, \|h\|_2^3 \cdot n^{-\frac{1}{2}}+ C\, n^{-\frac{1+3\beta_2}{2}}\cdot(\log n)^{\frac{(4+\beta_1)(3+3\beta_2)}{2\beta_1}}.
 \end{aligned}
\end{equation*}
For the term of $\frac{1}{2}\sum_{i=1}^n \big(\lambda(\theta)^T W_{\theta}^T g(X_i,\theta)\big)^2$, note that $X^{(n)}\in \m A_4$, so
\begin{equation*}
    \begin{aligned}
        &\Big|\sum_{i=1}^n \big(\lambda(\theta)^T W_{\theta}^T g(X_i,\theta)\big)^2-n \lambda(\theta)^T \Delta_0 \lambda(\theta)\Big|\\
        &\leq   \Big| \lambda(\theta)^T W_{\theta}^T \sum_{i=1}^n g(X_i,\theta)g(X_i,\theta)^T W_{\theta}\lambda(\theta)- n\cdot  \lambda(\theta)^T W_{\theta}^T \mb{E}[g(X,\theta)g(X,\theta)^T] W_{\theta}\lambda(\theta) \Big|\\
        &\qquad+ \Big|n\cdot  \lambda(\theta)^T W_{\theta}^T \mb{E}[g(X,\theta)g(X,\theta)^T] W_{\theta}\lambda(\theta) - n\cdot  \lambda(\theta)^T W_{\theta^*}^T \mb{E}[g(X,\theta^*)g(X,\theta^*)^T] W_{\theta^*}\lambda(\theta) \Big|\\
        &\leq n\cdot \|\lambda(\theta)\|^2\cdot \Bmnorm{\frac{1}{n}\sum_{i=1}^n g(X_i,\theta) g(X_i,\theta)^T-\mb{E}[g(X,\theta)g(X,\theta)^T]}_{\rm F}\\
        &\qquad+ n\cdot \|\lambda(\theta)\|^2\cdot\bmnorm{W_{\theta}^T\Delta_{\theta} W_{\theta}-W_{\theta^*}^T \Delta_{\theta^*}W_{\theta^*}}_{\rm F}\\
        & \lesssim \Big(\frac{(\log n)^{\frac{2}{\beta_1}+\frac{1}{2}}}{\sqrt{n}}+\frac{\|h\|}{\sqrt{n}}\Big)\cdot \Big(\|h\|^2+\frac{(\log n)^{\frac{(1+\beta_2)(4+\beta_1)}{2\beta_1}}}{n^{\beta_2}}\Big).
    \end{aligned}
\end{equation*}
For the term of $\frac{n}{2}\Big(\frac{1}{n}\sum_{i=1}^n\lambda(\theta)^T W_{\theta}^T g(X_i,\theta)\Big)^2$. Since $X^{(n)}\in \m A_1\cap \m A_3\cap \m A_5$, we have
\begin{equation*}
\begin{aligned}
\left\|\frac{1}{n}\sum_{i=1}^ng(X_i,\theta)\right\|&\leq  \Big\| \frac{1}{n}\sum_{i=1}^n  g(X_i,\theta)- \frac{1}{n}\sum_{i=1}^n  g(X_i,\theta^*)-\mathbb{E} [g(X,\theta)]+\mathbb{E} [ g(X,\theta^*)]\Big\|\\
&\quad+\big\|\frac{1}{n}\sum_{i=1}^n  g(X_i,\theta^*)\big\|+\big\|\mathbb{E} [g(X,\theta)]-\mathbb{E} [ g(X,\theta^*)]\big\|\lesssim \sqrt{\frac{\log n}{n}}+\frac{\|h\|}{\sqrt{n}},
\end{aligned}
\end{equation*}
and 
\begin{equation*}
\begin{aligned}
&\frac{n}{2}\Big(\frac{1}{n}\sum_{i=1}^n\lambda(\theta)^T W_{\theta}^T g(X_i,\theta)\Big)^2\\
&\leq\frac{n}{2}\cdot \|\lambda(\theta)\|^2\cdot  \Big\|\frac{1}{n}\sum_{i=1}^ng(X_i,\theta)\Big\|^2
&\lesssim \Big( \Big(\frac{(\log n)^{\frac{1}{2}+\frac{2}{\beta_1}}}{\sqrt{n}}\Big)^{1+\beta_2}+\frac{\|h\|_2}{\sqrt{n}}\Big)^2\cdot \big(\sqrt{\log n}+\|h\|\big).
\end{aligned}
\end{equation*}
Putting the pieces together, we obtain
\begin{equation*}
    \begin{aligned}
 &\Big|\log \frac{L(X^{(n)};\theta)}{(\frac{1}{n})^n}+n\cdot \lambda(\theta)^T \Delta_0 \lambda(\theta)\Big|\lesssim \Big(\frac{(\log n)^{\frac{2}{\beta_1}+\frac{1}{2}}}{\sqrt{n}}+\frac{\|h\|}{\sqrt{n}}\Big)\cdot \Big(\|h\|^2+\frac{(\log n)^{\frac{(1+\beta_2)(4+\beta_1)}{2\beta_1}}}{n^{\beta_2}}\Big).
    \end{aligned}
\end{equation*}
Together with~\eqref{difflambdatheta}, we obtain
\begin{equation*} 
 \begin{aligned}
 &\left|\log \frac{L\big(X^{(n)};\phi_{\theta^*}\big(\frac{W_{\theta^*}h}{\sqrt{n}}+\psi_{\theta^*}(\wh\theta^\diamond)\big)\big)}{(\frac{1}{n})^n}+\frac{1}{2} h^T\m H_0\Delta_0^{-1}\m H_0h\right|\\
 &\leq \Big|\log \frac{L(X^{(n)};\theta)}{(\frac{1}{n})^n}+n\cdot \lambda(\theta)^T \Delta_0 \lambda(\theta)\Big|+\left|\frac{1}{2} h^T\m H_0\Delta_0^{-1}\m H_0h-\frac{n}{2} \lambda(\theta)^T\Delta_0\lambda(\theta)\right|\\
 &\lesssim  \Big(\frac{(\log n)^{\frac{2}{\beta_1}+\frac{1}{2}}}{\sqrt{n}}+\frac{\|h\|}{\sqrt{n}}\Big)\cdot \Big(\|h\|^2+\frac{(\log n)^{\frac{(1+\beta_2)(4+\beta_1)}{2\beta_1}}}{n^{\beta_2}}\Big)\\
 &\qquad+ \Big((\log n)^{\frac{2}{\beta_1}} \sqrt{\frac{\log n}{n}} \Big(\frac{\|h\|+\sqrt{\log n}}{\sqrt{n}}\Big)^{\beta_2}\Big)\cdot\Big(\Big(\frac{(\log n)^{\frac{1}{2}+\frac{2}{\beta_1}}}{\sqrt{n}}\Big)^{1+\beta_2}+\frac{\|h\|_2}{\sqrt{n}}\Big)\\
 &\lesssim \frac{(\log n)^{\frac{2}{\beta_1}+1}}{n^{\frac{\beta_2}{2}}}\cdot(\|h\|^3+1).
 \end{aligned}
\end{equation*}
Furthermore, consider the transformed risk function $\wt {\m R}:B_r(0_d)\to \mb R$ defined by
$\wt {\m R}(z)=\m R(\phi_{\theta^*}(W_{\theta^*}z))=\mb{E}[\ell(X,\phi_{\theta^*}(W_{\theta^*}z))]$. By Assumption  1 and 3, $\wt {\m R}$ is thrice differentiable and $\nabla\wt{\m R}(0_d)=0$.  Therefore, 
\begin{equation*}
\begin{aligned}
&\Big|\wt{\mathcal{R}}\big(\frac{h}{\sqrt{n}}+W_{\theta^*}^T(\wh\theta^\diamond-\theta^*)\big)-\wt{\mathcal{R}}(0_d)\Big|\lesssim\Big\|\frac{h}{\sqrt{n}}+W_{\theta^*}^T(\wh\theta^\diamond-\theta^*)\Big\|^2\lesssim \frac{\log n}{n}+\frac{\|h\|^2}{n}.
 \end{aligned}
\end{equation*}
Then using $X^{(n)}\in \m A_6$, we can get
\begin{equation*} 
\begin{aligned}
&\Big|\mathcal{R}_n(\theta)-\mathcal{R}_n(\theta^*)\Big|\leq  \Big|n^{-1}\sum_{i=1}^n \ell(X_i,\theta)-n^{-1}\sum_{i=1}^n \ell(X_i,\theta^*)-\mathbb{E} [\ell(X,\theta)]+\mathbb{E} [\ell(X,\theta^*)]\Big|\\
&\qquad+\big|\wt{\mathcal{R}}\big(\frac{h}{\sqrt{n}}+W_{\theta^*}^T(\wh\theta^\diamond-\theta^*)\big)-\wt{\mathcal{R}}(0_d)\big|\lesssim \frac{\log n}{n}+\frac{\|h\|^2}{n}.
 \end{aligned}
\end{equation*}
 Collecting the above results and using $\alpha_n\lesssim \sqrt{n}$, we can finally obtain
        \begin{equation}\label{boundmsL}
            \begin{aligned}
          & \bigg| \log \ms L\Big(X^{(n)},\phi_{\theta^*}\Big(\frac{W_{\theta^*}h}{\sqrt{n}}+\psi_{\theta^*}(\wh\theta^\diamond)\Big)\Big)+ \frac{1}{2} h^T  \m H_0\Delta_0^{-1} \m H_0 h\bigg| \\
          &\leq \bigg|\log \frac{L\big(X^{(n)};\phi_{\theta^*}\big(\frac{W_{\theta^*}h}{\sqrt{n}}+\psi_{\theta^*}(\wh\theta^\diamond)\big)\big)}{(\frac{1}{n})^n}+\frac{1}{2} h^T\m H_0\Delta_0^{-1}\m H_0h\bigg|+\alpha_n \Big|\mathcal{R}_n(\theta)-\mathcal{R}_n(\theta^*)\Big|\\
          &\lesssim  \frac{(\log n)^{\frac{2}{\beta_1}+1}}{n^{\frac{\beta_2}{2}}}\cdot(\|h\|^3+1).
          \end{aligned}
          \end{equation}

\medskip
\noindent\textbf{Step 4: Show that when $c_2\log n\leq \alpha_n\leq c_3\sqrt{n}$ with sufficiently large $c_2$, for any constant $\delta>0$ and constant $c_1\geq 2$, it holds for sufficiently large $n$ that $\Pi_{\rm RP}(\|\theta-\wh\theta^\diamond\|\geq \delta\,|\,X^{(n)})\leq n^{-c_1}$.}

\medskip
\noindent Since $\|\wh\theta^\diamond-\theta^*\|\leq C\, \sqrt{\frac{\log n}{n}}$, for any positive constant $\delta$, there exists $N$ so that when $n\geq N$,
 \begin{equation*}
 \begin{aligned}
   \Pi_{\rm RP}  (\|\theta-\wh\theta^\diamond\|_2\geq \delta\,|\,X^{(n)})&\leq \Pi_{\rm RP} (\|\theta-\theta^*\|_2\geq \delta/2\,|\,X^{(n)})\\
  & =\frac{\int_{B_{\delta/2}(\theta^*)^c\cap S_{\Pi}} \exp(-\alpha_n \mathcal{R}_n(\theta))L(X^{(n)};\theta)\Pi(\dd\theta)}{\int  \exp(-\alpha_n \mathcal{R}_n(\theta))L(X^{(n)};\theta)\Pi (\dd\theta)}\\
  &\leq  \frac{\int_{B_{\frac{\delta}{2}}(\theta^*)^c\cap S_{\Pi}} \exp(-\alpha_n (\mathcal{R}_n(\theta)-\mathcal{R}_n(\theta^*)))L(X^{(n)};\theta)/(\frac{1}{n})^{\frac{1}{n}}\Pi (\dd\theta)}{\int_{  B_{{1}/{n}}(\wh\theta^\diamond)\cap S_{\Pi}}  \exp(-\alpha_n (\mathcal{R}_n(\theta)-\mathcal{R}_n(\theta^*)))L(X^{(n)};\theta)/(\frac{1}{n})^{\frac{1}{n}} \Pi (\dd\theta)}
  \end{aligned}
  \end{equation*}
For the denominator, by equation~\eqref{boundmsL}, there exists positive constant $c,c'$ such that  
\begin{equation*}
 \begin{aligned}
& \int_{\theta\in\mb  B_{{1}/{n}}(\wh\theta^\diamond)\cap S_{\Pi}}  \exp(-\alpha_n (\mathcal{R}_n(\theta)-\mathcal{R}_n(\theta^*)))L(X^{(n)};\theta)/(\frac{1}{n})^{\frac{1}{n}}\Pi(\dd\theta)\\
&\geq  c\int_{\theta\in B_{{1}/{n}}(\wh\theta^\diamond)\cap S_{\Pi}} \Pi(\dd\theta)\\
&\geq \exp(-c'\log n).
\end{aligned}
  \end{equation*}
   For the numerator, by the assumption that $\mathcal{R}(\theta)-\mathcal{R}(\theta^*)\geq L\, \|\theta-\theta^*\|$ and $X^{(n)}\in \m A_6$, when $n$ is large enough, it holds for any $\theta\in B_{\frac{\delta}{2}}(\theta^*)^c\cap S_{\Pi}$ that ,
    \begin{equation*}
\mathcal{R}_n(\theta)-\mathcal{R}_n(\theta^*) \geq \frac{L\delta}{2}-C\frac{(\log n)^{\frac{1}{2}+\frac{1}{\beta_1}}}{\sqrt{n}}\geq \frac{L\delta}{4}.
\end{equation*}
Therefore when $c_2\log n\leq \alpha_n\leq c_3\sqrt{n}$ with sufficiently large $c_2$,   it holds  that 
   \begin{equation*}
  \Pi_{\rm RP}  (\|\theta-\wh\theta^\diamond\|_2\geq \delta\,|\,X^{(n)})\leq \Pi_{\rm RP}  (\|\theta-\theta^*\|_2\geq \frac{\delta}{2}\,|\,X^{(n)})\leq \exp(c'\log n)\exp(-c_2 L\delta\log n/4)\leq n^{-c_1}.
  \end{equation*}

\medskip
\noindent\textbf{Step 5: Show that for any positive constant $c_1\geq 2$, with for sufficiently large $\delta_1$,  for any $h\in\mb R^d$ satisfying  $\|h\|\geq \delta_1(\log n)^{3/2}$ and $\frac{W_{\theta^*}h}{\sqrt{n}}+\psi_{\theta^*}(\wh\theta^\diamond)\in V_{\theta^*}$,  it holds that  $    \log\ms L\big(X^{(n)},\phi_{\theta^*}\big(\frac{W_{\theta^*}h}{\sqrt{n}}+\psi_{\theta^*}(\wh\theta^\diamond)\big)\big)\leq -2\,c_1 d\, \log n.$}

\medskip
\noindent Consider the transformed risk function $\wt {\m R}:B_r(0_d)\to \mb R$ defined by
$\wt {\m R}(z)=\m R(\phi_{\theta^*}(W_{\theta^*}z))=\mb{E}[\ell(X,\phi_{\theta^*}(W_{\theta^*}z))]$. Then it can be verified that $\nabla \wt {\m R}(z)=W_{\phi_{\theta^*}(W_{\theta^*}z)}^T \mb{E}[g(X,\phi_{\theta^*}(W_{\theta^*}z))]$ and the Hessian  matrix of $\wt {\m R}$ at $0_d$ is given by $\m H_0$. Now take an arbitrary $h\in \mb R^d$ with  $\|h\|\geq \delta_1(\log n)^{3/2}$ and $\frac{W_{\theta^*}h}{\sqrt{n}}+\psi_{\theta^*}(\wh\theta^\diamond)\in V_{\theta^*}$. Denote $\theta=\phi_{\theta^*}(\frac{W_{\theta^*}h}{\sqrt{n}}+\psi_{\theta^*}(\wh\theta^\diamond))$.    Then  when $\|\frac{W_{\theta^*}h}{\sqrt{n}}+\psi_{\theta^*}(\wh\theta^\diamond)\|\leq r$,  there exist  constant $C,C_1$ so that 
\begin{equation*}
    \begin{aligned}
      &\big\|W_{\theta}^T\mb{E}[g(X,\theta)]- W_{\theta^*}^T\mb{E}[g(X,\theta^*)]-\m H_0 W_{\theta^*}^T\psi_{\theta^*}(\theta)\big\|\\
      &=\big\| \nabla \wt {\m R}(W_{\theta^*}^T\psi_{\theta^*}(\theta))-\nabla \wt {\m R}(0_d)-\m H_0  W_{\theta^*}^T\psi_{\theta^*}(\theta)\big\|\\
      &\leq C\|W_{\theta^*}^T(\theta-\theta^*)\|^2\leq  C_1 (\frac{\|h\|^2}{n}+\frac{\log n}{n}).
    \end{aligned}
\end{equation*}
Moreover, since $\m H_0$ is positive definite, there exist positive  constants $\delta_1$, $\delta_2$ and $C_2$ such that when $\delta_1(\log n)^{1.5}\leq \|h\|\leq \delta_2 \sqrt{n}$ and $\theta=\phi_{\theta^*}(\frac{W_{\theta^*}h}{\sqrt{n}}+\psi_{\theta^*}(\wh\theta^\diamond))$, 
\begin{equation*}
\begin{aligned}
        &\big\|W_{\theta}^T\mb{E}[g(X,\theta)]\big\|\\
        &\geq\big\|\m H_0 W_{\theta^*}^T\psi_{\theta^*}(\theta)\big\|-\big\|W_{\theta}^T\mb{E}[g(X,\theta)]- W_{\theta^*}^T\mb{E}[g(X,\theta^*)]-\m H_0 W_{\theta^*}^T\psi_{\theta^*}(\theta)\big\|\\
        &\geq \|\m H_0(\frac{h}{\sqrt{n}}+W_{\theta^*}^T(\wh\theta^\diamond-\theta^*))\|-C_1 (\frac{\|h\|^2}{n}+\frac{\log n}{n})\geq C_2\frac{\|h\|}{\sqrt{n}}.
\end{aligned}
\end{equation*}
Then using $X^{(n)}\in \m A_1\cap\m A_2\cap \m A_5$, we have, for sufficiently large $n$ that, 
\begin{equation*}
    \begin{aligned}
           &\Big\|\frac{1}{n}\sum_{i=1}^nW_{\theta}^Tg(X_i,\theta)\Big\|\geq     \big\|W_{\theta}^T\mb{E}[g(X,\theta)]\big\|-\Big\|\frac{1}{n}\sum_{i=1}^n W_{\theta}^Tg(X_i,\theta^*)\Big\|\\
           &\qquad\qquad- \,\Big\| W_{\theta}^T\Big(\frac{1}{n}\sum_{i=1}^n  g(X_i,\theta)- n^{-1}\sum_{i=1}^n  g(X_i,\theta^*)-\mathbb{E} [g(X,\theta)]+\mathbb{E} [ g(X,\theta^*)]\Big)\Big\|\\
           &\qquad\qquad\qquad\geq\frac{C_2}{2}\frac{\|h\|}{\sqrt{n}}\geq \frac{C_2\delta_1}{2}\frac{(\log n)^{1.5}}{\sqrt{n}}.
    \end{aligned}
\end{equation*}
Then by 
\begin{equation*}
\begin{aligned}
&\log \frac{L(X^{(n)};  \theta)}{(\frac{1}{n})^{\frac{1}{n}}}=\sum_{i=1}^n p_i(\theta)-n\log \frac{1}{n};\\
&\sum_{i=1}^n p_i(\theta)W_{\theta}^Tg(X_i,\theta)=0_d.
\end{aligned}
\end{equation*}
It holds  that  when $\delta_1(\log n)^{1.5}\leq \|h\|_2\leq \delta_2 \sqrt{n}$, 
\begin{equation*}
\Big\|\sum_{i=1}^n\Big(p_i(\theta)-\frac{1}{n}\Big)W_{\theta}^Tg(X_i,\theta)\Big\|\geq \frac{C_2\delta_1}{2}\frac{(\log n)^{1.5}}{\sqrt{n}}.
\end{equation*}
Then using $X^{(n)}\in \m A_4$, there exists a positive constant $C_3$ such that 
\begin{equation*}
\sum_{i=1}^n\Big(p_i(\theta)-\frac{1}{n}\Big)^2\geq \frac{C_2\delta_1^2}{4}\frac{(\log n)^3}{n}\cdot \frac{1}{\sum_{i=1}^n \|W_{\theta}^T g(X_i,\theta)\|^2}\geq C_3 \delta_1^2 \frac{(\log n)^3}{n^2}.
\end{equation*}
Define $q(p_1, \cdots, p_{n-1})=\sum_{i=1}^{n-1} \log p_i +\log (1-\sum_{i=1}^{n-1} p_i)$. The Hessian matrix of function $q$ at point  $(p_1, \cdots, p_{n-1})$ is 
 \begin{equation*}
     \mathcal{H}_q|_{(p_1, \cdots, p_{n-1})}= \text{Diag}(-\frac{1}{p_1^2},\cdots,-\frac{1}{p_{n-1}^2})-\frac{1}{(1-\sum_{i=1}^{n-1}p_i)^2}\textbf{1}_{(n-1)\times(n-1)},
 \end{equation*}
 where $\textbf{1}_{(n-1)\times(n-1)}$ denotes the $(n-1)\times(n-1)$ matrix with all entries  being $1$. Let $p=(p_1, \cdots, p_{n})$ and $p_{-n}=(p_1, \cdots, p_{n-1})$. If  $\|p\|_{\infty} \geq 3c_1d \frac{\log n}{n}$, then 
 \begin{equation*}
 \sum_{i=1}^n \log p_i \leq \log \frac{3c_1d\log n}{n} +(n-1) \log \frac{1-3c_1d\frac{\log n}{n}}{n-1}.
 \end{equation*}
So,
 \begin{equation}
 \begin{aligned}
  -n\log n -\sum_{i=1}^n \log p_i  &\geq -\log(3c_1d\log n) -(n-1) \log\left( (1-3c_1d \frac{\log n}{n}) \frac{n}{n-1}\right)\\
  &\geq \frac{5}{2}c_1d \log n.
  \end{aligned}
 \end{equation}
 \quad\\
 If $\|p\|_{\infty} \leq 3c_1d \frac{\log n}{n}$, then when $\delta_1$ is large enough, we have $\sum_{i=1}^{n-1} (p_i-\frac{1}{n})^2\geq \frac{45c_1^3d^3(\log n)^3}{n^2}$, so by mean value theorem,  
 \begin{equation}
 \begin{aligned}
&q(\frac{1}{n}, \cdots ,\frac{1}{n})-q(p_{-n})\\
&=-\frac{1}{2}(p_{-n}-\frac{1}{n}\textbf{1}_{(n-1)})^T \mathcal{H}_q|_{(cp_{-n}+(1-c)\frac{1}{n}\textbf{1}_{(n-1)})}(p_{-n}-\frac{1}{n}\textbf{1}_{(n-1)})\\
&\geq \frac{5}{2}c_1d\log n.
  \end{aligned}
 \end{equation}
Therefore,  it holds that  when $\delta_1(\log n)^{1.5}\leq \|h\|\leq \delta_2 \sqrt{n}$, 
\begin{equation*}
\log \frac{L\Big(X^{(n)}; \phi_{\theta^*}\big(\frac{W_{\theta^*}h}{\sqrt{n}}+\psi_{\theta^*}(\wh\theta^\diamond)\big)\Big)}{(\frac{1}{n})^{\frac{1}{n}}}\leq -\frac{5}{2}c_1d\log n.
\end{equation*}
Moreover, by $X^{(n)}\in \m A_6$, it holds that
\begin{equation}\label{eqnRnminus}
\begin{aligned}
    &\mathcal{R}_n(\theta)-\mathcal{R}_n(\theta^*)-\mathcal{R}(\theta)-\mathcal{R}(\theta^*) \\
    &\geq -C\,\frac{\log n}{n} -C\, (\log n)^{\frac{1}{\beta_1}}\sqrt{\frac{\log n}{n}}\cdot \|\theta-\theta^* \|^{\beta_2},
    \end{aligned}
\end{equation}
then there exists a positive constant $c_3$ such that when $\alpha_n\leq c_3\sqrt{n}$ and  $\delta_1(\log n)^{1.5}\leq \|h\|\leq \delta_2 \sqrt{n}$, 

\begin{equation*}
\begin{aligned}
&\exp\bigg(\log \frac{L\Big(X^{(n)}; \phi_{\theta^*}(\frac{W_{\theta^*}h}{\sqrt{n}}+\psi_{\theta^*}(\wh\theta^\diamond))\Big)}{(\frac{1}{n})^{\frac{1}{n}}}-\alpha_n\bigg(\mathcal{R}_n\Big( \phi_{\theta^*}(\frac{W_{\theta^*}h}{\sqrt{n}}+\psi_{\theta^*}(\wh\theta^\diamond))\Big)-\mathcal{R}_n(\theta^*)\bigg)\bigg)\\
&\leq \exp\Bigg(\log \frac{L\big(X^{(n)}; \theta\big)}{(\frac{1}{n})^{\frac{1}{n}}}-\alpha_n\bigg(\mathcal{R}_n(\theta)-\mathcal{R}_n(\theta^*)-\mathcal{R}(\theta)+\mathcal{R}(\theta^*)\bigg)\Bigg)\leq \exp(-2c_1d\log n).
\end{aligned}
\end{equation*}
Then consider the case of $\|h\|>\delta_2 \sqrt{n}$.  Using equation~\eqref{eqnRnminus} and $\m R(\theta)-\m R(\theta^*)\geq L\|\theta-\theta^*\|^2$, there exists a constant $c_2$ such that when $\alpha_n\geq c_2\log n$, it holds that 
\begin{equation*}
\begin{aligned}
&\exp\bigg(\log \frac{L\Big(X^{(n)}; \phi_{\theta^*}(\frac{W_{\theta^*}h}{\sqrt{n}}+\psi_{\theta^*}(\wh\theta^\diamond))\Big)}{(\frac{1}{n})^{\frac{1}{n}}}-\alpha_n\bigg(\mathcal{R}_n\Big( \phi_{\theta^*}(\frac{W_{\theta^*}h}{\sqrt{n}}+\psi_{\theta^*}(\wh\theta^\diamond))\Big)-\mathcal{R}_n(\theta^*)\bigg)\bigg)\\
&\leq \exp\Big(-\alpha_n\big(\mathcal{R}(\theta)-\mathcal{R}(\theta^*)\big)+\alpha_n\big|
\mathcal{R}_n(\theta)-\mathcal{R}_n(\theta^*)-\mathcal{R}(\theta)+\mathcal{R}(\theta^*)\big|\Big)\leq  \exp(-2c_1d\log n).
\end{aligned}
\end{equation*}

\medskip
\noindent\textbf{Step 6: Show $\|\wh\theta-\wh\theta^\diamond\|\lesssim \frac{(\log n)^{\frac{1}{\beta_1}+1}}{n^{\frac{\beta_2}{2}+\frac{1}{2}}}$.}

\medskip
\noindent Denote $\wh\theta=\wh\theta(X^{(n)})$, using $X^{(n)}\in \m A_6$, we have 
\begin{equation*}
    \Big|n^{-1}\sum_{i=1}^n \ell(X_i,\wh\theta)-n^{-1}\sum_{i=1}^n \ell(X_i,\theta^*)-\mathbb{E} [\ell(X,\wh\theta)]+\mathbb{E} [\ell(X,\theta^*)]\Big| \leq C_1\,(\log n)^{\frac{1}{\beta_1}}\Big(\sqrt{\frac{\log n}{n}} \, \|\wh\theta-\theta^*\|_2+\frac{\log n}{n}\Big).
\end{equation*}
Moreover, since $n^{-1}\sum_{i=1}^n \ell(X_i,\wh\theta)<n^{-1}\sum_{i=1}^n \ell(X_i,\theta^*)$, we can get 
\begin{equation*}
    \frac{1}{L}\|\wh\theta-\theta^*\|^2\leq \mathbb{E} [\ell(X,\wh\theta)]-\mathbb{E} [\ell(X,\theta^*)]\leq  C_1\,(\log n)^{\frac{1}{\beta_1}}\Big(\sqrt{\frac{\log n}{n}} \, \|\theta-\theta^*\|_2+\frac{\log n}{n}\Big),
\end{equation*}
and therefore 
\begin{equation*}
    \|\wh\theta-\theta^*\|\lesssim \frac{(\log n)^{\frac{1}{2}+\frac{1}{\beta_1}}}{\sqrt{n}},
\end{equation*}
which implies $\|\wh\theta-\wh\theta^\diamond\|\leq   \|\wh\theta-\theta^*\|+  \|\wh\theta^\diamond-\theta^*\|\lesssim \frac{(\log n)^{\frac{1}{2}+\frac{1}{\beta_1}}}{\sqrt{n}}$.
Then using Lemma~\ref{lemmasmootharoundtheta}, there exist sets $V_{\wh\theta^\diamond}$, $U_{\wh\theta^\diamond}$ so that $T_{\wh\theta^\diamond}\m M\cap B_{r_1}(0_D)\subseteq V_{\wh\theta^\diamond}\subseteq T_{\wh\theta^\diamond}\m M$ and $\m M\cap B_{r_1}(\wh\theta^\diamond)\subset U_{\wh\theta^\diamond}\subset \m M$. The map $\psi_{\wh\theta^\diamond}: U_{\wh\theta^\diamond}\to V_{\wh\theta^\diamond}$ defined as $\psi_{\wh\theta^\diamond}(\theta)={\rm Proj}_{T_{\wh\theta^\diamond}\m M}(\theta-\wh\theta^\diamond)$, admits a inverse, denoted as $\phi_{\wh\theta^\diamond}$, which is thrice differentiable. 
% Then let $W_\diamond\in \mb R^{D\times d}$ be an arbitrary matrix whose columns form an orthonormal basis of $T_{\wh\theta^\diamond}\m M$, and define the map $\wt\phi_{\wh\theta^\diamond}: B_{r_1}(0_d)\to U_{\wh\theta^\diamond}$ as $\wt \phi_{\wh\theta^\diamond}(z)=\phi_{\wh\theta^\diamond}(W_{\diamond}z)$ and $\wt \psi_{\wh\theta^\diamond}:B_{r_1}(\wh\theta^\diamond)\cap \m M\to B_{r_1}(0_d)$ as $\wt \psi_{ \wh\theta^\diamond}(\theta)=W_{\diamond}^T(\theta-\wh\theta^\diamond)$. 
Now consider the curve 
\begin{equation*}
    c(t)=\phi_{ \wh\theta^\diamond}\Big(t\frac{ \psi_{ \wh\theta^\diamond}(\wh\theta)}{\|\psi_{ \wh\theta^\diamond}(\wh\theta)\|}\Big).
\end{equation*}
It holds that $c(0)=\wh\theta^\diamond$ and $c\big(\| \psi_{ \wh\theta^\diamond}(\wh\theta)\|\big)=\wh\theta$.  Let $\wt{\m H}_{\wh\theta^\diamond}$ be the Jacobian matrix of the map $\theta\mapsto {\rm Proj}_{T_{\theta}\m M} (\nabla {\ov{\m R}}(\theta))$ at $\theta=\wh\theta^\diamond$, where $\ov{\m R}(\cdot)$ is the ambient space extension of $\m R(\cdot)$. Then denote $P_{\wh\theta^\diamond}$ as the projection matrix onto $T_{\wh\theta^\diamond}\m M$ and $\m H_{\wh\theta^\diamond}=P_{\wh\theta^\diamond}\wt {\m H}_{\wh\theta^\diamond}P_{\wh\theta^\diamond}$. Using Assumption 1 and 3, it holds that 
\begin{equation*}
    \bmnorm{\m H_{\wh\theta^\diamond}-\m H_{\theta^*}}_{\rm F}\lesssim \sqrt{\frac{\log n}{n}}.
\end{equation*}
Moreover, it has been shown in Step 2 that there exists a positive constant $C$ so that $W_{\theta^*}^T \m H_{\theta^*} W_{\theta^*}\succcurlyeq C I_d$. For $W_{\wh{\theta^\diamond}}= \bold{J}_{{\phi}_{\theta^*}(W_{\theta^*}y)}(y=W_{\theta^*}^T {\psi}_{\theta^*}(\wh{\theta^\diamond}))$, when $n$ is sufficiently large, it holds that    $\frac{1}{2}I_d\preccurlyeq W_{\wh\theta^\diamond}^TW_{\wh\theta^\diamond}\preccurlyeq 2I_d $ and
\begin{equation*}
    W_{\wh\theta^\diamond}^T \m H_{\wh\theta^\diamond}    W_{\wh\theta^\diamond}\succcurlyeq \frac{C}{2}I_d.
\end{equation*}
Using second-order Taylor expansion on curves of manifold (see for example (5.26) of~\cite{boumal2020introduction}), it holds that 
\begin{equation*}
    \begin{aligned}
    &\bigg|\m R\Big(c\big(\| \psi_{ \wh\theta^\diamond}(\wh\theta)\|\big)\Big)-\m R(\wh\theta^\diamond)-\|\psi_{ \wh\theta^\diamond}(\wh\theta)\|\cdot \mb{E}[g(X,\wh\theta^\diamond)]^T c'(0)-\frac{\| \psi_{ \wh\theta^\diamond}(\wh\theta)\|^2}{2} \mb{E}[g(X,\wh\theta^\diamond)]^T c''(0)\\
    &\qquad\qquad-\frac{\|\psi_{ \wh\theta^\diamond}(\wh\theta)\|^2}{2} c'(0)^T \m H_{\wh\theta^\diamond} c'(0)\bigg| \leq C\, \| \psi_{ \wh\theta^\diamond}(\wh\theta)\|^3,
    \end{aligned}
\end{equation*}
Moreover, notice that $\|\psi_{\wh\theta^\diamond}(\wh\theta)\|\lesssim\|\wh\theta-\wh\theta^\diamond\|$ and $\|\mb{E}[g(X,\wh\theta^\diamond)]\|\lesssim \|\wh\theta^\diamond-\theta^*\|\lesssim \sqrt{\frac{\log n}{n}}$. Using $c'(0)=\frac{\psi_{\wh\theta^\diamond}(\wh\theta)}{\|\psi_{\wh\theta^\diamond}(\wh\theta)\|}$, it holds for a constant $C_1$  that
 \begin{equation}\label{eqntaylorR}
 \begin{aligned}
    &\bigg|\m R(\wh\theta)-\m R(\wh\theta^\diamond)-\mb{E}[ g(X,\wh\theta^\diamond)^T(\wh\theta-\wh\theta^\diamond)]-\frac{1}{2}(\wh\theta-\wh\theta^\diamond)^T{\m H}_{\wh\theta^\diamond}(\wh\theta-\wh\theta^\diamond)\bigg|\\
    &\leq C_1\, (\|\wh\theta-\wh\theta^\diamond\|^3+\|\wh\theta-\wh\theta^\diamond\|^2\sqrt{\frac{\log n}{n}}),
    \end{aligned}
 \end{equation}
Then using $X^{(n)}\in \m A_7$,  we have 
 \begin{equation*}
 \begin{aligned}
      &\Big|n^{-1}\sum_{i=1}^n \ell(X_i,\wh\theta)-n^{-1}\sum_{i=1}^n \ell(X_i,\wh\theta^\diamond) \\ &\qquad- \frac{1}{n}\sum_{i=1}^n   g(X_i,\wh\theta^\diamond)^T(\wh\theta-\wh\theta^\diamond)
     -\m R(\wh\theta)+\m R(\wh\theta^\diamond)+\mb{E}\big[  g(X,\wh\theta^\diamond)^T(\wh\theta-\wh\theta^\diamond)\big]\Big| \\
       &\leq C_1\,(\log n)^{\frac{1}{\beta_1}}\Big(\sqrt{\frac{\log n}{n}} \, \|\wh\theta-\wh\theta^\diamond\|^{\beta_2+1}+\frac{\log n}{n}\|\wh\theta-\wh\theta^\diamond\|+(\frac{\log n}{n})^2\Big).
 \end{aligned}
 \end{equation*}
Furthermore,   using $\frac{1}{2}I_d\preccurlyeq W_{\wh\theta^\diamond}^TW_{\wh\theta^\diamond}\preccurlyeq 2I_d $ and $    W_{\wh\theta^\diamond}^T \m H_{\wh\theta^\diamond}    W_{\wh\theta^\diamond}\succcurlyeq \frac{C}{2}I_d$,  there exists constants $C_1,C_2$ such that
\begin{equation*}
    \begin{aligned}
       & (\wh\theta-\wh\theta^\diamond)^T{\m H}_{\wh\theta^\diamond}(\theta-\wh\theta^\diamond)\\
       &\geq \psi_{\wh\theta^\diamond}(\wh \theta)^T {\m H}_{\wh\theta^\diamond}\psi_{\wh\theta^\diamond}(\wh \theta)-C_1\,\|\wh\theta-\wh\theta^\diamond\|^3\\
       &= \psi_{\wh\theta^\diamond}(\wh \theta)^T W_{\wh\theta^\diamond}(W_{\wh\theta^\diamond}^TW_{\wh\theta^\diamond})^{-1}W_{\wh\theta^\diamond}^T{\m H}_{\wh\theta^\diamond} W_{\wh\theta^\diamond}(W_{\wh\theta^\diamond}^TW_{\wh\theta^\diamond})^{-1}W_{\wh\theta^\diamond}^T\psi_{\wh\theta^\diamond}(\wh \theta)-C_1\,\|\wh\theta-\wh\theta^\diamond\|^3\\
       &\geq \frac{C}{4} \|\psi_{\wh\theta^\diamond}(\wh \theta)\|^2-C_1\,\|\wh\theta-\wh\theta^\diamond\|^3\geq C_2 \|\wh\theta-\wh\theta^\diamond\|^2.
    \end{aligned}
\end{equation*}
 We can get
\begin{equation*}
\begin{aligned}
    &    \Big|\frac{1}{n}\sum_{i=1}^n  g(X_i,\wh\theta^\diamond)^T(\wh\theta-\wh\theta^\diamond)\Big|\geq  -\frac{1}{n}\sum_{i=1}^n  g(X_i,\wh\theta^\diamond)^T(\wh\theta-\wh\theta^\diamond)\\
    &\geq \frac{1}{2}(\wh\theta-\wh\theta^\diamond)^T{\m H}_{\wh\theta^\diamond}(\wh\theta-\wh\theta^\diamond)-\Big|\frac{1}{2}(\wh\theta-\wh\theta^\diamond)^T{\m H}_{\wh\theta^\diamond}(\wh\theta-\wh\theta^\diamond)-\m R(\wh\theta)+\m R(\wh\theta^\diamond)+\mb{E}[ g(X,\wh\theta^\diamond)^T(\wh\theta-\wh\theta^\diamond)]\Big|\\
    &-\Big| \m R(\wh\theta)-\m R(\wh\theta^\diamond)-\mb{E}\big[  g(X,\wh\theta^\diamond)^T(\wh\theta-\wh\theta^\diamond)\big]\\
    &\qquad-n^{-1}\sum_{i=1}^n \ell(X_i,\wh\theta)+n^{-1}\sum_{i=1}^n \ell(X_i,\wh\theta^\diamond) + \frac{1}{n}\sum_{i=1}^n   g(X_i,\wh\theta^\diamond)^T(\wh\theta-\wh\theta^\diamond)
  \Big|\\
  &\qquad+n^{-1}\sum_{i=1}^n \ell(X_i,\wh\theta^\diamond)-n^{-1}\sum_{i=1}^n \ell(X_i,\wh\theta)\\
    & \geq C_2\, \| \wh\theta-\wh\theta^\diamond\|^2-C_1\,(\log n)^{\frac{1}{\beta_1}}\Big(\sqrt{\frac{\log n}{n}} \, \|\wh\theta-\wh\theta^\diamond\|^{\beta_2+1}+\frac{\log n}{n}\|\wh\theta-\wh\theta^\diamond\|+(\frac{\log n}{n})^2\Big)\\
    &\qquad\qquad-C_1 (\|\wh\theta-\wh\theta^\diamond\|^3+\|\wh\theta-\wh\theta^\diamond\|^2\sqrt{\frac{\log n}{n}})\\
     & \geq \frac{C_2}{2}\, \| \wh\theta-\wh\theta^\diamond\|^2-C_1\,(\log n)^{\frac{1}{\beta_1}}\Big(\sqrt{\frac{\log n}{n}} \, \|\wh\theta-\wh\theta^\diamond\|^{\beta_2+1}+\frac{\log n}{n}\|\wh\theta-\wh\theta^\diamond\|+(\frac{\log n}{n})^2\Big)\\
\end{aligned}
\end{equation*}
Moreover, using $X^{(n)}\in \m A_5$, we have
\begin{equation*}
    \begin{aligned}
      &  \Big\| n^{-1}\sum_{i=1}^n  g(X_i,\wh\theta^\diamond)- n^{-1}\sum_{i=1}^n   g(X_i,\theta^*)-\mathbb{E} [g(X,\wh\theta^\diamond)]+\mathbb{E} [   g(X,\theta^*)]\Big\|_2 \leq C_1\,\frac{(\log n)^{\frac{1}{\beta_1}+1}}{n^{\frac{\beta_2+1}{2}}}.
    \end{aligned}
\end{equation*}
Furthermore, 
\begin{equation*}
    \begin{aligned}
       & \Big\| n^{-1}\sum_{i=1}^n   g(X_i,\theta^*)+\mathbb{E} [g(X,\wh\theta^\diamond)]-\mathbb{E} [ g(X,\theta^*)]\Big\|\\
        &\leq \Big\|n^{-1}\sum_{i=1}^n  g(X_i,\theta^*)+{\m H}_{\theta^*}(\wh\theta^\diamond-\theta^*)\Big\|+C\,\frac{\log n}{n}\\
        &\leq  \Big\|n^{-1}\sum_{i=1}^n  g(X_i,\theta^*)-{\m H}_{\theta^*}W_{\theta^*}(W_{\theta^*}^T{\m H}_{\theta^*}W_{\theta^*})^{-1}\frac{1}{n}\sum_{i=1}^n W_{\theta^*}^Tg(X_i,\theta^*)\Big\|+C_1\frac{\log n}{n}\\
        &=          \Big\|n^{-1}\sum_{i=1}^nW_{\theta^*}W_{\theta^*}^T  g(X_i,\theta^*)-W_{\theta^*}W_{\theta^*}^T{\m H}_{\theta^*}W_{\theta^*}(W_{\theta^*}^T{\m H}_{\theta^*}W_{\theta^*})^{-1}\frac{1}{n}\sum_{i=1}^n W_{\theta^*}^Tg(X_i,\theta^*)\Big\|+C_1\frac{\log n}{n}\\
        &=C_1\frac{\log n}{n}.
    \end{aligned}
\end{equation*}
Therefore, we have 
\begin{equation*}
\begin{aligned}
&2C_1\,\frac{(\log n)^{\frac{1}{\beta_1}+1}}{n^{\frac{\beta_2+1}{2}}}\|
\wh\theta-\wh\theta^\diamond\|\\
    &   \geq  \Big\|\frac{1}{n}\sum_{i=1}^n g(X_i,\wh\theta^\diamond)^T(\wh\theta-\wh\theta^\diamond)\Big\|\\&\geq \frac{C_2}{2}\, \| \wh\theta-\wh\theta^\diamond\|^2 
    -C_3\,(\log n)^{\frac{1}{\beta_1}}\Big(\sqrt{\frac{\log n}{n}} \, \|\wh\theta-\wh\theta^\diamond\|^{\beta_2+1}+\frac{\log n}{n}\|\wh\theta-\wh\theta^\diamond\|+(\frac{\log n}{n})^2\Big).
\end{aligned}
\end{equation*}
So $\|\wh\theta-\wh\theta^\diamond\|\lesssim \frac{(\log n)^{\frac{1}{\beta_1}+1}}{n^{\frac{\beta_2+1}{2}}}$.

\medskip
\noindent\textbf{Step 7: Summarizing the results.}

\medskip
\noindent We have verified in Steps 1-6 the conditions of Lemma~\ref{lemma:generalresult}  with $\gamma_0=\frac{2}{\beta_1}+1$, $\gamma_1=\frac{\beta_2}{2}$, $\gamma_2=3$ and $\Sigma=\m H_0^{-1}\Delta_0\m H_0^{-1}$. Therefore, for any $X^{(n)}\in \m A$, and any positive constants $c_1$, there exists constants $C_0,C_1,C_2$ so that
    \begin{enumerate}
        \item $\Pi_{\rm RP}(\|\theta-\wh\theta^\diamond\|\geq C_0\sqrt{\frac{\log n}{n}}\,|\,X^{(n)})\leq C_1n^{-c_1}$,
        \item  $\|\wh\theta_p-\wh\theta^\diamond\|\leq C_1\,\frac{(\log n)^{\frac{2}{\beta_1}+1}}{n^{\frac{\beta_2}{2}+\frac{1}{2}}}$ and $\|\wh\theta-\wh\theta^\diamond\|\leq C_1\, \frac{(\log n)^{\frac{1}{\beta_1}+1}}{n^{\frac{\beta_2+1}{2}}}$.
        \item    for any $f: \m M\to \mb R^p$  satisfying Assumption B, it holds that
        \begin{equation}\label{resultsf}
              {\rm TV}\Big(f_{\#}\Pi_{\rm RP}(\cdot\,|\,X^{(n)}),\m N\big(f(\wh\theta^\diamond),n^{-1}J_fW_{\theta^*}\m H_0^{-1}\Delta_0\m H_0^{-1}W_{\theta^*}^TJ_f^T\big)\Big)\leq C_2 \frac{(\log n)^{\frac{2}{\beta_1}+1}}{n^{\frac{\beta_2}{2}}}.
        \end{equation}
 \end{enumerate}

\noindent Hence, $\|\sqrt{n}W_{\theta^*}^T\psi_{\theta^*}(\wh{\theta}^{\diamond})-\sqrt{n}W_{\theta^*}^T\psi_{\theta^*}(\wh{\theta})\|$ converge to $0$ in probability. 
Since $\sqrt{n}W_{\theta^*}^T\psi_{\theta^*}(\wh{\theta}^{\diamond})=\m H_0^{-1}\frac{1}{\sqrt{n}}\sum_{i=1}^n W_{\theta^*}^T g(X_i,{\theta^*})$,  by Slutsky’s Theorem and the central Limit theorem, we can obtain that   $\sqrt{n}W_{\theta^*}^T\psi_{\theta^*}(\wh{\theta})$ converges to $\mathcal{N}(0, \m H_0^{-1}W_{\theta^*}^T\ov\Delta_{\theta^*}W_{\theta^*}\m H_0^{-1})$ in distribution.    Using 
   \begin{equation*}
       {\m H}_{\theta^*}=W_{\theta^\ast} W_{\theta^\ast}^T{\m H}_{\theta^*}W_{\theta^\ast} W_{\theta^\ast}^T=W_{\theta^\ast} \m H_0 W_{\theta^\ast}^T,
   \end{equation*}
we conclude that  $\sqrt{n}\psi_{\theta^*}(\wh{\theta})=\sqrt{n}W_{\theta^*}W_{\theta^*}^T\psi_{\theta^*}(\wh{\theta})$ converges to $\mathcal{N}(0, {\m H}_{\theta^*}^{\dagger}\Delta_{\theta^*} {\m H}_{\theta^*}^{\dagger})$ in distribution. Moreover, choosing $f(\theta)=W_{\theta^*}^T(\theta-\theta^*)$, we obtain
\begin{equation*}
 {\rm TV}\bigg( \big(W_{\theta^*}^T(\cdot-\theta^*)\big)_{\#}[\Pi_{\rm RP}(\cdot\,|\,X^{(n)})], \mathcal{N}\big(W_{{\theta}^{\ast}}^T(\wh\theta^\diamond-\theta^*),n^{-1} \m H_0^{-1}\Delta_0\m H_0^{-1}\big)\bigg)  \leq C_2\,  \frac{(\log n)^{\frac{2}{\beta_1}+1}}{n^{\frac{\beta_2}{2}}}.
\end{equation*}
Using $\|\wh{\theta}-\wh\theta^\diamond\|\leq C_1\,\frac{(\log n)^{\frac{1}{\beta_1}+1}}{n^{\frac{\beta_2}{2}+\frac{1}{2}}}$, we further have
\begin{equation*}
 {\rm TV}\bigg( \big(W_{\theta^*}^T(\cdot-\theta^*)\big)_{\#}\Pi_{\rm RP}(\cdot\,|\,X^{(n)})], \mathcal{N}\big(W_{{\theta}^{\ast}}^T(\wh{\theta}-\theta^*),n^{-1} \m H_0^{-1}\Delta_0\m H_0^{-1}\big)\bigg)  \leq C_2\,  \frac{(\log n)^{\frac{2}{\beta_1}+1}}{n^{\frac{\beta_2}{2}}}.
\end{equation*}
 It follows that
\begin{equation*}
     {\rm TV}\bigg(\big({\rm Proj}_{T_{\theta^*}\m M}(\cdot-\theta^*\big)_{\#}[\Pi_{\rm RP}(\cdot\,|\,X^{(n)})], \mathcal{N}\big({\rm Proj}_{T_{\theta^*}\m M}(\wh{\theta}-\theta^*),n^{-1} {\m H}_{\theta^*}^{\dagger}{\Delta}_{\theta^*}{\m H}_{\theta^*}^{\dagger}\big)\bigg)  \leq C_2\, \frac{(\log n)^{\frac{2}{\beta_1}+1}}{n^{\frac{\beta_2}{2}}}.
\end{equation*}
Furthermore, using  $\Pi_{\rm RP}(\|\theta-\wh\theta^\diamond\|\geq C_0\sqrt{\frac{\log n}{n}}\,|\,X^{(n)})\leq n^{-1}$ and $B_{r}(\theta^*)\cap \m M\subseteq U_{\theta^*}$, $B_r(0_d)\cap \m M\subseteq V_{\theta^*}$, we have 
\begin{equation*}
\begin{aligned}
     & {\rm TV}\bigg( \psi_{\theta^*}{}_{\#}[\Pi_{\rm RP}(\cdot\,|\,X^{(n)})\big|_{U_{\theta^*}}], \mathcal{N}\big(\psi_{\theta^*}(\wh{\theta}),n^{-1} {\m H}_{\theta^*}^{\dagger}{\Delta}_{\theta^*}{\m H}_{\theta^*}^{\dagger}\big)\big|_{V_{\theta^*}}\bigg) \\
      &\leq     {\rm TV}\bigg(\big({\rm Proj}_{T_{\theta^*}\m M}(\cdot-\theta^*\big)_{\#}[\Pi_{\rm RP}(\cdot\,|\,X^{(n)})], \mathcal{N}\big({\rm Proj}_{T_{\theta^*}\m M}(\wh{\theta}-\theta^*),n^{-1} {\m H}_{\theta^*}^{\dagger}{\Delta}_{\theta^*}{\m H}_{\theta^*}^{\dagger}\big)\bigg) +\frac{2}{n}\\
&\lesssim  \frac{(\log n)^{\frac{w}{\beta_1}+1}}{n^{\frac{\beta_2}{2}}}.
     \end{aligned}
\end{equation*}
By the invertibility of $\psi_{\theta^*}: U_{\theta^*}\to V_{\theta^*}$, we have 
\begin{equation*}
\begin{aligned}
     & {\rm TV}\bigg(\Pi_{\rm RP}(\cdot\,|\,X^{(n)})\big|_{U_{\theta^*}}, \phi_{\theta^*}{}_{\#}\Big[\mathcal{N}\big(\psi_{\theta^*}(\wh{\theta}),n^{-1} {\m H}_{\theta^*}^{\dagger}{\Delta}_{\theta^*}{\m H}_{\theta^*}^{\dagger}\big)\big|_{V_{\theta^*}}\Big]\bigg)  \lesssim  \frac{(\log n)^{\frac{2}{\beta_1}+1}}{n^{\frac{\beta_2}{2}}}.
     \end{aligned}
\end{equation*}
Finally, using again  $\Pi_{\rm RP}(\|\theta-\wh\theta^\diamond\|\geq C_0\sqrt{\frac{\log n}{n}}\,|\, X^{(n)})\leq n^{-1}$, we conclude 
 \begin{equation*}
\begin{aligned}
     & {\rm TV}\bigg(\Pi_{\rm RP}(\cdot\,|\,X^{(n)}), \phi_{\theta^*}{}_{\#}\Big[\mathcal{N}\big(\psi_{\theta^*}(\wh{\theta}),n^{-1} {\m H}_{\theta^*}^{\dagger}{\Delta}_{\theta^*}{\m H}_{\theta^*}^{\dagger}\big)\big|_{V_{\theta^*}}\Big]\bigg)  \lesssim \frac{(\log n)^{\frac{w}{\beta_1}+1}}{n^{\frac{\beta_2}{2}}}.
     \end{aligned}
\end{equation*}

   \subsection{Proof of Corollary~\ref{co1:nonsmooth}}
Consider events $\m A$ defined in the proof of Theorem~\ref{th1}. Then we have $P(X^{(n)}\in \m A)\geq 1-n^{-2}$, and when $X^{(n)}\in \m A$, we have
$\|\wh{\theta}^{\diamond}-\theta^*\|_2\leq C_1\sqrt{\frac{\log n}{n}}$ and $\|\wh\theta_p-\wh{\theta}^{\diamond}\|_2\leq   C_1\frac{(\log n)^{\frac{1}{\beta_1}+1}}{n^{\frac{\beta_2}{2}+\frac{1}{2}}}$. Moreover, there exists a constant $C_0$ so that $\Pi_{\rm RP}(\|\theta-\wh\theta^\diamond\|\leq C_0\sqrt{\frac{\log n}{n}})\geq 1-n^{-2}$, and using analysis analogous to~\eqref{eqn:cov}, we can get
\begin{equation*}
    \Bmnorm{n\cdot\mb{E}_{\Pi_{\rm RP}}\big[W_{\theta^*}^T(\theta-\wh\theta^\diamond)(\theta-\wh\theta^\diamond)^TW_{\theta^*}\bold{1}\big(\|\theta-\wh\theta^\diamond\|\leq C_0\sqrt{\frac{\log n}{n}}\big)\big]-\m H_0^{-1}\Delta_0\m H_0}_{\rm F}\lesssim \frac{(\log n)^{\frac{2}{\beta_1}+1}}{n^{\frac{\beta_2}{2}}},
\end{equation*}
 and 
\begin{equation*}
\begin{aligned}
      & \Bmnorm{n\cdot\mb{E}_{\Pi_{\rm RP}}\big[W_{\wh\theta_p}^T(\theta-\wh\theta_p)(\theta-\wh\theta_p)^TW_{\wh\theta_p}\big]-\m H_0^{-1}\Delta_0\m H_0}_{\rm F}\\
       &\leq  \Bmnorm{n\cdot\mb{E}_{\Pi_{\rm RP}}\big[W_{\wh\theta_p}^T(\theta-\wh\theta_p)(\theta-\wh\theta_p)^TW_{\wh\theta_p}\bold{1}\big(\|\theta-\wh\theta^\diamond\|\leq C_0\sqrt{\frac{\log n}{n}}\big)\big]-\m H_0^{-1}\Delta_0\m H_0}_{\rm F}+\frac{C}{n}\\
       &\leq \Bmnorm{n\cdot\mb{E}_{\Pi_{\rm RP}}\big[W_{\wh\theta^\diamond}^T(\theta-\wh\theta^\diamond)(\theta-\wh\theta^\diamond)^TW_{\wh\theta^\diamond}\bold{1}\big(\|\theta-\wh\theta^\diamond\|\leq C_0\sqrt{\frac{\log n}{n}}\big)\big]-\m H_0^{-1}\Delta_0\m H_0}_{\rm F}+\frac{(\log n)^{\frac{1}{\beta_1}+2}}{n^{\frac{\beta_2}{2}+\frac{1}{2}}}\\
       &\lesssim  
\Bmnorm{n\cdot\mb{E}_{\Pi_{\rm RP}}\big[\big(W_{\wh\theta^\diamond}^T(\theta-\wh\theta^\diamond)(\theta-\wh\theta^\diamond)^TW_{\wh\theta^\diamond}-W_{\theta^*}^T(\theta-\wh\theta^\diamond)(\theta-\wh\theta^\diamond)^TW_{\theta^*}\big)  \cdot\bold{1}\big(\|\theta-\wh\theta^\diamond\|\leq C_0\sqrt{\frac{\log n}{n}}\big)\big]}_{\rm F}\\
&\qquad +\frac{(\log n)^{\frac{2}{\beta_1}+1}}{n^{\frac{\beta_2}{2}}}\\
&\lesssim \frac{(\log n)^{\frac{2}{\beta_1}+1}}{n^{\frac{\beta_2}{2}}}+\sqrt{\frac{\log n}{n}}\cdot n\cdot \mb{E}_{\Pi_{\rm RP}}\big[\big\|\theta-\wh\theta^\diamond\|^2\cdot\bold{1}\big(\|\theta-\wh\theta^\diamond\|\leq C_0\sqrt{\frac{\log n}{n}}\big)\big],
\end{aligned}
\end{equation*}
where we use the shorthand $\mb{E}_{\Pi_{\rm RP}}[f(\theta)]$ to denote the expectation of $f(\theta)$ with respect to $\theta\sim  \Pi_{\rm RP}(\cdot|X^{(n)})$.
Then using $W_{\theta^*}W_{\theta^*}{}^T(\theta-\wh\theta^{\diamond})=\psi_{\theta^*}(\theta)-\psi_{\theta^*}(\wh\theta^\diamond)$ and $\|\psi_{\theta^*}(\theta)-(\theta-\theta^*)\|\leq C\, \|\theta-\theta^*\|^2$, we have 
 \begin{equation}\label{eqn:squaredbound}
     \begin{aligned}
         &n\cdot \mb{E}_{\Pi_{\rm RP}}\big[\big\|\theta-\wh\theta^\diamond\|^2\cdot\bold{1}\big(\|\theta-\wh\theta^\diamond\|\leq C_0\sqrt{\frac{\log n}{n}}\big)\big]\\
         &\leq n\cdot   \mb{E}_{\Pi_{\rm RP}}\big[\big\|W_{\theta^*}(\theta-\wh\theta^\diamond)\|^2\cdot\bold{1}\big(\|\theta-\wh\theta^\diamond\|\leq C_0\sqrt{\frac{\log n}{n}}\big)+C\,\big(\frac{\log n}{n})^{\frac{3}{2}}\\
         &\leq C_1.
     \end{aligned}
 \end{equation}
Therefore, we have
\begin{equation*}
\begin{aligned}
&  \Bmnorm{n\cdot\mb{E}_{\Pi_{\rm RP}}\big[W_{\wh\theta_p}^T(\theta-\wh\theta_p)(\theta-\wh\theta_p)^TW_{\wh\theta_p}\big]-\m H_0^{-1}\Delta_0\m H_0}_{\rm F}
\lesssim \frac{(\log n)^{\frac{2}{\beta_1}+1}}{n^{\frac{\beta_2}{2}}}.
\end{aligned}
\end{equation*}
Furthermore, using~\eqref{resultsf} and choosing $f(\theta)=W_{\wh\theta_p}^T(\theta-\wh\theta_p)$, we have 
\begin{equation*}
\begin{aligned}
        &{\rm TV}(f_{\#}\Pi_{\rm RP}(\cdot\,|\,X^{(n)}), \m N(0, n^{-1} \m H_0^{-1}\Delta_0\m H_0^{-1}))\\
        &\leq {\rm TV}\Big(f_{\#}\Pi_{\rm RP}(\cdot\,|\,X^{(n)}), \m N\big(W_{\wh\theta_p}^T(\wh\theta^\diamond-\wh\theta_p), n^{-1}W_{\wh\theta_p}^TW_{\theta^*} \m H_0^{-1}\Delta_0\m H_0^{-1}W_{\theta^*}^TW_{\wh\theta_p}\big)\Big)\\
        &\qquad+{\rm TV}\Big(  \m N\big(W_{\wh\theta_p}^T(\wh\theta^\diamond-\wh\theta_p), n^{-1}W_{\wh\theta_p}^TW_{\theta^*} \m H_0^{-1}\Delta_0\m H_0^{-1}W_{\theta^*}^TW_{\wh\theta_p}\big), N(0, n^{-1} \m H_0^{-1}\Delta_0\m H_0^{-1})\Big)\\
        &\lesssim  \frac{(\log n)^{\frac{2}{\beta_1}+1}}{n^{\frac{\beta_2}{2}}}.
\end{aligned}
\end{equation*}
Therefore, let $\chi^2_{1-\alpha}$ denote the $1-\alpha$ quantile of $\chi^2(d)$, there exists a constant $C_3$ so that 
\begin{equation*}
\begin{aligned}
&\Pi_{\rm RP}\left(\psi_{\wh\theta_p}(\theta)^T{\Sigma}_p^{\dagger}\psi_{\wh\theta_p}(\theta)\leq \chi^2_{1-\alpha}(d)- C_3  \frac{(\log n)^{\frac{2}{\beta_1}+1}}{n^{\frac{\beta_2}{2}}}\,\Big|\, X^{(n)}\right)\\
&=\Pi_{\rm RP}\left(\psi_{\wh\theta_p}(\theta)^T{W}_{\wh\theta_p} \Big(\mb{E}_{\Pi_{\rm RP}}\big[W_{\wh\theta_p}^T(\theta-\wh\theta_p)(\theta-\wh\theta_p)^TW_{\wh\theta_p}\big]\Big)^{-1}{W}_{\wh\theta_p}^T\psi_{\wh\theta_p}(\theta)\leq \chi^2_{1-\alpha}(d)- C_3  \frac{(\log n)^{\frac{2}{\beta_1}+1}}{n^{\frac{\beta_2}{2}}}\,\Big|\, X^{(n)}\right)\\
     & \leq \Pi_{\rm RP}\left(n\cdot \psi_{\wh\theta_p}(\theta)^T{W}_{\wh\theta_p}\m H_0\Delta_0^{-1}\m H_0{W}_{\wh\theta_p}^T\psi_{\wh\theta_p}(\theta)\leq \chi^2_{1-\alpha}(d)-  \frac{C_3}{2}\frac{(\log n)^{\frac{2}{\beta_1}+1}}{n^{\frac{\beta_2}{2}}}\,\Big|\, X^{(n)}\right)\leq \alpha 
\end{aligned}
\end{equation*}
and similarly,
\begin{equation*}
\Pi_{\rm RP}\left(\psi_{\wh\theta_p}(\theta)^T{\Sigma}_p^{\dagger}\psi_{\wh\theta_p}(\theta)\leq \chi^2_{1-\alpha}(d)+ C_3  \frac{(\log n)^{\frac{1}{\beta_1}+1}}{n^{\frac{\beta_2}{2}}}\,\Big|\, X^{(n)}\right)\geq \alpha,
\end{equation*}
Which implies $|q_{1-\alpha}-\chi^2_{1-\alpha}(d)|\lesssim \frac{(\log n)^{\frac{2}{\beta_1}+1}}{n^{\frac{\beta_2}{2}}}$. Then let $\m P^*{}^{\otimes n}$ denote the  true distribution of the data $X^{(n)}$, it holds that
    \begin{equation*}
   \begin{aligned}
 & \m P^*{}^{\otimes n} \left(\left\{\psi_{\wh\theta_p}(\theta^*)^T{\Sigma}_p^{\dagger}\psi_{\wh\theta_p}(\theta^*)\leq q_{1-\alpha}\right\}\cap \mathcal{A}\right)\\
 &\leq  \m P^*{}^{\otimes n} \big(\theta^*\in\{\theta\in B_r(\wh\theta_p)\cap \m M\,:\,(\theta-\wh\theta_p)^T\Sigma_p^{\dagger}(\theta-\wh\theta_p)\leq q_{1-\alpha}\}\big)\\
 &\leq \m P^*{}^{\otimes n} \left(\left\{\psi_{\wh\theta_p}(\theta^*)^T {\Sigma}_p^{\dagger}\psi_{\wh\theta_p}(\theta^*)\leq q_{1-\alpha}\right\}\cap \mathcal{A}\right)+\frac{1}{n^2}.
    \end{aligned}
   \end{equation*}
  Furthermore, we can obtain that there exist  positive constants $c_1,c_2$ such that 
    \begin{equation*}
   \begin{aligned}
     &\m P^*{}^{\otimes n} \left(\left\{n\cdot\psi_{\theta^*}(\wh{\theta}^{\diamond})^T{W}_{\theta^*}\m H_0\Delta_0^{-1}\m H_0{W}_{\theta^*}^T\psi_{\theta^*}(\wh{\theta}^{\diamond})\leq \chi^2_{1-\alpha}(d)-c_2\frac{(\log n)^{\frac{2}{\beta_1}+1}}{n^{\frac{\beta_2}{2}}}\right\}\cap \mathcal{A}\right)\\
    &\leq\m P^*{}^{\otimes n} \left(\left\{n\cdot\psi_{\wh{\theta}^{\diamond}}(\theta^*)^T{W}_{\wh\theta_p}\m H_0\Delta_0^{-1}\m H_0{W}_{\wh\theta_p}^T\psi_{\wh{\theta}^{\diamond}}(\theta^*)\leq \chi^2_{1-\alpha}(d)-c_1 \frac{(\log n)^{\frac{2}{\beta_1}+1}}{n^{\frac{\beta_2}{2}}}\right\}\cap \mathcal{A}\right)\\
 &\leq \m P^*{}^{\otimes n}\left(\left\{\psi_{\wh\theta_p}(\theta^*)^T {\Sigma}_p^{\dagger}\psi_{\wh\theta_p}(\theta^*)\leq q_{1-\alpha}\right\}\cap \mathcal{A}\right)\\
 &\leq  \m P^*{}^{\otimes n} \left(\left\{n\cdot\psi_{\wh{\theta}^{\diamond}}(\theta^*)^T{W}_{\wh\theta_p}\m H_0\Delta_0^{-1}\m H_0{W}_{\wh\theta_p}^T\psi_{\wh{\theta}^{\diamond}}(\theta^*)\leq \chi^2_{1-\alpha}(d)+c_1 \frac{(\log n)^{\frac{2}{\beta_1}+1}}{n^{\frac{\beta_2}{2}}}\right\}\cap \mathcal{A}\right)\\
   &\leq\m P^*{}^{\otimes n} \left(\left\{n\cdot\psi_{\theta^*}(\wh{\theta}^{\diamond})^T{W}_{\theta^*}\m H_0\Delta_0^{-1}\m H_0{W}_{\theta^*}^T\psi_{\theta^*}(\wh{\theta}^{\diamond})\leq \chi^2_{1-\alpha}(d)+c_2\frac{(\log n)^{\frac{2}{\beta_1}+1}}{n^{\frac{\beta_2}{2}}}\right\}\cap \mathcal{A}\right)\\
    \end{aligned}
   \end{equation*}
   Then by $\wh{\theta}^{\diamond}=-\phi_{\theta^*}(W_{\theta^*}\m H_0^{-1}W_{\theta^*}^T\frac{1}{n}\sum_{i=1}^n g(X_i,\theta^*))$, we can obtain that 
     \begin{equation*}
   \begin{aligned}
 &\m P^*{}^{\otimes n} \left(\left\{n\cdot\psi_{\theta^*}(\wh{\theta}^{\diamond})^T{W}_{\theta^*}\m H_0\Delta_0^{-1}\m H_0{W}_{\theta^*}^T\psi_{\theta^*}(\wh{\theta}^{\diamond})\leq \chi^2_{1-\alpha}(d)+c_2\frac{(\log n)^{\frac{2}{\beta_1}+1}}{n^{\frac{\beta_2}{2}}}\right\}\cap \mathcal{A}\right)\\
 &=\m P^*{}^{\otimes n}\left(\left\{\left(W_{\theta^*}^T\frac{1}{n}\sum_{i=1}^n  g(X_i,\theta^*)\right)^T \Delta_0^{-1}W_{\theta^*}^T\frac{1}{n}\sum_{i=1}^n  g(X_i,\theta^*)\leq  \chi^2_{1-\alpha}(d)+c_2\frac{(\log n)^{\frac{2}{\beta_1}+1}}{n^{\frac{\beta_2}{2}}}\right\}\cap \mathcal{A}\right)\\
 &\leq \m P^*{}^{\otimes n}\left(\left(W_{\theta^*}^T\frac{1}{n}\sum_{i=1}^n  g(X_i,\theta^*)\right)^T \Delta_0^{-1}W_{\theta^*}^T\frac{1}{n}\sum_{i=1}^n  g(X_i,\theta^*)\leq  \chi^2_{1-\alpha}(d)+c_2\frac{(\log n)^{\frac{2}{\beta_1}+1}}{n^{\frac{\beta_2}{2}}}\right).
    \end{aligned}
   \end{equation*}
Since $\mathbb{E}_{\m P^*} [g(X,\theta)]=0$ and ${\rm Cov}_{\m P^*} (W_{\theta^*} g(X,\theta^*))=\Delta_0$, using Berry-Esseen theorem~\citep{Rai__2019},  we can obtain that
  \begin{equation*}
   \begin{aligned}
  & \m P^*{}^{\otimes n}\big(\theta^*\in\{\theta\in B_r(\wh\theta_p)\cap \m M\,:\,\psi_{\wh\theta_p}(\theta)^T{\Sigma}_p^{\dagger}\psi_{\wh\theta_p}(\theta)\leq q_{1-\alpha}\}\big)\\
&\leq \m P^*{}^{\otimes n}\left(\left(W_{\theta^*}^T\frac{1}{n}\sum_{i=1}^n   g(X_i,\theta^*)\right)^T \Delta_0^{-1}W_{\theta^*}^T\frac{1}{n}\sum_{i=1}^n  g(X_i,\theta^*)\leq  \chi^2_{1-\alpha}(d)+c_2\frac{(\log n)^{\frac{2}{\beta_1}+1}}{n^{\frac{\beta_2}{2}}}\right)\\
&\leq 1-\alpha+c_3 \frac{(\log n)^{\frac{2}{\beta_1}+1}}{n^{\frac{\beta_2}{2}}}.
     \end{aligned}
\end{equation*}
    Similarly, we can obtain that
      \begin{equation*}
   \begin{aligned}
& \m P^*{}^{\otimes n} \big(\theta^*\in\{\theta\in B_r(\wh\theta_p)\cap \m M\,:\,\psi_{\wh\theta_p}(\theta)^T{\Sigma}_p^{\dagger}\psi_{\wh\theta_p}(\theta)\leq q_{1-\alpha}\}\big)\\
&\geq 1-\alpha-c_3\frac{(\log n)^{\frac{1}{\beta_1}+1}}{n^{\frac{\beta_2}{2}}}.
     \end{aligned}
\end{equation*}
Proof is completed.

\subsection{Proof of Corollary~\ref{co:wellspecified}}

Denote
    \begin{equation*}
       \begin{aligned}
         &      \ell(x,\theta)=-\log p(x|\theta)\\
         &\ov {\m R}(\theta)=\mb{E}[\ell(x,\theta)], \, \theta\in \mb R^D\\
         &\m R=\ov {\m R}|_{\m M}\\
         &  g(x,\theta)={\rm grad}_{\theta}\ell(x,\theta)=P_{\theta^*}\nabla_{\theta} \ell(x,\theta).
           \end{aligned}
    \end{equation*}
  Since  
  \begin{equation*}
      \nabla \ov {\m R}(\theta^*)=\int -\nabla_{\theta} p(x|\theta^*)\, \dd x=0_D,
  \end{equation*}
using Corollary 5.47 of~\cite{boumal2022intromanifolds}, the Riemannian Hessian matrix ${\m H}_{\theta^*}$ of $\m R$ at $\theta^*$ satisfies
    \begin{equation}\label{eqnHessian}
{\m H}_{\theta^*}=P_{\theta^*}{\rm Hessian}(\ov {\m R}(\theta^*))P_{\theta^*}=P_{\theta^*}I_{\theta^*}P_{\theta^*},
    \end{equation}
    where ${\rm Hessian}(\ov{\m R}(\theta^*))$ denotes the Euclidean Hessian matrix of $\ov{\m R}(\cdot)$ at $\theta^*$, and $I_{\theta^*}$ is the Fisher Information matrix.  Then by Assumption 5, it is straightforward to verify that Assumption 3-4 holds with $\beta_2=1$. Moreover, 
    \begin{equation*}
        {\Delta}_{\theta^*}=\mb{E}[ g(X,\theta^*) g(X,\theta^*)^T]=P_{\theta^*}\mb{E}\left[\nabla_{\theta} \ell(x,\theta^*)\nabla_{\theta} \ell(x,\theta^*)^T\right]P_{\theta^*}=P_{\theta^*}I_{\theta^*}P_{\theta^*},
    \end{equation*}
    \begin{equation*}
         {\m H}_{\theta^*}^{\dagger}\Delta_{\theta^*}{\m H}_{\theta^*}^{\dagger}=(P_{\theta^*}I_{\theta^*}P_{\theta^*})^{\dagger}.
    \end{equation*}
    The desired result then directly follows from Theorem~\ref{th1}.

  \section{Proof for Gibbs Posterior}\label{proof:well-specified}
 In this section, we prove Theorem~\ref{th:classicbvm} and Corollary~\ref{co:classicbvm_region}. 
\subsection{Proof of Theorem~\ref{th:classicbvm}}
Similar as the proof of Theorem~\ref{th1}, we will use Lemma~\ref{lemma:generalresult} to show the desired result. We will start with verifying the conditions of  Lemma~\ref{lemma:generalresult}. Consider the set $\m A$ defined in the proof of Theorem~\ref{th1}, then $P(X^{(n)}\in \m A)\geq 1-n^{-2}$. Moreover, since $\m M$ is locally $C^3_{r,L}$ around $\theta^*$,  there exists $U_{\theta^*},V_{\theta*}$ with $B_{r}(\theta)\cap \m M\subseteq U_{\theta^*}\subseteq \m M$ and $B_{r}(0_D)\cap T_{\theta^*}\m M\subseteq V_{\theta^*}\subset T_{\theta^*}\m M$, so that $\psi_{\theta^*}: U_{\theta^*}\to V_{\theta^*}$ defined by $\psi_{\theta^*}(x)={\rm Proj}_{T_{\theta^*}\m M}(x-\theta^*)$ is a bijective, where the inverse, denoted by $\phi_{\theta^*}$ is thrice Fr\'{e}chet differentiable. Then it has been show in the proof of Theorem~\ref{th1} that, there exists a constant $C_2$ so that 
 \begin{equation*}
    \underset{v\in  V_{\theta^*} }{\sup} \frac{\|\phi_{\theta^*}(v)-(v+\theta^*)\|}{\|v\|^2}\leq C_2,
\end{equation*}
and
\begin{equation*}
\begin{aligned}
      \underset{\theta'\in U_{\theta^*}}{\sup}\frac{\|\psi_{\theta^*}(\theta')-(\theta'-\theta^*)\|}{\|\theta^*-\theta'\|^2}\leq C_2.
\end{aligned}
 \end{equation*}
Fix an arbitrary  $X^{(n)}\in \m A$,  and
define $$\wh\theta^{\diamond}(X^{(n)})=\phi_{\theta^*}
\big(-W_{\theta^*}(W_{\theta^*}^T{\m H}_{\theta^*}W_{\theta^*})^{-1}\frac{1}{n}\sum_{i=1}^n W_{\theta^*}^T g(X_i,\theta^*)\big).$$ It has been shown in the proof of Theorem~\ref{th1} that $\|\wh\theta^\diamond-\theta^*\|\lesssim \sqrt{\frac{\log n}{n}}$. Then it suffices to prove 
\begin{enumerate}
    \item Let  $\ms L(X^{(n)},\theta)=  \exp\big(-n \cdot(\m R_n(\theta)-\m R_n(\wh\theta^\diamond))\big)$. Show  that for any  $h\in B_{\delta_1(\log n)^{3/2}}(0_d)$, 
        \begin{equation*}
            \begin{aligned}
          & \bigg| \log \ms L\Big(X^{(n)},\phi_{\theta^*}\Big(\frac{W_{\theta^*}h}{\sqrt{n}}+\psi_{\theta^*}(\wh\theta^\diamond)\Big)\Big)+ \frac{1}{2} h^T  \m H_0 h\bigg| \leq C\,\frac{(\log n)^{\frac{1}{\beta_1}+1}}{n^{\frac{\beta_2}{2}}}\cdot(\|h\|^3+1).
            \end{aligned}
        \end{equation*}
    \item For any constant $\delta>0$ and constant $c_1\geq 2$, it holds for sufficiently large $n$ that $\Pi(\|\theta-\wh\theta^\diamond\|\geq \delta\,|\, X^{(n)})\leq n^{-c_1}$. 
    \item  For any positive constant $c_1\geq 2$, with for sufficiently large $\delta_1$,  for any $h\in\mb R^d$ satisfying  $\|h\|\geq \delta_1(\log n)^{3/2}$ and $\frac{W_{\theta^*}h}{\sqrt{n}}+\psi_{\theta^*}(\wh\theta^\diamond)\in V_{\theta^*}$,  it holds that  $    \log\ms L\big(X^{(n)},\phi_{\theta^*}\big(\frac{W_{\theta^*}h}{\sqrt{n}}+\psi_{\theta^*}(\wh\theta^\diamond)\big)\big)\leq -2\,c_1 d\, \log n.$
\end{enumerate}
The we can get the desired result by following the Step 7 of the proof of Theorem~\ref{th1}. Now we prove the above three claims. Then fix an arbitrary  $h\in B_{\delta_1(\log n)^{3/2}}(0_d)$, denote  $\theta=\phi_{\theta^*}\big(\frac{W_{\theta^*}h}{\sqrt{n}}+\psi_{\theta^*}(\wh\theta^\diamond)\big)$, it holds that
\begin{equation*}
    \begin{aligned}
        &\Big\|\theta-\wh\theta^\diamond-\frac{W_{\theta^*}h}{\sqrt{n}}\Big\|\\
        &\leq \Big\|\frac{W_{\theta^*}h}{\sqrt{n}}+\psi_{\theta^*}(\wh\theta^\diamond)+\theta^*-\theta)\Big\|+\big\|\wh\theta^\diamond-\theta^*-\psi_{\theta^*}(\wh\theta^\diamond)\big\|\\
        &\lesssim \frac{\log n}{n}+\frac{\|h\|^2}{n}.
    \end{aligned}
\end{equation*}
 Since $X^{(n)}\in \m A_7$,
\begin{equation*}
    \begin{aligned}
      &\Big| -\frac{1}{n}\ms L\big(X^{(n)},\theta\big)-\frac{1}{n}\sum_{i=1}^n  g(X_i,\wh\theta^\diamond)^T(\theta-\wh\theta^\diamond)-\big(\mb{E}[\ell(X,\theta)]-\mb{E}[\ell(X,\wh\theta^\diamond)]-\mb{E}[ g(X,\wh\theta^\diamond)^T(\theta-\wh\theta^\diamond)]\big)\Big|\\
       &\leq  C_1\,(\log n)^{\frac{1}{\beta_1}}\Big(\sqrt{\frac{\log n}{n}} \, \|\theta-\wh\theta^\diamond\|^{\beta_2+1}+\frac{\log n}{n}\|\theta-\wh\theta^\diamond\|+(\frac{\log n}{n})^2\Big)\bigg\}\\
       &\lesssim   \frac{(\log n)^{1+\frac{1}{\beta_1}}}{n^{1+\frac{\beta_2}{2}}}(\|h\|^{\beta_2+1}+1).
    \end{aligned}
\end{equation*}
Then notice that,
\begin{equation*}
    \begin{aligned}
 & \Big|\mb{E}[\ell(X,\theta)]-\mb{E}[\ell(X,\wh\theta^\diamond)]-\mb{E}[ g(X,\wh\theta^\diamond)^T(\theta-\wh\theta^\diamond)]-\frac{1}{2n} h^TW_{\theta^*}^T{\m H}_{\theta^*}W_{\theta^*}h\Big|\\
       &\leq  \Big|\mb{E}[\ell(X,\theta)]-\mb{E}[\ell(X,\wh\theta^\diamond)]-\mb{E}[ g(X,\wh\theta^\diamond)^T(\theta-\wh\theta^\diamond)]-\frac{1}{2} (\theta-\wh\theta^\diamond)^T{\m H}_{\theta^*}(\theta-\wh\theta^\diamond)\Big|\\
       &+\Big|\frac{1}{2n} h^TW_{\theta^*}^T{\m H}_{\theta^*}W_{\theta^*}h-\frac{1}{2} (\theta-\wh\theta^\diamond)^T{\m H}_{\theta^*}(\theta-\wh\theta^\diamond)\Big|\lesssim\frac{\log n}{n^{\frac{3}{2}}}(\|h\|^3+1).
    \end{aligned}
\end{equation*}
Moreover, since $X^{(n)}\in \m A_5$,  
\begin{equation*}
    \begin{aligned}
    &  \Big\| n^{-1}\sum_{i=1}^n  g(X_i,\wh\theta^\diamond)- n^{-1}\sum_{i=1}^n  g(X_i,\theta^*)-\mathbb{E} [ g(X,\wh\theta^\diamond)]\Big\| \\
        &=\Big\| n^{-1}\sum_{i=1}^n  g(X_i,\wh\theta^\diamond)- n^{-1}\sum_{i=1}^n  g(X_i,\theta^*)-\mathbb{E} [ g(X,\wh\theta^\diamond)]+\mathbb{E} [  g(X,\theta^*)]\Big\| \\
        &\leq C_2\,(\log n)^{\frac{1}{\beta_1}}(\frac{\log n}{n})^{\frac{1+\beta_2}{2}}.
    \end{aligned}
\end{equation*}
Furthermore, 
\begin{equation*}
    \begin{aligned}
        &\Big\|n^{-1}\sum_{i=1}^n   g(X_i,\theta^*)+\mathbb{E} [ g(X,\wh\theta^\diamond)]\Big\|=\Big\|n^{-1}\sum_{i=1}^n   g(X_i,\theta^*)+\mathbb{E} [ g(X,\wh\theta^\diamond)]-\mathbb{E} [g(X,\theta^*)]\Big\|\\
        &\leq \Big\|n^{-1}\sum_{i=1}^n g(X_i,\theta^*)+\m H_{\theta^*}(\wh\theta^\diamond-\theta^*)\Big\|+C\,\frac{\log n}{n}\\
        &\leq  \Big\|n^{-1}\sum_{i=1}^n  W_{\theta^*}W_{\theta^*}^T  g(X_i,\theta^*)-W_{\theta^*}W_{\theta^*}^T\m H_{\theta^*}W_{\theta^*}(W_{\theta^*}^T{\m H}_{\theta^*}W_{\theta^*})^{-1}\frac{1}{n}\sum_{i=1}^n W_{\theta^*}^T g(X_i,\theta^*)\Big\|+C_1\frac{\log n}{n}\\
        &=C_1\frac{\log n}{n}.
    \end{aligned}
\end{equation*}
 Therefore, we have 
 \begin{equation*}
     \begin{aligned}
         &\big\| n^{-1}\sum_{i=1}^n  g(X_i,\wh\theta^\diamond)(\theta-\wh\theta^\diamond)\big\|\leq\big\|n^{-1}\sum_{i=1}^n  g(X_i,\wh\theta^\diamond)\|\cdot \|\theta-\wh\theta^\diamond\|\\
         &\lesssim \frac{\|h\|+1}{\sqrt{n}}\cdot  \Big(\Big\| n^{-1}\sum_{i=1}^n  g(X_i,\wh\theta^\diamond)- n^{-1}\sum_{i=1}^n    g(X_i,\theta^*)-\mathbb{E} [ g(X,\wh\theta^\diamond)]\Big\|+\Big\|n^{-1}\sum_{i=1}^n  g(X_i,\theta^*)+\mathbb{E} [ g(X,\wh\theta^\diamond)]\Big\|\Big)\\
         &\lesssim (\log n)^{\frac{1}{\beta_1}}(\frac{\log n}{n})^{\frac{1+\beta_2}{2}}\frac{\|h\|+1}{\sqrt{n}}.
     \end{aligned}
 \end{equation*}
Therefore, combining all pieces, we can get 
        \begin{equation}\label{eqn:boundmsLlikelihood}
            \begin{aligned}
          & \bigg| \log \ms L\Big(X^{(n)},\phi_{\theta^*}\Big(\frac{W_{\theta^*}h}{\sqrt{n}}+\psi_{\theta^*}(\wh\theta^\diamond)\Big)\Big)+ \frac{1}{2} h^T  \m H_0 h\bigg| \\
          &= \bigg| \log \ms L\big(X^{(n)},\theta\big)+ \frac{1}{2} h^T  \m H_0 h\bigg|  \lesssim \frac{(\log n)^{\frac{1}{\beta_1}+1}}{n^{\frac{\beta_2}{2}}}\cdot(\|h\|^3+1).
            \end{aligned}
        \end{equation}
 For the second claim, Since $\|\wh\theta^\diamond-\theta^*\|\leq C\, \sqrt{\frac{\log n}{n}}$, for any positive constant $\delta$, there exists $N$ so that when $n\geq N$,
 \begin{equation*}
 \begin{aligned}
   \Pi (\|\theta-\wh\theta^\diamond\|_2\geq \delta\,|\, X^{(n)})&\leq \Pi (\|\theta-\theta^*\|_2\geq \delta/2\,|\, X^{(n)})\\
  & =\frac{\int_{B_{\frac{\delta}{2}}(\theta^*)^c\cap \mathcal{M}} \exp\big(-n\cdot(\mathcal{R}_n(\theta)-\mathcal{R}_n(\wh\theta^\diamond))\big) \Pi (\dd\theta)}{\int  \exp\big(-n\cdot(\mathcal{R}_n(\theta)-\mathcal{R}_n(\wh\theta^\diamond))\big)\Pi (\dd\theta)}\\
  &\leq  \frac{\int_{B_{\frac{\delta}{2}}(\theta^*)^c\cap \mathcal{M}}\exp\big(-n\cdot(\mathcal{R}_n(\theta)-\mathcal{R}_n(\wh\theta^\diamond))\big)\Pi (\dd\theta)}{\int_{  B_{{1}/{n}}(\wh\theta^\diamond)\cap \mathcal{M}} \exp\big(-n\cdot(\mathcal{R}_n(\theta)-\mathcal{R}_n(\wh\theta^\diamond))\big) \Pi (\dd\theta)}
  \end{aligned}
  \end{equation*}
For the denominator,  by equation~\eqref{eqn:boundmsLlikelihood}, there exists positive constant $c,c'$ such that  
\begin{equation*}
 \begin{aligned}
& \int_{\theta\in\mb  B_{{1}/{n}}(\wh\theta^\diamond)\cap \mathcal{M}}  \exp\big(-n\cdot(\mathcal{R}_n(\theta)-\mathcal{R}_n(\wh\theta^\diamond))\big)\Pi(\dd\theta)\\
&\geq  c\int_{\theta\in B_{{1}/{n}}(\wh\theta^\diamond)\cap \mathcal{M}} \Pi(\dd\theta)\\
&\geq \exp(-c'\log n).
\end{aligned}
  \end{equation*}
   For the numerator, by the assumption that $\mathcal{R}(\theta)-\mathcal{R}(\theta^*)\geq L\, \|\theta-\theta^*\|^2$ and $\|\theta^*-\wh\theta^\diamond\|\lesssim \sqrt{\frac{\log n}{n}}$, when $n$ is sufficiently large, there exists a positive constant $c_3$ so that for any $\theta\in B_{\frac{\delta}{2}}(\theta^*)^c\cap \mathcal{M}$, $\mathcal{R}(\theta)-\mathcal{R}(\wh\theta^\diamond)\geq c_3$. Moreover, since $X^{(n)}\in \m A_6$, for any $\theta\in B_{\frac{\delta}{2}}(\theta^*)^c\cap \mathcal{M}$,
    \begin{equation}\label{eqn:mRcons}
\mathcal{R}_n(\theta)-\mathcal{R}_n(\wh\theta^\diamond) \geq c_3/2.
\end{equation}
 Therefore, when $n$ is sufficiently large, 
   \begin{equation*}
  \Pi (\|\theta-\wh\theta^\diamond\|_2\geq \delta\,|\, X^{(n)})\leq \Pi (\|\theta-\theta^*\|_2\geq \frac{\delta}{2}\,|\, X^{(n)})\leq \exp(c'\log n)\exp(-c_3n/2)\leq n^{-c_1}.
  \end{equation*}
For the last claim,  consider any $h\in \mb R^d$ with $\|h\|\geq \delta_1(\log n)^{3/2}$ and $\frac{W_{\theta^*}h}{\sqrt{n}}+\psi_{\theta^*}(\wh\theta^\diamond)\in V_{\theta^*}$ denote $\theta=\phi_{\theta^*}(\frac{W_{\theta^*}h}{\sqrt{n}}+\psi_{\theta^*}(\wh\theta^\diamond))$.  If  $\theta\notin B_r(\theta^*)\cap \m M$, then using~\eqref{eqn:mRcons}, it is straightforward to show  that
\begin{equation*}
            \log\ms L\big(X^{(n)},\phi_{\theta^*}\big(\frac{W_{\theta^*}h}{\sqrt{n}}+\psi_{\theta^*}(\wh\theta^\diamond)\big)\big)\leq -2 c_1d\log n.
\end{equation*}
 If  $\theta\in B_r(\theta^*)\cap \m M$, using $X^{(n)}\in \m A_1\cap \m A_2\cap \m A_5\cap \m A_7$, we have 
\begin{equation*}
    \begin{aligned}
       &\log\ms L\big(X^{(n)},\phi_{\theta^*}\big(\frac{W_{\theta^*}h}{\sqrt{n}}
    +\psi_{\theta^*}(\wh\theta^\diamond)\big)\big)-\big(\mb{E}[\ell(X,\theta)]\\
    &\leq \m R(\wh\theta^\diamond)-\m R(\theta)+\Big|\m R(\theta)-\m R(\wh\theta^\diamond)-\mb{E}[g(X,\wh\theta^\diamond)](\theta-\wh\theta^\diamond)-\m R_n(\theta)+\m R_n(\wh\theta^\diamond)+\frac{1}{n}\sum_{i=1}^n g(X_i,\wh\theta^\diamond)(\theta-\wh\theta^\diamond)\Big|\\
    &+\|\theta-\wh\theta^\diamond\|\cdot \Big\|\mb{E}[g(X,\wh\theta^\diamond)]-\mb{E}[g(X,\theta^*)]-\frac{1}{n}\sum_{i=1}^n g(X_i,\wh\theta^\diamond)+\frac{1}{n}\sum_{i=1}g(X_i,\theta^*)\Big\|+\big\|\frac{1}{n}\sum_{i=1}g(X_i,\theta^*)\big\|\\
    &\leq  \m R(\wh\theta^\diamond)-\m R(\theta)+C_1 \frac{(\log n)^{\frac{1}{2}+\frac{1}{\beta_1}}}{\sqrt{n}} \big(\frac{\|h\|}{\sqrt{n}}\big)^{\beta_2+1}.
    \end{aligned}
\end{equation*}
Moreover, 
\begin{equation*}
    \begin{aligned}
    &\m R(\theta)- \m R(\wh\theta^\diamond) \\
    &\geq \m R(\theta)-\m R(\wh\theta^*)+\m R(\wh\theta^*)-\m R(\wh\theta^\diamond)\\
    &\geq L\, \frac{\|h\|^2}{n}-C_1{\frac{\log n}{n}}.
    \end{aligned}
\end{equation*}
Therefore, when $\delta_1$ is large enough, we have 
\begin{equation*}
    \begin{aligned}
        & \log\ms L\big(X^{(n)},\phi_{\theta^*}\big(\frac{W_{\theta^*}h}{\sqrt{n}}+\psi_{\theta^*}(\wh\theta^\diamond)\big)\big)\\
         &\leq n\cdot\Big(-L\frac{\|h\|^2}{n}+C_1{\frac{\log n}{n}}++C_1 \frac{(\log n)^{\frac{1}{2}+\frac{1}{\beta_1}}}{\sqrt{n}} \big(\frac{\|h\|}{\sqrt{n}}\big)^{\beta_2+1}\Big)\\
         &\leq -2c_1d\log n.
    \end{aligned}
\end{equation*}
 Proof is completed.
 \subsection{Proof of Corollary~\ref{co:classicbvm_region}}
Denote
    \begin{equation*}
       \begin{aligned}
         &      \ell(x,\theta)=-\log p(x|\theta)\\
         &\ov {\m R}(\theta)=\mb{E}[\ell(x,\theta)], \, \theta\in \mb R^D\\
         &\m R=\ov {\m R}|_{\m M}\\
         &  g(x,\theta)={\rm grad}_{\theta}\ell(x,\theta)=P_{\theta^*}\nabla_{\theta} \ell(x,\theta).
           \end{aligned}
    \end{equation*}
    Then  given Assumption 5, it is straightforward to verify that Assumption 3 and 4 hold with $\beta_2=1$. Moreover, using~\eqref{eqnHessian}, we have ${\m H}_{\theta^*}=P_{\theta^*}I_{\theta^*}P_{\theta^*}$ and $\m H_0=W_{\theta^*}^T I_{\theta^*}P_{\theta^*}$. 
 Consider events $\m A$ defined in the proof of Theorem~\ref{th1}. Then we have $P(X^{(n)}\in \m A)\geq 1-n^{-2}$. Fix an $X^{(n)}\in \m A$ and
define $$\wh\theta^\diamond=\wh\theta^{\diamond}(X^{(n)})=\phi_{\theta^*}
\big(-W_{\theta^*}(W_{\theta^*}^T{\m H}_{\theta^*}W_{\theta^*})^{-1}\frac{1}{n}\sum_{i=1}^n W_{\theta^*}^T g(X_i,\theta^*)\big).$$ Then it has been shown in the step 6 of the proof of Theorem~\ref{th1} that  $\|\wh\theta-\wh\theta^\diamond\|\lesssim  \frac{(\log n)^{\frac{1}{\beta_1}+1}}{n}$, where we have taken $\beta_2=1$. Then using  Theorem~\ref{th:classicbvm}, we have 
\begin{equation*}
    \begin{aligned}
            {\rm TV}(f_{\#}\Pi(\cdot|X^{(n)}), N(f(\wh\theta), n^{-1} \sigma_f^2)) 
     \leq C\,  \frac{(\log n)^{\frac{1}{\beta_1}+1}}{n^{\frac{\beta_2}{2}}},\quad \sigma_f^2=\nabla f(\theta^*)^T (P_{\theta^*}I_{\theta^*}P_{\theta^*})^{\dagger} \nabla f(\theta^*),
    \end{aligned}
\end{equation*}
and
\begin{equation*}
\begin{aligned}
        &{\rm TV}(f_{\#}\Pi(\cdot|X^{(n)}), N(f(\wh\theta^\diamond), n^{-1} \sigma_f^2)) 
     \leq C\,  \frac{(\log n)^{\frac{1}{\beta_1}+1}}{n^{\frac{\beta_2}{2}}}.
\end{aligned}
\end{equation*}
 Now denote
 \begin{equation*}
     P=I_{\theta^*} (P_{\theta^*}^TI_{\theta^*}P_{\theta^*})^{\dagger}=I_{\theta^*}W_{\theta^*}(W_{\theta^*}^TI_{\theta^*}W_{\theta^*})^{-1} W_{\theta^*}^T.
 \end{equation*}
Then we have  for any $\eta\in \mb R^D$,
\begin{equation*}
    \begin{aligned}
        \eta^TI_{\theta^*}^{-1}\eta&=(P\eta+(I_D-P)\eta)^TI_{\theta^*}^{-1}(P\eta+(I_D-P)\eta)\\
        &=\eta^TP^TI_{\theta^*}^{-1}P\eta+\eta^T(I_D-P)^TI_{\theta^*}^{-1}P\eta+\eta^TP^TI_{\theta^*}^{-1}(I_D-P)\eta+\eta^T(I_D-P)^TI_{\theta^*}^{-1}(I_D-P)\eta.\\
    \end{aligned}
\end{equation*}
\begin{equation*}
    \begin{aligned}
        P^TI_{\theta^*}^{-1}P&=W_{\theta^*}(W_{\theta^*}^TI_{\theta^*}W_{\theta^*})^{-1}W_{\theta^*}^TI_{\theta^*}I_{\theta^*}^{-1}I_{\theta^*}W_{\theta^*}(W_{\theta^*}^TI_{\theta^*}W_{\theta^*})^{-1} W_{\theta^*}^T\\
        &=W_{\theta^*}(W_{\theta^*}^TI_{\theta^*}W_{\theta^*})^{-1}W_{\theta^*}^T.
    \end{aligned}
\end{equation*}

\begin{equation*}
    \begin{aligned}
        (I_D-P)^TI_{\theta^*}^{-1}P&=(I_D-I_{\theta^*}W_{\theta^*}(W_{\theta^*}^TI_{\theta^*}W_{\theta^*})^{-1} W_{\theta^*}^T)^TI_{\theta^*}^{-1}I_{\theta^*}W_{\theta^*}(W_{\theta^*}^TI_{\theta^*}W_{\theta^*})^{-1} W_{\theta^*}^T\\
        &=W_{\theta^*}(W_{\theta^*}^TI_{\theta^*}W_{\theta^*})^{-1}W_{\theta^*}^T-W_{\theta^*}(W_{\theta^*}^TI_{\theta^*}W_{\theta^*})^{-1}W_{\theta^*}^T\\
        &=0.
    \end{aligned}
\end{equation*}
 Therefore,
 \begin{equation*}
    \begin{aligned}
        \eta^TI_{\theta^*}^{-1}\eta
        &=\eta^TW_{\theta^*}(W_{\theta^*}^TI_{\theta^*}W_{\theta^*})^{-1}W_{\theta^*}^T\eta+\eta^T(I_D-P)^TI_{\theta^*}^{-1}(I_D-P)\eta\\
        &\geq \eta^TW_{\theta^*}(W_{\theta^*}^TI_{\theta^*}W_{\theta^*})^{-1}W_{\theta^*}^T\eta=\eta^T (P_{\theta^*}^TI_{\theta^*}P_{\theta^*})^{\dagger}\eta.\\
    \end{aligned}
\end{equation*}
Hence  we have $(P_{\theta^*}^TI_{\theta^*}P_{\theta^*})\preccurlyeq I_{\theta^*}^{-1}$. 

We now prove the last statement. Let $z_{\frac{\alpha}{2}}$ to be the $\frac{\alpha}{2}$ quantile of $\m N(0,1)$ and let $Z\sim \m N(0,1)$, there exists a constant $C_1$ so that 
\begin{equation*}
    \begin{aligned}
       &\Pi\left( f(\theta)\geq f(\wh\theta^\diamond)+\frac{\sigma_f}{\sqrt{n}}z_{\frac{\alpha}{2}}+C_1\frac{(\log n)^{\frac{1}{\beta_1}+1}}{n^{\frac{1+\beta_2}{2}}}\,|\, X^{(n)}\right)\\
       &\leq P\Big(Z\geq z_{\frac{\alpha}{2}}+C_1\frac{(\log n)^{\frac{1}{\beta_1}+1}}{n^{\frac{\beta_2}{2}}\sigma_f}\Big)+C\,\frac{(\log n)^{\frac{1}{\beta_1}+1}}{n^{\frac{\beta_2}{2}}}\leq \frac{\alpha}{2},
    \end{aligned}
\end{equation*}
and 
\begin{equation*}
    \begin{aligned}
       &\Pi\left( f(\theta)\geq f(\wh\theta^\diamond)+\frac{\sigma_f}{\sqrt{n}}z_{\frac{\alpha}{2}}-C_1\frac{(\log n)^{\frac{1}{\beta_1}+1}}{n^{\frac{1+\beta_2}{2}}}\,|\, X^{(n)}\right)\\
       &\geq P\Big(Z\geq z_{\frac{\alpha}{2}}-C_1\frac{(\log n)^{\frac{1}{\beta_1}+1}}{n^{\frac{\beta_2}{2}}\sigma_f}\Big)-C\,\frac{(\log n)^{\frac{1}{\beta_1}+1}}{n^{\frac{\beta_2}{2}}}\geq \frac{\alpha}{2}.
    \end{aligned}
\end{equation*}
 So $$\Big|q_{\frac{\alpha}{2}}^f-f(\wh\theta^\diamond)-\frac{\sigma_f}{\sqrt{n}}z_{\frac{\alpha}{2}}\Big|\leq C_1\frac{(\log n)^{\frac{1}{\beta_1}+1}}{n^{\frac{1+\beta_2}{2}}}.$$ Similarly, we can show that
 $$\Big|q_{1-\frac{\alpha}{2}}^f-f(\wh\theta^\diamond)-\frac{\sigma_f}{\sqrt{n}}z_{1-\frac{\alpha}{2}}\Big|\leq C_1\frac{(\log n)^{\frac{1}{\beta_1}+1}}{n^{\frac{1+\beta_2}{2}}}.$$
Furthermore,  let $\m P^*{}^{\otimes n}$ be the distribution of the data $X^{(n)}$, then
 \begin{equation*}
 \begin{aligned}
  &\m P^*{}^{\otimes n}\left(\left\{ q_{\alpha/2}^f\leq f(\theta^*)\leq q_{1-\frac{\alpha}{2}}^f \right\}\cap \m A \right)\\
  &\leq\m P^*{}^{\otimes n}\left(  q_{\alpha/2}^f\leq f(\theta^*)\leq q_{z_{1-\frac{\alpha}/{2}}}^f  \right)\\
  &\leq \m P^*{}^{\otimes n}\left(\left\{ q_{\alpha/2}^f\leq f(\theta^*)\leq q_{z_{1-\frac{\alpha}/{2}}}^f  \right\}\cap \m A \right)+\frac{2}{n^2}.\\
 \end{aligned}
 \end{equation*}
For the term  $\m P^*{}^{\otimes n}\left(\left\{ q_{\alpha/2}^f\leq f(\theta^*)\leq q_{z_{1-\frac{\alpha}/{2}}}^f  \right\}\cap \m A \right)$,  using $\big|q_{\frac{\alpha}{2}}^f-f(\wh\theta^\diamond)-\frac{\sigma_f}{\sqrt{n}}z_{\frac{\alpha}{2}}\big|\leq C_1\frac{(\log n)^{\frac{1}{\beta_1}+1}}{n^{\frac{1+\beta_2}{2}}}$ and  $\big|q_{1-\frac{\alpha}{2}}^f-f(\wh\theta^\diamond)-\frac{\sigma_f}{\sqrt{n}}z_{1-\frac{\alpha}{2}}\big|\leq C_1\frac{(\log n)^{\frac{1}{\beta_1}+1}}{n^{\frac{1+\beta_2}{2}}}$, we can get
\begin{small}
 \begin{equation*}
    \begin{aligned}
  &\m P^*{}^{\otimes n}\left(\left\{z_{{\alpha}/{2}}+C_1\frac{(\log n)^{\frac{1}{\beta_1}+1}}{n^{\frac{\beta_2}{2}}\sigma_f}\leq \frac{\sqrt{n}}{\sigma_f}(f(\theta^*)-f(\wh\theta^\diamond)\leq z_{1-\frac{\alpha}{2}}- C_1\frac{(\log n)^{\frac{1}{\beta_1}+1}}{n^{\frac{\beta_2}{2}}\sigma_f}  \right\}\cap \m A \right)\\
    &\leq\m P^*{}^{\otimes n}\left(\left\{ q_{\alpha/2}^f\leq f(\theta^*)\leq q_{1-\alpha/2}^f  \right\}\cap \m A \right)\\
    &\leq\m P^*{}^{\otimes n}\left(\left\{z_{{\alpha}/{2}}- C_1\frac{(\log n)^{\frac{1}{\beta_1}+1}}{n^{\frac{\beta_2}{2}}\sigma_f}\leq \frac{\sqrt{n}}{\sigma_f}(f(\theta^*)-f(\wh\theta^\diamond)\leq z_{1-\frac{\alpha}{2}}+ C_1\frac{(\log n)^{\frac{1}{\beta_1}+1}}{n^{\frac{\beta_2}{2}}\sigma_f}  \right\}\cap \m A \right).
    \end{aligned}
\end{equation*}
\end{small}
Moreover, using  
 \begin{equation*}
 \sqrt{n}\cdot W_{\theta^*}^T\psi_{\theta^*}(\hat{\theta}^\diamond)=-\m H_0^{-1} \frac{1}{\sqrt{n}}\sum_{i=1}^n W_{\theta^*}^T  g(X_i,\theta^*),
\end{equation*}
 we have
\begin{equation*}
    \begin{aligned}
   &\Big\|f(\theta^*)-f(\wh\theta^\diamond)-\nabla f(\theta^*)^TW_{\theta^*}\m H_0^{-1} W_{\theta^*}^T\frac{1}{n}\sum_{i=1}^n  g(X_i,\theta^*)\Big\|\\
  &\leq \Big\|f(\theta^*)-f(\wh\theta^\diamond)-\nabla f(\theta^*)^T(\theta^*-\wh\theta^\diamond)\Big\|+C\,\|\theta^*-\wh\theta^\diamond\|^2\\
  &\lesssim \frac{\log n}{n}.
    \end{aligned}
\end{equation*}
Therefore, denote $\m J=\frac{1}{\sigma_f}\nabla f(\theta^*)^TW_{\theta^*} \m H_0^{-1}$,  there exists a constant $C_2$ such that 
\begin{small}
\begin{equation*}
    \begin{aligned}
   & \m P^*{}^{\otimes n}\left(\left\{z_{{\alpha}/{2}}- C_1\frac{(\log n)^{\frac{1}{\beta_1}+1}}{n^{\frac{\beta_2}{2}}\sigma_f}\leq \frac{\sqrt{n}}{\sigma_f}(f(\theta^*)-f(\wh\theta^\diamond)\leq z_{1-\frac{\alpha}{2}}+ C_1\frac{(\log n)^{\frac{1}{\beta_1}+1}}{n^{\frac{\beta_2}{2}}\sigma_f}  \right\}\cap \m A \right)\\
      & \leq\m P^*{}^{\otimes n}\left(\left\{z_{{\alpha}/{2}}- C_1\frac{(\log n)^{\frac{1}{\beta_1}+1}}{n^{\frac{\beta_2}{2}}\sigma_f}\leq \frac{1}{\sqrt{n}} \sum_{i=1}^n \m J W_{\theta^*}^Tg(X_i,\theta^*)\leq z_{1-\frac{\alpha}{2}}+ C_2\frac{(\log n)^{\frac{1}{\beta_1}+1}}{n^{\frac{\beta_2}{2}}\sigma_f}  \right\}\cap \m A \right)\\
  &\leq \m P^*{}^{\otimes n}\left(z_{{\alpha}/{2}}- C_1\frac{(\log n)^{\frac{1}{\beta_1}+1}}{n^{\frac{\beta_2}{2}}\sigma_f}\leq \frac{1}{\sqrt{n}} \sum_{i=1}^n \m J W_{\theta^*}^Tg(X_i,\theta^*)\leq z_{1-\frac{\alpha}{2}}+ C_2\frac{(\log n)^{\frac{1}{\beta_1}+1}}{n^{\frac{\beta_2}{2}}\sigma_f}   \right)\\
    \end{aligned}
\end{equation*}
\end{small}
Since $\mathbb{E} [W_{\theta^*}^Tg(X,\theta^*)]={0}_d$, ${\rm Cov}_{\m P^*}(W_{\theta^*}^Tg(X,\theta^*))=W_{\theta^*}^TI_{\theta^*}W_{\theta^*}$, we have 
\begin{equation*}
    \begin{aligned}
       &\mb E[\m J W_{\theta^*}^Tg(X,\theta^*)]={0}\\
     & {\rm Cov}_{\m P^*}( \m JW_{\theta^*}^Tg(X,\theta^*))=\m J\Delta_0\m J^T=1.\\
    \end{aligned}
    \end{equation*}
    Then by Berry-Esseen theorem~\citep{Rai__2019}, there exist constants $C_3,C_4$ such that
    \begin{equation*}
        \begin{aligned}
        &\m P^*\left( q_{\alpha/2}^f\leq f(\theta^*)\leq q_{1-\frac{\alpha}{2}}^f   \right)\\
           &\leq \m P^*\left(z_{{\alpha}/{2}}- C_1\frac{(\log n)^{\frac{1}{\beta_1}+1}}{n^{\frac{\beta_2}{2}}\sigma_f}\leq \frac{1}{\sqrt{n}} \sum_{i=1}^n \m J W_{\theta^*}^Tg(X_i,\theta^*)\leq z_{1-\frac{\alpha}{2}}+ C_2\frac{(\log n)^{\frac{1}{\beta_1}+1}}{n^{\frac{\beta_2}{2}}\sigma_f}   \right)+\frac{2}{n^2} \\
           &\leq   \m P^*\left(z_{{\alpha}/{2}}- C_1\frac{(\log n)^{\frac{1}{\beta_1}+1}}{n^{\frac{\beta_2}{2}}\sigma_f}\leq Z\leq z_{1-\frac{\alpha}{2}}+ C_2\frac{(\log n)^{\frac{1}{\beta_1}+1}}{n^{\frac{\beta_2}{2}}\sigma_f}   \right)+C_3\, \frac{(\log n)^{\frac{1}{\beta_1}+1}}{n^{\frac{\beta_2}{2}}} \\
           &=1-\alpha+C_4\frac{(\log n)^{\frac{1}{\beta_1}+1}}{n^{\frac{\beta_2}{2}}}. \\
        \end{aligned}
    \end{equation*}
Similarly, we can show that there exists a constant $C_5$ such that 
\begin{equation*}
        \begin{aligned}
        &\m P^*\left( q_{\alpha/2}^f\leq f(\theta^*)\leq q_{1-\alpha/2}^f   \right)\\
            &\geq 1-\alpha-C_5 \frac{(\log n)^{\frac{1}{\beta_1}+1}}{n^{\frac{\beta_2}{2}}} .\\
        \end{aligned}
    \end{equation*}
We then get the desired conclusion.
\section{Proof for Examples}\label{proof:example}
\subsection{Example 1: Reduced-Rank Multi-Response Regression}
We write the parameter matrix as 
\begin{equation*}
\theta=
    \begin{pmatrix}
        \theta_{11}&\theta_{12}\\
        \theta_{21}&\theta_{22}
    \end{pmatrix}, \text{ equivalently } \theta=(\theta_{11},\theta_{21},\theta_{12},\theta_{22}).
\end{equation*}
The squared error loss for one observation $(X,Y)$ is $$\ell((X,Y),\theta)=\frac{1}{2}(Y_1-\theta_{11}X_1-\theta_{21}X_2)^2+\frac{1}{2}(Y_2-\theta_{12}X_1-\theta_{22}X_2)^2, $$ where $Y=(Y_1,Y_2)^T$ and $X=(X_1,X_2)^T$. A direct calculation shows that the population risk $\m R(\theta)=\mb{E}[\ell((X,Y),\theta)]$ satisfies
\begin{equation*}
\begin{aligned}
        &\m R(\theta)- \m R(\theta^*)\\
        &=\frac{1}{2}\mb{E}[(\theta_{11}X_1-\theta_{21}X_2-\theta_{11}^*X_1-\theta_{21}^*X_2)^2]+\frac{1}{2}\mb{E}[(\theta_{12}X_1-\theta_{22}X_2-\theta_{12}^*X_1-\theta_{22}^*X_2)^2]\\
        &\geq \frac{1}{2}\|\theta-\theta^*\|^2.
\end{aligned}
\end{equation*}
which guarantees identifiability of the parameter. The gradient takes the form
 \begin{equation*}
    \nabla_{\theta} \ell((X,Y),\theta)=\begin{pmatrix}
        (\theta_{11}X_1+\theta_{21}X_2-Y_1)X_1\\
        (\theta_{11}X_1+\theta_{21}X_2-Y_1)X_2\\
        (\theta_{12}X_1+\theta_{22}X_2-Y_2)X_1\\
      (\theta_{12}X_1+\theta_{22}X_2-Y_2)X_2.
    \end{pmatrix}
\end{equation*}
Taking expectations, the gram matrix of the score at $\theta^*$ is give by:
\begin{equation*}
 \mb{E}[ \nabla_{\theta} \ell((X,Y),\theta^*)    \nabla_{\theta} \ell((X,Y),\theta)^T]=\begin{pmatrix}
       \Sigma_{11}&0&\Sigma_{12}&0\\
       0&\Sigma_{11}&0&\Sigma_{12}\\
        \Sigma_{21}&0&\Sigma_{22}&0\\
        0&\Sigma_{21}&0&\Sigma_{22}.
          \end{pmatrix}.
\end{equation*}
The Hessian of the loss at $\theta^*$ is given by 
\begin{equation*}
    {\rm Hess}_{\theta}\ell((X,Y),\theta^*)=\begin{pmatrix}
        X_1^2&X_1X_2&0&0\\
        X_1X_2&X_2^2&0&0\\
        0&0&X_1^2&X_1X_2\\
        0&0&X_1X_2&X_2^2
          \end{pmatrix}.
\end{equation*}
 So $\mb{E}[  {\rm Hess}_{\theta}\ell((X,Y),\theta)]=I_4$. Since loss function is smooth in $\theta$, when the parameter space is the Euclidean set of $\mb R^4$,  Assumption 1-4  are satisfied with $\beta_1=\beta_2=1$. 
Moreover,   the empirical risk minimizer corresponds to ordinary least squares:
 \begin{equation*}
 (\wh\beta_1^T,\wh\beta_2^T)^T=\arg\min_{\theta\in \mb R^4} n^{-1}\sum_{i=1}^n \ell((\wt X_i,\wt Y_i),\theta), \quad\wh\beta_j=(\wt X^T\wt X)\wt X^T\wt Y_{,j}.
 \end{equation*}
Because $\theta^*$ belongs to $S_{\Pi_{\rm E}}=\{\theta\in \mb R^4\,:\, \|\theta\|\leq 100\}$, the support of the Euclidean prior $\Pi_{\rm E}$ after flattening matrices into vectors. For sufficiently large $n$, it holds with probability at least $1-n^{-1}$ that the OLS estimator $(\wh\beta_1^T,\wh\beta_2^T)^T$ also lies in $S_{\Pi_{\rm E}}$. Hence the empirical risk minimizer over $S_{\Pi_{\rm E}}$  coincides with the OLS solution. Applying Theorem~\ref{th:classicbvm}, with high probability, for $f(\theta)=\theta_1-\theta_3$, the posterior satisfies
 \begin{equation*}
     {\rm TV}\Big(f_{\#}\Pi_{\rm E}(\cdot|(\wt X,\wt Y)), \m N\Big(\wh\beta_{11}-\wh\beta_{21},\frac{2}{n}\Big)\Big)\lesssim \frac{(\log n)^2}{\sqrt{n}}.
 \end{equation*}
In addition, the sampling distribution of the OLS estimator satisfies
\begin{equation*}
    \sqrt{n}(\wh\beta_{11}-\wh\beta_{21}-(\beta^*_{11}-\beta^*_{21}))\xrightarrow{d} \m N(0,\Sigma_{11}+\Sigma_{22}-2\Sigma_{12}).
\end{equation*}
We now turn to the manifold posterior. Note that the manifold $\ov{\m M}=\{\theta\in \mb R^{2\times 2}\,:\,{\rm Rank}(\theta)=1\}$ admits a global parametrization and can be written as 
\begin{equation*}
\ov{\m M}=\big\{\theta=\big(\begin{smallmatrix}
  \theta_{11} & a  \theta_{11}\\
  \theta_{12} & a   \theta_{12}
\end{smallmatrix}\big)\,:\,(\theta_{11},\theta_{12})\neq (0,0), a\in \mb R \big\}.
\end{equation*}
Flattening matrices into $\mb R^4$, this corresponds to 
\begin{equation*}
{\m M}=\big\{\theta=(\theta_{11},\theta_{21},\theta_{12},\theta_{22})^T\,:\, \big(\begin{smallmatrix}
  \theta_{11} &   \theta_{12}\\
  \theta_{21} &    \theta_{22}
\end{smallmatrix}\big)\in \ov{\m M}\big\}.
\end{equation*}
At a point $\theta=(\theta_{11},\theta_{21},a\theta_{11},a\theta_{21})^T\in \m M$, the projection matrix $P_{\theta}$ onto the tangent space $T_{\theta}\m M$ is given by 
\begin{equation*}
P_{\theta}=W_{\theta}(W_{\theta}^TW_{\theta})^{-1}W_{\theta}\text{ with }W_{\theta}=   \begin{pmatrix}
  1 & 0&0\\
0 & 1  &0\\
a&0&\theta_1\\
0&a&\theta_2
\end{pmatrix}.
\end{equation*}
At the true parameter $\theta^*=(1,1,2,2)^T$, this gives
\begin{equation*}
    P_{\theta^*}= \begin{pmatrix}
  0.6 & 0.4&0.2&0.2\\
 0.4&0.6&-0.2&0.2\\
0.2&-0.2&0.9&0.1\\
-0.2&0.2&0.1&0.9
\end{pmatrix}.
\end{equation*}
By Lemma~\ref{lemmasmootharoundtheta1}, the manifold $\m M$ is locally $C^3_{r,L}$-smooth around $\theta^*$ for  constants $r,L>0$. Consider the Riemannian gradient
\begin{equation*}
    g((X,Y),\theta)=P_{\theta}\nabla_{\theta} \ell((X,Y),\theta).
\end{equation*}
The corresponding covariance matrix at the truth is
\begin{equation*}
    \Delta_{\theta^*}=\mb{E}[ g((X,Y),\theta^*) g((X,Y),\theta^*)^T]=P_{\theta^*}\begin{pmatrix}
       \Sigma_{11}&0&\Sigma_{12}&0\\
       0&\Sigma_{11}&0&\Sigma_{12}\\
        \Sigma_{21}&0&\Sigma_{22}&0\\
        0&\Sigma_{21}&0&\Sigma_{22}
          \end{pmatrix}P_{\theta^*}.
\end{equation*}
Since $\mb{E}[\nabla_{\theta}\ell((X,Y),\theta^*)]=0_2$, the Riemannian Hessian matrix reduces to
\begin{equation*}
    \m H_{\theta^*}=P_{\theta^*}\mb{E}[  {\rm Hess}_{\theta}\ell((X,Y),\theta)] P_{\theta^*}=P_{\theta^*}.
\end{equation*}
So $\m H_{\theta^*}^\dagger=P_{\theta^*}$ and $\m H_{\theta^*}^\dagger\Delta_{\theta^*}\m H_{\theta^*}^\dagger=\Delta_{\theta^*}.$  Thus Assumption 1-4  can be verified for the manifold setting as well with $\beta_1=\beta_2=1$. Now we solve the empirical risk minimizer under the manifold:
 \begin{equation*}
 \arg\min_{\theta\in \m M} n^{-1}\sum_{i=1}^n \ell((\wt X_i,\wt Y_i),\theta).
 \end{equation*}
This is equivalent to solving for $(\wh\theta_{11},\wh\theta_{21},\wh a)$ satisfying
\begin{equation*}
    \begin{aligned}
     &\wt X^T(\wt Y_{,1}+\wt Y_{,2}-\wt X (\wh\theta_{11},\wh\theta_{21})^T (1+\wh a))=0_2\\
      &(\wh \theta_{11},\wh \theta_{21})\wt X^T(\wt Y_{,2}-\wt X(\wh \theta_{11},\wh \theta_{21})^T\wh  a)=0,
    \end{aligned}
\end{equation*}
which is
\begin{equation*}
    \begin{aligned}
       & \wh a=\frac{(\wt Y_{,1}+\wt Y_{,2})^T\wt X\wh\beta_2}{(\wt Y_{,1}+\wt Y_{,2})^T\wt X\wh\beta_1}\\
       &\wh\theta_{11}=\frac{\wh\beta_{11}+\wh\beta_{21}}{1+\wh a}\\
          &\wh\theta_{21}=\frac{\wh\beta_{12}+\wh\beta_{22}}{1+\wh a}.
    \end{aligned}
\end{equation*}
Letting $\wh s=\frac{1-\wh a}{1+\wh a}$. Similarly as the Euclidean case, applying Theorem~\ref{th:classicbvm} yields that,  with high probability, for $f(\theta)=\theta_1-\theta_3$,  
 \begin{equation*}
     {\rm TV}\Big(f_{\#}\Pi_{\rm M}(\cdot|(\wt X,\wt Y)), \m N\Big(\wh s (\wh\beta_{11}+\wh\beta_{21}),\frac{1.1}{n}\Big)\Big)\lesssim \frac{(\log n)^2}{\sqrt{n}},
 \end{equation*}
and 
\begin{equation*}
    \sqrt{n}(\wh s (\wh\beta_{11}+\wh\beta_{21})-(\beta^*_{11}-\beta^*_{21}))\xrightarrow{d} \m N(0,0.52\Sigma_{11}+0.58\Sigma_{22}-0.92\Sigma_{12}).
\end{equation*}
Applying Theorem~\ref{th1} then yields, 
 \begin{equation*}
     {\rm TV}\Big(f_{\#}\Pi_{\rm RP}(\cdot|(\wt X,\wt Y)), \m N\Big(\wh s (\wh\beta_{11}+\wh\beta_{21}),\frac{0.52\Sigma_{11}+0.58\Sigma_{22}-0.92\Sigma_{12}}{n}\Big)\Big)\lesssim \frac{(\log n)^2}{\sqrt{n}},
 \end{equation*}
 
\subsection{Example 2: Mean Direction of the Von Mises-Fisher Distribution}
Let $A(\kappa^*)=\coth \kappa^*-1/\kappa^*$, for $X\sim VMF3(\mu^*,\kappa^*)$, the standard moment formula (see for example~\cite{hillen2017moments}) are 
$$\mb{E}[X]=A(\kappa^*)\mu^*,$$ and $$\mb{E}[XX^T]=\frac{A(\kappa^*)}{\kappa^*}I_3+(1-\frac{3\coth \kappa^*}{\kappa^*}+\frac{3}{(\kappa^*)^2})\mu^*\mu^*{}^T.$$
Consider the loss function for one observation $\ell(X,\theta)=-\theta^TX$.   the risk function $\m R(\theta)=\mb{E}[\ell(X,\theta)]=-A(\kappa^*)\theta^T\mu^*$ satisfies for any $\theta\in \m M=\mb S_1^2$ that
\begin{equation*}
    \begin{aligned}
        \m R(\theta)-\m R(\mu^*)=A(\kappa^*)((\mu^*-\theta)^T\mu^*)=\frac{A(\kappa^*)}{2}\|\mu^*-\theta\|^2.
    \end{aligned}
\end{equation*}
Hence the population risk $\m R$ is uniquely minimized on $\m M$ at $\theta=\mu^*$. Moreover,  at a point $\theta\in \m M$, the projection matrix $P_{\theta}$ onto the tangent space $T_{\theta}\m M$ is given by $P_\theta=I_3-\theta\theta^T$. At $\mu^*=(\frac{1}{\sqrt{3}},\frac{1}{\sqrt{3}},\frac{1}{\sqrt{3}})$, this gives
\begin{equation*}
    P_{\mu^*}=\begin{pmatrix}
        2/3&-1/3&-1/3\\
        -1/3&2/3&-1/3\\
        1/3&-1/3&2/3
    \end{pmatrix}.
\end{equation*}
Consider the Riemannian gradient 
\begin{equation*}
    g(X,\theta)=P_{\theta}\nabla_{\theta}\ell(X,\theta)=-P_{\theta}X=-(I_3-\theta\theta^T)X,
\end{equation*}
 the Gram matrix at $\theta=\mu^*$ is  given by
\begin{equation*}
    \Delta_{\mu^*}=P_{\mu^*}\mb{E}[XX^T]P_{\mu^*}=\frac{A(\kappa^*)}{\kappa^*}P_{\mu^*}.
\end{equation*}
Moreover, the Riemannian Hessian matrix of the population risk at $\mu^*$ is
\begin{equation*}
    \m H_{\mu^*}=\mb{E}\big[P_{\mu^*}(\mu^* X^T+(X^T\mu^*)I_3)P_{\mu^*}\big]=A(\kappa^*)P_{\mu^*}.
\end{equation*}
So $\m H_{\theta^*}^\dagger=\frac{1}{A(\kappa^*)}P_{\mu^*}$ and $\m H_{\theta^*}^\dagger\Delta_{\theta}\m H_{\theta^*}^\dagger=\frac{1}{\kappa^*A(\kappa^*)}P_{\mu^*}$. Thus Assumption 1-4  hold with $\beta_2=1$, and any $\beta_1>0$, and here we choose $\beta_1=1$. The empirical risk minimizer on $\m M$ is given by 
\begin{equation*}
    \arg\min_{\theta\in\m M}-\theta^T\sum_{i=1}^n X_i=\frac{\sum_{i=1}^n X_i}{\|\sum_{i=1}^n X_i\|}=\frac{\ov X}{\|\ov X\|}.
\end{equation*}
 Applying Theorem~\ref{th:classicbvm} with $f(\theta)=\theta_1$ yields, with high probability,
 \begin{equation*}
     {\rm TV}\Big(f_{\#}\Pi(\cdot|X^{(n)}), \m N\Big(f\big( {\ov X}/{\|\ov X\|}\big),n^{-1}\frac{2}{3 A(\kappa^*)}\Big)\Big)\lesssim \frac{(\log n)^2}{\sqrt{n}},
 \end{equation*}
and 
\begin{equation*}
    \sqrt{n}(f\big({\ov X}/{\|\ov X\|}\big)-f(\mu^*))\xrightarrow{d} \m N(0,\frac{2}{3\kappa^* A(\kappa^*)}).
\end{equation*}
Applying Theorem~\ref{th1} yields,
 \begin{equation*}
     {\rm TV}\Big(f_{\#}\Pi_{\rm RP}(\cdot|X^{(n)}), \m N\Big(f\big( {\ov X}/{\|\ov X\|}\big),n^{-1}\frac{2}{3 \kappa^*A(\kappa^*)}\Big)\Big)\lesssim \frac{(\log n)^2}{\sqrt{n}}.
 \end{equation*}
 \section{Proof for Mixing time Bound}\label{Proofmix}
 
In this section, we prove Theorem~\ref{th:mixing} and Corollary~\ref{th:mixingbpetel} for the mixing time bound of sampling from $\Pi_{\rm RP}(\cdot|X^{(n)})$. 
 
\subsection{Proof of Theorem~\ref{th:mixing}}
 
 We first consider the mixing time for sampling from a truncated distribution $\mu^\ast|_{K_\theta}$, where recall $K_{\theta}=\{x=\phi_{\wt \theta}\big({W}_{\wt\theta}\frac{z}{\sqrt{n}}\Big)\,:\,\|(W_{\wt \theta}^T \wt I W_{\wt \theta})^{-\frac{1}{2}}z\|\leq R\}$.  Using Assumption B.1, there exists  a set of matrices $\wt{W}_\theta\in \mb R^{D\times d}$ indexed by $\theta\in \m M$ so that (1) the columns of $\wt  W_{\theta}$ form an orthonormal basis to $T_{\theta}\m M$; (2) $\wt W_{\wt\theta}=W_{\wt \theta}$  and $\mnorm{\wt W_{\theta}-\wt W_{ \wt\theta}}_{\rm F}\leq L_1\|\theta-\wt\theta\|$ for $\theta\in K_{\theta}$ and an $n$-independent constant $L_1$. Then we consider the following equivalent version of the $\zeta$-lazy RRWM algorithm described in Section~\ref{sec:Sampling}, which transforms the operations performed in each iteration into the $\mb R^d$ space for convenience of analysis.

 \begin{algorithm}[H]\label{algorithm_second_version}
\caption{Equivalent version of the $\zeta$-lazy  RRWM algorithm to sample from a density $\mu^\ast(\theta)$ on manifold}
\SetAlgoLined
\SetKwRepeat{Do}{do}{while}%
 \textbf{Input}: Number of iterations $L$, step size $\wt h$, initial distribution $\mu_0$, $\wt W_{\theta}\in \mb R^{D\times d}$ where for any $\theta\in \m M$, $\wt{W}_{\theta}$ forms an orthonormal basis for the tangent space of $T_{\theta}\m M$, covariance matrix $\wt I\in \mb R^{D\times D}$\;
 Sampling $\theta^0$ from $\mu_0$\;
 \For{$t \leftarrow 0 \,\,to\,\, L-1$}{
  Generate a uniform random number $u_0\in (0,1)$\;
  \eIf{ $u_0\leq \zeta$}
  {$\theta^{t+1}\leftarrow \theta^t$\;}
  {
   Sample $v$ from $\mathcal{N}(0,{2\wt h}\wt W_{\theta^t}^T\wt I \wt W_{\theta^t})$ and denote its density by $\wt{p}_{\theta^t}(v)$\;
    \eIf{$\wt W_{\theta^t}v\notin \wt V_{\theta^t}$}
   {$\theta^{t+1}\leftarrow \theta^t$\;}
   {  Let $y=\wt{\phi}_{\theta^t}(\wt{W}_{\theta^t}v)$\;
   \eIf{$\theta^t\notin\wt U_y$}
        {$\theta^{t+1}\leftarrow \theta^t$\;}
   { Let ${v}'=\wt{W}_y^T\wt\psi_y(\theta^t)$\;
   Generate a uniform random  number $u\in(0,1)$\;
  \eIf{$u> \frac{\mu^\ast(y)\cdot\wt{p}_y(v')\cdot\big({\rm det}\big(J_{\wt{\phi}_y(\wt W_{y}\cdot)}(v')^TJ_{\wt{\phi}_y(\wt W_{y}\cdot)}(v')\big)\big)^{-\frac{1}{2}}}{\mu^\ast(\theta^t)\cdot\wt{p}_{\theta^t}(v)\cdot\big({\rm det}\big(J_{\wt{\phi}_{\theta^t}(\wt W_{\theta^t}\cdot)}(v)^TJ_{\wt{\phi}_{\theta^t}(\wt W_{\theta^t}\cdot)}(v)\big)\big)^{-\frac{1}{2}}}$}
  { $\theta^{t+1}\leftarrow \theta^t$\;
   
   }{$\theta^{t+1}=y$\;
      
  }}
 }}}
 \end{algorithm}
 Note that the probability distribution obtained after $k$ steps of the Markov chain defined in Algorithm~\ref{algorithm_second_version} is same as the probability distribution obtained after $k$ steps of the Markov chain generated by the $\zeta$-lazy version of RRWM Algorithm defined in Section~\ref{sec:Sampling}. Then let $\wt\mu_k$ denotes the probability distribution obtained after $k$ steps of the Markov chain defined in Algorithm~\ref{algorithm_second_version} for sampling from $\mu^\ast|_{K_\theta}$ with initial distribution $\wt \mu_0=\mu_0|_{K_\theta}$.
Here,  $\mu^\ast|_{K_\theta}$ denotes  the normalized restriction of $\mu^*$  to $K_{\theta}$, i.e., $\mu^\ast|_{K_\theta}(A)=\frac{\mu^*(A\cap K_{\theta})}{\mu^*(K_{\theta}}$.   Define  $\ms Q=\{z=\sqrt{n}  W_{\wt\theta}^T\wt \psi_{\wt \theta}(\theta)\,:\, \theta\in  B_r(\wt\theta)\cap \m M\}$, and
  \begin{equation*}
      Q: B_r(\wt\theta)\cap \m M\to  \ms Q\,\text{ as }\, Q(\theta)=\sqrt{n}\cdot {W}^T_{\wt\theta}\wt\psi_{\wt\theta}(\theta).
  \end{equation*}
We can also define the inverse of $Q$ as 
  \begin{equation*}
      G: \ms Q\to B_r(\wt\theta)\cap \m M\,\text{ as }\, G(z)=Q^{-1}(z)=\wt\phi_{\wt\theta}\big( W_{\wt\theta}\frac{z}{\sqrt{n}}\big).
  \end{equation*}
  Then when $R\leq C\, \sqrt{n}$ with small enough $C$, it holds that $ K_{\theta}\subset B_r(\wt\theta)\cap \m M $. Define $\mu^\ast_{\rm loc}$ as the push-forward measure $Q_{\#}(\mu^\ast|_{K_\theta})$, which has a density (with respect to the Lebesgue measure of $\mb R^d$),
  \begin{equation*}
     \mu^\ast_{\rm loc}(z)= \mu^\ast|_{K_\theta}(G(z))\sqrt{{\rm det}\big(\bold{J}_G(z)^T\bold{J}_G(z)\big)},\quad z\in K=\{z\in \mb R^d\,:\, \|(  W_{\wt \theta}^T\wt I {W}_{\wt \theta})^{-\frac{1}{2}}z\|\leq R\}.
  \end{equation*}
  Similarly, denote $\nu_k=Q_{\#}\wt\mu_k$, it has a density (with respect to the Lebesgue measure on $\mb R^d$):
  \begin{equation*}
\nu_k(z)= \wt\mu_k(G(z))\sqrt{{\rm det}\big(\bold{J}_G(z)^T\bold{J}_G(z)\big)}, \quad z\in K,
  \end{equation*}
  where we abuse the notation to use $\wt\mu_k$ to denote its density function with respect to the volume measure of $\m M$.
%   So we have 
%  \begin{equation}\label{eqn:equivalence}
%      \begin{aligned}
%     \chi^2(\nu_k, \mu^\ast_{\rm loc})&=\mb{E}_{Q_{\#}(\mu^\ast|_{K_\theta})}\Bigg[\bigg(\frac{\mu_k(G(z))\sqrt{{\rm det}\big(\bold{J}_G(z)^T\bold{J}_G(z)\big)}}{\mu^\ast|_{K_\theta}(G(z))\sqrt{{\rm det}\big(\bold{J}_G(z)^T\bold{J}_G(z)\big)}}-1\bigg)^2\Bigg]\\
%     &=\mb{E}_{Q_{\#}(\mu^\ast|_{K_\theta})}\Bigg[\bigg(\frac{\mu_k(G(z))}{\mu^\ast|_{K_\theta}(G(z))}-1\bigg)^2\Bigg]\\
%     &=\mb{E}_{\mu^\ast|_{K_\theta}}\Bigg[\bigg(\frac{\mu_k(x)}{\mu^\ast|_{K_\theta}(x)}-1\bigg)^2\Bigg]\\
%     &=\chi^2(\mu_k,\mu^\ast|_{K_\theta}).
%      \end{aligned}
%  \end{equation}
%   So we only need to bound $\chi^2(\nu_k, \mu^\ast_{\rm loc})$ for $k\in \mb N$. 
 For $x\in \m M$, define $\Omega_x=\{v\in \mb R^d\,:\, \wt W_{x}v\in \wt V_{x}\}$ and
  \begin{equation*}
      \wt\eta_{x}=P(Z\in\Omega_x),\quad Z\sim \m N(0,\frac{2h}{n}\wt W_x^T\wt I\wt W_x).
  \end{equation*}
 Let   $\wt p_x$ denote the density function of $\m N(0,\frac{2h}{n}\wt W_x^T\wt I\wt W_x)$. Define $\wt\phi_x^*(\cdot)=\wt\phi_x(\wt W_{x}\cdot)$ and $\wt \psi^*_x(\cdot)=\wt W_x^T \wt \psi_x(\cdot)$. We write $p(x,\cdot)$ as the density of $(\wt{\phi}_x^*)_{\#}(\wt{p}_x|_{\Omega_x})=(\wt{\phi}_x^*)_{\#}(\m N(0,\frac{2h}{n}\wt W_x^T\wt I\wt W_x)|_{\Omega_x})$ with respect to the volume measure $\mu_{\m M}$ of $\m M$, then we have 
 \begin{equation}\label{defp}
     p(x,y)=\frac{1}{\wt\eta_x}\cdot\wt{p}_x(\wt{\psi}^*_x(y))\cdot\big({\rm det}\big(J_{\wt{\phi}^*_{x}}(\wt{\psi}^*_x(y))^TJ_{\wt{\phi}_{x}^*}(\wt{\psi}_x^*(y))\big)\big)^{-\frac{1}{2}}\cdot\bold{1}\big(y\in \wt{\phi}^*_x(\Omega_x)\big).
 \end{equation}
 So we can write the transition probability function $\m T$ associated with $\wt\mu_k\to \wt\mu_{k+1}$ as 
 \begin{equation*}
    \m T(x,\,\dd y)=(1-\zeta)\wt\eta_x\cdot\m A(x,y)p(x,y)\,\mu_{\m M}(\dd y)+\Big(1-\int(1-\zeta)\wt\eta_x\cdot\m A(x,y)p(x,y)\,{\mu_{\m M}(\dd y)}\Big)\cdot \delta_x(\dd y),
 \end{equation*}
 where $\delta_x(\cdot)$ denotes the Dirac measure at $x$, $\zeta$ is the lazy parameter and $\m A(x,y)$ is the acceptance ratio given by
 \begin{equation*}
     \m A(x,y)=1\wedge \frac{\wt \eta_y\cdot\mu^\ast|_{K_\theta}(y) p(y,x)}{\wt \eta_x\cdot\mu^\ast|_{K_\theta}(x)p(x,y)},
 \end{equation*}
 and note that when $x\notin \wt U_y$, the acceptance ratio $\m A(x,y)=0$, which is consistent with the algorithm that we will reject the update if $x\notin \wt U_y$. Next we will solve the transition probability function $\m T^\ast$ for $\nu_k \to \nu_{k+1}$. Denote $p^\ast(z_1,\cdot)$ as the density of $\big(Q\circ \wt{\phi}^*_{G(z_1)}\big)_{\#}(\wt p_{G(z_1)}|_{\Omega_{G(z_1)}})$ with respect to Lebesgue measure of $\mb R^d$, that is
 \begin{equation*}
   p^\ast(z_1,z_2)=\frac{1}{\wt\eta_{G(z_1)}}\cdot\wt{p}_{G(z_1)}(\wt{\psi}^*_{G(z_1)}\circ G(z_2))\cdot\Big|{\rm det}\big(\bold{J}_{\wt{\psi}^*_{G(z_1)}\circ G}(z_2)\big)\Big|\cdot \bold{1}\big(z_2\in [Q\circ \wt{\phi}^*_{G(z_1)}]\big(\Omega_{G(z_1)}\big) \big).
 \end{equation*}
 Then define 
 \begin{equation*}
     \alpha^\ast(z_1,z_2)=\frac{\wt \eta_{G(z_2)}\cdot p^\ast(z_2,z_1)\mu^\ast_{\rm loc}(z_2)}{\wt \eta_{G(z_1)}\cdot p^*(z_1,z_2)\mu^\ast_{\rm loc}(z_1)},
 \end{equation*}
 The next lemma shows that the acceptance ratio $\m A(G(z_1),G(z_2))$ is equal to $ \m A^*(z_1,z_2)=1\wedge \alpha^*(z_1,z_2)$.
\begin{lemma}\label{lemma:acceptance}
    For any $z_1\in K$ and $z_2\in [Q\circ \wt{\phi}^*_{G(z_1)}]\big(\Omega_{G(z_1)}\big)$, we have 
    \begin{equation*}
          \m A^*(z_1,z_2)=1\wedge \alpha^*(z_1,z_2)=1\wedge \frac{\wt \eta_{G(z_2)}\cdot\mu^\ast|_{K_\theta}(G(z_2))\cdot p(G(z_2),G(z_1))}{\wt \eta_{G(z_1)}\cdot\mu^\ast|_{K_\theta}(G(z_1))\cdot p(G(z_1),G(z_2))}=\m A(G(z_1),G(z_2)).
 \end{equation*}
\end{lemma}
\noindent Therefore, the  transition probability function $\m T^\ast$ for $\nu_k \to \nu_{k+1}$ is defined as 
  \begin{equation*}
  \begin{aligned}
       \m T^*(z_1,\,\dd z_2)&= (1-\zeta)\cdot\wt\eta_{G(z_1)}\cdot\m A^\ast(z_1,z_2)p^\ast(z_1,z_2)\,\dd z_2\\
       &\qquad+\Big(1-\int(1-\zeta)\cdot\wt\eta_{G(z_1)}\cdot\m A^\ast(z_1,z_2)p^\ast(z_1,z_2)\,\dd z_2\Big)\cdot \delta_{z_1}(\,\dd z_2).\\
  \end{aligned}
  \end{equation*}
   Now we show that given the choice of the step size stated in Theorem~\ref{th:mixing}, the $s$-conductance profile associated $\m T^\ast$ or $\m T$ can be lower bounded. 
  
   \begin{lemma}\label{lemma:conductance}
It holds that the $s$-conductance profile with $s=\frac{\varepsilon^2}{32M_0^2}$ satisfies 
 \begin{equation*}
 \begin{aligned}
    \Phi_{s}(v):&=\inf\left\{\frac{\int_S \m T(x,\m M \backslash S)\,  \mu^\ast|_{K_\theta}(\dd x)}{\mu^\ast|_{K_\theta}(S)-s}\,|\, S\subseteq \m M, s<\mu^\ast|_{K_\theta}(S)\leq v\right\}\\
    &\geq \inf\left\{\frac{\int_S \m T(x,\m M\backslash S)\,  \mu^\ast|_{K_\theta}(\dd x)}{\mu^\ast|_{K_\theta}(S)}\,|\, S\subseteq \m M, s<\mu^\ast|_{K_\theta}(S)\leq v\right\}\\
 & =  \inf\left\{\frac{\int_S \m T^\ast(z,\mb R^d\backslash S)\,\mu^\ast_{\rm loc}(z)\,\dd z}{\mu^\ast_{\rm loc}(S)}\,|\, S\subseteq \mb R^d, s<\mu^\ast_{\rm loc}(S)\leq v\right\}\\
    & \geq C\exp(-2\varepsilon_1)\, \min\Big\{1,\exp(-3\varepsilon_1)\sqrt{h\rho_1}\log^{\frac{1}{2}}\big(1+\frac{1}{v}\big)\Big\},
     \end{aligned}
 \end{equation*}
for $v\in [\frac{4}{M_0},\frac{1}{2}]$ and an $(n,D,d)$-independent constant $C$. 
\end{lemma}
We can then bound the mixing time using the following lemma,  whose directly follows the proof of Lemma 1 in~\cite{JMLR:v25:23-0783} by changing the Lebesgue measure to $\mu_{\m M}$.
\begin{lemma}[Lemma 1 of~\cite{JMLR:v25:23-0783}]\label{lemma:mixingtime}
Suppose we have a target distribution $\pi^*$ that is absolutely continuous with respect to $\mu_{\m M}$. Let $\varepsilon$ be an error tolerance, and  let $\pi_0$ be a $M_0$-warm distribution of $\pi^*$. Consider the Markov chain obtained from the $\zeta$-lazy version of the RRWM algorithm for sampling from $\pi^*$,  then its mixing time in $\chi^2$ divergence is bounded as
\begin{equation*}
    \tau_{\rm mix}(\varepsilon,\pi_0)\leq \int_{\frac{4}{M_0}}^{\frac{1}{2}} \frac{16\dd v}{\zeta\cdot v\Phi_s^2(v)}+\int_{\frac{1}{2}}^{\frac{4\sqrt{2}}{\varepsilon}} \frac{64\dd v}{\zeta\cdot v\Phi_s^2(\frac{1}{2})},
\end{equation*}
where $s=\frac{\varepsilon^2}{16M_0^2}$, $\Phi_s(v)=\inf\big\{\frac{\int_S \m T(x,\m M \backslash S)\,  \pi^\ast(\dd x)}{\pi^\ast(S)-s}\,|\, S\subseteq \m M, s<\pi^\ast(S)\leq v\big\}$, and $\m T(\cdot,\cdot)$ represents the transition probability function associated with the considered Markov chain.
\end{lemma}
 Then combining Lemmas~\ref{lemma:conductance} and~\ref{lemma:mixingtime}, follows equation (18) of~\cite{JMLR:v21:19-441}, we can get a mixing time bound 
\begin{equation*}
     \tau_{\rm mix}(\varepsilon,\mu_0|_{K_{\theta}})\lesssim \frac{\exp(2\varepsilon_1)}{\zeta} \,\bigg\{\bigg[\kappa\cdot \exp(3\varepsilon_1)\cdot\Big(d+\log \big(\frac{M_0d\kappa}{\varepsilon} \big)+\varepsilon_1\Big)\cdot \log \Big(\frac{\log M_0}{\varepsilon}\Big)\bigg]\vee\log \,(M_0)\bigg\},
 \end{equation*}
for sampling from $\mu^\ast|_{K_\theta}$ using the  $\zeta$-lazy version of the RRWM algorithm. \\
\quad\\Now we investigate the mixing time for sampling from the original distribution $\mu^\ast$. Let $\mu_k$ denote the probability distribution obtained after $k$ steps of the Markov chain obtained Algorithm~\eqref{algorithm_second_version} for sampling from $\mu^*$, with initial distribution $\mu_0$. Then transition probability function from $\mu_k$ to $\mu_{k+1}$ can be written as 
\begin{equation*}
    \ov{\m T}(x,\,\dd y)=(1-\zeta)\wt\eta_x \cdot\ov {\m A}(x,y)p(x,y)\,\mu_{\m M}(\dd y)+\Big(1-\int(1-\zeta)\wt\eta_x \cdot\ov {\m A}(x,y)p(x,y)\,{\mu_{\m M}}(\dd y)\Big)\cdot \delta_x(\dd y),
 \end{equation*}
 where $\zeta$ is the lazy parameter and $\ov {\m A}(x,y)$ is the acceptance ratio given by
 \begin{equation*}
     \ov{\m A}(x,y)=1\wedge \frac{\wt \eta_{y}\cdot\mu^\ast(y)p(y,x)}{\wt \eta_x\cdot\mu^\ast(x)p(x,y)}.
 \end{equation*}
Denote $K_{\theta}'=\{\theta=\wt\phi_{\wt \theta}\big(\wt{W}_{\wt\theta}\frac{z}{\sqrt{n}}\Big)\,:\,\|(W_{\wt \theta}^T \wt I W_{\wt \theta})^{-\frac{1}{2}}z\|\leq R/2\}$. Then for any $x\in K_{\theta}'$,  let $Z$ be a random variable follows $\m N(0,I_d)$, the acceptance probability satisfies that 
\begin{equation*}
\begin{aligned}
 &\Big| \int(1-\zeta)\wt\eta_x\cdot \ov{\m A}(x,y)p(x,y)\,{\mu_{\m M}}(\dd y)-\int(1-\zeta)\wt\eta_x\cdot {\m A}(x,y)p(x,y)\, {\mu_{\m M}}(\dd y)\Big|\\
 &= \int_{{K_\theta}^c}(1-\zeta)\wt\eta_x\cdot\ov{\m A}(x,y)p(x,y)\,{\mu_{\m M}}(\dd y)\\
 &\leq \int_{{K_{\theta}}^c}\wt \eta_x\cdot p(x,y)\,  \mu_{\m M}(\dd y)\\
  &\overset{(i)}{\leq} \int_{\{y\in \wt U_x\,:\,\|(W_{\wt \theta}^T \wt I W_{\wt \theta})^{-\frac{1}{2}}\cdot\wt{\psi}^*_x(y)\|\geq \frac{1}{4}R/\sqrt{n}\}}\wt \eta_x\cdot p(x,y)\,\mu_{\m M}(\dd y)\\
  &\leq P\Big(\|Z\|^2\geq \frac{R^2}{6h\|(W_{\wt \theta}^T\wt IW_{\wt \theta})^{-\frac{1}{2}}(\wt W_{x}^T\wt I\wt W_{x})^{\frac{1}{2}}\|_{\rm op}}\Big) \\
  &\overset{(ii)}{\leq }P\big(\|Z\|^2\geq \frac{R^2}{12h}\big) \\ 
  &\leq  \exp(-5\varepsilon_1)\cdot\frac{\varepsilon^2{h\rho_1}} {M_0^2},
   \end{aligned}
\end{equation*}
where $(i)$ is due to that, if $x\in K_{\theta}'$, $y\in \wt U_x$  and $\|(W_{\wt \theta}^T \wt I W_{\wt \theta})^{-\frac{1}{2}}\cdot\wt{\psi}^*_x(y)\|\leq \frac{1}{4}R/\sqrt{n}$, then there exist $n$-independent constants $C_0,C_1,C_2$ so that $\|y-x\|\leq C_0 R/\sqrt{n}$, $\|x-\wt\theta\|\leq C_0 R/\sqrt{n}$ and    
\begin{equation*}
\begin{aligned}
        \|(W_{\wt \theta}^T \wt I W_{\wt \theta})^{-\frac{1}{2}}\wt\psi^*_{\wt \theta}(y)\|&\leq   \|(W_{\wt \theta}^T \wt I W_{\wt \theta})^{-\frac{1}{2}}(\wt\psi^*_{\wt \theta}(y)-\wt\psi^*_{\wt \theta}(x))\|+\frac{R}{2\sqrt{n}}\\
        &\leq  \|(W_{\wt \theta}^T \wt I W_{\wt \theta})^{-\frac{1}{2}}(y-x)\|+C_1(\|y-\wt \theta\|^2+\|x-\wt\theta\|^2)+\frac{R}{2\sqrt{n}}\\
        &\leq  \|(W_{\wt \theta}^T \wt I W_{\wt \theta})^{-\frac{1}{2}}\wt \psi^*_x(y)\|+C_2\|y-x\|^2+C_1(\|y-\wt \theta\|^2+\|x-\wt\theta\|^2)+\frac{R}{2\sqrt{n}}\\
        &\leq R/\sqrt{n}
\end{aligned}
\end{equation*}
holds when $R\leq \frac{\sqrt{n}}{4(3C_1C_0^2+C_2C_0^2)}$; $(ii)$ uses  the Lipschitzness of $\wt W_x$ with respect to $x$ around $\wt \theta$, and the last inequality uses $R^2/h\geq \frac{36d}{\rho_1h}\geq 24\big(\sqrt{d}+\sqrt{\log \frac{M_0^2}{\varepsilon^2{h\rho_1}}+5\varepsilon_1}\big)^2$ given $h\leq c_0\rho_2^{-1}\big(d+\log (\frac{M_0d\rho_2}{\varepsilon \rho_1})+\varepsilon_1\big)^{-1}$ for small enough $c_0$. So for any $x\in K_\theta'$, and $S\subset K_{\theta}$, we have 
\begin{equation}\label{diffaccp}
    | \ov{\m T}(x,S)-{\m T}(x,S)|\leq   \exp(-5\varepsilon_1)\cdot\frac{\varepsilon^2h\rho_1} {M_0^2}.
\end{equation}
Then let $s=\frac{\varepsilon^2}{16M_0^2}$.  Fix any $S\subset \m M$ so that $s<\mu^\ast(S)\leq v$ for $v\in [\frac{4}{M_0},\frac{1}{2}]$. Then by $\mu^\ast(K_{\theta}')\geq 1-\exp(-5\varepsilon_1)\frac{\varepsilon^2h\rho_1}{M_0^2}$,  we have
\begin{equation*}
   \frac{s}{2} \leq s- \mu^\ast(K_\theta^c)<\mu^\ast(S\cap K_\theta)\leq  \mu^\ast|_{K_\theta}(S)=  \mu^\ast(S\cap K_\theta)/ \mu^\ast(K_\theta)\leq  \mu^\ast(S)/ \mu^\ast(K_\theta)\leq v+s< 2v.
\end{equation*}
 If $\frac{1}{2}\leq\mu^*|_{K_{\theta^*}}(S)\leq v+s\leq \frac{1}{2}+s$, then since $s=\frac{\varepsilon^2}{16M_0^2}\leq\frac{1}{16}$, we have 
 $s< \frac{1}{2}-s\leq \mu^*|_{K_{\theta^*}}(K_\theta\backslash S)\leq \frac{1}{2}<2v$.  Then by  Lemma~\ref{lemma:conductance}, 
 \begin{equation} 
 \begin{aligned}
      &\int_{K_\theta\cap S} \m T(x,K_\theta\backslash S)\,  \mu^\ast|_{ K_\theta}(\dd x)=\int_{K_\theta\backslash S}  \m T(x,K_\theta\cap S) \,   \mu^\ast|_{ K_\theta}(\dd x)\\
      &\geq  C \exp(-2\varepsilon_1) \min\Big\{1,\exp(-3\varepsilon_1)\sqrt{h\rho_1}\log^{\frac{1}{2}}\big(1+\frac{1}{2v}\big)\Big\}\mu^\ast|_{ K_\theta}(K_{\theta}\backslash S)\\
        &\overset{(i)}{\geq}  \frac{1}{2}C \exp(-2\varepsilon_1) \min\Big\{1,\exp(-3\varepsilon_1)\sqrt{h\rho_1}\log^{\frac{1}{2}}\big(1+\frac{1}{2v}\big)\Big\}\mu^\ast|_{ K_\theta}(S)\\
         &\geq  \frac{1}{4}C \exp(-2\varepsilon_1) \min\Big\{1,\exp(-3\varepsilon_1)\sqrt{h\rho_1}\log^{\frac{1}{2}}\big(1+\frac{1}{v}\big)\Big\}\mu^\ast|_{ K_\theta}(S),
      \end{aligned}
 \end{equation}
where $(i)$ uses that $\mu^\ast|_{ K_\theta}(K_{\theta}\backslash S)\geq \frac{1}{2}-s\geq \frac{7}{16}>\frac{1}{2}(\frac{1}{2}+s)\geq \frac{1}{2}\mu^\ast|_{ K_\theta}(S)$.
 If $\mu^\ast|_{ K_\theta}(K_{\theta}\backslash S)<\frac{1}{2}$, then combined with $\mu^\ast|_{ K_\theta}(K_{\theta}\backslash S)\leq 2v$, using  Lemma~\ref{lemma:conductance},  we have 
 \begin{equation}\label{lemma8conclusion}
 \begin{aligned}
     \int_{K_\theta\cap S} \m T(x,K_\theta\backslash S)\,  \mu^\ast|_{ K_\theta}(\dd x)&\geq C \exp(-2\varepsilon_1) \min\Big\{1,\exp(-3\varepsilon_1)\sqrt{h\rho_1}\log^{\frac{1}{2}}\big(1+\frac{1}{2v}\big)\Big\}\mu^\ast|_{ K_\theta}(S)\\
     &\geq \frac{1}{4}C \exp(-2\varepsilon_1) \min\Big\{1,\exp(-3\varepsilon_1)\sqrt{h\rho_1}\log^{\frac{1}{2}}\big(1+\frac{1}{v}\big)\Big\}\mu^\ast|_{ K_\theta}(S).
      \end{aligned}
 \end{equation}
For the right hand side, we further have 
\begin{equation*}
   \mu^\ast|_{K_\theta}(S)\geq   \mu^\ast(S)-\frac{s}{2}\geq \frac{1}{2} \mu^\ast(S).
\end{equation*}
For the left hand side, we further have
\begin{equation*}
\begin{aligned}
    & \int_{S} \ov{\m T}(x, \m M\backslash S)\,  \mu^*(\dd x)-  \int_{K_\theta\cap S} \m T(x,K_\theta\backslash S)\,  \mu^\ast|_{ K_\theta}(\dd x)\\
    &\geq \int_{ S\cap K_\theta'} \ov{\m T}(x, K_\theta\backslash S)\,  \mu^*(\dd x)-  \int_{K_\theta\cap S} \m T(x,K_\theta\backslash S)\, \mu^\ast|_{ K_\theta}(\dd x)\\
    &\geq \int_{ S\cap K_\theta'} \ov{\m T}(x, K_\theta\backslash S)\,  \mu^*(\dd x)-  \int_{S\cap K_\theta'} \m T(x,K_\theta\backslash S)\, \mu^\ast|_{ K_\theta}(\dd x)-\mu^\ast|_{K_\theta}(K_{\theta}\backslash K_\theta')\\
    &\geq -\exp(-5\varepsilon_1)\cdot\frac{3\varepsilon^2{h\rho_1}} { M_0^2},
      \end{aligned}
\end{equation*}
where the last inequality uses equation~\eqref{diffaccp}
 and $\mu^\ast(K_\theta')\geq 1-\exp(-5\varepsilon_1)\cdot\frac{\varepsilon^2{h\rho_1}} {M_0^2}$. Then by equation~\eqref{lemma8conclusion}, we have
 \begin{equation*}
      \int_{S} \ov{\m T}(x, \m M\backslash S)\,  \mu^*(\dd x)+ \exp(-5\varepsilon_1)\cdot\frac{3\varepsilon^2{h\rho_1}} {M_0^2}\geq  \frac{C\exp(-2\varepsilon_1)}{8} \min\Big\{1,\exp(-3\varepsilon_1)\sqrt{h\rho_1}\log^{\frac{1}{2}}\big(1+\frac{1}{v}\big)\Big\}\mu^\ast(S).
 \end{equation*}
 Using $\mu^\ast(S)\geq s=\frac{\varepsilon^2}{16M_0^2}$, we have 
\begin{equation*}
      \int_{S} \ov{\m T}(x, \m M\backslash S)\,\dd \mu^*(x)\geq  \frac{C\exp(-2\varepsilon_1)}{16} \min\Big\{1,\exp(-3\varepsilon_1)\sqrt{h\rho_1}\log^{\frac{1}{2}}\big(1+\frac{1}{v}\big)\Big\}\mu^\ast(S).
 \end{equation*}
Taking infimum over $S$,  we can obtain that
 \begin{equation*}
 \begin{aligned}
&\inf\left\{\frac{\int_S  \ov{\m T}(x,\m M \backslash S)\,\dd \mu^*(x)}{\mu^\ast(S)-s}\,|\, S\subseteq \m M, s<\mu^\ast(S)\leq v\right\}\\
    & \geq C_2\exp(-2\varepsilon_1)\, \min\Big\{1,\exp(-3\varepsilon_1)\sqrt{h\rho_1}\log^{\frac{1}{2}}\big(1+\frac{1}{v}\big)\Big\},
     \end{aligned}
 \end{equation*}
for $v\in [\frac{4}{M_0},\frac{1}{2}]$ and $(n,d,D)$-independent constant $C_2$.  The desired result then follows from Lemma~\ref{lemma:mixingtime}.

   \subsection{Proof of Lemma~\ref{lemma:conductance}}
 Denote $ I^{\Delta}=\wt W_{\wt\theta}^T \wt I  \wt W_{\wt\theta}$ and $J^{\Delta}=(I^{\Delta})^{\frac{1}{2}}J (I^{\Delta})^{\frac{1}{2}}$. We first present the following lemma for bounding the $s$-conductance profile for sampling from $\mu_{\rm loc}^*=\sqrt{n}\cdot W_{\wt \theta}^T\wt\psi_{\wt\theta}(\cdot)_{\#}(\mu^*|_{K_{\theta}})$ with $K_{\theta}=\{x=\wt\phi_{\wt \theta}\big({W}_{\wt\theta}\frac{z}{\sqrt{n}}\Big)\,:z\in K\}$ and $K=\{z\in \mb R^d\,:\, \|(W_{\wt \theta}^T \wt I W_{\wt \theta})^{-\frac{1}{2}}z\|\leq R\}$. 
\begin{lemma}\label{lemma:conductance1}
Suppose (1) $\underset{\xi \in K}{\sup} \big|\log \big(\mu^*_{\rm loc}(\xi)\big)-\log\big((2\pi{\rm det}(J^{-1}))^{-\frac{d}{2}}\exp(-\frac{1}{2}\xi^TJ\xi)\big)\big|\leq \varepsilon_1$; (2) there exists a set $E$  satisfying $\mu^\ast_{\rm loc}(E)\geq 1-\exp(-3\varepsilon_1)\frac{4\varepsilon^2h\rho_1}{M_0^2}$, so that for any $x,z\in E$ with $\| (I^{\Delta})^{-\frac{1}{2}}(x-z)\|\leq \frac{1}{8}\sqrt{2h}$, ${\rm TV}(\m T^\ast(x,\cdot),\m T^\ast(z,\cdot))< 1-\omega$. Then it holds that the $s$-conductance profile of the Markov chain  under transition probability  $\m T^\ast$ with $s=\frac{\varepsilon^2}{32M_0^2}$ satisfies 
 \begin{equation*}
 \begin{aligned}
 & \inf\left\{\frac{\int_S \m T^\ast(x,\mb R^d\backslash S)\mu^\ast_{\rm loc}(x)\,\dd x}{\mu^\ast_{\rm loc}(S)}\,|\, S\subseteq \mb R^d, s<\mu^\ast_{\rm loc}(S)\leq v\right\} \\
 &\geq \frac{\omega}{4}\, \min\Big\{1,\frac{\exp(-3\varepsilon_0)}{64}\sqrt{\rho_1h}\log^{\frac{1}{2}}\big(1+\frac{1}{v}\big)\Big\},
     \end{aligned}
 \end{equation*}
for $v\in [\frac{4}{M_0},\frac{1}{2}]$. 
\end{lemma}
  By Assumption B.2 and lemma~\ref{lemma:conductance1}, it only remains to show that when $\| (I^{\Delta})^{-\frac{1}{2}}(x-z)\|\leq \frac{1}{8}\sqrt{2h}$, ${\rm TV}(\m T^\ast(x,\cdot),\m T^\ast(z,\cdot))$ can be controlled with high probability. Define $E=\big\{\xi\in K\,:\,\big|\|(I^{\Delta})^{\frac{1}{2}}J\xi\|^2-{\rm tr}(J^{\Delta})\big|\leq r_d\big\}$, where recall $ K=\{z\in \mb R^d\,:\, \|(W_{\wt \theta}^T\wt I\wt{W}_{\wt \theta})^{-\frac{1}{2}}z\|\leq R\}$. Note that by Assumption B.2, 
 \begin{equation}\label{eqn:Assc}
   \underset{\xi \in K}{\sup} \Big|\log \big(\mu^*_{\rm loc}(\xi)\big)-\log\Big((2\pi{\rm det}(J^{-1}))^{-\frac{d}{2}}\exp(-\frac{1}{2}\xi^TJ\xi)\Big)\Big|\leq\varepsilon_1.
\end{equation}
Moreover,
   \begin{equation*}
   \begin{aligned}
       &1-\mu^\ast_{\rm loc}(E)=\mu^\ast_{\rm loc}\big(\big|\|(I^{\Delta})^{\frac{1}{2}}J\xi\|^2-{\rm tr}(J^{\Delta})\big|> r_d\big)\\
       &=\int_{\big\{\xi\in K\,:\, \big|\|(I^{\Delta})^{\frac{1}{2}}J\xi\|^2-{\rm tr}(J^{\Delta})\big|> r_d\big\}}\mu^\ast_{\rm loc}(\xi)\,\dd \xi\\
       &\leq\int_{\big\{\xi\in K\,:\, \big|\|(I^{\Delta})^{\frac{1}{2}}J\xi\|^2-{\rm tr}(J^{\Delta})\big|> r_d\big\}}\frac{\sqrt{{\rm det}(J)}}{(2\pi)^{\frac{d}{2}}}\exp(-\frac{\xi^TJ \xi}{2})\,\dd \xi\cdot\underset{\xi\in K}{\inf}\frac{\mu^\ast_{\rm loc}(\xi)}{(2\pi{\rm det}((J)^{-1}))^{-\frac{d}{2}}\exp(-\frac{1}{2}\xi^TJ\xi)}\\
       &\leq \exp(\varepsilon_1)\cdot\int_{ \big\{\xi\in \mb R^d\,:\,\big|\|(I^{\Delta})^{\frac{1}{2}}J\xi\|^2-{\rm tr}(J^{\Delta})\big|> r_d\big\}}\frac{\sqrt{{\rm det}(J)}}{(2\pi)^{\frac{d}{2}}}\exp(-\frac{\xi^TJ\xi}{2})\,\dd \xi\\
       &= \exp(\varepsilon_1)\cdot\int_{ \big\{\xi\in \mb R^d\,:\,\big|\xi^T(J^{\Delta})^2\xi-{\rm tr}(J^{\Delta})\big|> r_d\big\}}\frac{\sqrt{{\rm det}(J^{\Delta})}}{(2\pi)^{\frac{d}{2}}}\exp(-\frac{\xi^TJ^{\Delta}\xi}{2})\,\dd \xi.
          \end{aligned}
   \end{equation*}
   Furthermore, by Bernstein's inequality (see for example, Theorem 2.8.2 of~\cite{vershynin_2018}), for $Z\sim \m N(0,\Sigma)$, it holds that 
  \begin{equation}\label{Bernstein}
    \mb P(\left|\|Z\|^2-{\rm tr}(\Sigma)\right|\geq t)\leq 2\exp(-\frac{1}{c'}(\frac{t^2}{\mnorm{\Sigma}_{\scriptsize  \rm F}^2}\wedge \frac{t}{\mnorm{\Sigma}_{\scriptsize  \rm  op}}))
  \end{equation}
   for a positive $(n,d,D)$-independent constant $c'$. 
When $r_d=\Big(\sqrt{c'\big(\log \frac{M_0^2}{\varepsilon^2 h\rho_1}+4\varepsilon_1\big)}\mnorm{J^\Delta}_{\scriptsize  \rm F} \Big)\vee \Big(c'\big(\log \frac{M_0^2}{\varepsilon^2 h\rho_1}+4\varepsilon_1\big)\rho_2\Big)$,  we can obtain
   \begin{equation*}
   \begin{aligned}
       \mu^\ast_{\rm loc}(E)&\geq  1- \exp(\varepsilon_1)\cdot\int_{ \big\{|\xi^T(J^{\Delta})^2\xi-{\rm tr}(J^{\Delta})|> r_d\big\}}\frac{\sqrt{{\rm det}(J^{\Delta})}}{(2\pi)^{\frac{d}{2}}}\exp(-\frac{\xi^TJ^{\Delta}\xi}{2})\,\dd \xi\\
       &\geq 1- 2\cdot\exp(\varepsilon_1)\cdot P_{Z\sim \m N(0,J^{\Delta})}\Big(\left|\|Z\|^2-{\rm tr}(J^{\Delta})\right|\geq r_d\Big)\\
       &\geq 1-4\frac{\varepsilon^2 h\rho_1}{M_0^2}\exp(-3\varepsilon_1).
          \end{aligned}
   \end{equation*}
 Recall 
   \begin{equation*}
     \m T^*(x,\dd y)= (1-\zeta)\cdot\wt\eta_{G(x)}\cdot\m A^\ast(x,y)p^\ast(x,y)\,\dd y+\Big(1-\int(1-\zeta)\cdot\wt\eta_{G(x)}\cdot\m A^\ast(x,y)p^\ast(x,y)\,\dd y\Big)\cdot \delta_{x}(\dd y).
  \end{equation*}
   For any $x, z\in E$, we consider the following decomposition:
   \begin{equation} 
    \begin{aligned}
     &{\rm TV}(\m T^\ast(x,\cdot),\m T^\ast(z,\cdot))\\
     &=\frac{1}{2}\int |\m T^\ast(x,y)-\m T^\ast(z,y)|\,\dd y\\
     &=\frac{1}{2}\m T^\ast_{x}(\{x\})+\frac{1}{2}\m T^\ast_{z}(\{z\})+\frac{1}{2}\int_{\mathbb{R}^d\backslash\{x,z\}}|\m T^\ast(x,y)-\m T^\ast(z,y)|\,\dd y\\
     &=\frac{1}{2}-\frac{1-\zeta}{2}\cdot\wt\eta_{G(x)}\cdot\int_K p^\ast(x,y)\m A^*(x,y)\,\dd y+\frac{1}{2}-\frac{1-\zeta}{2}\cdot\wt\eta_{G(z)}\cdot\int_K p^\ast(z,y)\m A^*(z,y)\,\dd y\\
     &\qquad-\frac{1-\zeta}{2}\int_{K}\big|\wt\eta_{G(x)}\cdot p^\ast(x,y)\m A^*(x,y)-\wt\eta_{G(z)}\cdot p^\ast(z,y)\m A^*(z,y)\big|\,\dd y\\
     &=1-(1-\zeta)\int_K\min\left(\wt\eta_{G(x)}\cdot p^\ast(x,y)\m A^*(x,y),\wt\eta_{G(z)}\cdot p^\ast(z,y)\m A^*(z,y)\right)\,\dd y
    \end{aligned}
\end{equation}
 where we use $\m T^*_x$ to denote $\m T^*(x,\cdot)$. Recall that 
 \begin{equation*}
     \m A^*(x,y)=1\wedge\frac{\wt\eta_{G(y)}\cdot p^\ast(y,x)\mu^*_{\rm loc}(y)}{\wt\eta_{G(x)}\cdot p^\ast(x,y)\mu^*_{\rm loc}(x)}.
 \end{equation*}
 Define $\overline{\mu}:\,=\m N(0,J^{-1})$, we have 
 \begin{equation*}
     \underset{\xi \in K}{\sup}\left|\log(\mu^*_{\rm loc}(\xi))-\log(\ov \mu(\xi)\right|\leq \varepsilon_1,
 \end{equation*}
 and thus 
\begin{equation*}
    \frac{\mu^*_{\rm loc }(y)}{\mu^*_{\rm loc }(x)}\geq \exp(-2\varepsilon_1)\frac{\ov\mu(y)}{\ov\mu(x)}.
    \end{equation*}
 Therefore, denote 
 \begin{equation*}
     \ov{\m A}^*(x,y)=1\wedge\frac{\wt\eta_{G(y)}\cdot p^\ast(y,x)\ov \mu(y)}{\wt\eta_{G(x)}\cdot p^\ast(x,y)\ov \mu(x)}.
 \end{equation*}
 We have 
 \begin{equation}\label{decompTXz}
    \begin{aligned}
     &{\rm TV}(\m T^\ast(x,\cdot),\m T^\ast(z,\cdot))\\
    &\leq 1-(1-\zeta)\exp(-2\varepsilon_1)\int_K\min\left(\wt\eta_{G(x)}\cdot p^\ast(x,y)\ov{\m A}^*(x,y),\wt\eta_{G(z)}\cdot p^\ast(z,y) \ov{\m A}^*(z,y)\right)\,\dd y\\
    &=1-\frac{1-\zeta}{2}\exp(-2\varepsilon_1)\cdot\Big(\wt\eta_{G(x)}\cdot\int_K p^\ast(x,y)\ov{\m A}^*(x,y)\,\dd y+\wt\eta_{G(z)}\cdot\int_K p^\ast(z,y)\ov{\m A}^*(z,y)\,\dd y\\
     &\qquad-\int_{K}\big|\wt\eta_{G(x)}\cdot p^\ast(x,y)\ov{\m A}^*(x,y)-\wt\eta_{G(z)}\cdot p^\ast(z,y)\ov{\m A}^*(z,y)\big|\,\dd y\Big).\\
    \end{aligned}
\end{equation}
Then consider the inequality,
\begin{equation*}
\begin{aligned}
 \int_{K}|&\wt \eta_{G(x)}\cdot p^\ast(x,y)\ov{\m A}^*(x,y)-\wt \eta_{G(z)}\cdot p^\ast(z,y)\ov{\m A}^*(z,y)|\,\dd y\leq \int_K \wt\eta_{G(x)}\cdot p^\ast(x,y)(1-\ov{\m A}^*(x,y))\,\dd y\\
 &+ \int_K \wt\eta_{G(z)}\cdot p^\ast(z,y)(1-\ov{\m A}^*(z,y))\,\dd y+\int_K  |\wt\eta_{G(x)}\cdot p^\ast(x,y)-\wt\eta_{G(z)}\cdot p^\ast(z,y)|\,\dd y.
 \end{aligned}
\end{equation*}
Moreover,  consider the proposal distribution of RWM,
  \begin{equation*}
 p^{\Delta}_x(\cdot)=p^{\Delta}(x,\cdot)=\m N(x,2hI^{\Delta}),     
 \end{equation*} 
 we have the equation,
 \begin{equation}\label{eqn:Tx}
    \begin{aligned}
  &\wt\eta_{G(x)}\cdot\int_K p^\ast(x,y)\ov{\m A}^*(x,y)\,\dd y\\
   &=1-\Big(\int p^\Delta(x,y)\,\dd y-\int_K \wt \eta_{G(x)}\cdot p^\ast(x,y)\ov{\m A}^*(x,y)\,\dd y\Big) \\
     &=1-\Big(\int_{K} \big(p^\Delta(x,y)-\wt \eta_{G(x)}\cdot p^\ast(x,y)\big)\,\dd y\\
     &\qquad\qquad+\int_{K}\wt \eta_{G(x)}\cdot(1-\ov{\m A}^*(x,y))p^\ast(x,y)\,\dd y+\int_{K^c}p^\Delta(x,y)\,\dd y\Big)\\
    \end{aligned}
\end{equation}
Combined with~\eqref{decompTXz}, we can then obtain 
\begin{equation*}
    \begin{aligned}
     &{\rm TV}(\m T^\ast(x,\cdot),\m T^\ast(z,\cdot))\\
     &\leq 1-(1-\zeta)\exp(-2\varepsilon_1)\bigg(1-\int_{K}\wt \eta_{G(x)}\cdot(1-\ov{\m A}^*(x,y))p^\ast(x,y)\,\dd y-\int_{K}\wt \eta_{G(z)}\cdot(1-\ov{\m A}^*(z,y))p^\ast(z,y)\,\dd y\\
     &\qquad-\frac{1}{2}\int_K  |\wt\eta_{G(x)}\cdot p^\ast(x,y)-\wt\eta_{G(z)}\cdot p^\ast(z,y)|\,\dd y-\frac{1}{2}\int_{K^c}p^\Delta(x,y)\,\dd y-\frac{1}{2}\int_{K^c}p^\Delta(z,y)\,\dd y\\
     &\qquad-\frac{1}{2}\int_{K} \big(p^\Delta(x,y)-\wt \eta_{G(x)}\cdot p^\ast(x,y)\big)\,\dd y-\frac{1}{2}\int_{K} \big(p^\Delta(z,y)-\wt \eta_{G(z)}\cdot p^\ast(z,y)\big)\,\dd y\bigg).
    \end{aligned}
\end{equation*}
 Notice that 
 \begin{equation*}
 \begin{aligned}
&\int_K  |\wt\eta_{G(x)}\cdot p^\ast(x,y)-\wt\eta_{G(z)}\cdot p^\ast(z,y)|\,\dd y \\
&\leq \int_K |\wt\eta_{G(x)}\cdot p^\ast(x,y)-p^{\Delta}(x,y)|\,\dd y+ 2\,{\rm TV}(p^{\Delta}(x,\cdot),p^{\Delta}(z,\cdot)) + \int_K |\wt\eta_{G(z)} \cdot p^\ast(z,y)-p^{\Delta}(z,y)|\,\dd y.
 \end{aligned}
  \end{equation*}
 The term ${\rm TV}(p^{\Delta}(x,\cdot),p^{\Delta}(z,\cdot))$
can be upper bounded by Pinsker's inequality, that is, for any $x,z \in K$,
  \begin{equation*}
    {\rm TV}(p^{\Delta}(x,\cdot),p^{\Delta}(z,\cdot))\leq \frac{\|(I^{\Delta})^{-\frac{1}{2}}(x-z)\|}{2\sqrt{2h}}.
 \end{equation*}
 For the term  $\int_K |\wt \eta_{G(x)}\cdot p^\ast(x,y)-p^{\Delta}(x,y)|\,\dd y$, recall
 \begin{equation*}
   p^\ast(x,y)=\frac{1}{\wt\eta_{G(x)}}\cdot\wt{p}_{G(x)}(\wt{\psi}^*_{G(x)}\circ G(y))\cdot\Big|{\rm det}\big(\bold{J}_{\wt{\psi}^*_{G(x)}\circ G}(y)\big)\Big|\cdot \bold{1}\big(y\in (Q\circ \wt{\phi}^*_{G(x)})(\Omega_{G(x)}) \big).
 \end{equation*}
 We have the following lemma.
\begin{lemma}\label{lemma:diffpast}
There exists a constant $C$ independent of $n$ so that for any $x,y\in K$
\begin{equation*}
  \Big|1-\frac{\wt \eta_{G(x)} \cdot p^\ast(x,y)}{p^\Delta(x,y)}\Big|\leq   C\frac{R^3}{h\sqrt{n}}.
\end{equation*}
\end{lemma}
\noindent So by Lemma~\ref{lemma:diffpast}, for any $x\in K$, $\int_K |\wt \eta_{G(x)}\cdot p^\ast(x,y)-p^{\Delta}(x,y)|\,\dd y\leq  C\frac{R^3}{h\sqrt{n}}$ holds for an $n$-independent constant $C$. For the term of $\int_{K}\wt \eta_{G(x)}\cdot  p^\ast(x,y)(1-\ov{\m A}^*(x,y))\,\dd y$, we use Assumption B.2 by comparing $\mu^*_{\rm loc}$ with $\ov \mu$, leading to the following decomposition:
\begin{equation*}
    \begin{aligned}
     &\int_{K}\wt \eta_{G(x)}\cdot p^\ast(x,y)(1-\ov{\m A}^*(x,y))\,\dd y\\
     &\leq \int_K\Big|\wt \eta_{G(x)}\cdot p^\ast(x,y)-\frac{\wt \eta_{G(y)}\cdot\ov\mu(y)p^\ast(y,x)}{\ov\mu(x)}\Big|\,\dd y\\
     &\leq \int_K |\wt \eta_{G(x)}\cdot p^\ast(x,y)-p^{\Delta}(x,y)|\,\dd y+\underbrace{\int\left|p^{\Delta}(x,y)-\frac{\ov\mu(y)p^{\Delta}(y,x)}{\ov\mu(x)}\right|\dd y}_{\rm (A)}\\
     &\quad+\underbrace{\int_{K}\left|\frac{\ov\mu(y)p^{\Delta}(y,x)}{\ov\mu(x)}-\frac{\wt \eta_{G(y)}\cdot\ov\mu(y)p^\ast(y,x)}{\ov\mu(x)}\right|\,\dd y}_{\rm (B)}.
    \end{aligned}
\end{equation*}
   We then state the following lemma for bounding the term $(A)$.
\begin{lemma}\label{lemma:termA}
There exists a small enough $(n,d,D)$-independent positive constant $c_0$  so that when $h \leq c_0\,\rho_2^{-1}
 \Big(d+\log \big(\frac{M_0d\kappa}{\varepsilon} \big)\Big)^{-1}$, for any $x\in E$, it holds that 
\begin{equation*}
    \int\left|p^{\Delta}(x,y)-\frac{\ov\mu(y)p^{\Delta}(y,x)}{\ov\mu(x)}\right|\,\dd y\leq \frac{1}{24}.
\end{equation*}
\end{lemma}
\noindent For the term $(B)$, by Lemma~\ref{lemma:diffpast} and~\ref{lemma:termA}, we have 
\begin{equation*}
\begin{aligned}
    &\int_K \left|p^{\Delta}(y,x)-\wt \eta_{G(y)}\cdot p^\ast(y,x)\right|\frac{\ov{\mu}(y)}{\ov{\mu}(x)}\,\dd y\\
    &=\int_K \left|1-\frac{\wt \eta_{G(y)}\cdot p^\ast(y,x)}{p^{\Delta}(y,x)}\right|\frac{\ov{\mu}(y)p^{\Delta}(y,x)}{\ov{\mu}(x)}\,\dd y\\
    &\leq \frac{25}{24}\cdot\frac{CR^3}{h\sqrt{n}}.\\
    \end{aligned}
\end{equation*}
Then it remains to bound $\int_{K^c}p^{\Delta}(x,y)\,\dd y$, which is captured by the following lemma.
\begin{lemma}\label{lemmaboundTx}
 When $R\geq 6\sqrt{d/\lambda_{\min}(J^{\Delta})}$, there exists a $(n,d,D)$-independent constant $c_0$ so that when $h\leq c_0\rho_2^{-1}d^{-1}$,  for any $x\in K$, it holds that 
\begin{equation*}
    \int_{K^c} \m N(x,2h I^{\Delta})\,\dd y\leq \frac{13}{24}.
\end{equation*}
\end{lemma}
\noindent Then by combining all the pieces, we can obtain that
\begin{equation*}
    \begin{aligned}
         &{\rm TV}(\m T^\ast(x,\cdot),\m T^\ast(z,\cdot))\leq 1-(1-\zeta)\exp(-2\varepsilon_1)\Big(\frac{3}{8}-\frac{73}{12}\frac{CR^3}{h\sqrt{n}}-\frac{\|(I^{\Delta})^{-\frac{1}{2}}(x-z)\|}{2\sqrt{2}h}\Big).
    \end{aligned}
\end{equation*}
So when $R\leq C_1 (h\sqrt{n})^{\frac{1}{3}}$ for a small enough $n$-independent constant $C_1$, for any $x,z\in E$ with $\| (I^{\Delta})^{-\frac{1}{2}}(x-z)\|\leq \frac{1}{8}\sqrt{2h}$ and $\zeta\in (0,\frac{1}{2}]$, it holds that ${\rm TV}(\m T^\ast(x,\cdot),\m T^\ast(z,\cdot))\leq 1-\frac{1-\zeta}{4}\exp(-2\varepsilon_1)\leq  1-\frac{\exp(-2\varepsilon_1)}{8}$. The desired result then directly follows from Lemma~\ref{lemma:conductance1}.
  \subsection{Proof of Corollary~\ref{th:mixingbpetel}}
 
%We introduce the following Lemma to verify Assumption C.

Fix an $X^{(n)}\in \m A$ where $\m A$ is defined in the proof of Theorem~\ref{th1}. Let $R=C\, \sqrt{\log n}$, where $C$ is a large enough constant that will be chosen later. Define $$\wh\theta^{\diamond}=\phi_{\theta^*}
\big(-W_{\theta^*}(W_{\theta^*}^T{\m H}_{\theta^*}W_{\theta^*})^{-1}\frac{1}{n}\sum_{i=1}^n W_{\theta^*}^T g(X_i,\theta^*)\big),$$ as in the proof of Theorem~\ref{th1}. Then we have shown $\|\theta^*-\wh\theta^{\diamond}\|\lesssim \sqrt{\frac{\log n}{n}}$.  Let $W_{\theta^*}\in \mb R^{D\times d}$ be an orthonormal basis of $T_{\theta^*}\m M$,  and define $\wt W_{\wh\theta^\diamond}=\m U_{\wh\theta^\diamond}\m W_{\wh\theta^\diamond}^T$ with $\m U_{\wh\theta^\diamond}S_{\wh\theta^\diamond}\m W_{\wh\theta^\diamond}^T$ being the singular value decomposition of $\bold{J}_{{\phi}_{\theta^*}(W_{\theta^*}y)}(y=W_{\theta^*}^T {\psi}_{\theta^*}({\wh\theta^\diamond}))$. We will show the desired result using  Theorem~\ref{th:mixing}. We start by verifying the four conditions in Theorem~\ref{th:mixing}.
\begin{enumerate}
    \item   Using Lemma~\ref{lemmasmootharoundtheta}, and $\|\theta^*-\wh\theta^{\diamond}\|\lesssim \sqrt{\frac{\log n}{n}}$, there exists a positive constant $r$ so that $B_r(0_D)\cap T_{\wh\theta^\diamond}\m M\subseteq V_{\wh\theta^\diamond}$.  Moreover, since $R\lesssim \sqrt{\log n}$, for  $z\in K=\{z\in \mb R^d\,: \|(\wt W_{\wh\theta^\diamond}\wt I \wt W_{\wh\theta^\diamond})^{-\frac{1}{2}}z\|\leq R\}$, we have $\|z\|\leq R\cdot \bmnorm{\wt W_{\wh\theta^\diamond}\wt I\wt  W_{\wh\theta^\diamond}}_{\rm op}\lesssim \sqrt{\log n}$.  Thus $\{v=\wt W_{\wh \theta^\diamond}\frac{z}{\sqrt{n}}\,:\, z\in K\}\subseteq B_r(0_D)\cap T_{\wh\theta^\diamond}\m M\subseteq V_{\wh\theta^\diamond}$.
    \item
Define $K_{\theta}=\{x=\phi_{\wh \theta^\diamond}(\wt W_{\wh \theta^\diamond}\frac{z}{\sqrt{n}})\,:\,z\in K\}$ and  let  $\ms L(X^{(n)},\theta)= \frac{L(X^{(n)},\theta)}{(\frac{1}{n})^n}\exp\big(-\alpha_n \cdot(\m R_n(\theta)-\m R_n(\theta^*))\big)$. Denote $G(z)=\sqrt{n}\cdot\phi_{\wh \theta^\diamond}\big(\wt W_{\wh\theta^\diamond}\frac{z}{\sqrt{n}}\big)$. The density function $\mu^\ast_{\rm loc}(\cdot)$ of $\big[\big(\sqrt{n}\cdot \wt W_{\wh \theta^\diamond}^T\psi_{\wh \theta^\diamond}\big)_{\#}(\Pi_{\rm RP}(\cdot|X^{(n)})|_{K_\theta})\big]$ is given by
\begin{equation*}
\begin{aligned}
& \mu^\ast_{\rm loc}(z) =\frac{\pi(G(z)) \ms L\big(X^{(n)}, G(z)\big)\sqrt{\operatorname{det}(\mathbf{J}_{G}(z)^{T} \mathbf{J}_{G}(z))}}{\int_K \pi(G(z)) \ms L\big(X^{(n)}, G(z)\big)\sqrt{\operatorname{det}(\mathbf{J}_{G}(z)^{T} \mathbf{J}_{G}(z))} \,\dd z},\quad z\in K.
\end{aligned}
\end{equation*}
Since $\|\wh\theta^\diamond-\theta^\ast\|\leq C_1\,\sqrt{\frac{\log n}{n}}$, we have $\bmnorm{\wt W_{\wh \theta^\diamond}-W_{\theta^*}}_{\rm F}\leq  C_2\,\sqrt{\frac{\log n}{n}}$.
Define  $h=\sqrt{n}\cdot W_{\theta^*}^T\Big(\phi_{\wh \theta^\diamond}\big(\frac{\wt W_{\wh \theta^\diamond}z
 }{\sqrt{n}}\big)-\wh\theta^\diamond\Big)$,  we can get
 \begin{equation*}
     \begin{aligned}
         \big\|\frac{h}{\sqrt{n}}-\frac{z}{\sqrt{n}}\big\|
        &=\Big\|W_{\theta^*}^T\Big(\phi_{\wh \theta^\diamond}\big(\frac{\wt W_{\wh \theta^\diamond}z
 }{\sqrt{n}}\big)-\wh\theta^\diamond\Big)-\frac{z}{\sqrt{n}}\Big\|\\
 &\leq \Big\|W_{\theta^*}^T\frac{\wt W _{\wh \theta^\diamond}z
 }{\sqrt{n}}-\frac{z}{\sqrt{n}}\Big\|+C_3\,\frac{\|z\|^2}{n}\\
 &\leq \frac{\|z\|}{\sqrt{n}} \bmnorm{W_{\theta^*}^T\wt W_{\wh \theta^\diamond}-W_{\theta^*}^TW_{\theta^*}}_{\rm op}+C_3\,\frac{\|z\|^2}{n}\lesssim  \frac{R^2}{n}
     \end{aligned}
\end{equation*}
and by Step 3 of the proof of Theorem~\ref{th1}, 
\begin{equation*} 
 \begin{aligned}
 &\bigg|\log \ms L(X^{(n)},G(z))+\frac{1}{2} z^T\m H_0\Delta_0^{-1}\m H_0z\bigg|\\
 &\leq \bigg|\log \ms L\Big(X^{(n)}, \phi_{\theta^*}\big(\frac{W_{\theta^*}h}{\sqrt{n}}+\psi_{\theta^*}(\wh\theta^\diamond)\big)\Big)+\frac{1}{2} h^T\m H_0\Delta_0^{-1}\m H_0h\bigg|+\Big|\frac{1}{2} h^T\m H_0\Delta_0^{-1}\m H_0h-\frac{1}{2} z^T\m H_0\Delta_0^{-1}\m H_0z\Big|\\
 &\lesssim \frac{R^3(\log n)^{1+\frac{2}{\beta_1}}}{n^{\frac{\beta_2}{2}}}.
 \end{aligned}
\end{equation*}
 Moreover, by Assumption 1 and Lemma~\ref{lemmasmootharoundtheta}, we can verify
\begin{equation*}
        \big|\sqrt{n}\cdot \sqrt{\operatorname{det}(\mathbf{J}_{G}(z)^{T} \mathbf{J}_{G}(z))}-1\big|=\big|\sqrt{n}\cdot \sqrt{\operatorname{det}(\mathbf{J}_{G}(z)^{T} \mathbf{J}_{G}(z))}-\sqrt{n}\cdot \sqrt{\operatorname{det}(\mathbf{J}_{G}(0_d)^{T} \mathbf{J}_{G}(0_d))}\big|\leq C\,\frac{R}{\sqrt{n}}.
\end{equation*}
So there exists a constant $C_1$ so that for any $z\in K$,
 \begin{equation*}
     \begin{aligned}
         &\Big|\log\Big(\pi(G(z)) \ms L(X^{(n)},G(z))\sqrt{n\cdot\operatorname{det}(\mathbf{J}_{G}(z)^{T} \mathbf{J}_{G}(z))}\Big)-\log\Big(\pi(\theta^\ast)\exp(-\frac{1}{2}z^T\m H_0\Delta_0^{-1}\m H_0z)\Big)\Big|\\
         &\leq C_1\,\frac{R^3(\log n)^{1+\frac{1}{\beta_1}}}{n^{\frac{\beta_2}{2}}}. 
     \end{aligned}
 \end{equation*}
Furthermore,  when $R=C\, \sqrt{\log n}$ with large enough $C$, there exists a constant $C_2$ so that
\begin{equation*}
     \begin{aligned}
         &\Big|\int_{K}\pi(G(z)) \ms L(X^{(n)},G(z))\sqrt{n\cdot\operatorname{det}(\mathbf{J}_{G}(z)^{T} \mathbf{J}_{G}(z))}\,\dd z-\int\pi(\theta^\ast)\exp(-\frac{1}{2}z^T\m H_0\Delta_0^{-1}\m H_0z)\,\dd z\Big|\\
         &\leq \Big|\int_{K}\pi(G(z)) \ms L(X^{(n)},G(z))\sqrt{n\cdot\operatorname{det}(\mathbf{J}_{G}(z)^{T} \mathbf{J}_{G}(z))}\,\dd z\\
         &\qquad-\int_K\pi(\theta^\ast)\exp(-\frac{1}{2}z^T\m H_0\Delta_0^{-1}\m H_0z)\,\dd z\Big|+\int_{K^c}\pi(\theta^\ast)\exp(-\frac{1}{2}z^T\m H_0\Delta_0^{-1}\m H_0z)\,\dd z\\
         &\leq C_2\frac{R^3(\log n)^{1+\frac{1}{\beta_1}}}{n^{\frac{\beta_2}{2}}}.
         \end{aligned}
 \end{equation*}
We can then obtain by combining all pieces that there exists a constant $C_3$ so that
\begin{equation*} 
  \underset{z\in K}{\sup}  \Big|\log\big(\mu_{\rm loc}^*(z)\big)-\log\Big((2\pi{\rm det}(J^{-1}))^{-\frac{d}{2}}\exp(-\frac{1}{2}z^TJz)\Big)\Big|\leq C_3\,\frac{R^3(\log n)^{1+\frac{1}{\beta_1}}}{n^{\frac{\beta_2}{2}}},
\end{equation*}
with $J=\m H_0\Delta_0^{-1}\m H_0$.

    \item  Using $\rho_1I_d\preccurlyeq(W_{\theta^*}^T\wt IW_{\theta^*})^{\frac{1}{2}}J(W_{\theta^*}^T\wt I W_{\theta^*})^{\frac{1}{2}}\preccurlyeq\rho_2 I_d$,  and  $\bmnorm{W_{\theta^*}-\wt W_{\wh\theta^\diamond}}_{\rm F}\lesssim\sqrt{\frac{\log n}{n}}$, when $n$ is large enough, we have 
 \begin{equation*}
     \frac{\rho_1}{2}I_d\preccurlyeq(\wt W_{\wh\theta^\diamond}^T\wt I\wt W_{\wh\theta^\diamond})^{\frac{1}{2}}J(\wt W_{\wh\theta^\diamond}^T\wt I\wt W_{\wh\theta^\diamond})^{\frac{1}{2}}\preccurlyeq 2\rho_2 I_d.
 \end{equation*}
 \item Notice that for sufficiently large $n$, we have  
 \begin{equation*}
     \begin{aligned}
          \varepsilon_1&=\underset{z\in K}{\sup}  \Big|\log\big(\mu_{\rm loc}^*(z)\big)-\log\Big((2\pi{\rm det}(J^{-1}))^{-\frac{d}{2}}\exp(-\frac{1}{2}z^TJz)\Big)\Big|\\
          &\leq C_3\,\frac{R^3(\log n)^{1+\frac{1}{\beta_1}}}{n^{\frac{\beta_2}{2}}}\leq 1
     \end{aligned}
 \end{equation*}
 Then using $\frac{\varepsilon}{M_0}\geq n^{-c}$, we have  for sufficiently large $n$ that $\exp(-5\varepsilon_1)\frac{\varepsilon^2h\rho_1}{2M_0^2}\geq n^{-2c-1}$. Furthermore, using the results from Step 6 of the proof of Theorem~\ref{th1}, there exists a constant $C_0$ so that $\Pi_{\rm RP}(\|\theta-\wh\theta^\diamond\|\leq C_0\sqrt{\frac{\log n}{n}}\,|\, X^{(n)})\geq 1-n^{-2c-1}$. Therefore,  when $R=C\, \sqrt{\log n}$ with large enough $C$, when have 
 \begin{equation*}
     \begin{aligned}
         &\Pi_{\rm RP}\Big(\theta\in \big\{\theta=\phi_{\wh\theta^\diamond}(\wt W_{\wh\theta^\diamond}\frac{z}{\sqrt{n}})\,:\, \|(\wt W_{\wh\theta^\diamond}\wt I \wt W_{\wh\theta^\diamond})^{-\frac{1}{2}}z\|\leq R/2\big\}\,|\, X^{(n)}\Big)\\
         &\geq \Pi_{\rm RP}(\|\theta-\wh\theta^\diamond\|\leq C_0\sqrt{\frac{\log n}{n}}\,|\, X^{(n)})\geq 1-n^{-2c-1}.
     \end{aligned}
 \end{equation*} 
\end{enumerate}
The desired result then follows from Theorem~\ref{th:mixing}.

 \section{Proof of Technical Results}
\subsection{Proof of Lemma~\ref{RGradientpotential}}
 Recall that
 \begin{equation*}
\prod_{i=1}^n p_i(\theta)=\prod_{i=1}^n\frac{\exp\left(\ov\lambda(\theta)^T  g(X_i,\theta)\right)}{\sum_{i=1}^n\exp\left(\ov\lambda(\theta)^T  g(X_i,\theta)\right)},
  \end{equation*}
  with $\ov\lambda(\theta)=\underset{\xi\in T_{\theta}\m M}{\arg\min}\sum_{i=1}^n \exp(\xi^T g(X_i,\theta))$. Therefore, we can write 
  \begin{equation*}
      f(\theta)=\frac{\alpha_n}{n}\sum_{i=1}^n\ell(X_i,\theta)-\sum_{i=1}^n \ov{\lambda}(\theta)^T g(X_i,\theta)-n\cdot\log\Big(\sum_{i=1}^n\exp\big(\ov\lambda(\theta)^T  g(X_i,\theta)\big)\Big).
  \end{equation*}
So the Riemannian gradient of $f$ is given by 
 \begin{equation*}
     \begin{aligned}
       {\rm grad} \,f(\theta)&=\frac{\alpha_n}{n}\sum_{i=1}^n g(X_i,\theta)-\sum_{i=1}^n [{\rm grad}_{\theta}\, g(X_i,\theta)]\cdot\ov\lambda(\theta)-\sum_{i=1}^n [{\rm grad}\, \ov\lambda(\theta)]\cdot g(X_i,\theta)\\
       &\,-n\cdot\frac{\sum_{i=1}^n\exp\big(\ov\lambda(\theta)^T  g(X_i,\theta)\big)
       \cdot\big([{\rm grad}_{\theta}\, g(X_i,\theta)]\cdot\ov\lambda(\theta)+ [{\rm grad}\, \ov\lambda(\theta)]\cdot g(X_i,\theta)\big)}{\sum_{i=1}^n\exp\big(\ov\lambda(\theta)^T  g(X_i,\theta)\big)}\\
       &=\frac{\alpha_n}{n}\sum_{i=1}^n g(X_i,\theta)-\sum_{i=1}^n [{\rm grad}_{\theta}\, g(X_i,\theta)]\cdot\ov\lambda(\theta)-\sum_{i=1}^n [{\rm grad}\, \ov\lambda(\theta)]\cdot g(X_i,\theta)\\
       &\,-n\cdot\frac{\sum_{i=1}^n\exp\big(\ov\lambda(\theta)^T  g(X_i,\theta)\big)
       \cdot[{\rm grad}_{\theta}\, g(X_i,\theta)]\cdot\ov\lambda(\theta)}{\sum_{i=1}^n\exp\big(\ov\lambda(\theta)^T  g(X_i,\theta)\big)},
     \end{aligned}
 \end{equation*}
 where ${\rm grad}\,\ov\lambda(\cdot)$ denotes the Riemannian gradient of  $\ov\lambda(\cdot)$ on $\m M$ and the last equality is due to the fact that 
 \begin{equation}\label{lambdacondition}
     \sum_{i=1}^n \frac{\exp(\ov\lambda(\theta)^Tg(X_i,\theta))}{{\sum_{i=1}^n\exp\big(\ov\lambda(\theta)^T  g(X_i,\theta)\big)}}\cdot g(X_i,\theta)=\sum_{i=1}^n p_i(\theta)g(X_i,\theta)={0}_D.
 \end{equation}
 So it remains to show that $g_{\ov\lambda}(\theta)=[{\rm grad}\,\ov\lambda(\theta)]^T$. Recall for any $k\in [D]$ and $\theta\in \m M$, 
 \begin{equation*}
     \sum_{i=1}^n \exp(\ov\lambda(\theta)^Tg(X_i,\theta))g_k(X_i,\theta)=0.
 \end{equation*}
 Therefore, we have for any $k\in [D]$, the Riemannian gradient of $\sum_{i=1}^n \exp(\ov\lambda(\theta)^Tg(X_i,\theta))g_k(X_i,\theta)$ with respect to $\theta$ on $\m M$ satisfies that
 \begin{equation*}
     {\rm grad}_{\theta}\, \Big( \sum_{i=1}^n \exp(\ov\lambda(\theta)^Tg(X_i,\theta))g_k(X_i,\theta)\Big)=0_D.
 \end{equation*}
 We can then obtain that  for any $k\in [D]$,
 \begin{equation*}
     \begin{aligned}
       &\sum_{i=1}^n \exp(\ov\lambda(\theta)^Tg(X_i,\theta))\cdot\Big({\rm grad}_{\theta}\, g_k(X_i,\theta)+ g_k(X_i,\theta)[{\rm grad}\, \ov\lambda(\theta)]g(X_i,\theta)+g_k(X_i,\theta)[{\rm grad}_{\theta}\, g(X_i,\theta)]\ov\lambda(\theta)\Big)\\
       &=0_D,
     \end{aligned}
 \end{equation*}
which further lead to 
\begin{equation*}
\begin{aligned}
    &\sum_{i=1}^n \exp(\ov\lambda(\theta)^Tg(X_i,\theta)) g(X_i,\theta) g^T(X_i,\theta) [{\rm grad}\, \ov\lambda(\theta)]^T\\
    &=- \sum_{i=1}^n \exp(\ov\lambda(\theta)^Tg(X_i,\theta))\cdot\Big([{\rm grad}_{\theta}\, g(X_i,\theta)]^T+g(X_i,\theta)\ov\lambda(\theta)^T[{\rm grad}_{\theta}\, g(X_i,\theta)]^T\Big).
    \end{aligned}
    \end{equation*}
    Therefore, we have
 \begin{equation*}
     \begin{aligned}
         [{\rm grad}\, \ov\lambda(\theta)]^T
         &=-\left(\sum_{i=1}^n \exp(\ov\lambda(\theta)^T g(X_i,\theta))g(X_i,\theta)g(X_i,\theta)^T\right)^{+}\\
       &\cdot\left(\sum_{i=1}^n \exp(\ov\lambda(\theta)^T g(X_i,\theta))\left(I_D+g(X_i,\theta)\ov\lambda(\theta)^T\right)[{\rm grad}_{\theta} \,g(X,\theta)]^T\right)=g_{\ov\lambda}(\theta).
     \end{aligned}
 \end{equation*}
 \subsection{Proof of Lemma~\ref{lemma1}}

Using $\mb{E}[\exp\big((\frac{b(X)}{L})^{\beta_1}\big)]\leq 1$ and Markov's inequality,  we obtain,  for $C\geq L 4^{\frac{1}{\beta_1}}$,  that
\begin{equation}\label{eqnprobA1}
\begin{aligned}
      P(X^{(n)}\notin \m A_1)&\leq n\cdot P(b(X)> C(\log n)^{\frac{1}{\beta_1}})\\
      &\leq n \frac{1}{\exp((\frac{C}{L})^{\beta_1}\log n)}\leq \frac{1}{n^3}.
\end{aligned}
\end{equation}
Furthermore,  for any $X\in \m X$, 
\begin{equation*}
  \big\| g(X,\theta^*)\cdot\bold{1}\big(b(X)\leq  C(\log n)^{\frac{1}{\beta_1}}\big)\big\|\leq b(X) \cdot\bold{1}\big(b(X)\leq  C(\log n)^{\frac{1}{\beta_1}}\big)\leq  C(\log n)^{\frac{1}{\beta_1}}.
  \end{equation*}
Moreover, there exists a constant $C_2$ so that 
\begin{equation*}
\begin{aligned}
       & \mb{E}\Big[ \big\|g(X,\theta^*)\cdot\bold{1}\big(b(X)\leq  C(\log n)^{\frac{1}{\beta_1}}\big)\big\|^2\Big]\\
       &\leq \mb{E}\big[  \|  g(X,\theta^*) \|^2\big]\leq  \mb{E}\big[ (b(X))^2\big]\leq C_2.
\end{aligned}
\end{equation*}
In addition, using $\mb{E}[g(X,\theta^*)]={\rm grad}_{\theta} \m R(\theta^*)=0_D$, we have
\begin{equation*}
\begin{aligned}
    &  \Big\|\mb{E}\Big[ g(X,\theta^*)\cdot\bold{1}\big(b(X)\leq  C(\log n)^{\frac{1}{\beta_1}}\big)\Big]\Big\|\\
    &= \Big\|\mb{E}\Big[g(X,\theta^*)\cdot\bold{1}\big(b(X)>C(\log n)^{\frac{1}{\beta_1}}\big)\Big]\Big\|\\
    &\leq \sqrt{d} \sqrt{\mb{E}\Big[\Big\|g(X,\theta^*)\Big\|^2\Big]}\cdot\sqrt{P\big(b(X)>C(\log n)^{\frac{1}{\beta_1}}\big)}\\
    &\leq \frac{C_3}{n^{2}}.
\end{aligned}
\end{equation*}
 We can  then obtain via Bernstein's inequality that, when $C_1$ is sufficiently large, 
\begin{equation*}
\begin{aligned}
       P(X^{(n)}\notin \m A_2)&\leq P\Big(\Big\|\frac{1}{n}\sum_{i=1}^n g(X_i,\theta^*)\cdot\bold{1}\big(b(X_i)\leq  C(\log n)^{\frac{1}{\beta_1}}\big)\\
       & \quad\qquad-\mb{E}\Big[g(X,\theta^*)\cdot\bold{1}\big(b(X)\leq  C(\log n)^{\frac{1}{\beta_1}}\big)\Big]\Big\|\geq C_1 \sqrt{\frac{\log n}{n}}-\frac{C_3}{n^{\frac{3}{2}}}\Big)\\
       &\leq  P\Big(\Big\|\frac{1}{n}\sum_{i=1}^n g(X_i,\theta^*)\cdot\bold{1}\big(b(X_i)\leq  C(\log n)^{\frac{1}{\beta_1}}\big)\\
       & \quad\qquad-\mb{E}\Big[g(X,\theta^*)\cdot\bold{1}\big(b(X)\leq  C(\log n)^{\frac{1}{\beta_1}}\big)\Big]\Big\|\geq \frac{C_1}{2} \sqrt{\frac{\log n}{n}}\Big)\leq \frac{1}{n^3}.
\end{aligned}
\end{equation*}
 Similarly, since $\mb{E}[\|g(X,\theta^*)\|^6]\leq \mb{E}[b(X)^6]<\infty$, we can apply Bernstein inequality to obtain  $P(X^{(n)}\notin \m A_3)\leq \frac{1}{n^3}$. Define
\begin{equation*}
    \wt g(x,\theta)=\frac{1}{ C(\log n)^{\frac{1}{\beta_1}}}\cdot g(x,\theta)\cdot\bold{1}\big(b(x)\leq C(\log n)^{\frac{1}{\beta_1}}\big)
\end{equation*}
 and
 \begin{equation*}
     \wt \ell(x,\theta)=\frac{1}{ C(\log n)^{\frac{1}{\beta_1}}}\cdot \ell(x,\theta)\cdot\bold{1}(b(x)\leq C(\log n)^{\frac{1}{\beta_1}}).
 \end{equation*}
Then, for any $x\in \m X$ and $\theta\in S_{\Pi}$, $\|\wt g(x,\theta)\|\leq \frac{b(x)}{ C(\log n)^{\frac{1}{\beta_1}}}\cdot \bold{1}(b(x)\leq C(\log n)^{\frac{1}{\beta_1}})\leq 1$ and for any $x\in \m X$, $\theta,\theta'\in S_{\Pi}$, 
\begin{equation*}
    | \wt \ell(x,\theta)-\wt   \ell(x,\theta')|\leq \frac{1}{ C(\log n)^{\frac{1}{\beta_1}}} b(x)\cdot \bold{1}(b(x)\leq C(\log n)^{\frac{1}{\beta_1}})\cdot \|\theta-\theta'\|\leq \|\theta-\theta'\|.
\end{equation*}
Now define the pseudo-metric $d_n^{\wt g}:S_{\Pi}\times S_{\Pi}\to \mb R$ as $$d_n^{\wt g}(\theta,\theta')=\sqrt{\frac{1}{n}\sum_{i=1}^n \| \wt g(X_i,\theta)-\wt g(X_i,\theta')\|^2}.$$ Then for any $\theta,\theta'\in S_{\Pi}$, 
 \begin{equation*}
     \begin{aligned}
         &d_n^{\wt g}(\theta,\theta')=\frac{1}{C(\log n)^{\frac{1}{\beta_1}}}\sqrt{\frac{1}{n}\sum_{i=1}^n \| g(X_i,\theta)-g(X_i,\theta')\|^2\cdot \bold{1}(b(X_i)\leq C(\log n)^{\frac{1}{\beta_1}})}\\
         &=\frac{1}{C(\log n)^{\frac{1}{\beta_1}}}\sqrt{\frac{1}{n}\sum_{i\in [n], \,b(X_i)\leq C(\log n)^{\frac{1}{\beta_1}}} \| g(X_i,\theta)-g(X_i,\theta')\|^2}.
     \end{aligned}
 \end{equation*}
Therefore 
\begin{equation*}
    \log \bold{N}(B_r(\theta^*)\cap\m M, d_n^{\wt g}, \varepsilon)= \log \bold{N}(B_r(\theta^*)\cap S_{\Pi}, d_n^{\wt g}, C(\log n)^{\frac{1}{\beta_1}}\varepsilon)\leq   
 L_2\log n+L_3\log(\frac{1}{\varepsilon}).
\end{equation*}
Furthermore, for any $\theta,\theta'\in S_{\Pi}$, it holds that 
\begin{equation*}
    \begin{aligned}
       & \sqrt{\frac{1}{n}\sum_{i=1}^n \bmnorm{ \wt g(X_i,\theta)\wt g(X_i,\theta)^T-\wt g(X_i,\theta')\wt g(X_i,\theta')^T}_{\rm F}^2}\leq 2 d_n^{\wt g}(\theta,\theta').
    \end{aligned}
\end{equation*}
Using the uniform low via Rademacher complexity (see for example Theorem 4.10 of~\cite{wainwright_2019}) and Dudley’s entropy integral bound (see for example Theorem 5.22 of~\cite{wainwright_2019}),  there exist constants $C_2,C_3$ so that it holds with probability at least $1-n^{-3}$ that  for any  $\theta \in B_{r}(\theta^*)\cap \m M$,
 \begin{equation*}
     \begin{aligned}
         &\Bmnorm{n^{-1}\sum_{i=1}^n \wt g(X_i,\theta)\wt g(X_i,\theta)^T-\mathbb{E} \big[\wt g(X,\theta)\wt g(X,\theta)^T\big]}_{\rm F}\\
         &\leq C_2\sqrt{\frac{\log n}{n}}+\frac{C_2}{\sqrt{n}}\mb{E}\Big[\int_0^2 \sqrt{ \log \bold{N}(B_r(\theta^*)\cap\m M, d_n^{\wt g}, \varepsilon)}\,\dd \varepsilon\Big]\\
         &\leq C_3 \sqrt{\frac{\log n}{n}}.
     \end{aligned}
 \end{equation*}
In addition, for any $\theta\in \m M$,  
\begin{equation*}
    \mb{E}[\|\wt g(X,\theta)-\wt g(X,\theta^*)\|^2]\leq \frac{1}{C^2 (\log n)^{\frac{2}{\beta_1}}} \mb{E}[\|g(X,\theta)-g(X,\theta^*)\|^2] \leq \frac{L}{C^2 (\log n)^{\frac{2}{\beta_1}}}\|\theta-\theta^*\|^{2\beta_2}.
\end{equation*}
Moreover, for any $\theta,\theta'\in B_r(\theta^*)\cap \m M$, 
\begin{equation*}
\begin{aligned}
    \mathbb{E} \big[\big(\wt\ell(X,\theta)-\wt\ell(X,\theta')-\wt g(X,\theta')(\theta-\theta')\big)^2\big]&\leq  \frac{1}{C^2 (\log n)^{\frac{2}{\beta_1}}}  \mathbb{E} \big[\big(\ell(X,\theta)-\ell(X,\theta')-g(X,\theta')(\theta-\theta')\big)^2\big]\\
     &\leq \frac{L}{C^2 (\log n)^{\frac{2}{\beta_1}}}\|\theta-\theta^*\|^{2+2\beta_2}.
\end{aligned}
\end{equation*}
 Following the proof of Lemma 24 in~\cite{JMLR:v25:23-0783}, we can obtain that there exists a constant $C_4$ so that it holds with probability at least $1-n^{-3}$ that 
 \begin{enumerate}
     \item  
 For any  $\theta\in B_r(\theta^*)\cap \m M$, 
 \begin{equation*}
     \begin{aligned}
         \Big\| n^{-1}\sum_{i=1}^n  \wt g(X_i,\theta)- n^{-1}\sum_{i=1}^n  \wt g(X_i,\theta^*)-\mathbb{E} [\wt g(X,\theta)]+\mathbb{E} [ \wt g(X,\theta^*)]\Big\|_2 \leq C_4\,\Big(\sqrt{\frac{\log n}{n}} \, \|\theta-\theta^*\|^{\beta_2}+\frac{\log n}{n}\Big).
     \end{aligned}
 \end{equation*}
 \item 
 For any $\theta,\theta'\in S_{\Pi}$, 
  \begin{equation*}
     \begin{aligned}
        \Big|n^{-1}\sum_{i=1}^n \wt\ell(X_i,\theta)-n^{-1}\sum_{i=1}^n \wt\ell(X_i,\theta')-\mathbb{E} [\wt\ell(X,\theta)]+\mathbb{E} [\wt\ell(X,\theta')]\Big| \leq C_4\,\Big(\sqrt{\frac{\log n}{n}} \, \|\theta-\theta'\|+\frac{\log n}{n}\Big).
     \end{aligned}
 \end{equation*}
 \item  For any $\theta,\theta'\in B_r(\theta^*)\cap \m M$, 
   \begin{equation*}
     \begin{aligned}
& \Big|n^{-1}\sum_{i=1}^n \wt\ell(X_i,\theta)-n^{-1}\sum_{i=1}^n \wt\ell(X_i,\theta')- \frac{1}{n}\sum_{i=1}^n \wt g(X_i,\theta')(\theta-\theta')
-\mathbb{E} [\wt \ell(X,\theta)]+\mathbb{E} [\wt \ell(X,\theta')]\\
&\qquad+\mb{E}\big[\wt g(X,\theta')(\theta-\theta')\big]\Big| \leq C_4\,\Big(\sqrt{\frac{\log n}{n}} \, \|\theta-\theta'\|^{\beta_2+1}+\frac{\log n}{n}\|\theta-\theta'\|+(\frac{\log n}{n})^2\Big).
      \end{aligned}
 \end{equation*}
 \end{enumerate}
 Furthermore, by H\"{o}lder's inequality, there exist constants $C_5, C_6$ so that for any  $\theta \in S_{\Pi}$,
 \begin{equation*}
\begin{aligned}
         &\Bmnorm{\mathbb{E} \big[g(X,\theta)g(X,\theta)^T\cdot \bold{1}\big(b(X)> C(\log n)^{\frac{1}{\beta_1}}\big)\big]}_{\rm F}\\
         &\leq  C_5\, \sqrt{\mathbb{E} \big[b(X)^4]}\sqrt{P(b(X)> C(\log n)^{\frac{1}{\beta_1}})}\\
         &\leq C_6 \frac{1}{n^2}.
\end{aligned}
 \end{equation*}
Similarly, for any $\theta,\theta'\in S_{\Pi}$,
\begin{equation*}
    \begin{aligned}
          &\Big|\mathbb{E} \big[(\ell(X,\theta)-\ell(X,\theta'))\cdot \bold{1}\big(b(X)> C(\log n)^{\frac{1}{\beta_1}}\big)\big]\Big|\\
          &\leq \mathbb{E} \big[b(X)\cdot \bold{1}\big(b(X)> C(\log n)^{\frac{1}{\beta_1}}\big)\big] \cdot\|\theta-\theta'\|\\
          &\leq C_6 \frac{1}{n^2},
    \end{aligned}
\end{equation*}
and
\begin{equation*}
    \begin{aligned}
          &\Big|\mathbb{E} \big[(\ell(X,\theta)-\ell(X,\theta')-g(X,\theta')(\theta-\theta'))\cdot \bold{1}\big(b(X)> C(\log n)^{\frac{1}{\beta_1}}\big)\big]\Big|\\
          &\leq 2\cdot\mathbb{E} \big[b(X)\cdot \bold{1}\big(b(X)> C(\log n)^{\frac{1}{\beta_1}}\big)\big] \cdot\|\theta-\theta'\|\\
          &\leq C_6 \frac{1}{n^2}.
    \end{aligned}
\end{equation*}
Furthermore, notice that given $X^{(n)}\in \m A_1$ and  $\bmnorm{n^{-1}\sum_{i=1}^n \wt g(X_i,\theta)\wt g(X_i,\theta)^T-\mathbb{E} \big[\wt g(X,\theta)\wt g(X,\theta)^T\big]}_{\rm F}\leq C_3 \sqrt{\frac{\log n}{n}}$, it holds that
 \begin{equation*}
\begin{aligned}
  &\Bmnorm{n^{-1}\sum_{i=1}^n g(X_i,\theta)g(X_i,\theta)^T-\mathbb{E} \big[g(X,\theta)g(X,\theta)^T\big]}_{\rm F}\\
     &= \Bmnorm{ C^2(\log n)^{\frac{2}{\beta_1}}\cdot n^{-1}\sum_{i=1}^n\wt g(X_i,\theta)\wt g(X_i,\theta)^T-\mathbb{E} \big[g(X,\theta)g(X,\theta)^T\big]}_{\rm F}\\
          &\leq C^2(\log n)^{\frac{2}{\beta_1}}\cdot \Bmnorm{ n^{-1}\sum_{i=1}^n\wt g(X_i,\theta)\wt g(X_i,\theta)^T-\mathbb{E} \big[\wt g(X,\theta)\wt g(X,\theta)^T\big]}_{\rm F} \\
          &\qquad+  \Bmnorm{\mathbb{E} \big[g(X,\theta)g(X,\theta)^T\cdot \bold{1}\big(b(X)> C(\log n)^{\frac{1}{\beta_1}}\big)\big]}_{\rm F}\\
     &\leq C_1 \frac{(\log n)^{\frac{2}{\beta_1}+\frac{1}{2}}}{\sqrt{n}}.
\end{aligned}
\end{equation*}
Therefore,  $P(X^{(n)}\in  \m A_4)\geq P(X^{(n)}\in \m A_1\cap \m A_4)\geq 1-2n^{-3}$. Proceeding analogously, we obtain the same type of bounds for  $\m A_5,\m A_6,\m A_7$, and hence conclude the desired result.

  \subsection{Proof of Lemma~\ref{lemma2}}
 Write $\wh\theta^\diamond=\wh\theta^\diamond(X^{(n)})=\phi_{\theta^*}
\big(-W_{\theta^*}\m H_0^{-1}\frac{1}{n}\sum_{i=1}^n W_{\theta^*}^Tg(X_i,\theta^*)\big)$. Recall $\wt{\lambda}(\theta)=-\Delta_0^{-1}\m H_0W_{\theta^*}^T (\psi_{\theta^*}(\theta)-\psi_{\theta^*}(\wh\theta^\diamond))$, we have 
 \begin{equation*}
  \begin{aligned}
  &\big\|\Delta_0^{-1} W_{\theta}^T\frac{1}{n}\sum_{i=1}^n g(X_i, \theta)+\wt{\lambda}(\theta)\big\|\\
  &= \left\|\Delta_0^{-1} \left(\frac{1}{n}\sum_{i=1}^n  W_{\theta}^T g(X_i, \theta)-\frac{1}{n}\sum_{i=1}^n W_{\theta^*}^Tg(X_i,\theta^*)-\m H_0W_{\theta^*}^T\psi_{\theta^*}(\theta)\right)\right\|.
  \end{aligned}
   \end{equation*}
Since $X^{(n)}\in \m A$ and $\mathbb{E}[g(X,\theta^*)]=0_D$,  there exists a constant $C$ so that it holds for any $\theta\in B_{r}(\theta^*)\cap \mathcal{M}$ that,
   \begin{equation*}
  \begin{aligned}
  &\Big\|\frac{1}{n}\sum_{i=1}^n W_{\theta}^Tg\big(X_i,\theta\big)-\frac{1}{n}\sum_{i=1}^n W_{\theta^*}^Tg(X_i,\theta^*)-\mathbb{E}[W_{\theta}^Tg\big(X,\theta\big)]+\mathbb{E}[W_{\theta^*}^Tg(X,\theta^*)]\Big\|\\
  &\leq \Big\|W_{\theta}^T\cdot\big(\frac{1}{n}\sum_{i=1}^n g\big(X_i,\theta\big)-\frac{1}{n}\sum_{i=1}^n g(X_i,\theta^*)-\mathbb{E}[g\big(X,\theta\big)]+\mathbb{E}[g(X,\theta^*)]\big)\Big\|\\
  &+\|(W_{\theta}-W_{\theta^*})\cdot\big(\frac{1}{n}\sum_{i=1}^n g(X_i,\theta^*)-\mathbb{E}[g(X,\theta^*)]\big)\|\\
  &\leq  C\,(\log n)^{\frac{1}{\beta_1}}\Big(\sqrt{\frac{\log n}{n}} \, \|\theta-\theta^*\|^{\beta_2}+\frac{\log n}{n}\Big).
  \end{aligned}
  \end{equation*}
Consider the transformed risk function $\wt {\m R}:B_r(0_d)\to \mb R$ defined by
$\wt {\m R}(z)=\m R(\phi_{\theta^*}(W_{\theta^*}z))=\mb{E}[\ell(X,\phi_{\theta^*}(W_{\theta^*}z))]$. By Assumption  1 and 3, $\wt {\m R}$ is thrice differentiable.  
% \begin{enumerate}
%     \item $|\wt{\mathcal{R}}(y)|+\big|\m D\wt{\mathcal{R}}(y)[\eta_1]\big|+\big|\m D\big(\m D\wt{\mathcal{R}}(y)[\eta_1]\big)[\eta_1]\big|\leq L_1$;
%     \item $\m D\big(\m D\wt{\mathcal{R}}(0_D)[\eta_1]\big)[\eta_1]\geq a^2\|\eta_1\|^2$;
%     \item $\big|\m D\big(\m D\wt{\mathcal{R}}(y)[\eta_1]\big)[\eta_1]-\m D\big(\m D\wt{\mathcal{R}}(y')[\eta_1]\big)[\eta_1]\big|\leq L_2\|y-y'\|$.
% \end{enumerate}
Moreover, by proposition 5.44 of~\citep{boumal2022intromanifolds}, we have $\m H_0=W_{\theta^\ast}^T{\m H}_{\theta^*}W_{\theta^\ast}={\rm Hessian}(\wt R(z)|_{z=0_d})$. Let $\m H_z$ denote the Hessian matrix of $\wt {\m R}(\cdot)$ at $z$, then $\m H_z$ is uniformly Lipshitz-continuous over $B_r(0_d)$. Hence, there exists a constant $C$ so that for any $\theta\in B_{r}(\theta^*)\cap\m M$, 
  \begin{equation*}
  \begin{aligned}
  &\Big\|\mathbb{E}[W_{\theta}^Tg(X, \theta)]-W_{\theta^*}^T\mathbb{E}[ g(X,\theta^*)]-\m H_0W_{\theta^*}^T\psi_{\theta^*}(\theta)\Big\|\\
  &\leq \underset{\eta\in \mb S_1^{d-1}}{\sup}\Big|\eta^T\nabla \wt {\m R}( W_{\theta^*}^T\psi_{\theta^*}(\theta))-\eta^T\nabla \wt {\m R}(0_d)-\eta^T\m H_0 W_{\theta^*}^T\psi_{\theta^*}(\theta)\Big|\\
  & \leq\underset{\eta\in \mb S_1^{d-1}}{\sup}\underset{t\in (0,1)}{\sup} \Big|\eta^T(\m H_{t W_{\theta^*}^T\psi_{\theta^*}(\theta)}-\m H_0) W_{\theta^*}^T\psi_{\theta^*}(\theta)\Big|\\
  &\leq C\,\|\theta-\theta^*\|_2^2.
  \end{aligned}
 \end{equation*}
   Therefore, for any $\theta\in B_{r}(\theta^*)\cap\m M$,  it holds that
    \begin{equation}\label{eqn:deltalambda}
  \begin{aligned}
   &\big\|\Delta_0^{-1} W_{\theta}^T\frac{1}{n}\sum_{i=1}^n  g(X_i, \theta)+\wt{\lambda}(\theta)\big\|\\
   &\leq  C\,(\log n)^{\frac{1}{\beta_1}} \Big(\frac{\log n}{n}+\sqrt{\frac{\log n}{n}}\|\theta-\theta^*\|_2^{\beta_2}\Big)+C\,\|\theta-\theta^*\|_2^2.
   \end{aligned}
   \end{equation}
Moreover, notice that  
\begin{equation}\label{eqn:Wbound}
\begin{aligned}
      & \|\frac{1}{n}\sum_{i=1}^n W_{\theta}^T g\big(X_i,\theta\big)\|\leq \Big\|\frac{1}{n}\sum_{i=1}^n W_{\theta}^Tg\big(X_i,\theta\big)-\frac{1}{n}\sum_{i=1}^n W_{\theta^*}^Tg(X_i,\theta^*)-\mathbb{E}[W_{\theta}^Tg\big(X,\theta\big)]+\mathbb{E}[W_{\theta^*}^Tg(X,\theta^*)]\Big\|\\
      &+\Big\|\mathbb{E}[W_{\theta}^Tg(X, \theta)]-W_{\theta^*}^T\mathbb{E}[ g(X,\theta^*)]-\m H_0W_{\theta^*}^T\psi_{\theta^*}(\theta)\Big\|+
      \|\m H_0W_{\theta^*}^T(\theta-\theta^*)\|+ \|\frac{1}{n}\sum_{i=1}^n W_{\theta^*}^T g\big(X_i,\theta^*\big)\|\\
      &\leq   C\,(\log n)^{\frac{1}{\beta_1}}\Big(\sqrt{\frac{\log n}{n}} \, \|\theta-\theta^*\|^{\beta_2}+\frac{\log n}{n}\Big)+ C\,\|\theta-\theta^*\|_2^2+\|\m H_0W_{\theta^*}^T(\theta-\theta^*)\|+C_1\sqrt{\frac{\log n}{n}}\\
      &\lesssim \|\theta-\theta^*\|+\sqrt{\frac{\log n}{n}}. 
\end{aligned}
\end{equation}
Consequently, there exist constant $C_1,C_2,C_3$ so that for any $\theta\in \m M$ with $\|\theta-\theta^*\|\leq \delta\frac{(\log n)^{\frac{3}{2}}}{\sqrt{n}}$, we have
 \begin{equation}\label{eqn:le2.1}
     \begin{aligned}
&  \bigg \|\frac{1}{n}\sum_{i=1}^n \exp\left(\wt{\lambda}(\theta)^T  W_{\theta}^Tg(X_i,\theta)\right) W_{\theta}^Tg(X_i,\theta)\\
&-\frac{1}{n}\sum_{i=1}^n \exp\Big(-\big(\Delta_0^{-1}W_{\theta}^T\frac{1}{n}\sum_{i=1}^n g(X_i, \theta)\big)^TW_{\theta}^Tg(X_i,\theta)\Big)W_{\theta}^Tg(X_i,\theta))\bigg\|\\
&= \bigg \|\frac{1}{n}\sum_{i=1}^n \exp\left(\wt{\lambda}(\theta)^T  W_{\theta}^Tg(X_i,\theta)\right) W_{\theta}^Tg(X_i,\theta)\cdot \textbf{1}(b(X_i)\leq C(\log n)^{\frac{1}{\beta_1}})\\
&\qquad-\frac{1}{n}\sum_{i=1}^n \exp\Big(-\big(\Delta_0^{-1}W_{\theta}^T\frac{1}{n}\sum_{i=1}^n g(X_i, \theta)\big)^TW_{\theta}^Tg(X_i,\theta)\Big)W_{\theta}^Tg(X_i,\theta) \textbf{1}(b(X_i)\leq C(\log n)^{\frac{1}{\beta_1}})\bigg\|\\
&\leq C_1\Big( (\log n)^{\frac{1}{\beta_1}} \Big(\frac{\log n}{n}+\sqrt{\frac{\log n}{n}}\|\theta-\theta^*\|_2^{\beta_2}\Big)+\|\theta-\theta^*\|_2^2\Big) \cdot \big(\frac{1}{n}\sum_{i=1}^n \|g(X_i,\theta)\|^2\big)\\\
&\leq C_1\Big( (\log n)^{\frac{1}{\beta_1}} \Big(\frac{\log n}{n}+\sqrt{\frac{\log n}{n}}\|\theta-\theta^*\|_2^{\beta_2}\Big)+\|\theta-\theta^*\|_2^2\Big) \cdot \big({\rm tr}(\ov\Delta_{\theta})+C_2 \frac{(\log n)^{\frac{2}{\beta_1}+\frac{1}{2}}}{\sqrt{n}}\Big) \\
&\leq  C_3\Big( (\log n)^{\frac{1}{\beta_1}} \Big(\frac{\log n}{n}+\sqrt{\frac{\log n}{n}}\|\theta-\theta^*\|_2^{\beta_2}\Big)+\|\theta-\theta^*\|_2^2\Big).
        \end{aligned}
   \end{equation}
Furthermore, using $|\exp(x)- (1+x)|\leq x^2$ for $x\in [-1,1]$, and  the Lipschitz continuity of $\ov\Delta_\theta$  around $\theta^*$, there exist constant $C_1,C_2,C_3$ so that for any $\theta\in \m M$ with $\|\theta-\theta^*\|\leq \delta\frac{(\log n)^{\frac{3}{2}}}{\sqrt{n}}$, 
  \begin{equation*}
     \begin{aligned}
& \bigg\|\frac{1}{n}\sum_{i=1}^n \exp\Big(-\big(\Delta_0^{-1}W_{\theta}^T \frac{1}{n}\sum_{j=1}^n g(X_j, \theta)\big)^T W_{\theta}^Tg(X_i, \theta)\Big)W_{\theta}^Tg(X_i, \theta)\\
&-\frac{1}{n}\sum_{i=1}^n  W_{\theta}^Tg(X_i, \theta)+\frac{1}{n}\sum_{j=1}^n  W_{\theta}^Tg(X_j, \theta)g(X_j, \theta)^T W_{\theta}\cdot \Delta_0^{-1}
\frac{1}{n}\sum_{i=1}^n W_{\theta}^Tg(X_i, \theta)\bigg\|\\
&\leq C_1  \big\|\Delta_0^{-1} W_{\theta}^T\frac{1}{n}\sum_{i=1}^n g(X_i, \theta)\big\|^2 \cdot \frac{1}{n}\sum_{i=1}^n \|g(X_i, \theta)\|^3 \cdot \bold{1}(b(X_i)\leq C(\log n)^{\frac{1}{\beta_1}})\\
&\leq C_2  \Big(\|\theta-\theta^*\|+\sqrt{\frac{\log n}{n}}\Big)^2,
\end{aligned}
   \end{equation*}
and
\begin{equation*}
\begin{aligned}
    &\Bmnorm{I_d-n^{-1}\sum_{j=1}^n  W_{\theta}^Tg(X_j, \theta)g(X_j, \theta)^T W_{\theta}\Delta_0^{-1}}_{\rm F}\\
        &=\Bmnorm{W_{\theta^*}^T \Delta_{\theta^*} W_{\theta^*} \Delta_0^{-1}-n^{-1}\sum_{j=1}^n  W_{\theta}^Tg(X_j, \theta)g(X_j, \theta)^T W_{\theta}\Delta_0^{-1}}_{\rm F}\\
    &  \leq C_1\Big(\bmnorm{\Delta_{\theta^*}-\Delta_{\theta}}_{\rm F}+\bmnorm{W_{\theta^*}-W_{\theta}}_{\rm F}+\big\vert\kern-0.25ex\big\vert\kern-0.25ex\big\vert n^{-1}\sum_{j=1}^n   g(X_j, \theta)g(X_j, \theta)^T-\Delta_{\theta}\big\vert\kern-0.25ex\big\vert\kern-0.25ex\big\vert_{\rm F}\Big)\\
    &\leq C_2\,\|\theta-\theta^*\|+C_2\, \frac{(\log n)^{\frac{2}{\beta_1}+\frac{1}{2}}}{\sqrt{n}},
    \end{aligned}
\end{equation*}
which leads to
\begin{equation*}
     \begin{aligned}
&\Big \|\frac{1}{n}\sum_{i=1}^n \exp\Big(-\big(\Delta_0^{-1}W_{\theta}^T \frac{1}{n}\sum_{j=1}^n g(X_j, \theta)\big)^T W_{\theta}^Tg(X_i, \theta)\Big)W_{\theta}^Tg(X_i, \theta)\Big\|\\
&\leq \bigg\|\Big(I_d-n^{-1}\sum_{j=1}^n  W_{\theta}^Tg(X_j, \theta)g(X_j, \theta)^T W_{\theta}\Delta_0^{-1}\Big)\frac{1}{n}\sum_{i=1}^n W_{\theta}^Tg(X_i, \theta)\bigg\|\\
&\quad+C_2  \Big(\|\theta-\theta^*\|+\sqrt{\frac{\log n}{n}}\Big)^2,\\
&\leq  C_3\Big(\|\theta-\theta^*\|+\sqrt{\frac{\log n}{n}}\Big)\cdot \Big(\|\theta-\theta^*\|+\frac{(\log n)^{\frac{2}{\beta_1}+\frac{1}{2}}}{\sqrt{n}}\Big).
\end{aligned}
   \end{equation*}
Together with~\eqref{eqn:le2.1},  we can obtain  that there exists a constant $C$ such that, for any $\theta\in \m M$ with $\|\theta-\theta^*\|\leq \delta\frac{(\log n)^{\frac{3}{2}}}{\sqrt{n}}$,
   \begin{equation*}
     \begin{aligned}
&  \Big \|\frac{1}{n}\sum_{i=1}^n \exp\left(\wt{\lambda}(\theta)^T  W_{\theta}^Tg(X_i,\theta)\right) W_{\theta}^Tg(X_i,\theta)\Big\|\\
&\leq C\, (\log n)^{\frac{2}{\beta_1}} \Big(\frac{\log n}{n}+\sqrt{\frac{\log n}{n}}\|\theta-\theta^*\|_2^{\beta_2}\Big) +C\,\|\theta-\theta^*\|^2.
        \end{aligned}
   \end{equation*}
   Fix any $\theta\in \m M$ with $\|\theta-\theta^*\|\leq \delta\frac{(\log n)^{\frac{3}{2}}}{\sqrt{n}}$. Set $v=\frac{\wt \lambda(\theta)-\lambda(\theta)}{\|\wt \lambda(\theta)-\lambda(\theta)\|}$ and $a=\|\wt \lambda(\theta)-\lambda(\theta)\|$.  Define the function $f:\mb R_{\geq 0}\to \mb R$ by
$$f(t)=\frac{1}{n}\sum_{i=1}^n \exp\Big(\big(\lambda(\theta)+  t\cdot v\big)^TW_{\theta}^T{g}(X_i,\theta)\Big),$$   then $f$ has a first-order derivative
 \begin{equation*}
\begin{aligned}
 &f^{'}(t)=\frac{1}{n}\sum_{i=1}^n \exp\Big(\big(\lambda(\theta)+t\cdot v\big)^TW_{\theta}^T{g}(X_i,\theta)\Big)\cdot v^T W_{\theta}^T{g}(X_i,\theta),
    \end{aligned}
  \end{equation*}
  and  a second-order derivative
  \begin{equation*}
\begin{aligned}
 &f^{''}(t)=\frac{1}{n}\sum_{i=1}^n \exp\Big(\big(\lambda(\theta)+  t\cdot v\big)^TW_{\theta}^T{g}(X_i,\theta)\Big)\cdot v^T W_{\theta}^T{g}(X_i,\theta){g}(X_i,\theta)^TW_{\theta}v.
   \end{aligned}
  \end{equation*}
Then it holds that $f^{'}(0)=0$, and $f^{''}(t)>0$ for $t>0$, so $f^{'}(\cdot)$ is non-decreasing on $[0,\infty)$.  Moreover,  it holds that 
\begin{equation*}
\begin{aligned}
        &f'(a)= \frac{1}{n}\sum_{i=1}^n \exp\left(\wt{\lambda}(\theta)^T  W_{\theta}^Tg(X_i,\theta)\right) \frac{(\wt \lambda(\theta)-\lambda(\theta))^T}{\|\wt \lambda(\theta)-\lambda(\theta)\|}W_{\theta}^Tg(X_i,\theta)\\
&\leq C\, (\log n)^{\frac{2}{\beta_1}} \Big(\frac{\log n}{n}+\sqrt{\frac{\log n}{n}}\|\theta-\theta^*\|_2^{\beta_2}\Big) +C\,\|\theta-\theta^*\|^2\\
&\lesssim  \frac{(\log n)^3}{n}+\frac{(\log n)^{\frac{1}{2}+\frac{2}{\beta_1}+\frac{3\beta_2}{2}}}{n^{\frac{1}{2}+\frac{\beta_2}{2}}}.
\end{aligned}
\end{equation*}
We will first show $a=\|\wt \lambda(\theta)-\lambda(\theta)\|\lesssim  (\log n)^{-\frac{1}{\beta_1}}$. Let's suppose $a> C (\log n)^{-\frac{1}{\beta_1}}$ with a positive constant $C$.  Then
\begin{equation*}
    f^{'}(a)\geq f^{'}\Big(a-\frac{C}{2} (\log n)^{-\frac{1}{\beta_1}} \Big).
\end{equation*}
On the other hand,  we have $\|\wt \lambda(\theta)\|\lesssim \frac{(\log n)^{3/2}}{\sqrt{n}}$, and $\|{g}(X_i,\theta)\|\leq b(X_i)\leq C(\log n)^{\frac{1}{\beta_1}}$ holds for any $i\in [n]$.
Thus, there exists a positive constant $C_1$ so that for any  $t\in [a-\frac{C}{2} (\log n)^{-\frac{1}{\beta_1}},a]$ and any $i\in [n]$
\begin{equation*}
    \exp\Big(\big(\lambda(\theta)+  t\cdot v\big)^TW_{\theta}^T{g}(X_i,\theta)\Big)\geq     C_1.
\end{equation*}
Furthermore, note that for $W_{{\theta}}= \bold{J}_{{\phi}_{\theta^*}(W_{\theta^*}y)}(y=W_{\theta^*}^T {\psi}_{\theta^*}({\theta}))$, there exists a positive constant $C_2$ so that for any $\theta\in B_r(\theta^*)\cap \m M$, 
\begin{equation*}
    \bmnorm{W_{\theta}-W_{\theta^*}}_{\rm F}\leq C_2\, \|\theta-\theta^*\|.
\end{equation*}
Combined with $\|  \Delta_{\theta}- \Delta_{\theta^*}\|\leq L\|\theta-\theta^*\|$  and $W_{\theta^*}^T {\Delta}_{\theta^*} W_{\theta^*}\succcurlyeq L_1 I_d$ for $L_1>0$, there exists a small enough positive constant $r_1$ so that  for any $\theta\in B_{r_1}(\theta^*)\cap \m M$, 
\begin{equation*}
W_{\theta}^T   {\Delta}_{\theta} W_{\theta}\succcurlyeq \frac{L_1}{2} I_d,
\end{equation*}
and 
\begin{equation*}
W_{\theta}^T  \frac{1}{n}\sum_{i=1}^n  {g}(X_i,\theta) {g}(X_i,\theta)^TW_{\theta}\succcurlyeq \frac{L_1}{4} I_d.
\end{equation*}
Therefore, for any  $t\in [a-\frac{C}{2} (\log n)^{-\frac{1}{\beta_1}},a]$, it holds that $f^{''}(t)\geq \frac{L_1}{4}C_1:=C_2>0$. Thus, when $n$ is sufficiently large,
\begin{equation*}
\begin{aligned}
        &f(a-\frac{C}{2} (\log n)^{-\frac{1}{\beta_1}})-f(a)\geq -f^{'}(a)\cdot \frac{C}{2} (\log n)^{-\frac{1}{\beta_1}}+\frac{C_2}{2}(\frac{C}{2} (\log n)^{-\frac{1}{\beta_1}})^2>0,
\end{aligned}
\end{equation*}
which cause contradiction.  Therefore, we have $a\leq C (\log n)^{-\frac{1}{\beta_1}}$, and there exists a positive constant $C_3$ so that  for any $t\in [0,a]$, it holds that 
\begin{equation*}
    f^{''}(t)\geq  C_3
\end{equation*}
 So we can get 
  \begin{equation*}
\begin{aligned}
 f^{'}(a)= f^{'}(a)- f^{'}(0) \geq C_3 a \Rightarrow a\leq \frac{f^{'}(a)}{C_3}.
 \end{aligned}
  \end{equation*}
 Thus there exists a constant $C$ so that for any  $\theta\in \m M$ with $\|\theta-\theta^*\|\leq \delta\frac{(\log n)^{\frac{3}{2}}}{\sqrt{n}}$,
   \begin{equation*}
   \|\wt{\lambda}(\theta)-\lambda(\theta)\|   \leq C\, (\log n)^{\frac{2}{\beta_1}} \Big(\frac{\log n}{n}+\sqrt{\frac{\log n}{n}}\|\theta-\theta^*\|_2^{\beta_2}\Big) +C\,\|\theta-\theta^*\|^2.\\
  \end{equation*}

\subsection{Proof of Lemma~\ref{lemma:acceptance}}
 
Firstly if $z_2\notin K$, then we have $G(z_2)\notin K_{\theta}$, and thus $ \m A^*(z_1,z_2)=\m A(G(z_1),G(z_2))=0$. 
So it remains to show that for any $z_1\in K$ and $z_2\in [Q\circ \wt{\phi}^*_{G(z_1)}]\big(\Omega_{G(z_1)}\big)\cap K$,
  \begin{equation}\label{eqnalpha}
      \alpha^*(z_1,z_2)=\frac{\wt \eta_{G(z_2)}\cdot\mu^\ast|_{K_\theta}(G(z_2))\cdot p(G(z_2),G(z_1))}{\wt \eta_{G(z_1)}\cdot\mu^\ast|_{K_\theta}(G(z_1))\cdot p(G(z_1),G(z_2))}.
  \end{equation}
  We claim that it suffices to show that for any $z\in Q(B_r(\wt\theta)\cap \m M)$ and $x\in B_r(\wt\theta)\cap \m M$,
  \begin{equation}\label{eqn:relationjacobian}
      {\rm det}\big(J_G(z)^TJ_G(z)\big)= \Big({\rm det}\big(J_{\wt{\psi}^*_{x}\circ G}( z)\big)\Big)^2\cdot{\rm det}\big(J_{\wt\phi^*_{x}}(\wt\psi^*_{x}\circ G(z))^TJ_{\wt\phi^*_{x}}(\wt\psi^*_{x}\circ G(z))\big).
  \end{equation}
  Indeed, under~\eqref{eqn:relationjacobian}, when  $z_1\in (Q\circ \wt{\phi}^*_{G(z_2)})(\Omega_{G(z_2)})$ and $z_2\in K$, we have 
  \begin{equation*}
  \begin{aligned}
      \alpha^*(z_1,z_2)&=\frac{\wt \eta_{G(z_2)}\cdot p^*(z_2,z_1)\mu^\ast_{\rm loc}(z_2)}{\wt \eta_{G(z_1)}\cdot p^*(z_1,z_2)\mu^\ast_{\rm loc}(z_1)}\\
      &=\frac{ \wt{p}(\wt{\psi}^*_{G(z_2)}\circ G(z_1))\cdot\Big|{\rm det}\big(\bold{J}_{\wt{\psi}^*_{G(z_2)}\circ G}(z_1)\big)\Big|\cdot\mu^\ast|_{K_\theta}(G(z_2))\cdot\sqrt{{\rm det}\big(\bold{J}_G(z_2)^T\bold{J}_G(z_2)\big)}}{\wt{p}(\wt{\psi}^*_{G(z_1)}\circ G(z_2))\cdot\Big|{\rm det}\big(\bold{J}_{\wt{\psi}^*_{G(z_1)}\circ G}(z_2)\big)\Big|\cdot\mu^\ast|_{K_\theta}(G(z_1))\cdot\sqrt{{\rm det}\big(\bold{J}_G(z_1)^T\bold{J}_G(z_1)\big)}}\\
      &\overset{(i)}{=} \frac{ \wt{p}(\wt{\psi}^*_{G(z_2)}\circ G(z_1))\cdot\mu^\ast|_{K_\theta}(G(z_2))\cdot \Big({\rm det}\big(J_{\wt\phi^*_{G(z_2)}}(\wt\psi^*_{G(z_2)}\circ G(z_1))^TJ_{\wt\phi^*_{G(z_2)}}(\wt\psi^*_{G(z_2)}\circ G(z_1))\big)\Big)^{-\frac{1}{2}}}{ \wt{p}(\wt{\psi}^*_{G(z_1)}\circ G(z_2))\cdot \mu^\ast|_{K_\theta}(G(z_1))\cdot  \Big({\rm det}\big(J_{\wt\phi^*_{G(z_1)}}(\wt\psi^*_{G(z_1)}\circ G(z_2))^TJ_{\wt\phi^*_{G(z_1)}}(\wt\psi^*_{G(z_1)}\circ G(z_2))\big)\Big)^{-\frac{1}{2}}}\\
      &=\frac{\wt \eta_{G(z_2)}\cdot\mu^\ast|_{K_\theta}(G(z_2))\cdot p(G(z_2),G(z_1))}{\wt \eta_{G(z_1)}\cdot\mu^\ast|_{K_\theta}(G(z_1))\cdot p(G(z_1),G(z_2))},
       \end{aligned} 
  \end{equation*}
  where $(i)$ uses equation~\eqref{eqn:relationjacobian} and  $K_{\theta}\subseteq B_r(\wt\theta)\cap \m M$; the last inequality uses the definition of $p(\cdot,\cdot)$ in~\eqref{defp}. On the other hand, when $z_1\notin (Q\circ \wt{\phi}^*_{G(z_2)})(\Omega_{G(z_2)})$, we have $G(z_1)\notin \wt{\phi}^*_{G(z_2)}(\Omega_{G(z_2)})$, and thus 
  \begin{equation*}
      \alpha^*(z_1,z_2)=0=\frac{\wt \eta_{G(z_2)}\cdot\mu^\ast|_{K_\theta}(G(z_2))\cdot p(G(z_2),G(z_1))}{\wt \eta_{G(z_1)}\cdot\mu^\ast|_{K_\theta}(G(z_1))\cdot p(G(z_1),G(z_2))}.
  \end{equation*}
  Now we show claim~\eqref{eqn:relationjacobian}. For any  $z\in Q(B_r(\wt\theta)\cap \m M)$ and $x\in B_r(\wt\theta)\cap \m M$, we have 
  \begin{equation*}
      \wt{\phi}^*_x\circ \wt\psi^*_{x}\circ G(z)=G(z),
  \end{equation*}
  thus 
  \begin{equation*}
  J_{\wt\phi^*_{x}}(\wt\psi^*_{x}\circ G(z))\cdot J_{\wt{\psi}^*_{x}\circ G}( z)
  =J_G(z).
  \end{equation*}
Therefore,
  \begin{equation*}
  \begin{aligned}
     &\quad\quad J_G(z)^TJ_G(z)= J_{\wt{\psi}^*_{x}\circ G}( z)^TJ_{\wt\phi^*_{x}}(\wt\psi^*_{x}\circ G(z))^TJ_{\wt\phi^*_{x}}(\wt\psi^*_{x}\circ G(z))\cdot J_{\wt{\psi}^*_{x}\circ G}(z)\\
     &\Longleftrightarrow{\rm det}\big(J_G(z)^TJ_G(z)\big)=\Big({\rm det}\big(J_{\wt{\psi}^*_{x}\circ G}( z)\big)\Big)^2\cdot{\rm det}\big(J_{\wt\phi^*_{x}}(\wt\psi^*_{x}\circ G(z))^TJ_{\wt\phi^*_{x}}(\wt\psi^*_{x}\circ G(z))\big),
     \end{aligned}
   \end{equation*}
 which proves claim~\eqref{eqn:relationjacobian}.

 \subsection{Proof of Lemma~\ref{lemma:diffpast}}
Since $B_r(0_D)\cap T_{\theta}\m M\subset V_{\theta}$ holds for $\theta\in B_{r}(\wt \theta)\cap \m M$, when $R\leq c\sqrt{n}$ for a small enough $n$-independent constant $c$ and $x\in K$, we have $K\subseteq (Q\circ \wt{\phi}_{G(x)})(\Omega_{G(x)})$, where recall $\Omega_{\theta}=\{v\in \mb R^d\,:\, \wt W_{\theta}v\in V_{\theta}\}$. Therefore, when $x,y\in K$,
 \begin{equation*}
  \wt\eta_{G(x)}\cdot  p^\ast(x,y)=\wt{p}_{G(x)}(\wt{\psi}^*_{G(x)}\circ G(y))\cdot\Big|{\rm det}\big(\bold{J}_{\wt{\psi}^*_{G(x)}\circ G}(y)\big)\Big|.
 \end{equation*}
 Moreover,  for any $x,y\in K$, we have 
 \begin{equation*}
 \begin{aligned}
   \wt{\psi}^*_{G(x)}\circ G(y)&=  \wt{W}_{G(x)}^T\wt\psi_{G(x)}\big(\wt\phi_{\wt\theta}\big(  W_{\wt\theta}\frac{y}{\sqrt{n}}\big)\big)\\
   &=\wt{W}_{G(x)}^T\Big(\wt \phi_{\wt\theta}\big( W_{\wt\theta}\frac{y}{\sqrt{n}}\big)-G(x)\Big)+O\Big(\big\|\wt\phi_{\wt\theta}\big(  W_{\wt\theta}\frac{y}{\sqrt{n}}\big)-G(x)\big\|^2\Big)\\
   &=\wt{W}_{G(x)}^T\big(\wt\theta+W_{\wt\theta}\frac{y}{\sqrt{n}}-G(x)\big)+O\Big(\big\|\wt\theta+W_{\wt\theta}\frac{y}{\sqrt{n}}-G(x)\big\|^2\Big)+O\Big(\frac{\|y\|^2}{n}\Big)\\
   &=\wt{W}_{G(x)}^T\big(W_{\wt\theta}\frac{y}{\sqrt{n}}-W_{\wt\theta}\frac{x}{\sqrt{n}}\big)+O\Big(\big\|W_{\wt\theta}\frac{y}{\sqrt{n}}-W_{\wt\theta}\frac{x}{\sqrt{n}}\big\|^2\Big)+O\Big(\frac{\|y\|^2+\|x\|^2}{n}\Big)\\
   &=\wt{W}_{G(x)}^TW_{\wt\theta}\frac{y-x}{\sqrt{n}}+O\Big(\frac{\mnorm{\wt I}_{\rm op}R^2}{n}\Big).
    \end{aligned}
 \end{equation*}
Therefore,
 \begin{equation*}
     \begin{aligned}
     & n\cdot  (\wt{\psi}^*_{G(x)}\circ G(y))^T(\wt W_{G(x)}^T\wt I\wt W_{G(x)})^{-1}(\wt{\psi}^*_{G(x)}\circ G(y))\\
      &=(y-x)^TW_{\wt \theta}^T\wt W_{G(x)}(\wt W_{G(x)}^T\wt I\wt W_{G(x)})^{-1}\wt W_{G(x)}^TW_{\wt\theta}(y-x)+O\Big(\frac{R^3}{\sqrt{n}}\mnorm{\wt I}_{\rm op}^{\frac{3}{2}}\mnorm{\wt I^{-1}}_{\rm op}\Big)\\
    &=(y-x)^T(W_{\wt \theta}^T\wt IW_{\wt \theta})^{-1}(y-x)+O\Big(\frac{R^3}{\sqrt{n}}\mnorm{\wt I}_{\rm op}^{\frac{3}{2}}\mnorm{\wt I^{-1}}_{\rm op}+\frac{R}{\sqrt{n}}\mnorm{\wt I}_{\rm op}\mnorm{\wt I^{-1}}_{\rm op}+\frac{R}{\sqrt{n}}\mnorm{\wt I^{-1}}_{\rm op}\Big).
     \end{aligned}
 \end{equation*}
 Furthermore, since for any $z\in B_r(0_d)$,
 \begin{equation*}
     \wt\psi_{G(x)}^*\circ \wt\phi_{G(x)}^*(z)=z,
 \end{equation*}
 we have 
  \begin{equation*}
     \bold{J}_{\wt\psi^*_{G(x)}}\big(\wt\phi^*_{G(x)}(z)\big)\bold{J}_{\wt\phi^*_{G(x)}}(z)= I_d.
 \end{equation*}
 For $x,y\in K$, choose $z=  \wt{\psi}^*_{G(x)}(G(y))$, we have
 \begin{equation*}
      \bold{J}_{\wt\psi^*_{G(x)}}\big(G(y)\big)\bold{J}_{\wt\phi^*_{G(x)}}\big( \wt{\psi}^*_{G(x)}(G(y))\big)=I_d.
 \end{equation*}
 Then by Assumption B.1, there exist some $n$-independent constants $C_1,C_2,C_3,C_4$ so that,
 \begin{equation}\label{eqn:VGx}
     \mnorm{\bold{J}_{\wt\phi^*_{G(x)}}\big( \wt{\psi}^*_{G(x)}(G(y))\big)-\wt W_{G(x)}}_{\rm op}\leq C_1\,\|\wt{\psi}_{G(x)}(G(y))\|\leq C_2\frac{L}{\sqrt{n}}\|y-x\|+C_2\,\frac{R^2}{n}\leq C_3 \frac{R}{\sqrt{n}},
 \end{equation}
 and
 \begin{equation}\label{eqn:Vhattheta}
     \mnorm{\sqrt{n}\cdot\bold{J}_{G}(y)-  W_{\wt\theta}}_{\rm op}= \mnorm{\bold{J}_{\wt{\phi}^*_{\wt\theta}}(y/\sqrt{n})-  W_{\wt\theta}}_{\rm op} \leq C_4\,\frac{R}{\sqrt{n}}.
 \end{equation}
We can obtain that for any $x,y\in K$,
 \begin{equation*}
 \begin{aligned}
  \Big|{\rm det}\Big(\sqrt{n}\cdot\bold{J}_{\wt{\psi}^*_{G(x)}\circ G}(y)\Big)-1\Big|&\leq \Big|{\rm det}\Big(\sqrt{n}\cdot\bold{J}_{\wt{\psi}^*_{G(x)}\circ G}(y)\Big)-
     {\rm det}\Big( \bold{J}_{\wt\psi^*_{G(x)}}\big(G(y)\big)\cdot W_{\wt\theta}\Big)\Big|\\
     &+\Big| {\rm det}\Big( \bold{J}_{\wt\psi^*_{G(x)}}\big(G(y)\big)\cdot W_{\wt\theta}\Big)- {\rm det}\Big( \bold{J}_{\wt\psi^*_{G(x)}}\big(G(y)\big)\cdot\wt W_{G(x)}\Big)\Big|\\
     &+\Big|  {\rm det}\Big( \bold{J}_{\wt\psi^*_{G(x)}}\big(G(y)\big)\cdot\wt W_{G(x)}\Big)- {\rm det}\Big( \bold{J}_{\wt\psi^*_{G(x)}}\big(G(y)\big)\bold{J}_{\wt\phi^*_{G(x)}}\big( \wt{\psi}^*_{G(x)}(G(y))\big)\Big)\Big|\\
    & \lesssim \frac{R}{\sqrt{n}},
      \end{aligned}
 \end{equation*}
 where the last step uses equation~\eqref{eqn:Vhattheta},~\eqref{eqn:VGx} and the fact that $\|\wt\theta-G(x)\|\lesssim R/\sqrt{n}$ for $x\in K$. We can then obtain the desired result by combining all pieces. 

  \subsection{Proof of Lemma~\ref{lemma:termA}}
Recall $\overline{u}=\m N(0,J^{-1})$ and $p^{\Delta}(x,\cdot)=\m N(x,2hI^{\Delta})$, we have
 \begin{equation*}
     \begin{aligned}
&\int\left|p^{\Delta}(x,y)-\frac{\ov\mu(y)p^{\Delta}(y,x)}{\ov\mu(x)}\right|\,\dd y\\
&=\int \frac{1}{(4\pi h)^{\frac{d}{2}}}\left|\exp\left(-\frac{\|(I^{\Delta})^{-\frac{1}{2}}(y-x)\|^2}{4h}\right)-\exp\left(\frac{x^TJ x-y^TJy}{2}\right)\exp\left(-\frac{\|(I^{\Delta})^{-\frac{1}{2}}\cdot(x-y)\|^2}{4h}\right)\right|\,\dd y\\
&=\int  \frac{1}{(4\pi h)^{\frac{d}{2}}}\exp\left(-\frac{\|(I^{\Delta})^{-\frac{1}{2}}\cdot(y-x)\|^2}{4h}\right)\left|1-\exp\left(\frac{x^TJx-y^TJy}{2}\right)\right|\,\dd y.
\end{aligned}
 \end{equation*}
 Let $u=\frac{(I^{\Delta})^{-\frac{1}{2}}(y-x)}{\sqrt{2h}}$ in the above integral, then consider $u\sim \m N(0,I_d)$, we have 
\begin{equation*}
     \begin{aligned}
&\int\left|p^{\Delta}(x,y)-\frac{\ov\mu(y)p^{\Delta}(y,x)}{\ov\mu(x)}\right|\,\dd y\\
&=\mb{E}_{u\sim \m N(0,I_d)}\bigg[\Big|1-\exp\Big(\frac{x^TJx-(x+\sqrt{2h}(I^{\Delta})^{\frac{1}{2}}u)^TJ(x+\sqrt{2h}(I^{\Delta})^{\frac{1}{2}}u)}{2}\Big)\Big|\bigg]\\
&=\mb{E}_{u\sim \m N(0,I_d)}\bigg[\Big|1-\exp\Big(-h\cdot u^TJ^{\Delta}u-\sqrt{2h}\cdot u^T(I^{\Delta})^{\frac{1}{2}}Jx\Big)\Big|\bigg].
\end{aligned}
 \end{equation*}
 Let
$$\m B=\left\{u\in \mb R^d: \big|h\cdot u^TJ^{\Delta}u+\sqrt{2h}\cdot u^T(I^{\Delta})^{\frac{1}{2}}Jx\big|\leq\frac{1}{49}\right\},$$   
then by H\"{o}lder inequality,  we have
\begin{equation*}
    \begin{aligned}
     &\int\left|p^{\Delta}(x,y)-\frac{\ov\mu(y)p^{\Delta}(y,x)}{\ov\mu(x)}\right|\,\dd y\\
     &\leq \exp(1/49)-1+\mathbb{E}_{u}\left[\bold{1}_{\m B^c}(u)\right]\\
     &\quad+\sqrt{\mathbb{E}_{u}\left[\bold{1}_{\m B^c}(u)\right]}\cdot \Big(\mathbb{E}_{u} \left[\exp(-4h\cdot u^TJ^{\Delta}u)\right]\cdot\mathbb{E}_{u} \left[\exp(-4\sqrt{2h}u^T(I^{\Delta})^{\frac{1}{2}}Jx)\right]\Big)^{\frac{1}{4}}\\
     &\leq 1/48+\mathbb{E}_{u}\left[\bold{1}_{\m B^c}(u)\right]+\sqrt{\mathbb{E}_{u}\left[\bold{1}_{\m B^c}(u)\right]}\cdot \Big(\mathbb{E}_{u} \left[\exp(-4h\cdot u^TJ^{\Delta}u)\right]\cdot\mathbb{E}_{u} \left[\exp(-4\sqrt{2h}u^T(I^{\Delta})^{\frac{1}{2}}Jx)\right]\Big)^{\frac{1}{4}},\\
    \end{aligned}
\end{equation*}
where we use the shorthand $\mb {E}_u$ to denote $\mb E_{u\sim \m N(0,I_d)}$. Then by (1) $h\leq\sqrt{c_0}\,({\rm tr}(J^{\Delta})+r_d)^{-1}$  with  $r_d= \bigg\{\Big(\sqrt{c'\log \frac{M_0^2}{\varepsilon^2 h\rho_1}}\mnorm{J^{\Delta}}_{\scriptsize  \rm F} \Big)\vee \Big(c'\log \frac{M_0^2 }{\varepsilon^2 h\rho_1}\rho_2^2\Big)\bigg\}\wedge  (\rho_2^2R^2)$ and $R\geq C(\frac{d}{\rho_1})^{\frac{1}{2}}$; (2) $x\in E=\{x\in K: \big|\|(I^{\Delta})^{\frac{1}{2}}Jx\|^2-{\rm tr}(J^{\Delta})\big|\leq r_d\}$,   it holds that for $0\leq t\leq\frac{1}{4h\lambda_{\rm max}(J^{\Delta})}$,
 \begin{equation*}
 \begin{aligned}
 \mathbb{E}_{u}\left[\exp(th\cdot u^T(I^{\Delta})^{\frac{1}{2}}J(I^{\Delta})^{\frac{1}{2}}u)\right]&=\prod_{j=1}^{d}\frac{1}{\sqrt{1-2th\cdot\lambda_j(J^{\Delta})}}\\
 &\leq\exp(2th\cdot{\rm tr}(J^{\Delta}))\leq \exp(2t\sqrt{c_0}),
      \end{aligned}
 \end{equation*}
 where $\{\lambda_j(J^{\Delta})\}_{j=1}^d$ denotes the eigenvalues of $J^{\Delta}$, and $\lambda_{\rm max}(J^{\Delta})$ denotes the maximum eigenvalues of $J^{\Delta}$. Moreover, for $t\in \mb R$, 
 \begin{equation*}
     \mathbb{E}_{u} \left[\exp(t\sqrt{h}u^T(I^{\Delta})^{\frac{1}{2}}Jx)\right]=\exp(\frac{1}{2}t^2h\|(I^{\Delta})^{\frac{1}{2}}Jx\|^2)\leq \exp(\frac{1}{2}t^2h(r_d+{\rm tr}(J^{\Delta})))\leq \exp(\frac{1}{2}t^2\sqrt{c_0}).
 \end{equation*}
 Thus we have 
 \begin{equation*}
    \begin{aligned}
     \mathbb{E}_{u}\left[\bold{1}_{\m B^c}(u)\right]&\leq  P\left(h\cdot u^TJ^{\Delta}u\geq \frac{1}{96}\right)+P\left(\sqrt{h}\cdot |u^T(I^{\Delta})^{\frac{1}{2}}Jx|\geq \frac{1}{96\sqrt{2}}\right)\\
     &\leq \underset{0\leq t\leq\frac{1}{4h\lambda_{\rm max}(J^{\Delta})}}{\inf} \exp\left(2t\sqrt{c_0}-\frac{t}{96}\right)+2\, \underset{t>0}{\inf} \exp\left(\frac{1}{2}t^2\sqrt{c_0}-\frac{t}{96\sqrt{2}}\right)\\
     &\leq \exp\left(\frac{1}{2}-\frac{1}{384\sqrt{c_0}}\right)+2\,\exp\left(-\frac{1}{2\cdot(96\sqrt{2})^2\sqrt{c_0}}\right),
    \end{aligned} 
 \end{equation*}
and  
 \begin{equation*}
     \Big(\mb {E}_u\left[\exp(-4h\cdot u^TJ^{\Delta}u)\right]\cdot\mathbb{E}_{u} \left[\exp(-4\sqrt{2h}u^T(I^{\Delta})^{\frac{1}{2}}Jx)\right]\Big)^{\frac{1}{4}}\leq \exp(4\sqrt{c_0}).
 \end{equation*}
So when $c_0$ is small enough, we have 
 \begin{equation*}
     \mathbb{E}_{u}\left[\bold{1}_{\m B^c}(u)\right]+\sqrt{\mathbb{E}_{u}\left[\bold{1}_{\m B^c}(u)\right]}\cdot \Big(\mathbb{E}_{u} \left[\exp(-4h\cdot u^TJ^{\Delta}u)\right]\cdot\mathbb{E}_{u} \left[\exp(-4\sqrt{2h}u^T(\wt I^{\Delta})^{\frac{1}{2}}Jx)\right]\Big)^{\frac{1}{4}}\leq \frac{1}{48},
 \end{equation*}
 which leads to the desired result. 
   
 \subsection{Proof of Lemma~\ref{lemmaboundTx}}
 Since
\begin{equation*}
\begin{aligned}
    &\int_{K^c} \m N(x,2hI^{\Delta})\,\dd y
    \\
    &=\mb{E}_{u\sim \m N(0,I_d)}\Big[\bold{1}\big(\|(I^{\Delta})^{-\frac{1}{2}}x+\sqrt{2h}u\|\geq R\big)\Big]\\
    &=\mb{E}_{u\sim \m N(0,I_d)}\Big[\bold{1}\big(2\sqrt{2h}x^T(I^{\Delta})^{-\frac{1}{2}}u\geq 2h\|u\|^2+R^2-\|(I^{\Delta})^{-\frac{1}{2}}x\|^2\big)\Big]\\
    &=\mb{E}_{u\sim \, \m N(0,I_d)}\bigg[\bold{1}\Big(\frac{x^T(I^{\Delta})^{-\frac{1}{2}}}{\|(I^{\Delta})^{-\frac{1}{2}}x\|}u\geq -\frac{\sqrt{2h}}{{2}}\frac{\|u\|^2}{\|(I^{\Delta})^{-\frac{1}{2}}x\|}+\frac{R^2-\|(I^{\Delta})^{-\frac{1}{2}}x\|^2}{2\sqrt{2h}\|(I^{\Delta})^{-\frac{1}{2}}x\|}\Big)\bigg]\\
    &\leq \exp(-4)\\
    &\quad + \mb{E}_{u\sim \m N(0,I_d)}\bigg[\bold{1}\Big(\frac{x^T(I^{\Delta})^{-\frac{1}{2}}}{\|(I^{\Delta})^{-\frac{1}{2}}x\|}u\geq -\frac{\sqrt{2h}}{{2}}\frac{\|u\|^2}{\|(I^{\Delta})^{-\frac{1}{2}}x\|}+\frac{R^2-\|(I^{\Delta})^{-\frac{1}{2}}x\|^2}{2\sqrt{2h}\|(I^{\Delta})^{-\frac{1}{2}}x\|}\Big)\cdot \bold{1}\big(\|u\|^2\leq d+4\sqrt{d}+8\big)\bigg]\\
     &\leq \exp(-4)+  \mb{E}_{u\sim \m N(0,1)}\bigg[\bold{1}\Big(u\geq -\frac{\sqrt{2h}}{{2}}\cdot\frac{d+4\sqrt{d}+8}{\|(I^{\Delta})^{-\frac{1}{2}}x\|}+\frac{R^2-\|(I^{\Delta})^{-\frac{1}{2}}x\|^2}{2\sqrt{2h}\|(I^{\Delta})^{-\frac{1}{2}}x\|}\Big)\bigg].\\
    \end{aligned}
\end{equation*}
 Then when $h\leq c_0 \rho_2^{-1} d^{-1}$ for a small enough $c_0$, by $R\geq 6\sqrt{d/\lambda_{\min}(J^{\Delta})}$, we can obtain
 \begin{equation*}
\begin{aligned}
    \int_{K^c} \m N(x,2hI_d)\,\dd y&\leq \exp(-4)+  \mb{E}_{u\sim\m N(0,1)}\bigg[\bold{1}\Big(u\geq -\frac{\sqrt{2h}}{{2}}\frac{d+4\sqrt{d}+8}{\|(I^{\Delta})^{-\frac{1}{2}}x\|}+\frac{R^2-\|(I^{\Delta})^{-\frac{1}{2}}x\|^2}{2\sqrt{2h}\|(I^{\Delta})^{-\frac{1}{2}}x\|}\Big)\bigg]\\
    &\leq \exp(-4)+  \mb{E}_{u\sim \m N(0,1)}\bigg[\bold{1}\Big(u\geq -\frac{\sqrt{2h\lambda_{\min}(J^{\Delta})}}{{12}}\frac{d+4\sqrt{d}+8}{\sqrt{d}}\Big)\bigg]\\
    &\leq \frac{13}{24}.
    \end{aligned}
\end{equation*}

  \section{Proof Related to Conductance Profile}
 \subsection{Proof of Lemma~\ref{lemma:conductance1}}
    Recall $s= \frac{\varepsilon^2}{32M_0^2}$,  and let $S$ be any measurable set of $\mb R^d$ with $s\leq \mu^\ast_{\rm loc}(S)\leq v\leq \frac{1}{2}$. Define the following subsets:
 \begin{equation*}
     \begin{aligned}
     & S_1:=\{x\in S|\m T^\ast(x,S^c)\leq \frac{\omega}{2}\},\\
      & S_2:=\{x\in S^c| \m T^\ast(x,S)\leq\frac{\omega}{2}\},\\
      &  S_3:= (S_1\cup S_2)^c,
     \end{aligned}
 \end{equation*}
  Then if $\mu^\ast_{\rm loc}(S_1)\leq \mu^\ast_{\rm loc}(S)/2$ or $\mu^\ast_{\rm loc}(S_2)<\mu^\ast_{\rm loc}(S^c)/2$,  by the fact that $\mu^\ast_{\rm loc}$ is stationary w.r.t the transition kernel $\m T^*$, we have 
  \begin{equation*}
  \begin{aligned}
      \int_S  \m T^\ast(x,S^c)\mu^\ast_{\rm loc}(x)\,\dd x&=\int  \m T^\ast(x,S)\mu^\ast_{\rm loc}(x)\,\dd x-\int_{S} \m T^\ast(x,S)\mu^\ast_{\rm loc}(x)\,\dd x\\
      &=\int_{S^c}  \m T^\ast(x,S)\mu^\ast_{\rm loc}(x)\,\dd x\geq \frac{\omega}{2}\cdot \max\{\mu^\ast_{\rm loc}(S\cap S_1^c),\mu^\ast_{\rm loc}(S^c\cap S_2^c)\}\\
      &\geq \frac{\mu^\ast_{\rm loc}(S)\omega}{4}.
        \end{aligned}
  \end{equation*}
Moreover, when $\mu^\ast_{\rm loc}(S_1)\wedge \mu^\ast_{\rm loc}(S_2)\geq \frac{\mu^\ast_{\rm loc}(S)}{2}$, consider $x\in E\cap S_1$ and $z\in E\cap S_2$, then $\|\m T^*(x,\cdot)-\m T^*(z,\cdot)\|_{\scriptsize  \rm TV}\geq  \m T^\ast(z,S^c)- \m T^\ast(x,S^c)\geq1-\omega$, thus $\|(I^{\Delta})^{\frac{1}{2}}(x-z)\|\geq \frac{\sqrt{2h}}{8}$, which implies that $\inf_{x\in E\cap S_1,z\in E \cap S_2}\|(I^{\Delta})^{\frac{1}{2}}(x-z)\|\geq  \frac{\sqrt{2h}}{8}$. We then state the following log-isoperimetric inequality to lower bound $\mu^\ast_{\rm loc}(S_3)$.
  \begin{lemma}\label{lemmalogiso}
Suppose $\underset{\xi \in K}{\sup} \big|\log \big(\mu^*_{\rm loc}(\xi)\big)-\log\big((2\pi{\rm det}(J^{-1}))^{-\frac{d}{2}}\exp(-\frac{1}{2}\xi^TJ\xi)\big)\big|\leq \varepsilon_1$. Consider any measurable partition form $K=S_1\cup S_2\cup S_3$ such that $\inf_{x\in S_1,z\in  S_2}\|(I^{\Delta})^{\frac{1}{2}}(x-z)\|\geq t$, we have 
  \begin{equation*}
      \mu^\ast_{\rm loc}(S_3)\geq \frac{\sqrt{\lambda_{\min}(J^{\Delta})}}{2}t\exp(-3{\varepsilon_1}) \min\{\mu^\ast_{\rm loc}(S_1),\mu^\ast_{\rm loc}(S_2)\}\log^{\frac{1}{2}}\Big(1+\frac{1}{\min\{\mu^\ast_{\rm loc}(S_1),\mu^\ast_{\rm loc}(S_2)\}}\Big),
  \end{equation*}
  where $\lambda_{\min}(J^{\Delta})$ denotes the minimum eigenvalue of $J^{\Delta}$.
  \end{lemma}
  \noindent Then take $S_1$ as $E\cap S_1$, $S_2$ as $E\cap S_2$, and $t=\frac{\sqrt{2h}}{8}$ in Lemma~\ref{lemmalogiso}, we can obtain that
  \begin{equation*}
  \begin{aligned}
      \mu^\ast_{\rm loc}(((E\cap S_1)\cup (E\cap S_2))^c)&\geq \exp(-3\varepsilon_1) \frac{\sqrt{2h\rho_1}}{16}\min\{\mu^\ast_{\rm loc}(E\cap S_1),\mu^\ast_{\rm loc}(E\cap S_2)\}\\
      &\quad\cdot \log^{\frac{1}{2}}\Big(1+\frac{1}{\min\{\mu^\ast_{\rm loc}(E\cap S_1),\mu^\ast_{\rm loc}(E \cap S_2)\}}\Big).
        \end{aligned}
  \end{equation*}
Without loss of generality, we can assume $\mu^\ast_{\rm loc}(E\cap S_1)\leq\mu^\ast_{\rm loc}(E\cap S_2)$, then by $((E\cap S_1)\cup (E\cap S_2))^c\subseteq E^c\cup S_3$ and $\mu^\ast_{\rm loc}(E^c)\leq \exp(-3\varepsilon_1)\frac{4\varepsilon^2h\rho_1}{M_0^2}= 128s h\rho_1\exp(-3\varepsilon_1)$, when $c_0$ is small enough, we can obtain
  \begin{equation*}
      \begin{aligned}
      &\mu^\ast_{\rm loc}(S_3)+ 128s h\rho_1\exp(-3\varepsilon_1)\\
      &\geq  \exp(-3\varepsilon_1) \frac{\sqrt{2h\rho_1}}{16} \mu^\ast_{\rm loc}(E\cap S_1)\log^{\frac{1}{2}}\Big(1+\frac{1}{ \mu^\ast_{\rm loc}(E\cap S_1)}\Big) \\
      &\overset{\rm {(i)}}{\geq} \exp(-3\varepsilon_1) \frac{\sqrt{2h\rho_1}}{16}\big(\frac{\mu^\ast_{\rm loc}(S)}{4}+\frac{s}{4}-128s h\rho_1\big)
      \log^{\frac{1}{2}}\Big(1+\frac{1}{\frac{\mu^\ast_{\rm loc}(S)}{4}+\frac{s}{4}-128s h\rho_1}\Big)\\
      &\geq \exp(-3\varepsilon_1) \frac{\sqrt{2h\rho_1}}{16} \frac{\mu^\ast_{\rm loc}(S)}{4}
      \log^{\frac{1}{2}}\Big(1+\frac{4}{{\mu^\ast_{\rm loc}(S)}}\Big),
      \end{aligned}
  \end{equation*}
  where (i) uses $ \mu^\ast_{\rm loc}(E\cap S_1)\geq \pi_{\rm loc }(S_1)-\mu^\ast_{\rm loc}(E^c)$, $\pi_{\rm loc }(S_1)\geq \frac{\mu^\ast_{\rm loc}(S)}{2}\geq \frac{s}{2}$ and the function $x\log^{\frac{1}{2}} (1+\frac{1}{x})$ is an increasing function. Then when $h\leq c_0\rho_1^{-1}$ with small enough $c_0$, we have 
  \begin{equation*}
      \mu^\ast_{\rm loc}(S_3)\geq \exp(-3\varepsilon_1) \frac{\sqrt{h\rho_1}\mu^\ast_{\rm loc}(S)}{64} 
      \log^{\frac{1}{2}}\Big(1+\frac{4}{{\mu^\ast_{\rm loc}(S)}}\Big).
  \end{equation*}
  hence 
  \begin{equation*}
  \begin{aligned}
      \int_S  \m T^\ast(x,S^c) \mu^\ast_{\rm loc}(x)\,\dd x&\geq \frac{1}{2}\left(  \int_S \m T^\ast(x,S^c) \mu^\ast_{\rm loc}(x)\,\dd x+  \int_{S^c} \m T^\ast \mu^\ast_{\rm loc}(x)\, \dd x\right)\\
      &\geq \frac{\omega}{4}\mu^\ast_{\rm loc}(S_3)\geq \frac{\omega}{256}\exp(-3\varepsilon_1)\, \sqrt{\rho_1h}\mu^\ast_{\rm loc}(S)\log^{\frac{1}{2}}\Big(1+\frac{4}{{\mu^\ast_{\rm loc}(S)}}\Big),
       \end{aligned}
  \end{equation*}
  which leads to 
  \begin{equation*}
  \begin{aligned}
     \frac{\int_S \m T^\ast(x,S^c) \mu^\ast_{\rm loc}(x)}{\mu^\ast_{\rm loc}(S)}\,\dd x\geq  \frac{\omega}{256}\exp(-3\varepsilon_1)\, \sqrt{\rho_1h}\log^{\frac{1}{2}}\Big(1+\frac{4}{{\mu^\ast_{\rm loc}(S)}}\Big)\geq \frac{\omega}{4}\exp(-3\varepsilon_1)\, \sqrt{\rho_1h}\log^{\frac{1}{2}}\Big(1+\frac{1}{v}\Big).
       \end{aligned}
  \end{equation*}
   Then combining with the result for the first case, we can obtain a lower bound of $$\frac{\omega}{4}\,\min\big\{1,\frac{\exp(-3\varepsilon_1)}{64}\sqrt{\rho_1h\log\big(1+\frac{1}{v}\big)}\big\}$$ on $s$-conductance profile $\Phi_s(v)$ with $s=\frac{\varepsilon^2}{32M_0^2}$.
 \subsection{Proof of Lemma~\ref{lemmalogiso}}
 To begin with, we consider the following lemma stated in~\cite{JMLR:v21:19-441}.
 \begin{lemma}\label{lemmalogiso1}
 (Lemma 16 of~\cite{JMLR:v21:19-441}) Let \(\gamma\) denote the density of the standard Gaussian distribution \(\mathcal{N}\left(0, \sigma^{2} {I}_{d}\right)\), and let \(\mu^\ast\) be a distribution with density \(\mu^\ast=q \cdot \gamma\), where \(q\) is a log-concave function. Then for any partition \(S_{1}, S_{2}, S_{3}\) of \(\mathbb{R}^{d}\), we have
$$
\mu^\ast\left(S_{3}\right) \geq \frac{d\left(S_{1}, S_{2}\right)}{2 \sigma} \min \left\{\mu^\ast\left(S_{1}\right), \mu^\ast\left(S_{2}\right)\right\} \log ^{\frac{1}{2}}\left(1+\frac{1}{\min \left\{\mu^\ast\left(S_{1}\right), \mu^\ast\left(S_{2}\right)\right\}}\right) .
$$
 \end{lemma}
 We first consider the case $J=I_d$. Define $\overline{\mu}=N(0,I_d)$ and $\overline{\mu}|_K=N(0,I_d)|_{K}$, by the fact that $K$ is a convex set and $\bold{1}_K$ is a log-concave function, using lemma~\ref{lemmalogiso1}, we can obtain that for any partition  \(S_{1}, S_{2}, S_{3}\) of \(K\), we have 
 $$
\overline\mu|_K \left(S_{3}\right) \geq \frac{d\left(S_{1}, S_{2}\right)}{2} \min \left\{\overline\mu|_K\left(S_{1}\right), \overline\mu|_K\left(S_{2}\right)\right\} \log ^{\frac{1}{2}}\left(1+\frac{1}{\min \left\{\overline\mu|_K\left(S_{1}\right), \overline\mu|_K\left(S_{2}\right)\right\}}\right) .
$$
Then since $\underset{\xi \in K}{\sup} \big|\log \big(\mu^*_{\rm loc}(\xi)\big)-\log\big((2\pi{\rm det}(J^{-1}))^{-\frac{d}{2}}\exp(-\frac{1}{2}\xi^TJ\xi)\big)\big|=\varepsilon_1$, we have 

\begin{equation*}
    1\geq \int_{K}\ov \mu(\xi)\,\dd \xi= \int_{K}\mu^*_{\rm loc}(\xi)\,\dd \xi\cdot\frac{ \int_{K}\ov \mu(\xi)\,\dd \xi}{\int_{K}\mu^*_{\rm loc}(\xi)\,\dd \xi}\overset{(i)}{\geq}  \underset{\xi\in K}{\min}\frac{ \ov \mu(\xi)}{\mu^*_{\rm loc}(\xi)}\geq  \exp(-\varepsilon_1),
\end{equation*}
where $(i)$ uses $\int_{K}\mu^*_{\rm loc}(\xi)\,\dd \xi=1$.
 Furthermore, we can obtain that for any measurable set $S\subseteq K$, 
\begin{equation*}
\exp(-2 {\varepsilon}_1)\leq \frac{\mu^\ast_{\rm loc}(S)}{\overline\mu|_{K}(S)}=\frac{\int_{S}\mu^\ast_{\rm loc}(\xi)\dd \xi \int_K \ov{\mu}(\xi)\dd \xi}{\int_{S}\ov{\mu}(\xi)\dd \xi}\leq  \exp( {\varepsilon}_1).
\end{equation*}
Thus
\begin{equation}\label{eqn:caseI}
    \begin{aligned}
        &\mu^\ast_{\rm loc}(S_3)\geq  \exp(-2 {\varepsilon}_1)\overline{\mu}|_{K}(S_3)\\
        &\geq \frac{d(S_1,S_2)}{2} \exp(-2 {\varepsilon}_1)\min \left\{\overline\mu|_{K}\left(S_{1}\right), \overline\mu|_{K}\left(S_{2}\right)\right\} \log ^{\frac{1}{2}}\left(1+\frac{1}{\min \left\{\overline\mu|_{K}\left(S_{1}\right), \overline\mu|_{K}\left(S_{2}\right)\right\}}\right)\\
        &\overset{(i)}{\geq} \frac{d(S_1,S_2)}{2} \exp(-3\varepsilon_1)\min \left\{ \mu^\ast_{\rm loc}\left(S_{1}\right),  \mu^\ast_{\rm loc}\left(S_{2}\right)\right\} \log ^{\frac{1}{2}}\left(1+\frac{1}{\exp(-\varepsilon_1)\min \left\{ \mu^\ast_{\rm loc}\left(S_{1}\right),  \mu^\ast_{\rm loc}\left(S_{2}\right)\right\}}\right)\\
        &\geq  \frac{d(S_1,S_2)}{2} \exp(-3\varepsilon_1)\min \left\{ \mu^\ast_{\rm loc}\left(S_{1}\right),  \mu^\ast_{\rm loc}\left(S_{2}\right)\right\} \log ^{\frac{1}{2}}\left(1+\frac{1}{\min \left\{ \mu^\ast_{\rm loc}\left(S_{1}\right),  \mu^\ast_{\rm loc}\left(S_{2}\right)\right\}}\right),\\
    \end{aligned}
\end{equation}
where $(i)$ uses the fact that $x\log^{\frac{1}{2}}(1+\frac{1}{x})$ is an increasing function.  For the general case where $J$ is not necessary an identity matrix,  we can define $K'=J^{\frac{1}{2}}K=\{x=J^{\frac{1}{2}}y\,: \,y\in K\}$, and $\lambda=J^{\frac{1}{2}}\xi$, where $\xi$ is a random variable with density $\mu^\ast_{\rm loc}$. Thus $\lambda$ has a density
\begin{equation*}
    \widetilde\mu^\ast(\lambda)=\mu^\ast_{\rm loc}(J^{-\frac{1}{2}}\lambda)({\rm det}(J))^{\frac{d}{2}}.
\end{equation*}
Then for any $\lambda\in K'$, it holds that
\begin{equation*}
    \Big|\log( \widetilde\mu^\ast(\lambda))-\log\big((2\pi)^{-\frac{d}{2}}\exp(-\frac{1}{2}\lambda^T\lambda)\big)\Big|\leq  \varepsilon_1.
\end{equation*}
Consider any partition  \(S_{1}, S_{2}, S_{3}\) of \(K\), let 
 \begin{equation*}
     \begin{aligned}
         &\widetilde{S_1}=J^{\frac{1}{2}}S_1;\\
          &\widetilde{S_2}=J^{\frac{1}{2}}S_2;\\
           &\widetilde{S_3}=J^{\frac{1}{2}}S_3.\\
     \end{aligned}
 \end{equation*}
  Then for any point $x\in S_1$ and $y\in S_2$, we have 
  \begin{equation*}
  \begin{aligned}
        \|J^{\frac{1}{2}}x-J^{\frac{1}{2}}y\|^2&=(x-y)^T(I^{\Delta})^{-\frac{1}{2}}J^{\Delta}(I^{\Delta})^{-\frac{1}{2}}(x-y)\\
        &\geq \lambda_{\min}(J^{\Delta})\|(I^{\Delta})^{-\frac{1}{2}}(x-y)\|^2
  \end{aligned}
  \end{equation*}
 So by applying $\widetilde{\mu}^\ast$ to statement~\eqref{eqn:caseI}, we can obtain
 \begin{equation*}
     \begin{aligned}
         &\mu^\ast_{\rm loc}(S_3)=\widetilde{\mu}^\ast(\widetilde S_3)
         \geq    \frac{d(\widetilde S_1,\widetilde S_2)}{2} \exp(-3\varepsilon_1)\min \left\{ \widetilde\mu^\ast (\widetilde S_{1} ),  \widetilde\mu^\ast(\widetilde S_{2} )\right\} \log ^{\frac{1}{2}}\bigg(1+\frac{1}{\min \left\{ \widetilde\mu^\ast (\widetilde S_{1} ),  \widetilde\mu^\ast (\widetilde S_{2})\right\}}\bigg)\\
         &\geq \frac{\sqrt{\lambda_{\min}(J^{\Delta})}}{2} t\exp(-3\varepsilon_1)\min \left\{ \mu^\ast_{\rm loc}\left(S_{1}\right),  \mu^\ast_{\rm loc}\left(S_{2}\right)\right\} \log ^{\frac{1}{2}}\left(1+\frac{1}{\min \left\{ \mu^\ast_{\rm loc}\left(S_{1}\right),  \mu^\ast_{\rm loc}\left(S_{2}\right)\right\}}\right),
     \end{aligned}
 \end{equation*}
 where the last inequality uses $\inf_{x\in S_1,z\in  S_2}\|(I^{\Delta})^{\frac{1}{2}}(x-z)\|\geq t$. Proof is completed.
 
\end{document}